# Quantized resistance limitations for graphene *p-n* junction devices

T. M. Padwa,[1,2*] D. Henry,[1,2*] D. S. Scaletta,[3] H. M. Hill,[1] A. L. Levy,[4] D. G. Jarrett,[1] D. B. Newell,[1] and A. F. Rigosi[1,a)]

[1]*Physical Measurement Laboratory, National Institute of Standards and Technology (NIST), Gaithersburg, Maryland 20899, USA*

[2]*Department of Physics, Georgetown University, Washington, District of Columbia 20057, USA*

[3]*Department of Physics, Mount San Jacinto College, Menifee, California 92584, USA*

[4]*Department of Physics and Astronomy, Howard University, Washington, District of Columbia 20059, USA*

Developing high-resistance quantized Hall resistance standards requires accessing values beyond the robust value seen in epitaxial graphene ($R_{\mathrm{H}} = \frac{1}{2}\frac{h}{e^2}$). This study evaluates the available parameter space of the coefficient of effective resistance across various epitaxial graphene *p-n* junction topologies, including rectangular, Corbino, and triangular layouts. Because mapping desired resistances to device configurations is mathematically degenerate, an automated Landauer-Büttiker Python solver was developed for all three topologies and was validated by cryogenic multi-terminal transport measurements and LTspice simulations. By analyzing the Shannon entropy of available states, it is found that rectangular checkerboards maximize topological complexity and tunable outputs, whereas Corbino geometries exhibit highly constrained resistance availabilities.

[a)] Author to whom correspondence should be addressed. email: afr1@nist.gov

[b)] *The authors have contributed equally to this work.

Epitaxial graphene (EG) has been a key component of developing quantum electrical standards, particularly for applications that require expertise in electrical metrology [1]. The quantum Hall effect continues to serve as the primary realization of the SI unit of electrical resistance (ohm), with some standards relying on the stable $\nu = 2$ Landau level for dissemination ($R_{\mathrm{H}} = \frac{1}{2}\frac{h}{e^2} \approx 12906.4037\ \Omega$, where $h$ is the Planck constant, and $e$ is the elementary charge) [2-6]. However, modern instrumentation found in industry typically requires calibration at higher resistance values, reaching into the MΩ regime and beyond, and so it would be ideal to transcend the limitations of conventional single quantum Hall devices that only output one quantized resistance.

To access other quantized resistances, recent work in graphene-based devices for metrology have focused on either constructing arrays of many elements, requiring metallic electrical contacts between each element [7-12], and engineering *p*-*n* junction (*pn*Js) devices [13-17], which do not require any metallic contacts to exhibit rational multiples of $R_{\mathrm{H}}$, especially when designs allow the use of multiple source and drain terminals [18-19]. To avoid complications in fabrication of the latter case, one may take advantage of localized chemical doping or functionalization to create adjoining regions of electron-like (*n*-type) and hole-like (*p*-type) carrier concentrations. The boundaries between these distinct regions act as physical *p*-*n* interfaces where chiral edge channels from adjacent domains co-propagate [13-17]. The measurable resistances take the form $qR_{\mathrm{H}}$, where $q$ is defined as the *coefficient of effective resistance* (CER) and should be restricted to the set of positive rational numbers ($q: q \in \mathbb{Q}^{+}$) for reasons pertaining to how one utilizes Landauer-Büttiker formalism.

Though the mechanics of a single row of *pn*Js are well-established, the scaling of *pn*Js into complex, multi-domain topologies with a clear understanding of the accessible quantized resistance parameter space remains an open and challenging problem. Such an understanding would give necessary clarity to the extent of utility of *pn*J devices for electrical metrology. Earlier work demonstrated the viability of scalable designs through gateless quantum Hall checkerboard devices and concentric Corbino dartboard geometries [20-21], but did not fully lay out how one can reverse engineer a desired value, and though Landauer-Büttiker formalism can lead one to calculate a resulting quantized resistance in a complex *pn*J device, it cannot easily map a desired value to a configuration of source and drain terminals for a given configuration.

This manuscript comprehensively evaluates various parameter spaces of available CERs across diverse *pn*J topologies. Empirical quantum Hall transport data acquired from *pn*J devices establish a baseline verification for multi-terminal equilibration when compared with simulations performed with the circuit simulator LTspice. From that baseline, one extends this analysis to various checkerboard, Corbino, and triangular topologies by writing and utilizing Python code to solve specific configurations for all unique CERs via Landauer-Büttiker formalism. Lastly, an analysis on the parameter space of available quantized resistances is conducted for each configuration and topology to reveal statistically meaningful representations of a numerical density of states, Farey gap visualization, and Shannon entropy.

To establish baseline verification, several pnJ devices were fabricated with EG, with the full process described in the Supplementary Material, but that can be summarized as fabrication of devices on SiC substrate via sublimation, photolithography, and functionalization (to stabilize carrier density values in a way that outperforms other encapsulation methods) [9, 22-29]. An example completed device containing 8 regions is shown in Fig. 1 (a). Simulations for *pn*J devices were performed with LTspice to verify acquired data and custom code that emulates Landauer-Büttiker formalism [30], with a sample layout provided in Fig. 1 (b). As with similar work [18-21], the circuit uses both $p$-type and $n$-type quantum Hall elements. Devices were measured in a cryogenic system at 1.6 K and magnetic flux densities of up to 9 T. To confirm that the device was quantized after setting the necessary doping levels, several orthogonal pairs of voltages were measured while using 1 μA of current and shown in Fig. 1 (c). This current level ensures operation well within the linear quantum Hall regime, preventing critical breakdown, inter-Landau-level scattering, or thermal dissipation along the $p$-$n$ interfaces.

Despite the $n$-type region exhibiting longitudinal and Hall resistance mixing at low magnetic flux densities, the high flux density cases still subdued traces of longitudinal resistance. With established alternating regions, one may now select from a large set of terminal placements to generate quantized resistances that are rational multiples of $R_H$. Three such selections are shown in Fig. 1 (d), with their corresponding color-coded illustrations present to the lower left of the relevant curve. The measured plateaus closely match theoretically predicted fractions 20/9, 32/23, and 18/35 (indicated by the dashed horizontal lines), calculated by using the algebraic framework set forth in Ref. [18] for a device containing *N* total terminals (each of which may take on either a positive or negative polarity):

$$q_{N-1}(n_{N-1}) = \frac{q_{N-2}(n_{N-1}+1)}{n_{N-1} + \frac{q_{N-2}}{q_{N-1}^{(0)}}} \tag{1}$$

Equation 1 predicts a CER given a particular configuration, with $q$ being the CER (or a CER from a subset of the configuration – see [18] for precise use cases) and $n$ being the number of junctions between each adjacent pair of terminals. LTspice was also used to confirm the values in Fig. 1 (d).

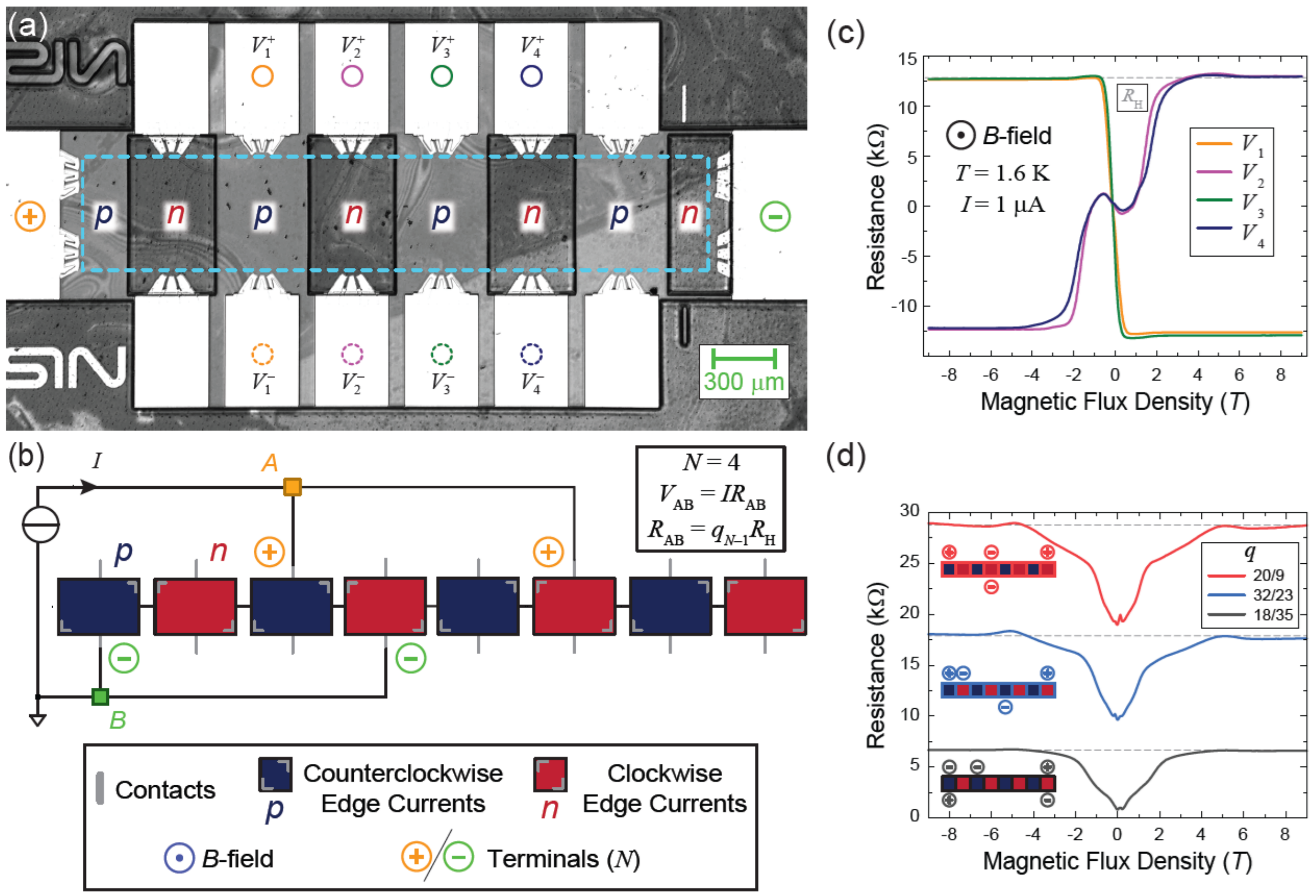

FIG. 1. (a) Optical image of the fabricated linear device, showcasing the alternating *p*-type and *n*-type regions. Electrical contacts are labeled along the top ($V^+$) and bottom ($V^-$) edges. (b) An LTspice schematic of the device shown in (a), modeling the Landauer-Büttiker chiral edge state network. The schematic shows an example $N = 4$ terminal configuration used to measure CER ($q$). (c) Standard four-terminal magnetoresistances were measured at 1.6 K and 1 µA as a function of magnetic flux density, verifying the integer quantum Hall plateaus across the *n*-type or *p*-type regions. (d) Experimental fractional quantized resistance plateaus achieved by configuring different multi-terminal sources and drains. The measured plateaus closely match the theoretically predicted fractions 20/9, 32/23, and 18/35 (indicated by the dashed horizontal lines). Color-coded insets show the exact configuration of terminals.

While LTspice simulations yield rapid numerical outputs, they lack structural insight into CER evolution [18, 30]. Conversely, while the analytical framework of Eq. 1 provides physical clarity, it is strictly restricted to one-dimensional *pn*J chains [18]. Introducing a second interface creates parallel chiral pathways, requiring full two-dimensional Landauer-Büttiker directed graph calculations [20, 21].

In a two-row checkerboard configuration, like the one in Fig. 2 (b), one now introduces parallel vertical chiral pathways, breaking the strict linear sequencing of Eq. 1, necessitating treatment of the *pn*J device as a two-dimensional directed graph subject to Landauer-Büttiker formalism, where $V_i$ is the voltage at terminal $i$ and $T_{ij}$ is the transmission probability from terminal $j$ to terminal $i$. In the quantum Hall regime with perfectly equilibrated chiral edge states, $T_{ij} = 1$ if there is a direct edge path from $j$ to $i$, and 0 otherwise:

$$I_i = \frac{1}{R_H} \sum_j \left( T_{ji} V_i - T_{ij} V_j \right) \quad ; \quad \mathbf{I} = \frac{1}{R_H} \mathbf{G}\,\mathbf{V}$$

(2)

Equation 2 sets the conditions for solving for a CER. One must invert the matrix **G** to solve algebraically for the voltage vector **V**. The CER ($q$) between the entry terminals (point A prior to splitting multiple entries) and exit terminals (point B where multiple exits meet again) is simply:

$$q = \frac{V_A - V_B}{I_{total} \cdot R_H}$$

Directly reverse-engineering a device layout from a target CER is mathematically intractable due to three factors: (1) high mapping degeneracy, where multiple topologies and terminal allocations yield identical CERs; (2) discrete parameter space gaps, which offer no guarantee that an arbitrary target value exists; and (3) factorial scaling of matrix permutations as grid dimensions expand.

The most effective approach to realizing the vision of a user-selected CER prior to designing a *pn*J device is to construct a bounded algorithmic library; that is, to write and use Python code to procedurally generate thousands of physically valid Landauer-Büttiker matrices, calculate their exact CERs, and index them into a searchable database for reverse-lookup. This searchable database directly addresses the issue of output arbitrariness by allowing metrologists to deterministically look up and select the exact multi-terminal arrangement required to yield a specific target resistance from the dense spectrum of available states. To that end, a discrete network model maps a user-defined device geometry onto a rectangular grid of alternating *n*-doped and *p*-doped regions and employs an edge-node topology to describe the one-dimensional chiral channels at the perimeters of each region (for more details on Python code development, see the Supplementary Material).

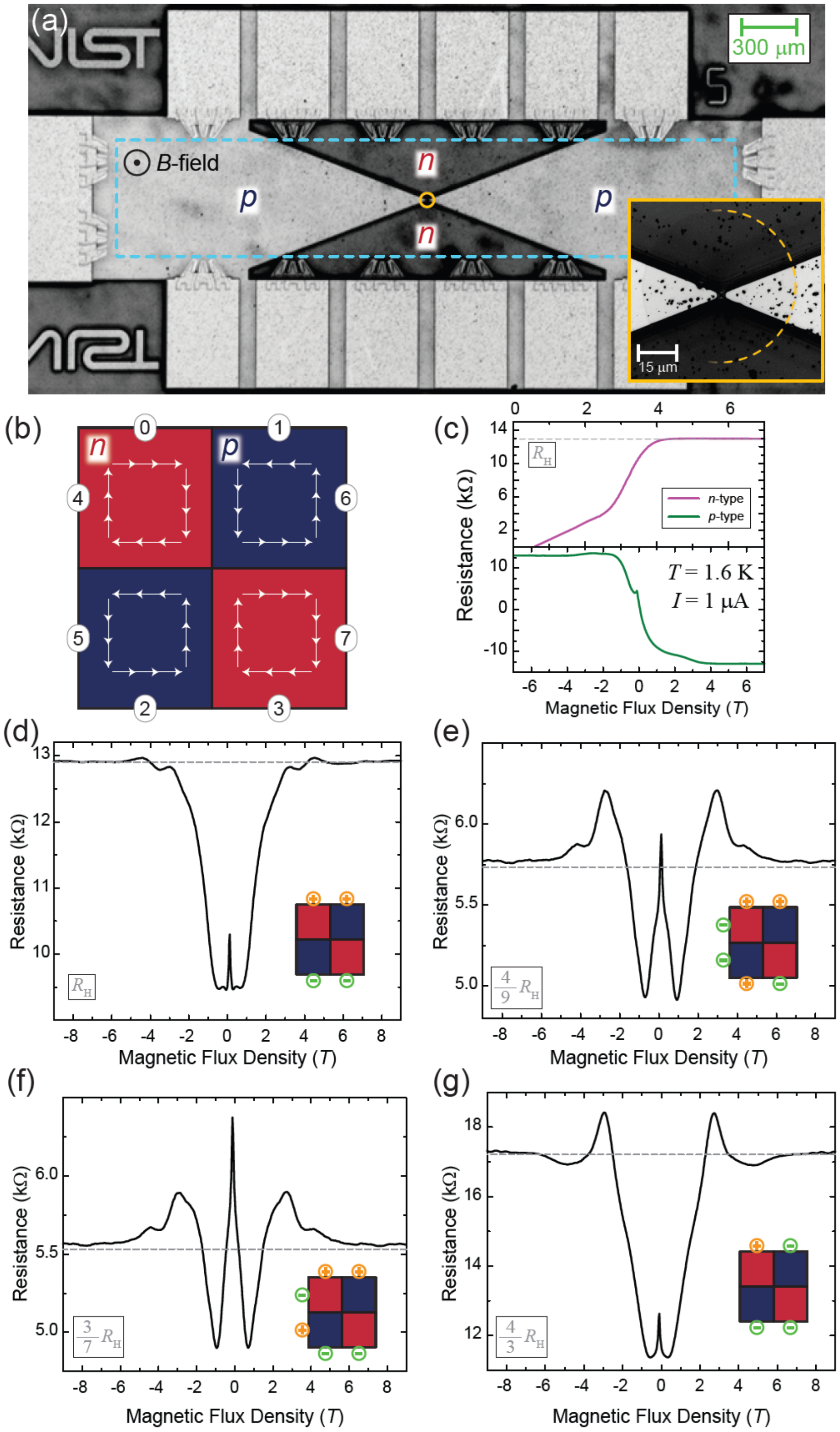

FIG. 2. (a) Optical image of the fabricated **2** × **2** checkerboard matrix showing alternating *n*-type and *p*-type EG regions, with a circular cutout region to avoid disordered edge state propagation at the central vertex. (b) Schematic mapping the external perimeter contacts (labeled 0 through 7), serving as the topological basis for the Landauer-Büttiker matrix solver written in Python. (c) Magnetoresistance validation sweeps for isolated *n*-type and *p*-type regions measured at 1.6 K and 1 μA as a function of magnetic flux density. (d - g) Multi-terminal quantized resistance sweeps exhibiting stable fractional plateaus. The solid experimental curves are plotted against horizontal

dashed lines representing the CERs (including $\frac{4}{9}\mathbf{R}_{\mathrm{H}}$, $\frac{3}{7}\mathbf{R}_{\mathrm{H}}$, and $\frac{4}{3}\mathbf{R}_{\mathrm{H}}$) generated by the Python simulation code (matching the LTspice simulation results).

This model was implemented with Python code to perform an exhaustive combinatorial search across all possible boundary contact allocations, allowing the user to vary the number of active terminals up to a maximum limit, $N_{max}$. All additional results not presented herein are provided in the supplementary material. With this approach, an example $2 \times 2$ quantum Hall checkerboard *pn*J device was selected for device testing and prediction verification, with Fig. 2 showing an image of the device, a schematic of the external perimeter contacts serving as the basis for the Landauer-Büttiker matrix solver written in Python, magnetoresistance sweeps as a function of magnetic flux density, and multi-terminal quantized resistance sweeps exhibiting stable fractional plateaus. The takeaway experimental results are plotted against horizontal dashed lines representing the CERs (including $R_H$, $\frac{4}{9}$ $R_H$, $\frac{3}{7}$ $R_H$, and $\frac{4}{3}$ $R_H$) generated by the Python code, which also match LTspice simulations. This agreement across physical multi-terminal measurements confirms that the Landauer-Büttiker network solver accurately captures the transport physics without requiring ad hoc correction factors at these grid scales.

Having validated the functionality of the Python code, one can now expand on practicability of various *pn*J topologies. In graph theory, one may design a grid where *p* and *n* regions alternate without identical regions merging, with the requirement that the dual graph of the tiling is bipartite, meaning that an even number of tiles meet at every vertex of the grid. By this mathematical constraint, only several topologies become accessible for fabrication. Three distinct topologies are shown in Fig. 3, namely the rectangular, Corbino, and triangular layouts. The numbered contacts along the perimeters correspond to possible source and drain terminals. Example calculations of the Landauer-Büttiker matrices for different multi-terminal configurations are also provided and include $\frac{1776}{881}$ $R_H$, $\frac{1901}{811}$ $R_H$, $\frac{1947}{817}$ $R_H$, $\frac{1411}{196}$ $R_H$, $\frac{2222}{761}$ $R_H$, $\frac{577}{239}$ $R_H$, and $\frac{1349}{777}$ $R_H$, demonstrating the extensive and highly tunable parameter space made accessible by imposing varied boundary conditions on EG *pn*J devices.

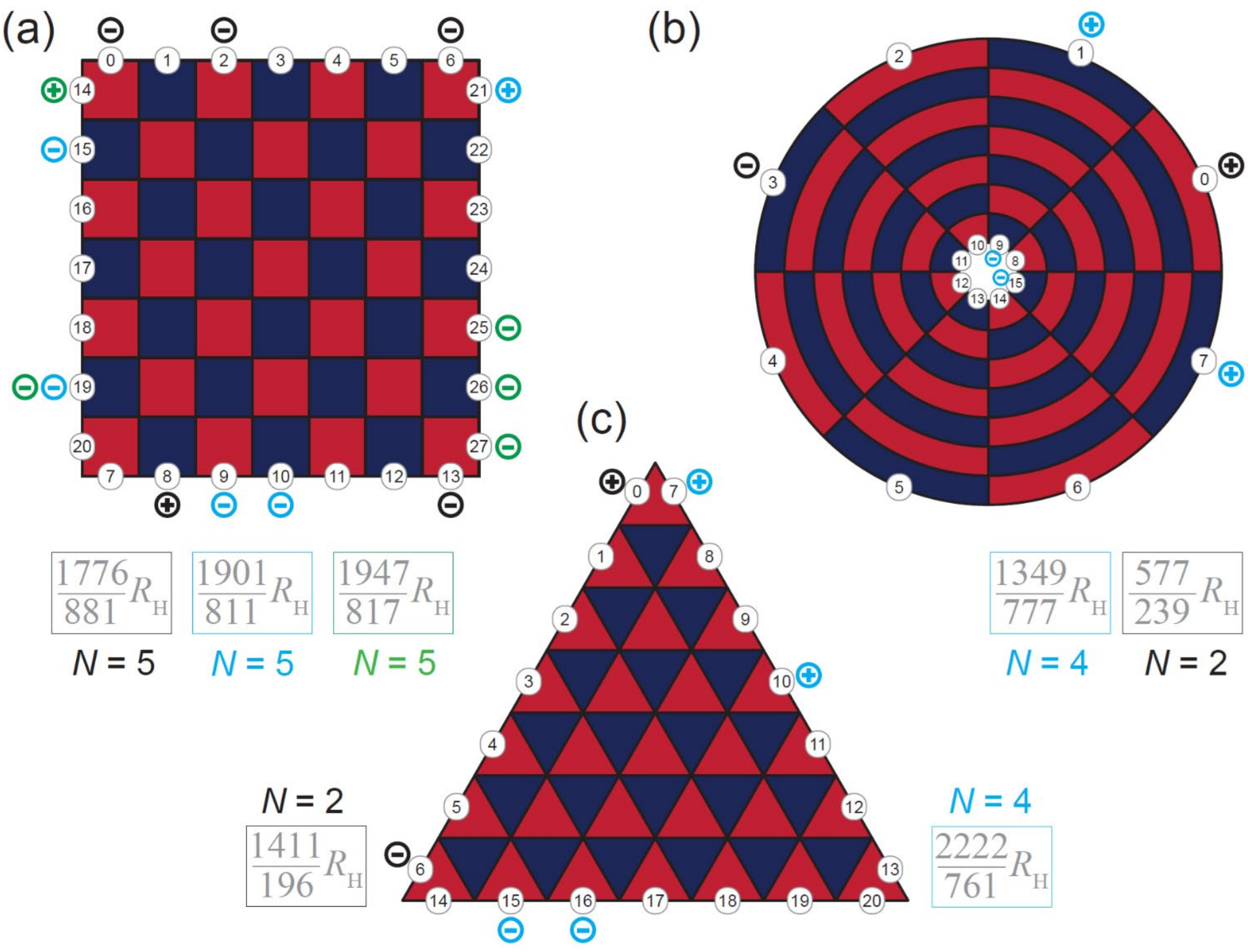

FIG. 3. Graphs and CERs for three distinct *pn*J topologies. Number labels correspond to metallic contacts situated on the boundaries, with red and blue regions representing *n*-type and *p*-type regions, respectively. (a) Rectangular topology shown with three example calculations and color-coded terminal placements including $\frac{1776}{881}$ $R_H$ (black), $\frac{1901}{811}$ $R_H$ (cyan), and $\frac{1947}{817}$ $R_H$ (green). (b) Corbino topology shown with two examples calculated analytically from the Landauer-Büttiker matrix solving Python code: $\frac{1411}{196}$ $R_H$ (black) and $\frac{2222}{761}$ $R_H$ (cyan). (c) Triangular topology shown with two examples similarly calculated with modified Python code: $\frac{577}{239}$ $R_H$ (black) and $\frac{1349}{777}$ $R_H$ (cyan). These example calculations demonstrate the extensive tunable parameter space made accessible by switching topologies.

To successfully calculate predicted CERs for the Corbino and triangular topologies, minor Python code modifications were required, such as periodic boundary conditions for the former and a switch from Cartesian indexing to oblique coordinate indexing. Furthermore, it became evident that the number of unique CERs was highly dependent on the topology and configuration, as was the clustering of those solutions within the number line, which is an important property to understand if one wishes to engineer a device with highly programmable, but local, outputs. By tracking the degeneracies of the CERs for a given configuration, the Python code was modified to calculate the Shannon entropy (H) of the accessible states, providing a way to differentiate between high clustering, where CERs heavily collapse into primary attractors, and wide distributions spanning the Farey sequence, projecting results as the one-dimensional density of states:

$$H = -\sum_{i=1}^{K} P(q_i) \log_2 P(q_i)$$

(4)

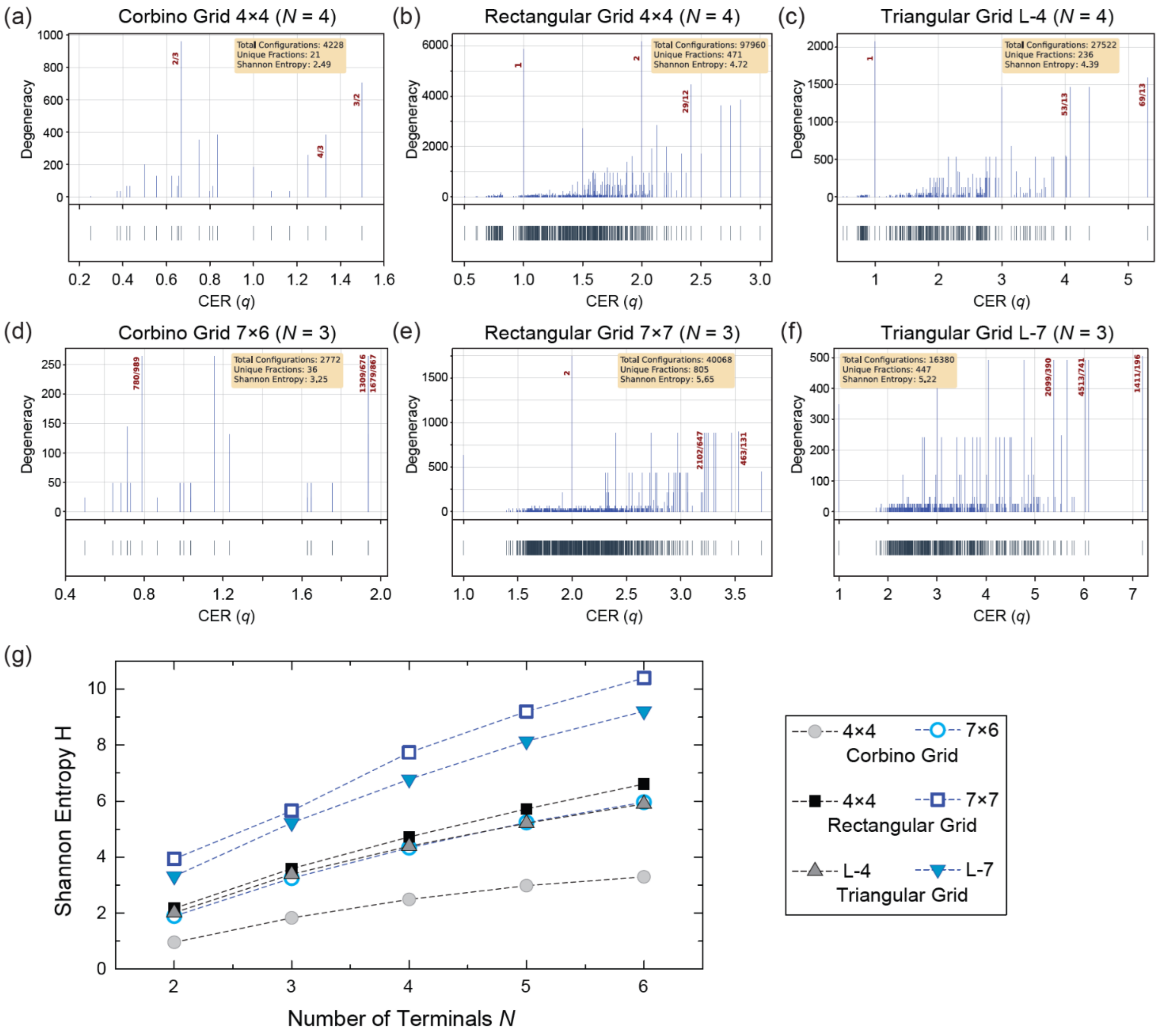

FIG. 4. (a)-(c) The full set of CERs were calculated for each of the three systems listed, all with four maximum allowable terminals and all topologies of similar dimensions. The one-dimensional density of states was plotted as a function of CER, along with Farey sequence visualizations to intuit the relationship between the Shannon entropy and CER clustering along the number line. (d)-(f) Another full set of CERs were calculated for each of the three topologies of similar dimension allowing only three terminals. The same quantities of the previous three panels were also computed. For all six panels, the three largest CER degeneracies are indicated in red, and an inset indicates the total number of possible configurations, unique fractions, and Shannon entropy. (g) The Shannon entropy (H) was determined as a function of the number of total terminals for checkerboards (square data points), Corbino rings (circular data points), and triangular layouts (triangular data points). Black and gray points are for the first set of configurations, and the blue and cyan points are for the second set of configurations.

The Shannon entropy in Eq. 4 serves as a statistical measure of topological complexity and diversity of states. It quantifies how evenly distributed or clustered CERs ($q_i$) are for a given device geometry and terminal count, using $K$ as the number of unique fractions and $P(q_i)$ as the probability of measuring a specific value $q_i$ when choosing a valid terminal arrangement at random (that is, $P(q_i)$ = (degeneracy of $q_i$)/(number of total valid configurations)).

The distribution of accessible CERs reveals a distinct, topology-dependent hierarchy in the Shannon entropy of the *pn*J layouts. Figure 4 shows degeneracy counts for systems with four terminals (a-c) and three terminals (d-f). The Corbino layout yields a highly degenerate spectrum (21 unique fractions for (a)), corresponding to the lowest calculated Shannon entropy (2.49 in Fig. 4 (a)). The topological complexity expands moderately in the triangular layout (Fig. 4 (c)), and the rectangular checkerboard maximizes the available parameter space, achieving the highest Shannon entropy of 4.72 and generating 471 unique fractions (Fig. 4 (b)). This clear progression from low, to intermediate, to high entropy is similarly maintained at larger grid scales, as demonstrated by Fig. 4 (g).

This hierarchy of topological complexity is further validated by the scaling behavior of the Shannon entropy as a function of the number of terminals ($N$). The data demonstrate that regardless of the specific terminal count, the Corbino geometry persistently maintains the lowest overall entropy across both small and large grid sizes. This constrained parameter space is a consequence of the continuous azimuthal boundary conditions and high rotational symmetry inherent to the Corbino geometry, which forces significant path degeneracy. The triangular grids trace an intermediate entropy behavior, balancing structural symmetry with a diversity of routing pathways, and lastly, the rectangular grids consistently reach the highest Shannon entropies across all tested terminal counts. This maximized entropy confirms that the open boundaries of the checkerboard minimize degeneracy and produce the richest spectrum of fractional quantized resistances. For all cases, it should be noted that quantized resistances never go beyond an order of magnitude of $R_H$, highlighting the range restriction expected from these kinds of *pn*J devices.

Although individual *pn*J devices are bounded within an order of magnitude of $R_{\mathrm{H}}$, their true metrological strength emerges when integrated as tunable, modular nodes within star-mesh resistance networks [9, 10]. Traditional star-mesh quantum Hall arrays rely on uniform, single-value elements ($R_{\mathrm{H}} \approx 12.9\ \mathrm{k\Omega}$) linked by metallic interconnects, where synthesizing high resistances (MΩ to GΩ) requires large numbers of identical nodes and complex graph topologies. Replacing standard single-value elements with multi-terminal *pn*J nodes introduces adjustable local coefficients ($q_i R_{\mathrm{H}}$) directly into the star-mesh matrix transformations. Because *pn*J devices achieve fractional quantization entirely via chiral edge-state equilibration along carrier-density domain boundaries, without internal metallic contacts [18, 19], they eliminate the parasitic contact resistances and thermal EMFs that typically accumulate in multi-element arrays. Furthermore, the ability to select from a dense spectrum of $q$-values via the Landauer-Büttiker solver allows star-mesh network branches to be fine-tuned deterministically, significantly reducing deviations from nominal decadic values of resistance typically preferred in standards laboratories.

In conclusion, this work establishes a comprehensive approach to evaluate CER parameter spaces across several EG *pn*J devices and proposed topologies. Rather than directly reverse-engineering specific device layouts from a desired CER, automated Python codes based on the Landauer-Büttiker formalism were developed to procedurally generate and index valid transmission matrices, a method that demonstrated excellent agreement with both LTspice modeling and baseline multi-terminal cryogenic measurements. The Shannon entropies of the configurations showed a clear topological hierarchy, with Corbino, triangular, and rectangular layouts exhibiting low, intermediate, and high values, respectively. Furthermore, this insight indicates that Shannon entropy could be increased even further by deploying asymmetric polygon tessellations, Penrose tilings, or randomized Voronoi grids. Such aperiodic layouts would further break structural boundary symmetries, forcing chiral edge channels into non-redundant pathways and expanding the density of unique quantized states. Coupled with knowledge of the full range of available CERs for any given configuration, this systematic evaluation defines the operational boundaries for any quantized Hall resistance standards based on *pn*J devices.

## SUPPLEMENTARY MATERIAL

Additional details for device fabrication, Python code development, and Landauer-Büttiker matrix calculations are provided in the Supplementary Material.

## ACKNOWLEDGMENTS

The authors would like to express thanks to Mattias Kruskopf of Physikalisch-Technische Bundesanstalt (PTB) for assistance with graphene growth during his postdoctoral guest researchership at NIST and to Paola Barbara at Georgetown University for fruitful discussions and advising. Unless otherwise noted, NIST work was funded solely by the United States Government.

The authors declare no competing interests. Commercial equipment, instruments, and materials are identified in this paper in order to specify the experimental procedure adequately. Such identification is not intended to imply recommendation or endorsement by the National Institute of Standards and Technology or the United States government, nor is it intended to imply that the materials or equipment identified are necessarily the best available for the purpose.

# Supplementary Material

T. M. Padwa,[1,2*] D. Henry,[1,2*] D. S. Scaletta,[3] H. M. Hill,[1] A. L. Levy,[4] D. G. Jarrett,[2] D. B. Newell,[1] and A. F. Rigosi[1,a)]

[1]*Physical Measurement Laboratory, National Institute of Standards and Technology (NIST), Gaithersburg, Maryland 20899, USA*

[2]*Department of Physics, Georgetown University, Washington, District of Columbia 20057, USA*

[3]*Department of Physics, Mount San Jacinto College, Menifee, California 92584, USA*

[4]*Department of Physics and Astronomy, Howard University, Washington, District of Columbia 20059, USA*

Table of Contents:

*All references numbered in main text.

## 1. Additional details for device fabrication

For the fabrication process, chips must first be diced from 4*H*-SiC(0001) wafers (CREE - see main text acknowledgments) and chemically cleaned with a 5:1 diluted solution of hydrofluoric acid and deionized water, followed then by a coating with AZ-5214-E photoresist to utilize polymer-assisted sublimation [22]. Chips were then placed on a polished graphite substrate (SPI Glas 22 – see acknowledgments) silicon-face down and subjected to an ambient argon environment at 1900 °C inside of a graphite-lined furnace (Materials Research Furnaces Inc. – see main text acknowledgments). Next, superconducting NbTiN electrical contacts were deposited, with specific process details available in other work [9, 23]. Visual inspection and non-invasive testing preceded the last segments of fabrication [24], which involved $Cr(CO)_3$ functionalization to lower the electron density to about $10^{10}$ $cm^{-2}$ [25-26] and deposition of S1813 photoresist as a protective layer for intended *n*-type regions; *p*-type doping could be done controllably with chip exposure to nitric acid vapors [23].

## 2. Additional details for Python code development

For a checkerboard of *r* rows and *c* columns, the network comprises $\mathrm{N_{perimeter}} = 2(\mathrm{r} + \mathrm{c})$ external boundary segments and $\mathrm{N_{internal}} = (\mathrm{r}-1)\mathrm{c} + \mathrm{r}(\mathrm{c}-1)$ internal junction segments, yielding a total of $\mathrm{N}$ node segments.

At each individual domain, the local routing of the chiral edge states is dictated by the carrier type. In *n*-type domains, carriers circulate clockwise, prompting directed transmissions between adjacent boundary segments in the order of North → East → South → West. Conversely, in *p*-type domains, the circulation is counterclockwise, reversing the routing sequence to North → West → South → East. The superposition of these local circulations across all regions yields the transmission matrix, $G \in R^{N \times N}$, which satisfies current conservation requirements. To calculate the CER for an arbitrary multi-terminal configuration, the system is solved for all floating node (full set *F*, with specific labeled nodes *f*) potentials, with drains set to zero voltage and source nodes (full set *S,* with indices *s* and *s'*) constrained to a summed unit injected current. This yields a reduced linear system:

$$\begin{bmatrix} \sum_{s\in S}\sum_{s'\in S} G_{s,s'} & \sum_{s\in S} G_{s,f_1} & \cdots \\ \sum_{s\in S} G_{k,s} & G_{k,f_1} & \cdots \\ \vdots & \vdots & \ddots \end{bmatrix} \begin{bmatrix} V_{source} \\ V_{f_1} \\ \vdots \end{bmatrix} = \begin{bmatrix} 1 \\ 0 \\ \vdots \end{bmatrix} \tag{S1}$$

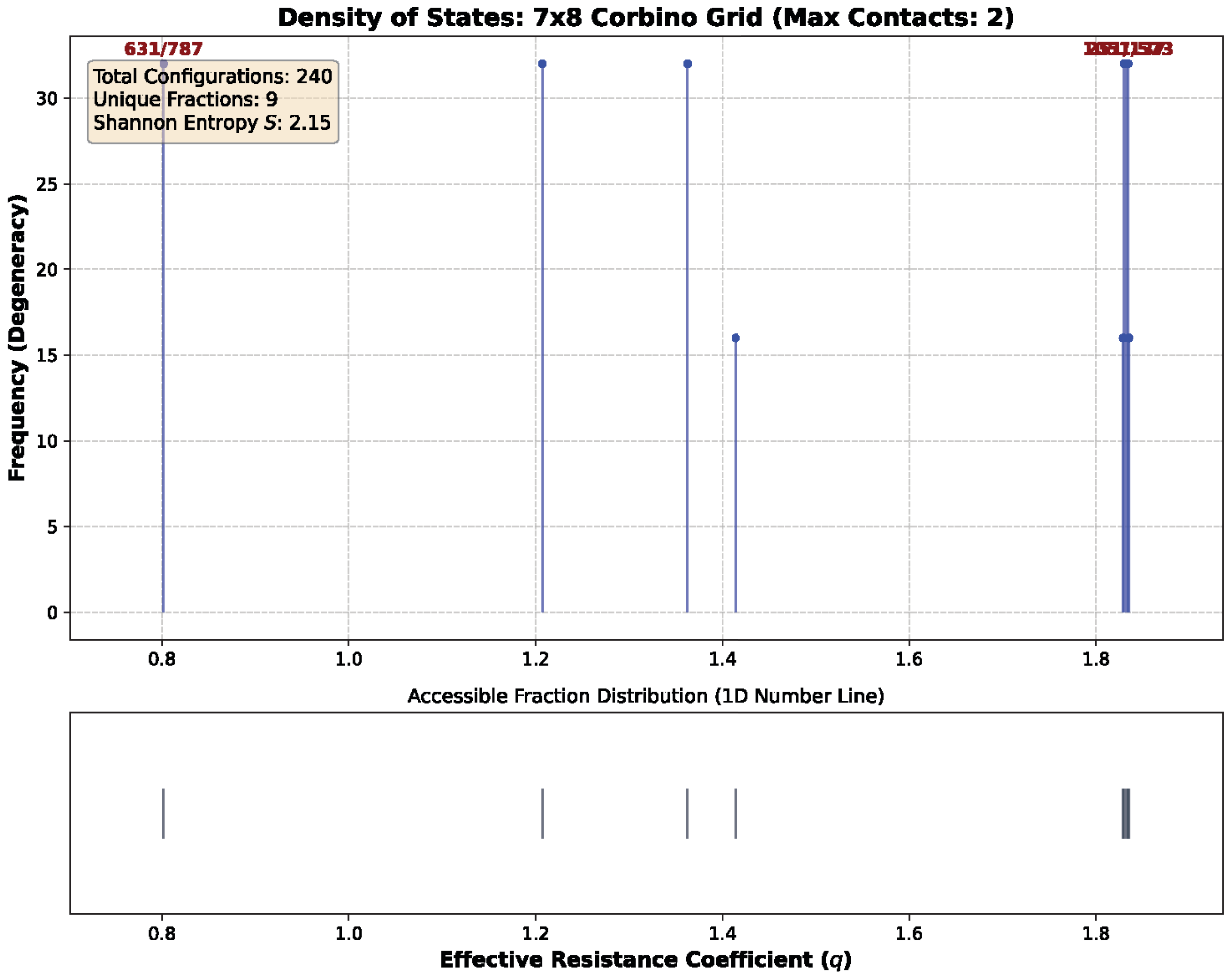

Density of States: 7x8 Corbino Grid (Max Contacts: 2)
631/787
Total Configurations: 240
Unique Fractions: 9
Shannon Entropy S: 2.15
Frequency (Degeneracy)
0
5
10
15
20
25
30
0.8
1.0
1.2
1.4
1.6
1.8
Accessible Fraction Distribution (1D Number Line)
Effective Resistance Coefficient (q)

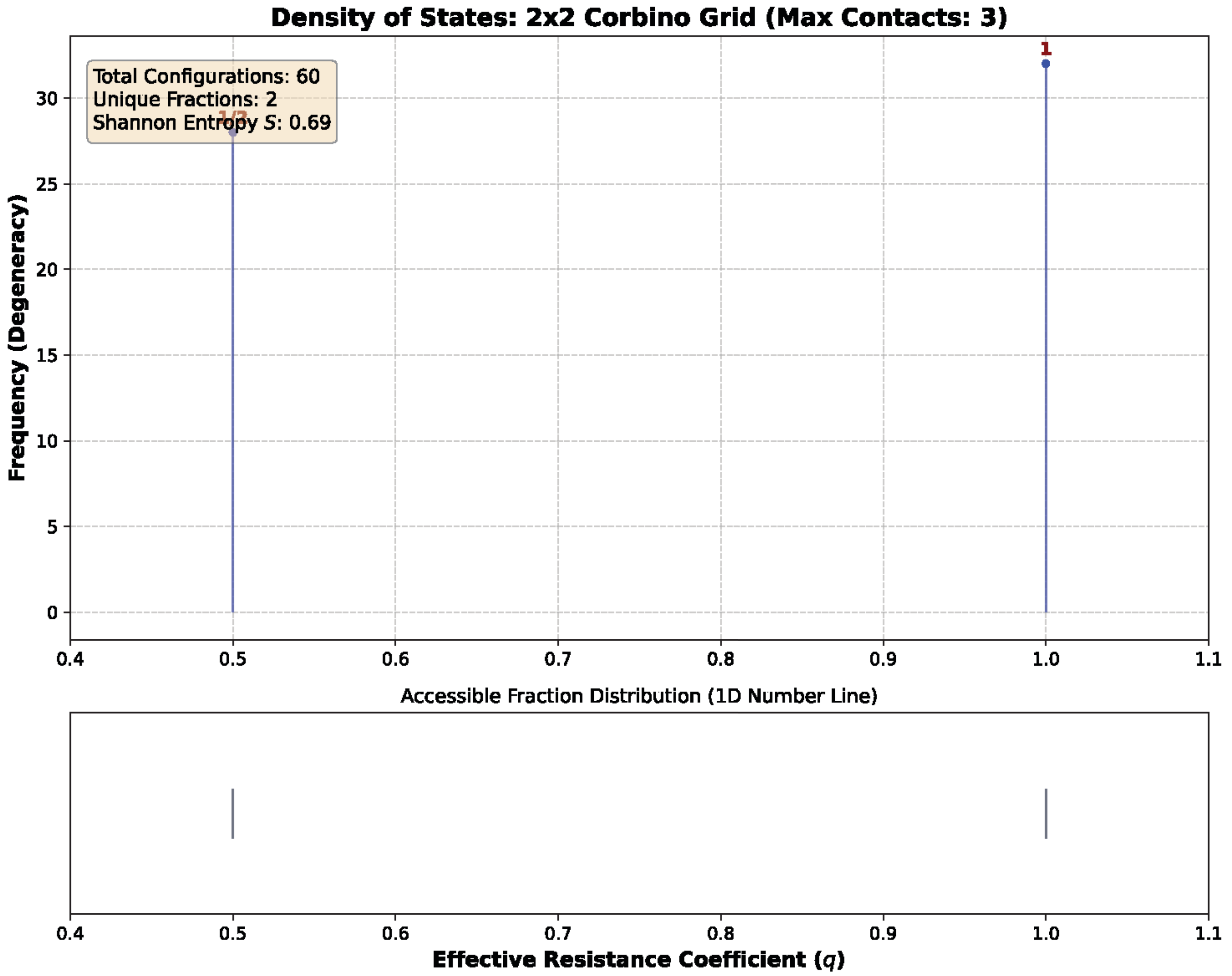
Density of States: 2x2 Corbino Grid (Max Contacts: 3)
Total Configurations: 60
Unique Fractions: 2
Shannon Entropy S: 0.69
1/2
1
Frequency (Degeneracy)
0
5
10
15
20
25
30
0.4
0.5
0.6
0.7
0.8
0.9
1.0
1.1
Accessible Fraction Distribution (1D Number Line)
Effective Resistance Coefficient (q)

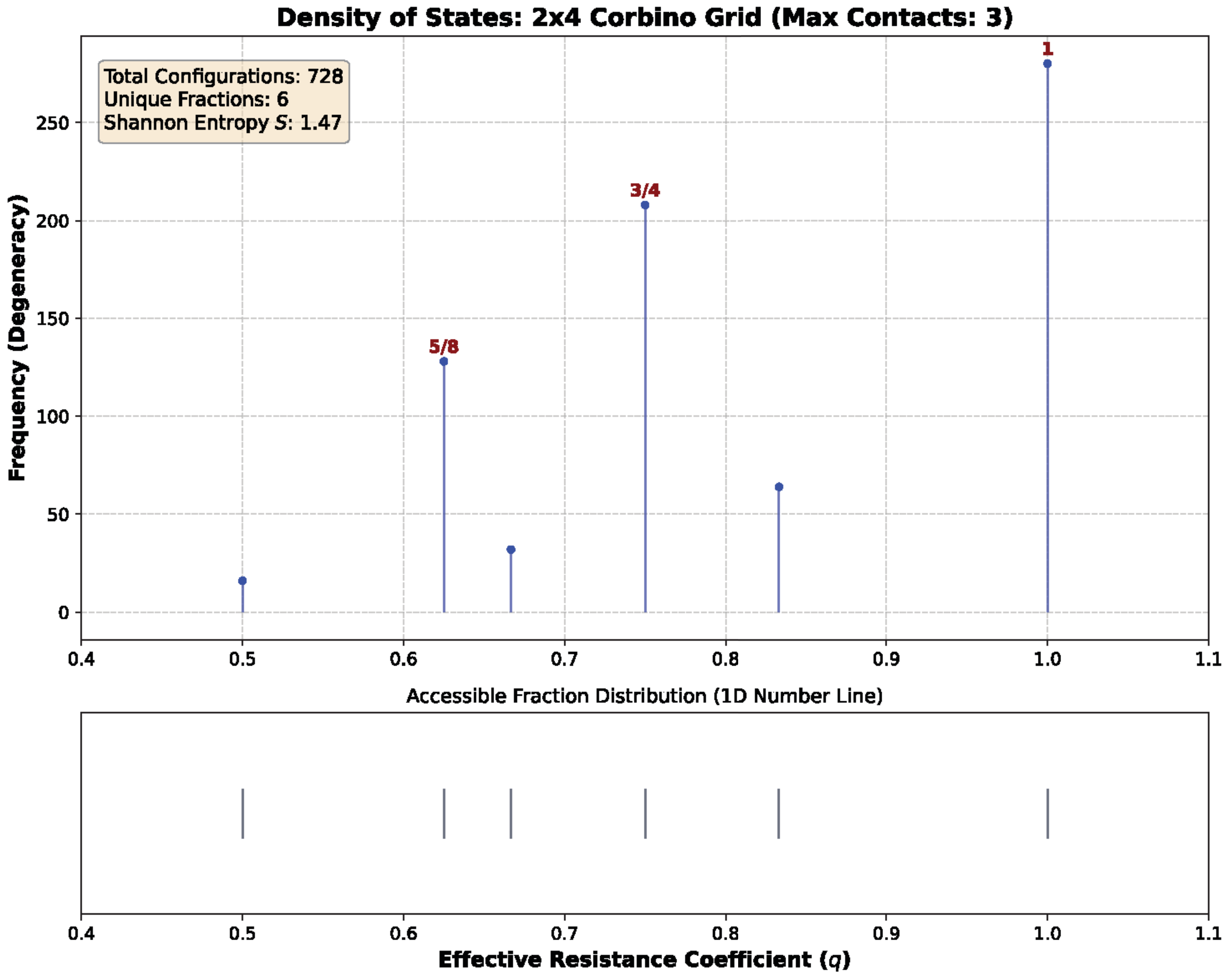

Density of States: 2x4 Corbino Grid (Max Contacts: 3)
Total Configurations: 728
Unique Fractions: 6
Shannon Entropy S: 1.47
1
3/4
5/8
Frequency (Degeneracy)
0
50
100
150
200
250
0.4
0.5
0.6
0.7
0.8
0.9
1.0
1.1
Accessible Fraction Distribution (1D Number Line)
Effective Resistance Coefficient (q)

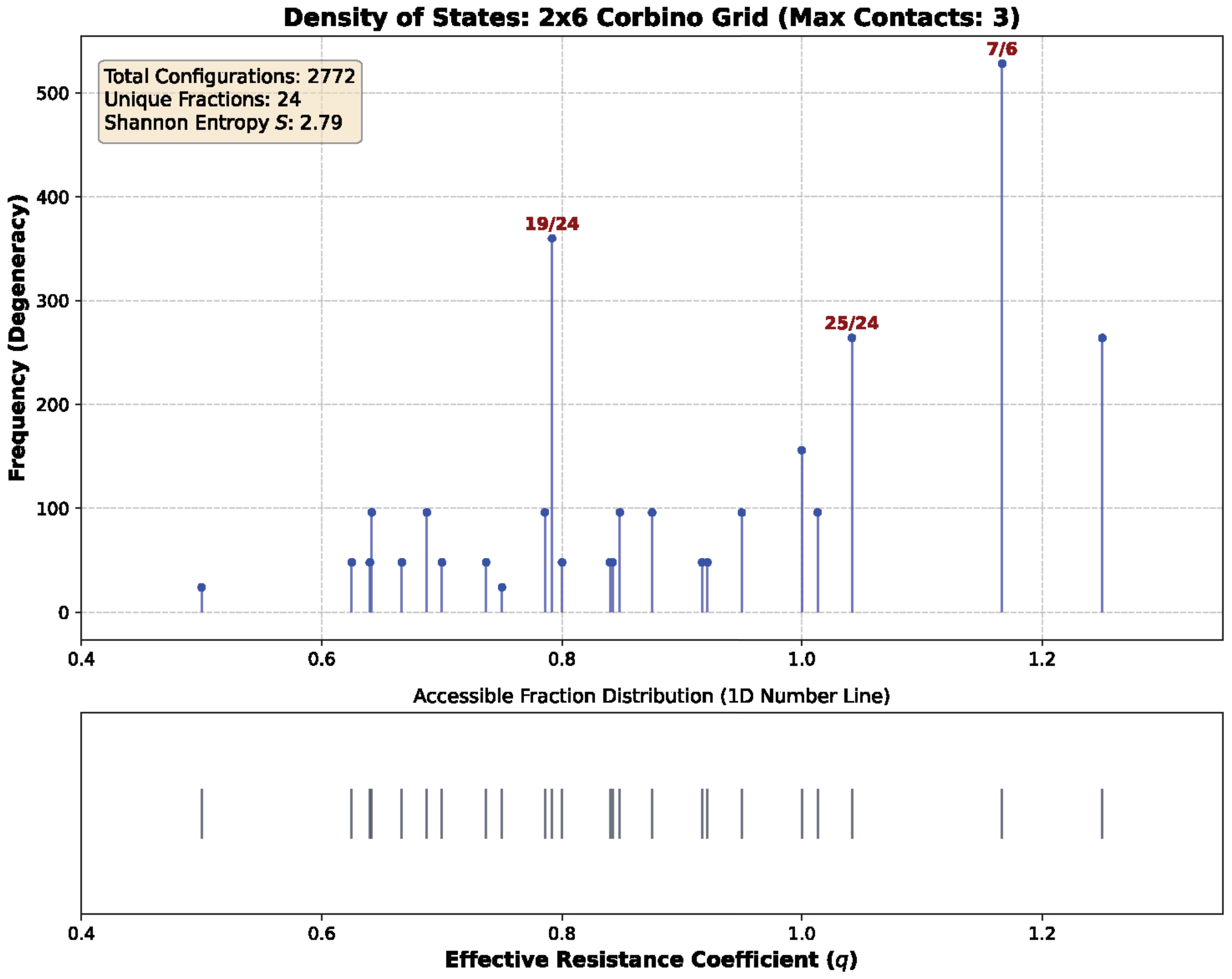

Density of States: 2x6 Corbino Grid (Max Contacts: 3)
Total Configurations: 2772
Unique Fractions: 24
Shannon Entropy S: 2.79
7/6
19/24
25/24
Frequency (Degeneracy)
0
100
200
300
400
500
0.4
0.6
0.8
1.0
1.2
Accessible Fraction Distribution (1D Number Line)
Effective Resistance Coefficient (q)

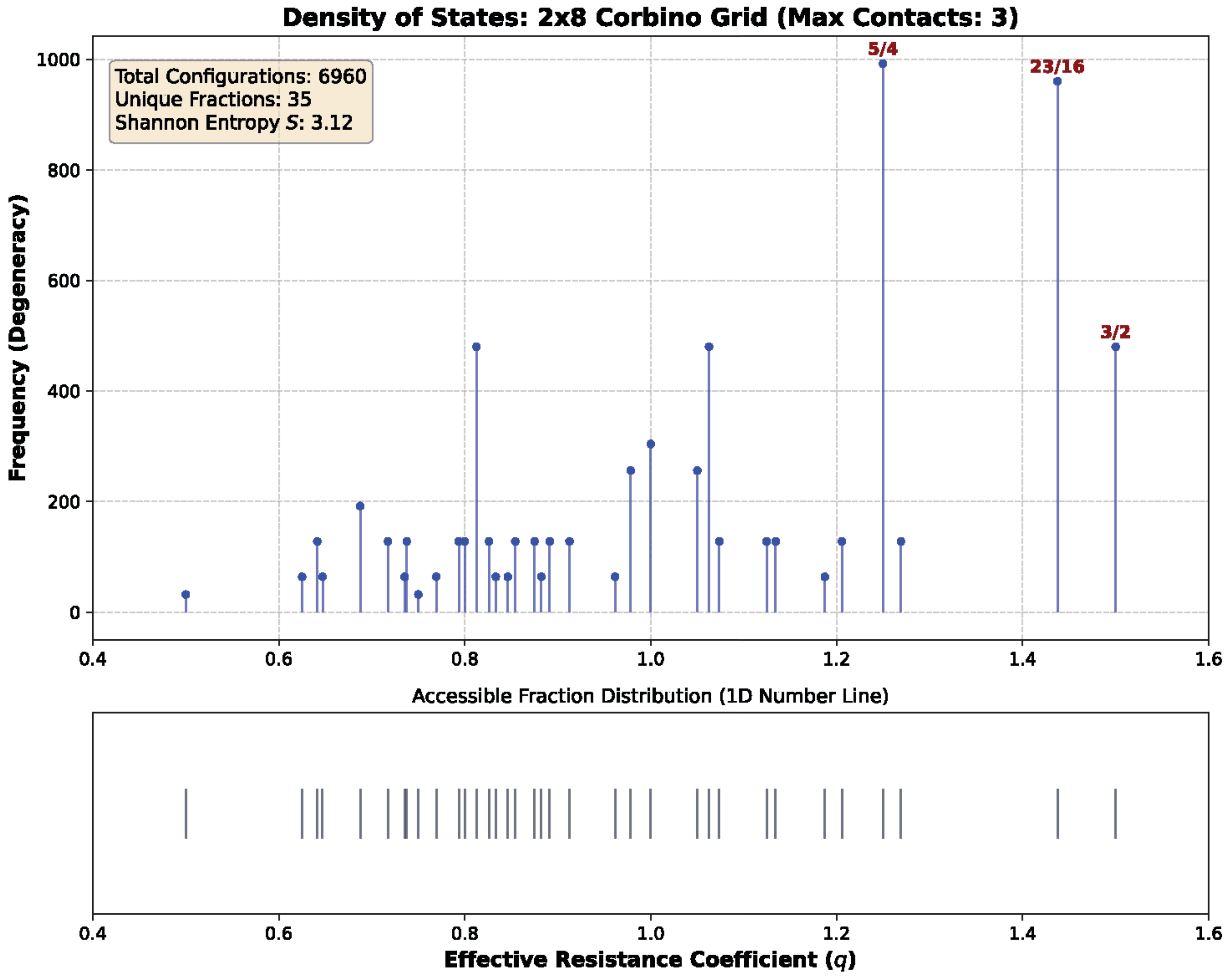

Density of States: 2x8 Corbino Grid (Max Contacts: 3)
Total Configurations: 6960
Unique Fractions: 35
Shannon Entropy S: 3.12
5/4
23/16
3/2
Frequency (Degeneracy)
1000
800
600
400
200
0
0.4
0.6
0.8
1.0
1.2
1.4
1.6
Accessible Fraction Distribution (1D Number Line)
0.4
0.6
0.8
1.0
1.2
1.4
1.6
Effective Resistance Coefficient (q)

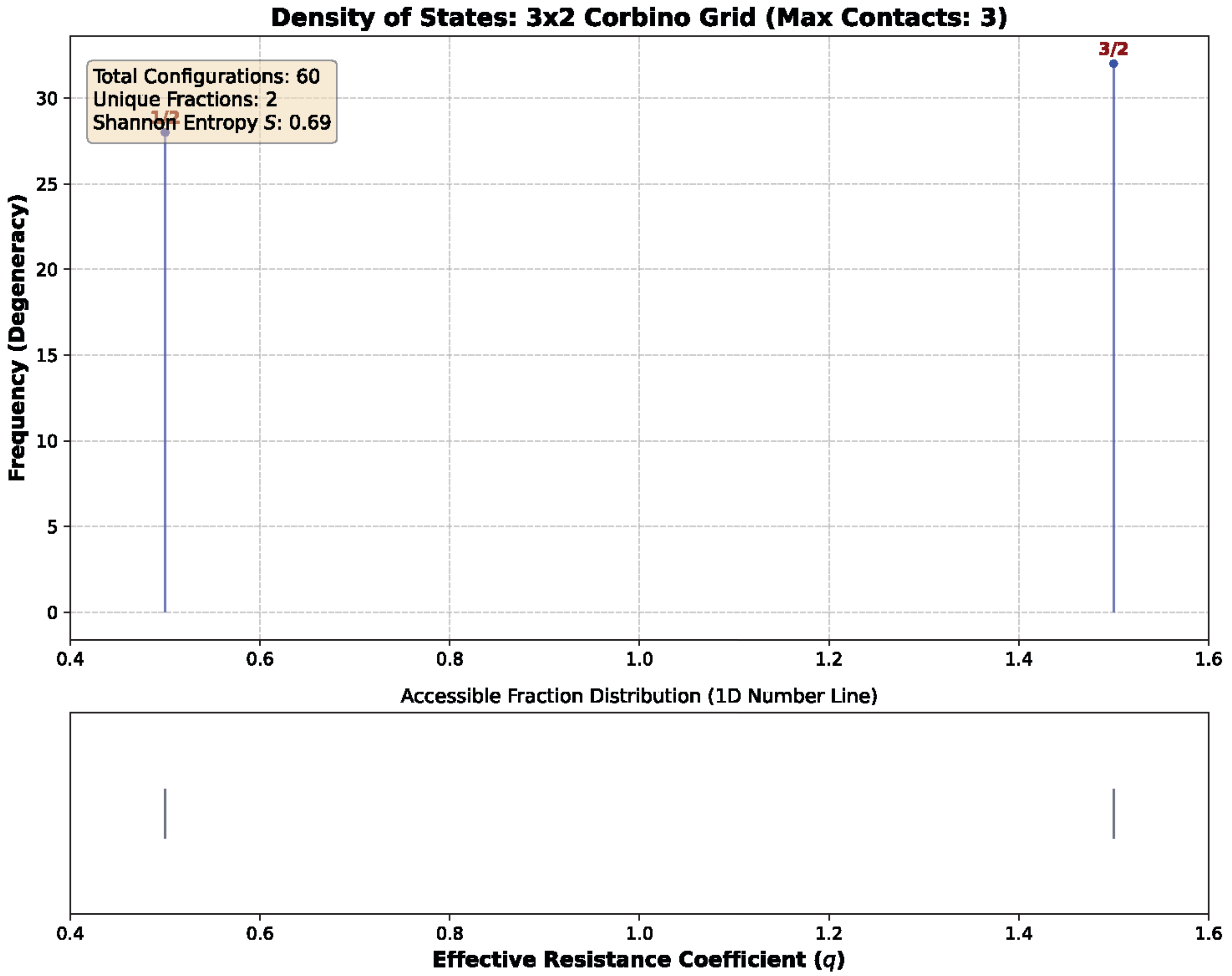

Density of States: 3x2 Corbino Grid (Max Contacts: 3)
Total Configurations: 60
Unique Fractions: 2
Shannon Entropy S: 0.69
1/2
3/2
Frequency (Degeneracy)
0
5
10
15
20
25
30
0.4
0.6
0.8
1.0
1.2
1.4
1.6
Accessible Fraction Distribution (1D Number Line)
Effective Resistance Coefficient (q)

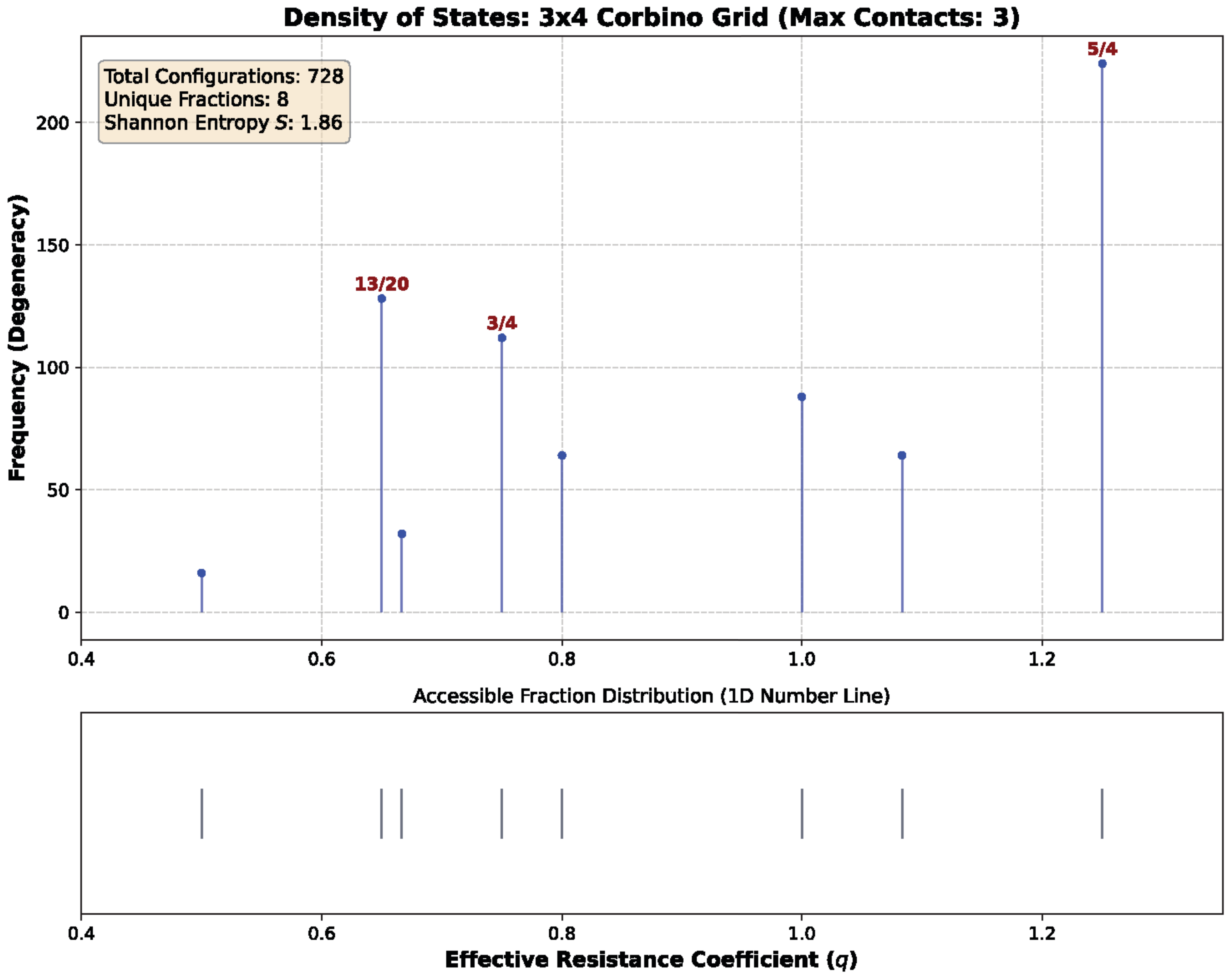
Density of States: 3x4 Corbino Grid (Max Contacts: 3)
Total Configurations: 728
Unique Fractions: 8
Shannon Entropy S: 1.86
Frequency (Degeneracy)
13/20
3/4
5/4
0
50
100
150
200
0.4
0.6
0.8
1.0
1.2
Accessible Fraction Distribution (1D Number Line)
Effective Resistance Coefficient (q)

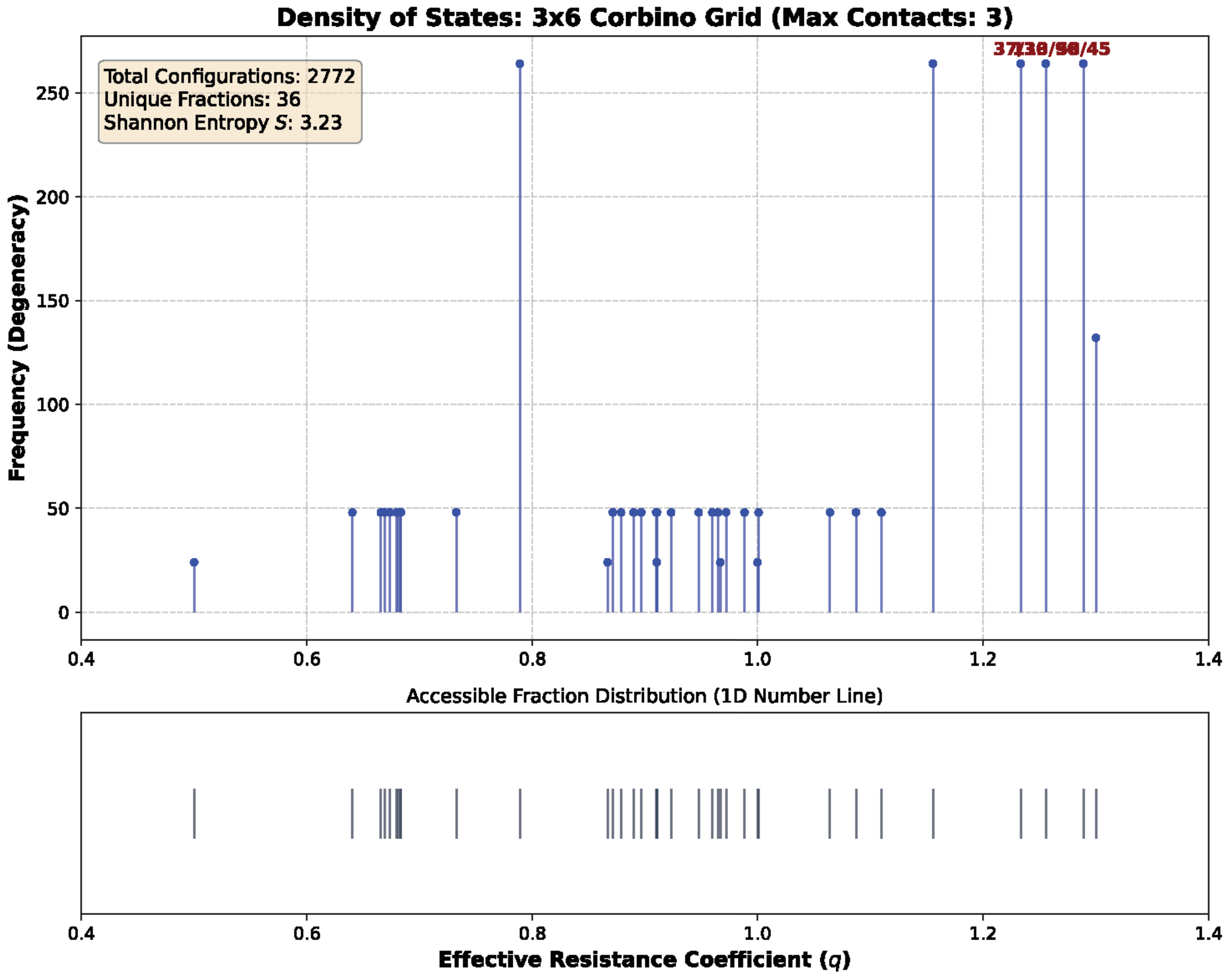

Density of States: 3x6 Corbino Grid (Max Contacts: 3)
Total Configurations: 2772
Unique Fractions: 36
Shannon Entropy S: 3.23
Frequency (Degeneracy)
0
50
100
150
200
250
0.4
0.6
0.8
1.0
1.2
1.4
Accessible Fraction Distribution (1D Number Line)
Effective Resistance Coefficient (q)

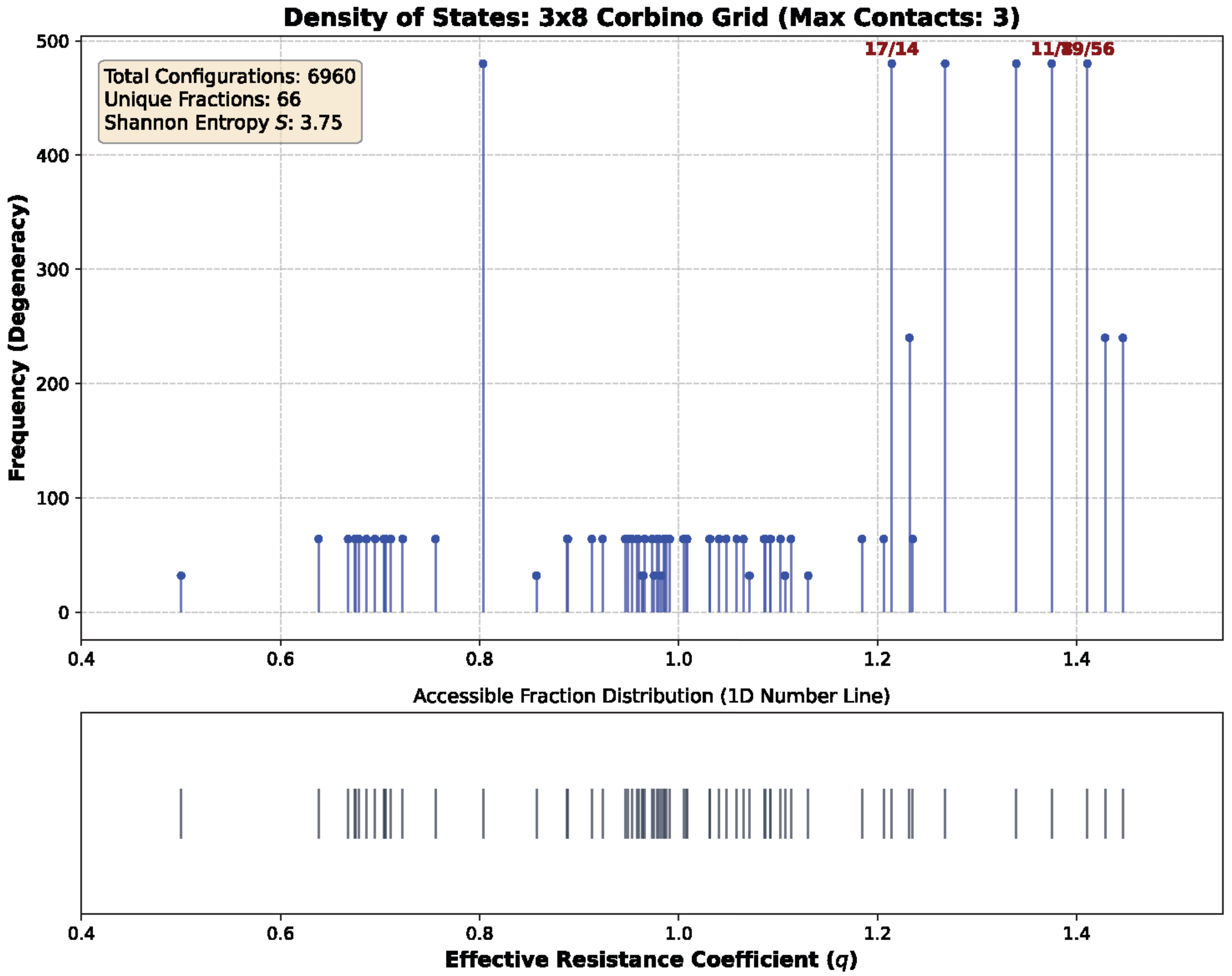

Density of States: 3x8 Corbino Grid (Max Contacts: 3)
Total Configurations: 6960
Unique Fractions: 66
Shannon Entropy S: 3.75
17/14
11/8
39/56
Frequency (Degeneracy)
0
100
200
300
400
500
0.4
0.6
0.8
1.0
1.2
1.4
Accessible Fraction Distribution (1D Number Line)
Effective Resistance Coefficient (q)

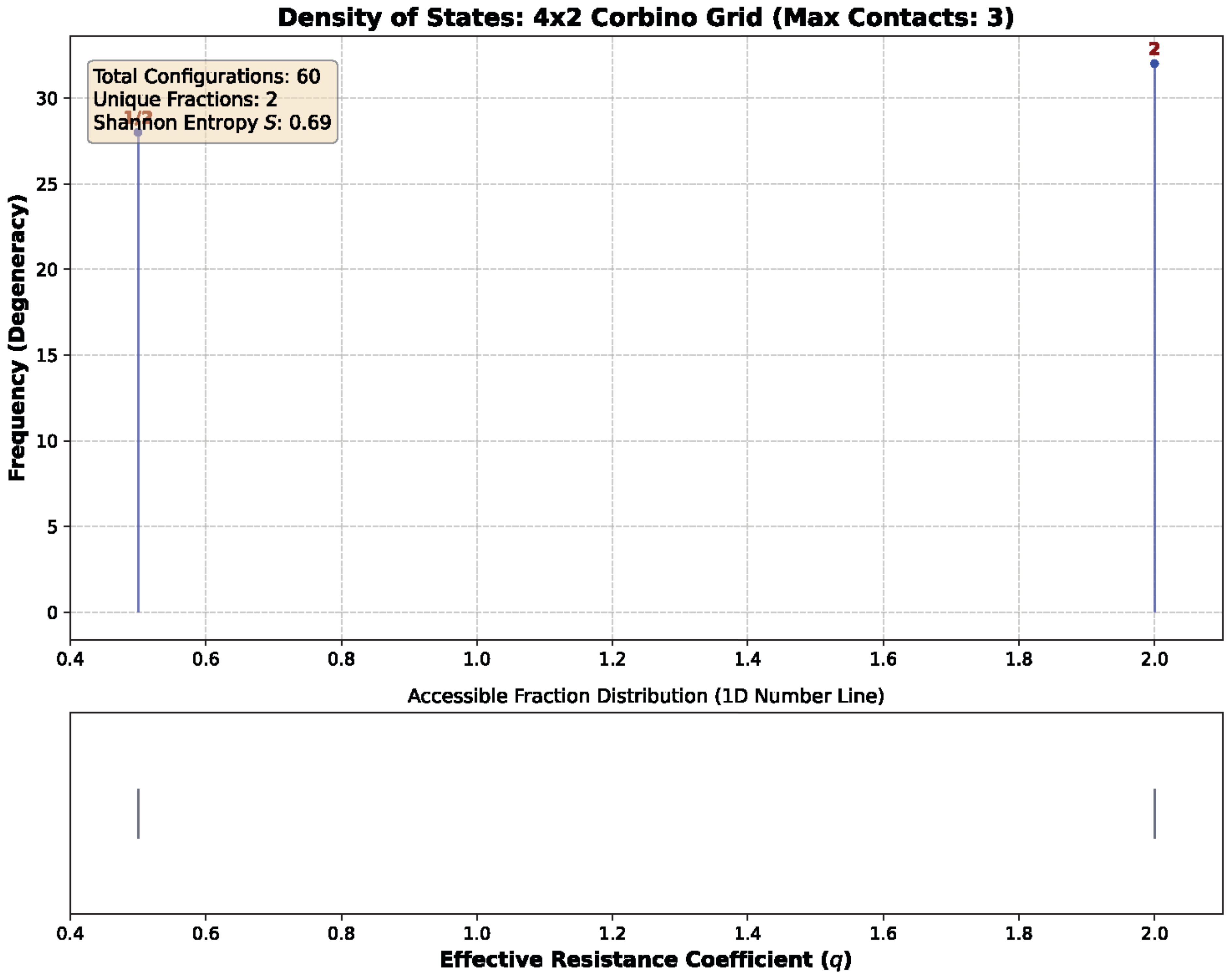

Density of States: 4x2 Corbino Grid (Max Contacts: 3)
Total Configurations: 60
Unique Fractions: 2
Shannon Entropy S: 0.69
1/2
2
Frequency (Degeneracy)
0
5
10
15
20
25
30
0.4
0.6
0.8
1.0
1.2
1.4
1.6
1.8
2.0
Accessible Fraction Distribution (1D Number Line)
Effective Resistance Coefficient (q)

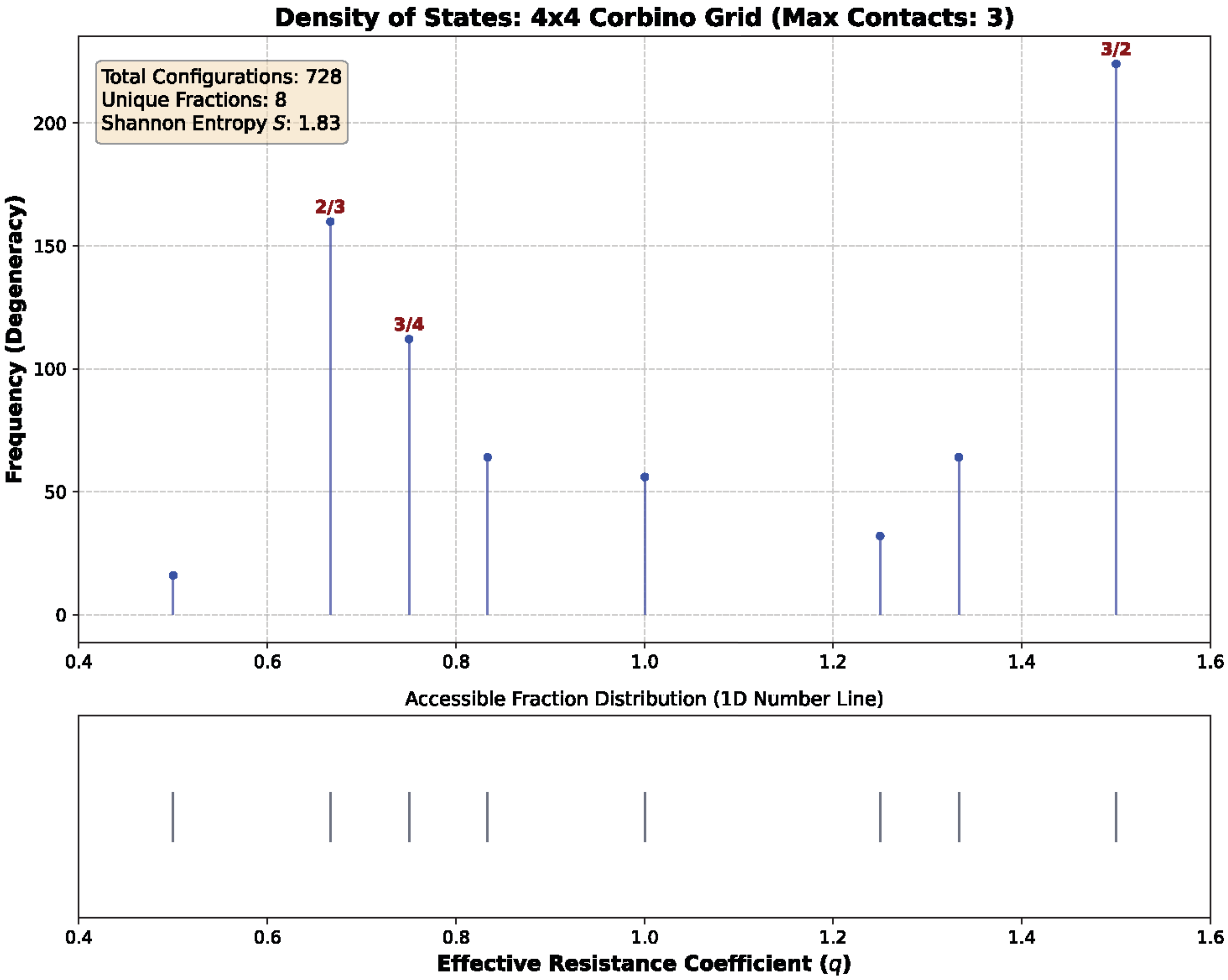
Density of States: 4x4 Corbino Grid (Max Contacts: 3)
Total Configurations: 728
Unique Fractions: 8
Shannon Entropy S: 1.83
2/3
3/4
3/2
Frequency (Degeneracy)
0
50
100
150
200
0.4
0.6
0.8
1.0
1.2
1.4
1.6
Accessible Fraction Distribution (1D Number Line)
Effective Resistance Coefficient (q)

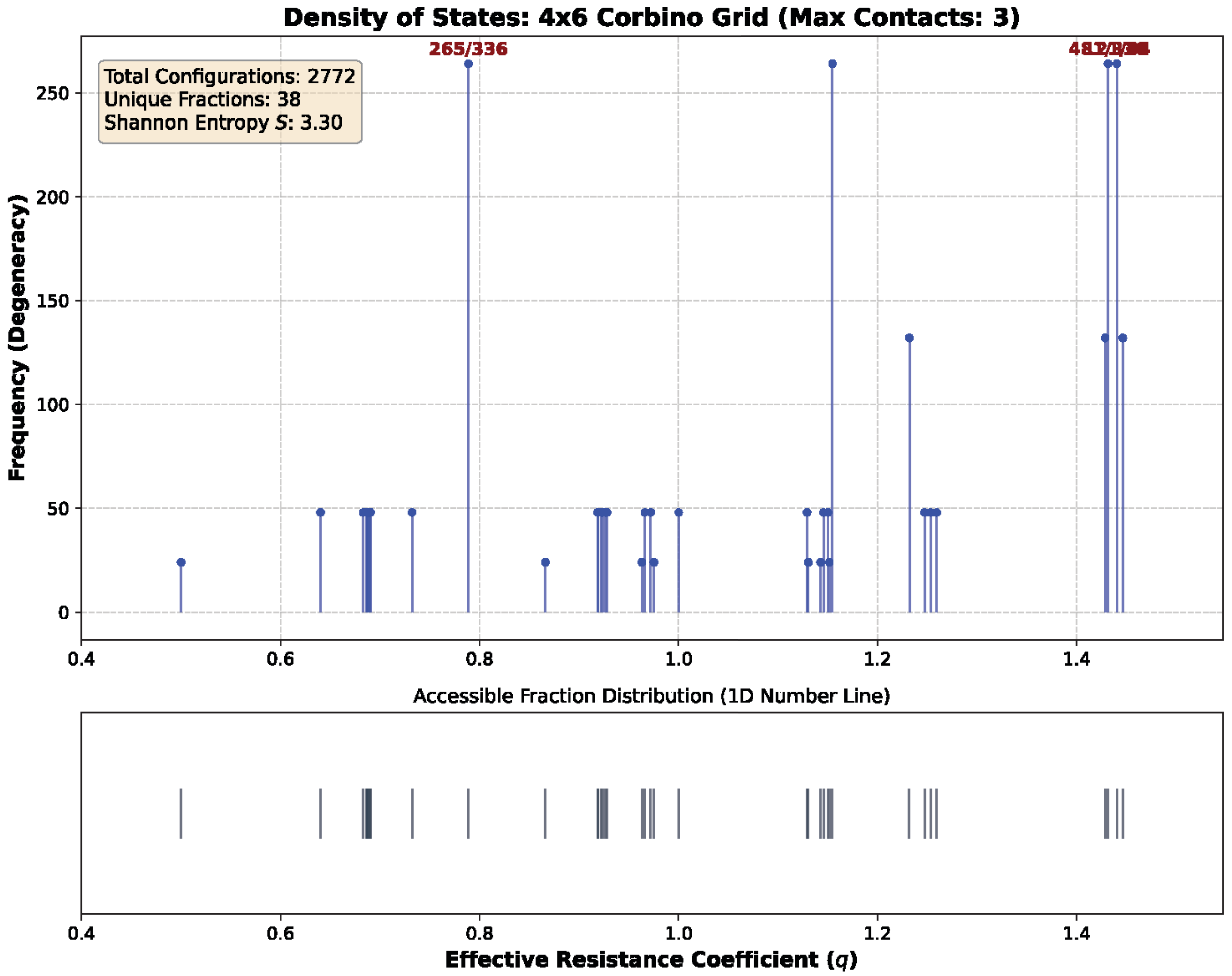
Density of States: 4x6 Corbino Grid (Max Contacts: 3)
265/336
Total Configurations: 2772
Unique Fractions: 38
Shannon Entropy $S$: 3.30
Frequency (Degeneracy)
0
50
100
150
200
250
0.4
0.6
0.8
1.0
1.2
1.4
Accessible Fraction Distribution (1D Number Line)
0.4
0.6
0.8
1.0
1.2
1.4
Effective Resistance Coefficient ($q$)

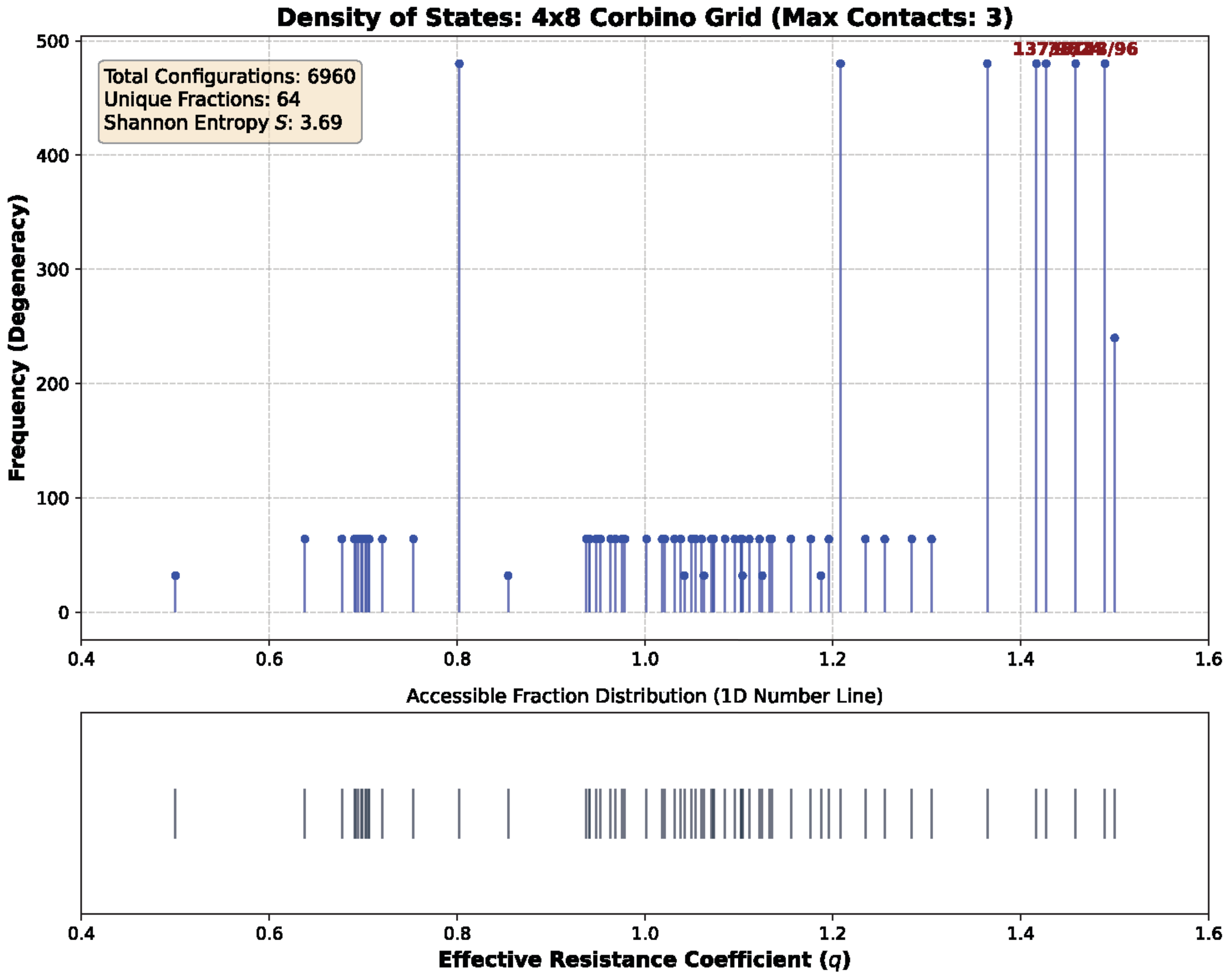
Density of States: 4x8 Corbino Grid (Max Contacts: 3)
Total Configurations: 6960
Unique Fractions: 64
Shannon Entropy S: 3.69
Frequency (Degeneracy)
0
100
200
300
400
500
0.4
0.6
0.8
1.0
1.2
1.4
1.6
Accessible Fraction Distribution (1D Number Line)
Effective Resistance Coefficient (q)

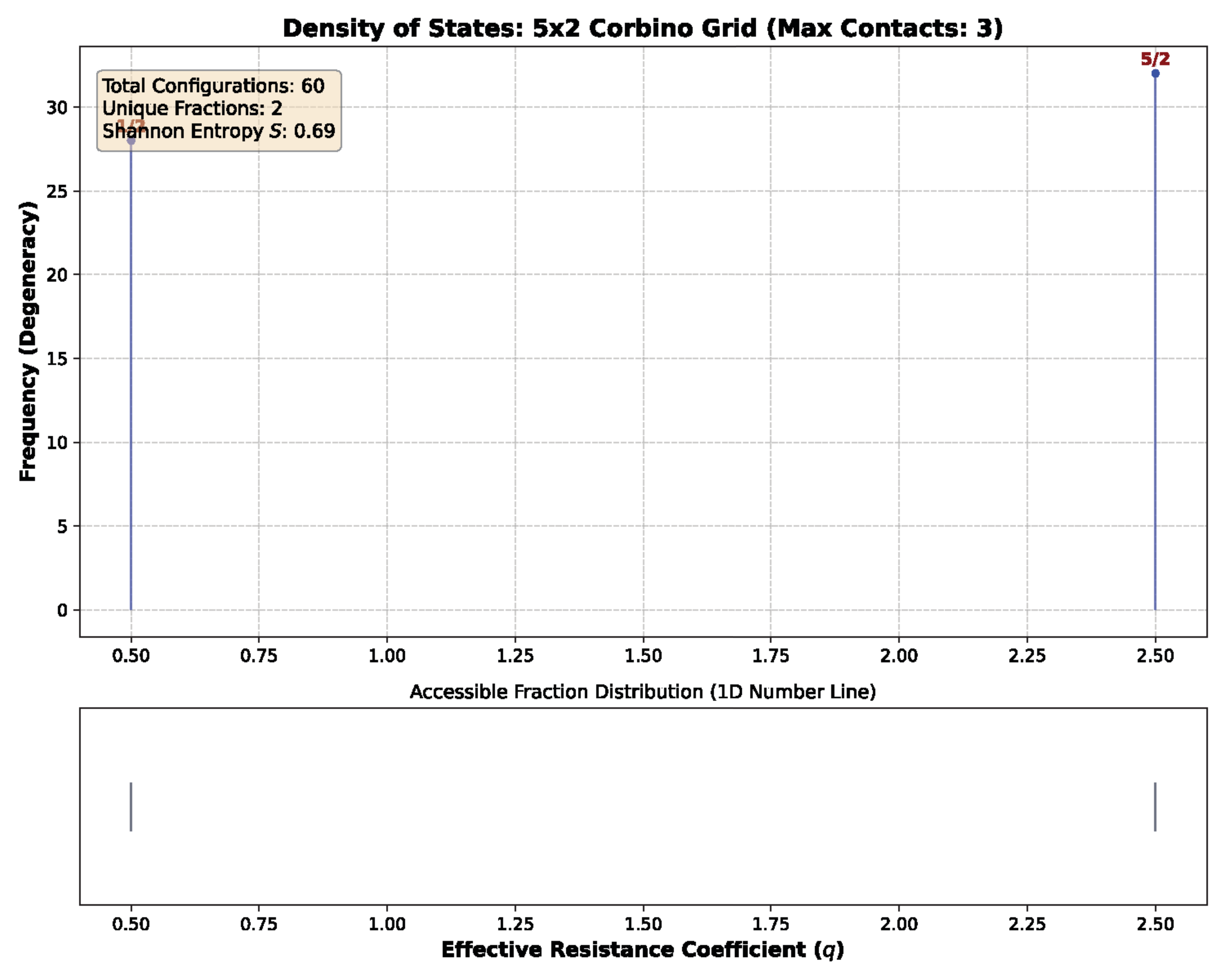

Density of States: 5x2 Corbino Grid (Max Contacts: 3)
Total Configurations: 60
Unique Fractions: 2
Shannon Entropy S: 0.69
1/2
5/2
Frequency (Degeneracy)
0
5
10
15
20
25
30
0.50
0.75
1.00
1.25
1.50
1.75
2.00
2.25
2.50
Accessible Fraction Distribution (1D Number Line)
Effective Resistance Coefficient (q)

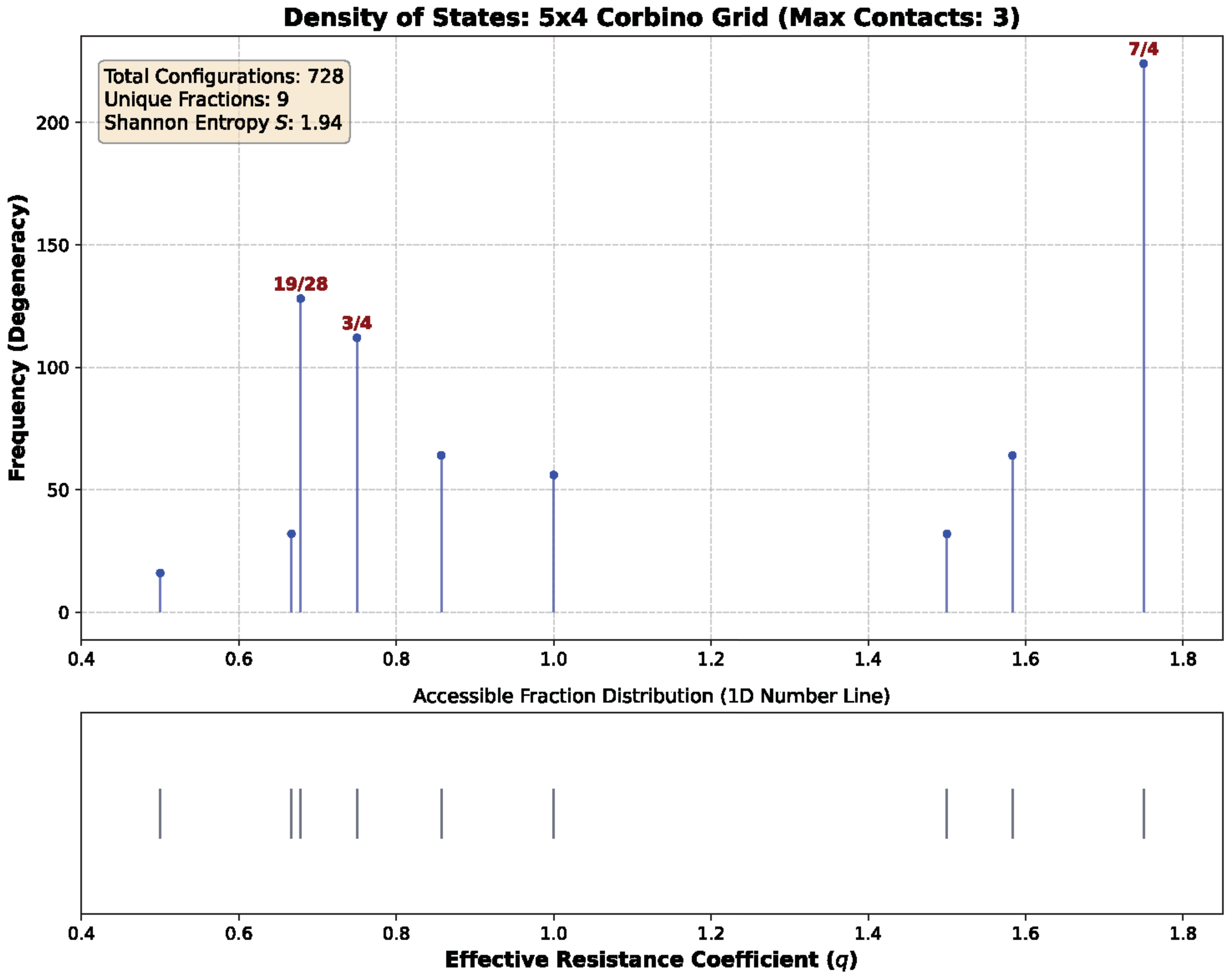

Density of States: 5x4 Corbino Grid (Max Contacts: 3)
Total Configurations: 728
Unique Fractions: 9
Shannon Entropy S: 1.94
Frequency (Degeneracy)
7/4
19/28
3/4
0
50
100
150
200
0.4
0.6
0.8
1.0
1.2
1.4
1.6
1.8
Accessible Fraction Distribution (1D Number Line)
Effective Resistance Coefficient (q)

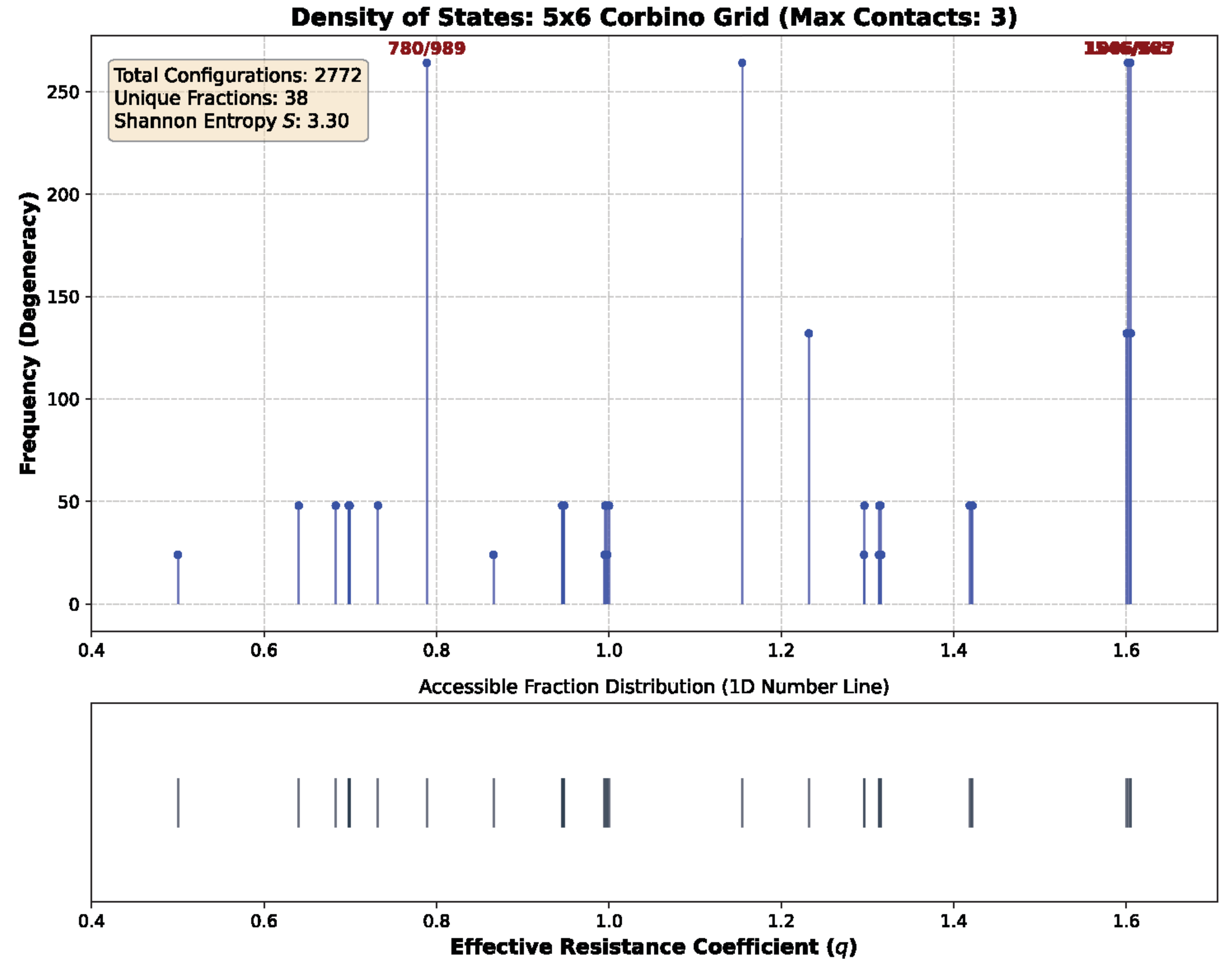
Density of States: 5x6 Corbino Grid (Max Contacts: 3)
780/989
Total Configurations: 2772
Unique Fractions: 38
Shannon Entropy S: 3.30
Frequency (Degeneracy)
0
50
100
150
200
250
0.4
0.6
0.8
1.0
1.2
1.4
1.6
Accessible Fraction Distribution (1D Number Line)
Effective Resistance Coefficient (q)

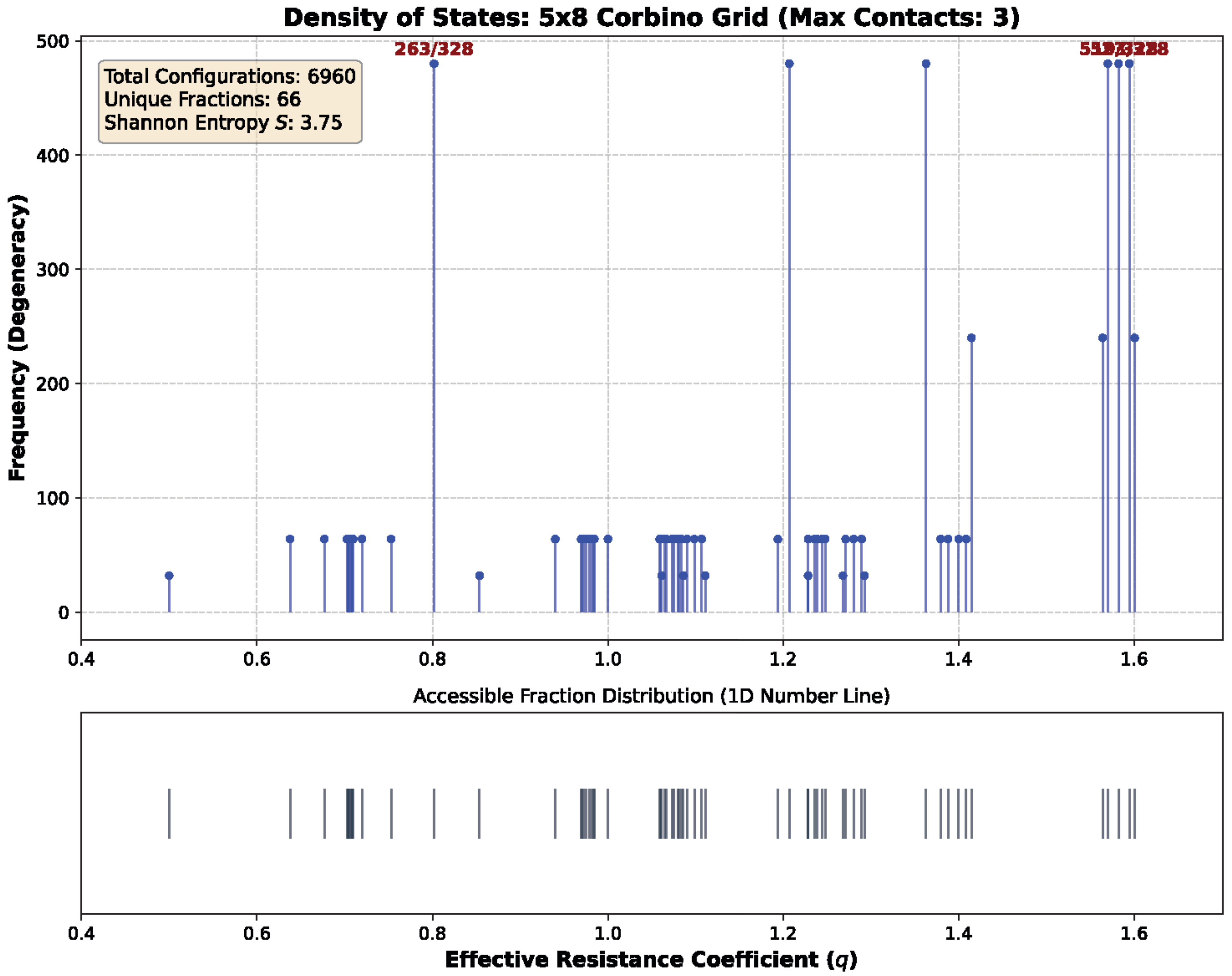

Density of States: 5x8 Corbino Grid (Max Contacts: 3)
Total Configurations: 6960
Unique Fractions: 66
Shannon Entropy S: 3.75
263/328
Frequency (Degeneracy)
0
100
200
300
400
500
0.4
0.6
0.8
1.0
1.2
1.4
1.6
Accessible Fraction Distribution (1D Number Line)
Effective Resistance Coefficient (q)

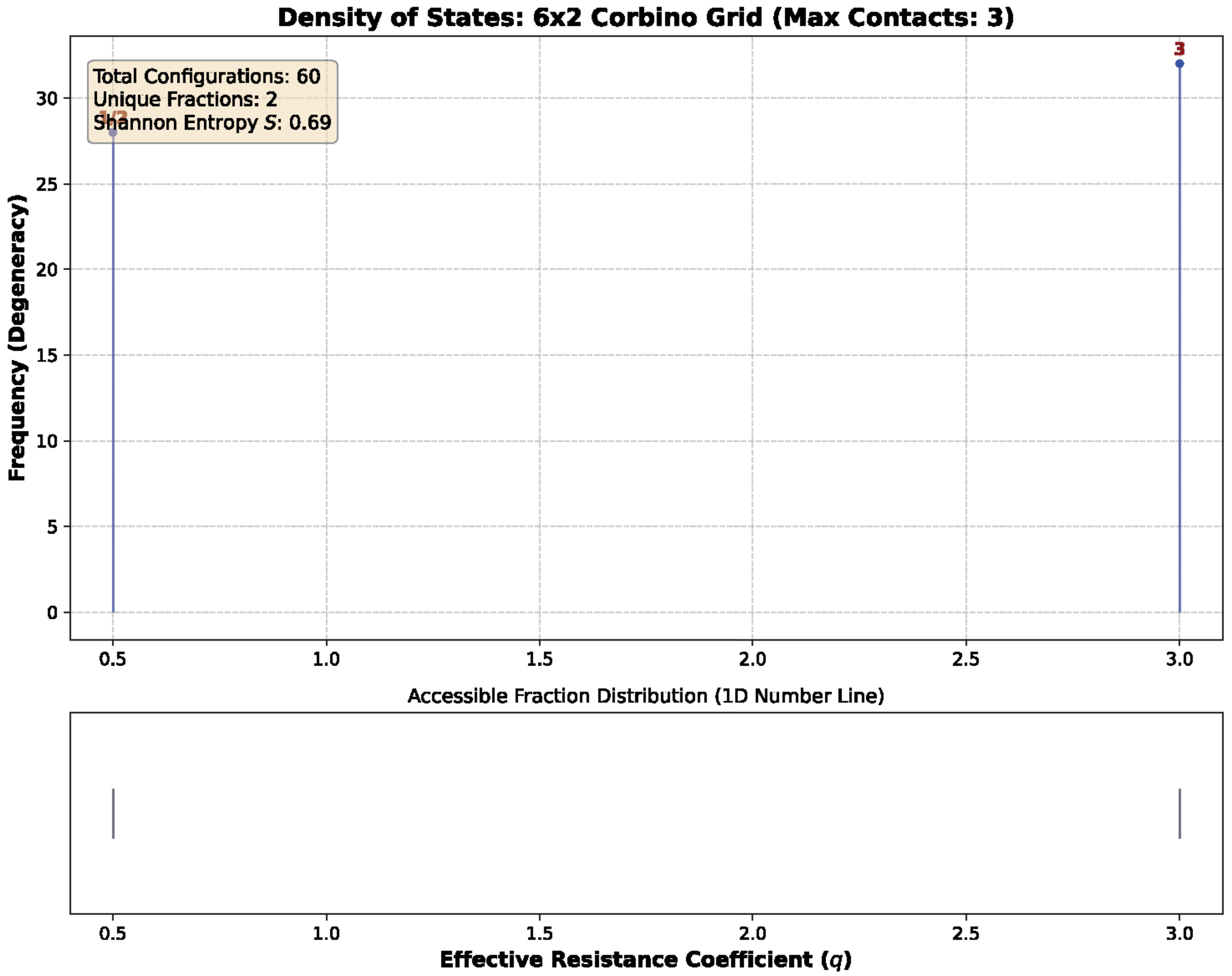

Density of States: 6x2 Corbino Grid (Max Contacts: 3)
Total Configurations: 60
Unique Fractions: 2
Shannon Entropy S: 0.69
1/2
3
Frequency (Degeneracy)
0
5
10
15
20
25
30
0.5
1.0
1.5
2.0
2.5
3.0
Accessible Fraction Distribution (1D Number Line)
Effective Resistance Coefficient (q)

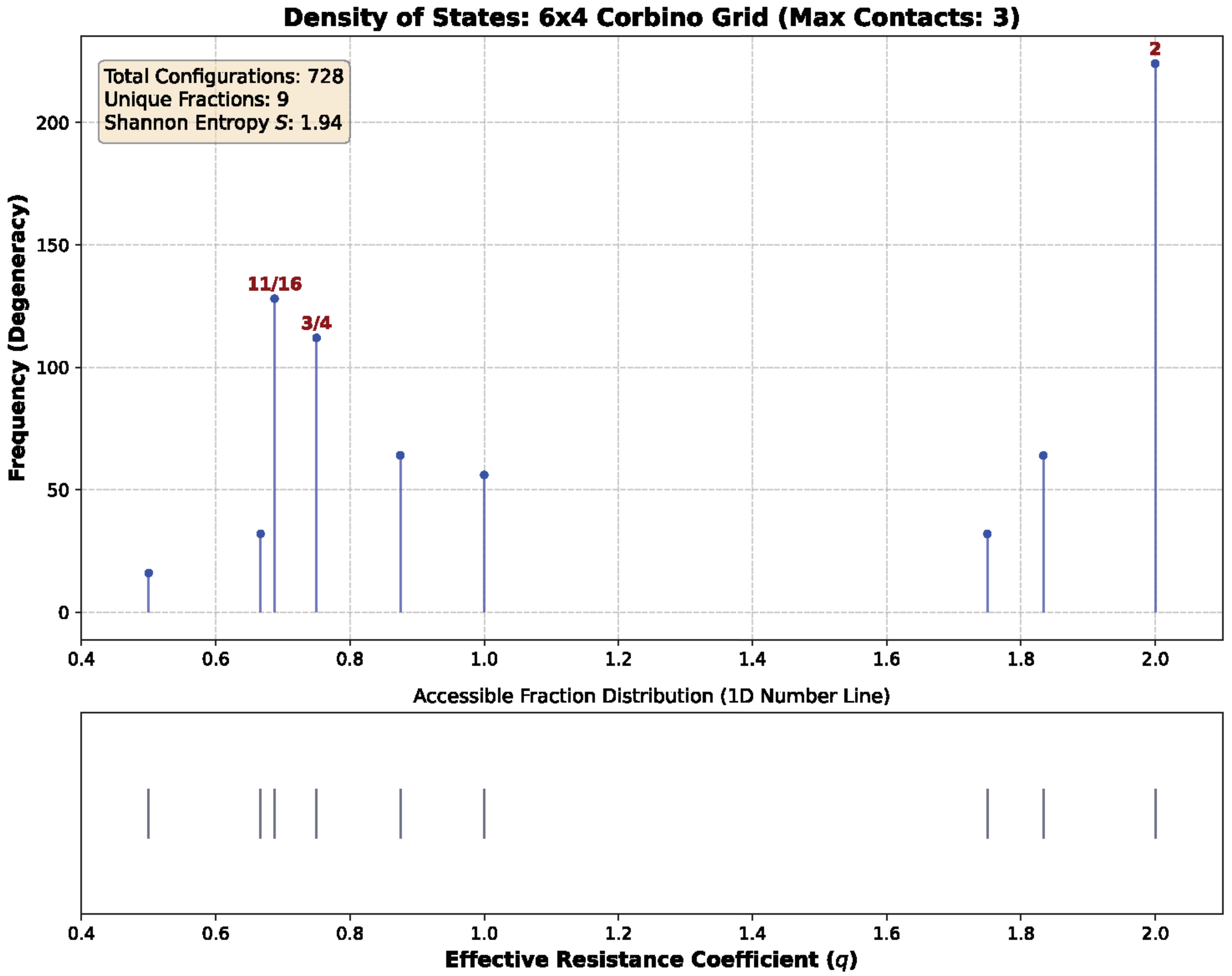

Density of States: 6x4 Corbino Grid (Max Contacts: 3)
Total Configurations: 728
Unique Fractions: 9
Shannon Entropy S: 1.94
Frequency (Degeneracy)
11/16
3/4
2
0
50
100
150
200
0.4
0.6
0.8
1.0
1.2
1.4
1.6
1.8
2.0
Accessible Fraction Distribution (1D Number Line)
Effective Resistance Coefficient (q)

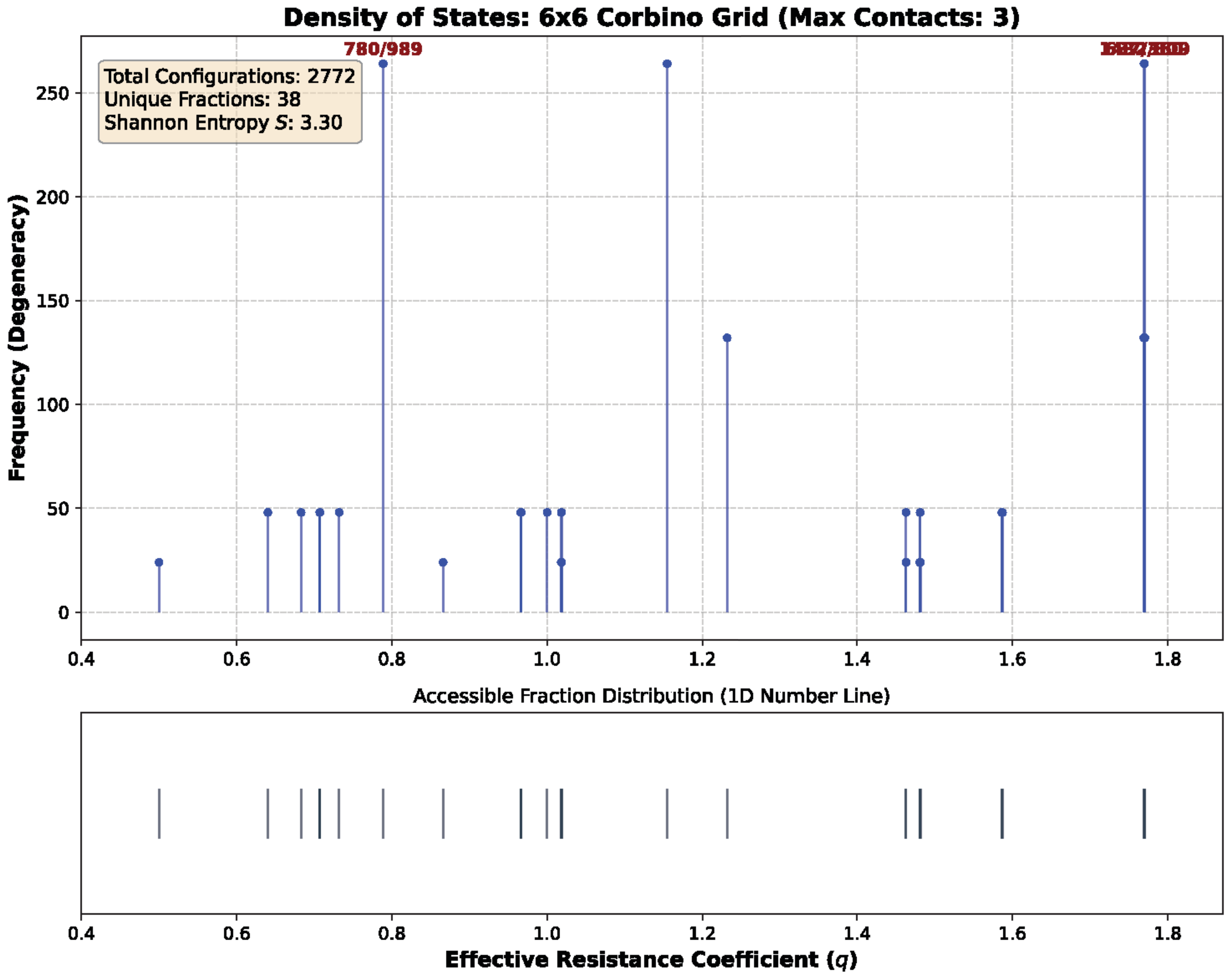
Density of States: 6x6 Corbino Grid (Max Contacts: 3)
Total Configurations: 2772
Unique Fractions: 38
Shannon Entropy S: 3.30
780/989
Frequency (Degeneracy)
Accessible Fraction Distribution (1D Number Line)
Effective Resistance Coefficient (q)

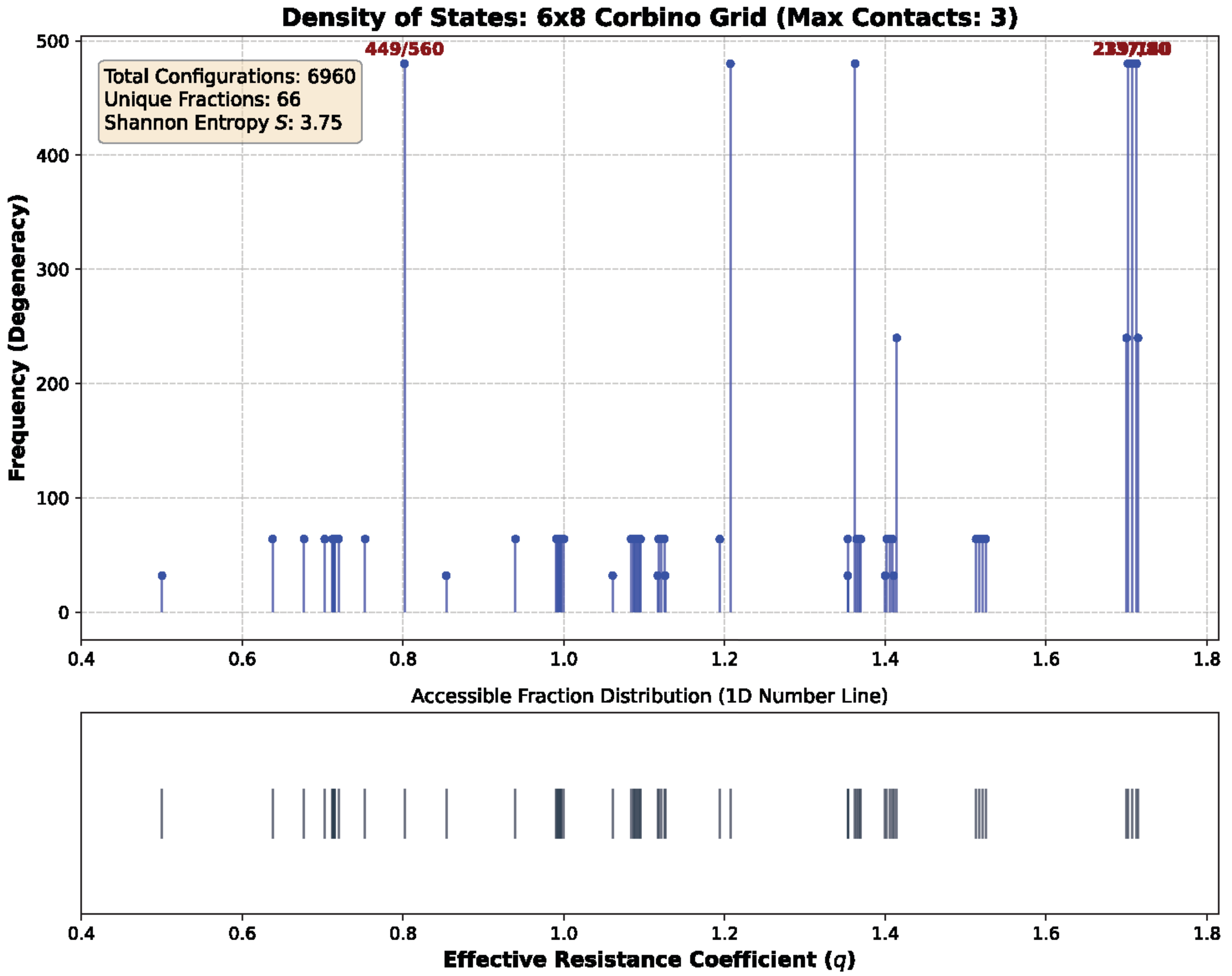

Density of States: 6x8 Corbino Grid (Max Contacts: 3)
Total Configurations: 6960
Unique Fractions: 66
Shannon Entropy S: 3.75
449/560
Frequency (Degeneracy)
0
100
200
300
400
500
0.4
0.6
0.8
1.0
1.2
1.4
1.6
1.8
Accessible Fraction Distribution (1D Number Line)
Effective Resistance Coefficient (q)

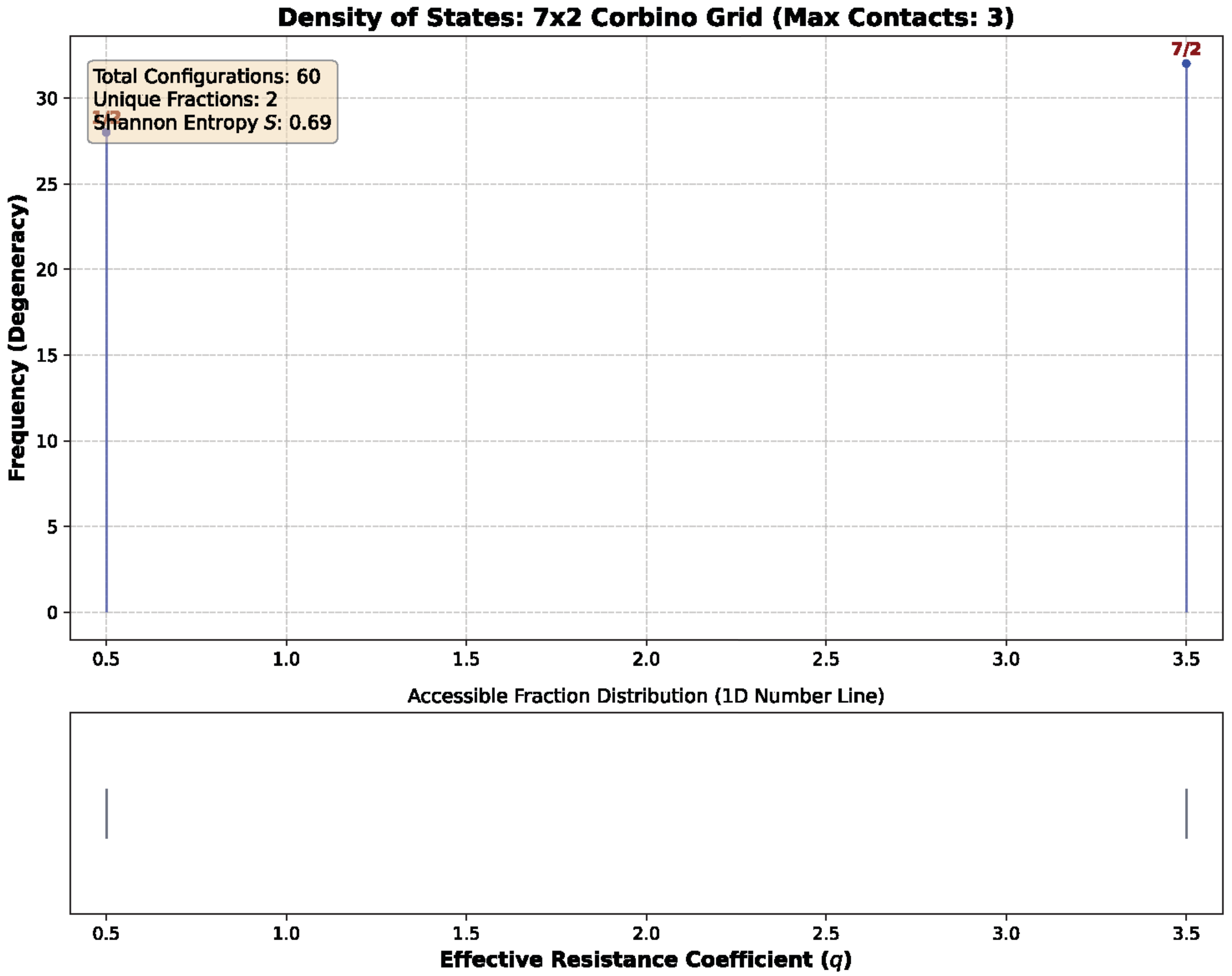
Density of States: 7x2 Corbino Grid (Max Contacts: 3)
Total Configurations: 60
Unique Fractions: 2
Shannon Entropy S: 0.69
1/2
7/2
Frequency (Degeneracy)
0
5
10
15
20
25
30
0.5
1.0
1.5
2.0
2.5
3.0
3.5
Accessible Fraction Distribution (1D Number Line)
Effective Resistance Coefficient (q)

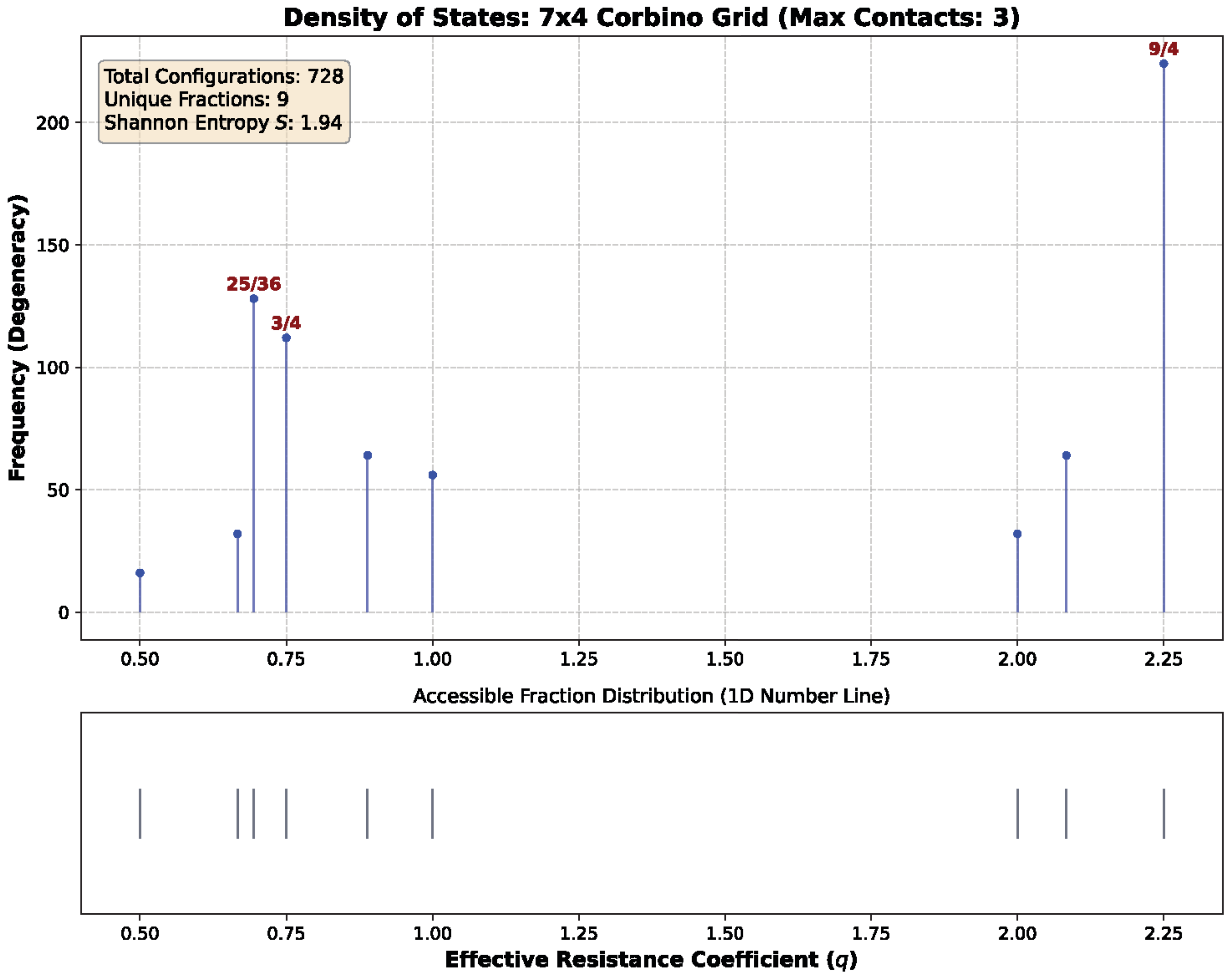

Density of States: 7x4 Corbino Grid (Max Contacts: 3)
Total Configurations: 728
Unique Fractions: 9
Shannon Entropy S: 1.94
Frequency (Degeneracy)
25/36
3/4
9/4
0
50
100
150
200
0.50
0.75
1.00
1.25
1.50
1.75
2.00
2.25
Accessible Fraction Distribution (1D Number Line)
Effective Resistance Coefficient (q)

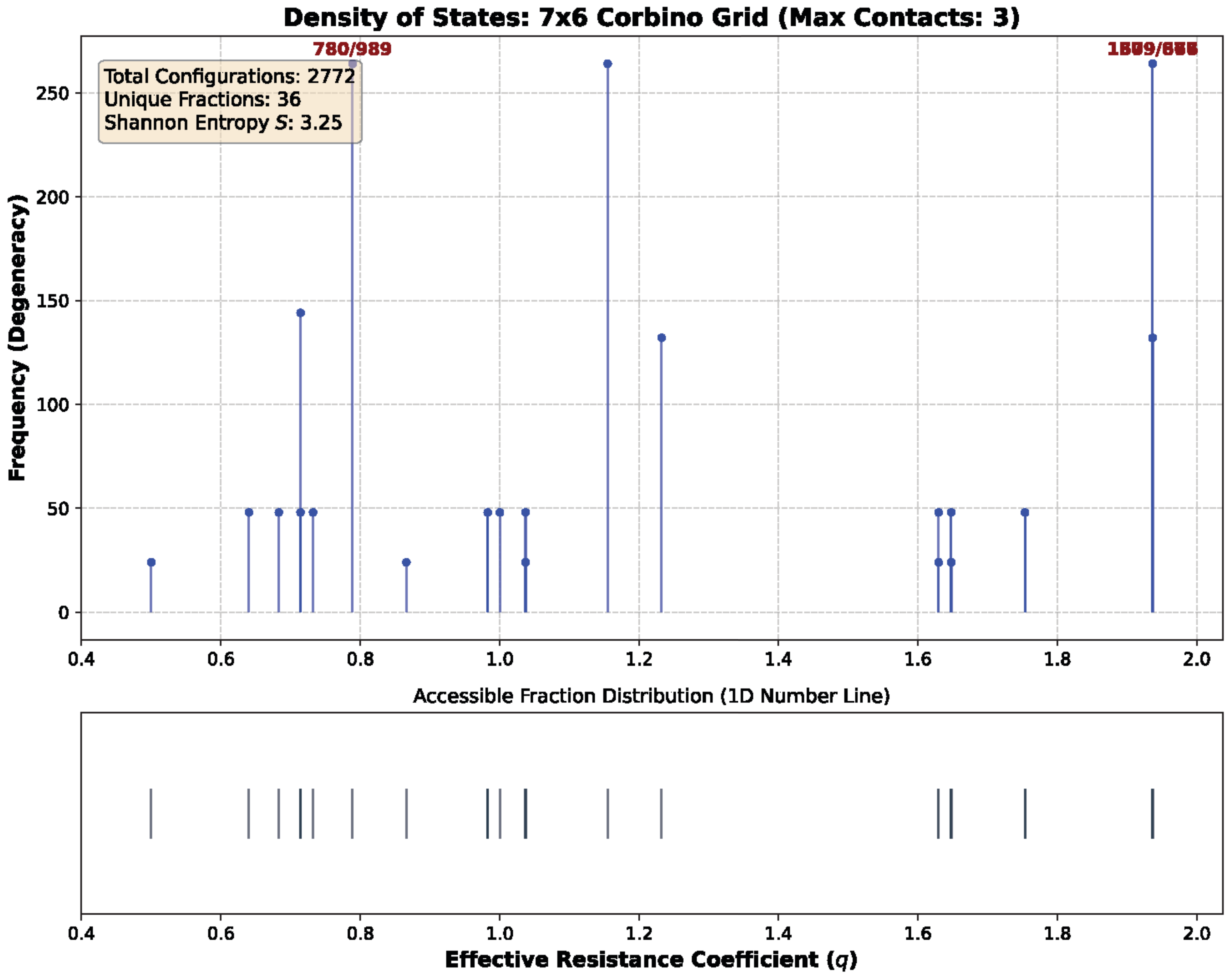

Density of States: 7x6 Corbino Grid (Max Contacts: 3)
780/989
Total Configurations: 2772
Unique Fractions: 36
Shannon Entropy S: 3.25
Frequency (Degeneracy)
0
50
100
150
200
250
0.4
0.6
0.8
1.0
1.2
1.4
1.6
1.8
2.0
Accessible Fraction Distribution (1D Number Line)
0.4
0.6
0.8
1.0
1.2
1.4
1.6
1.8
2.0
Effective Resistance Coefficient (q)

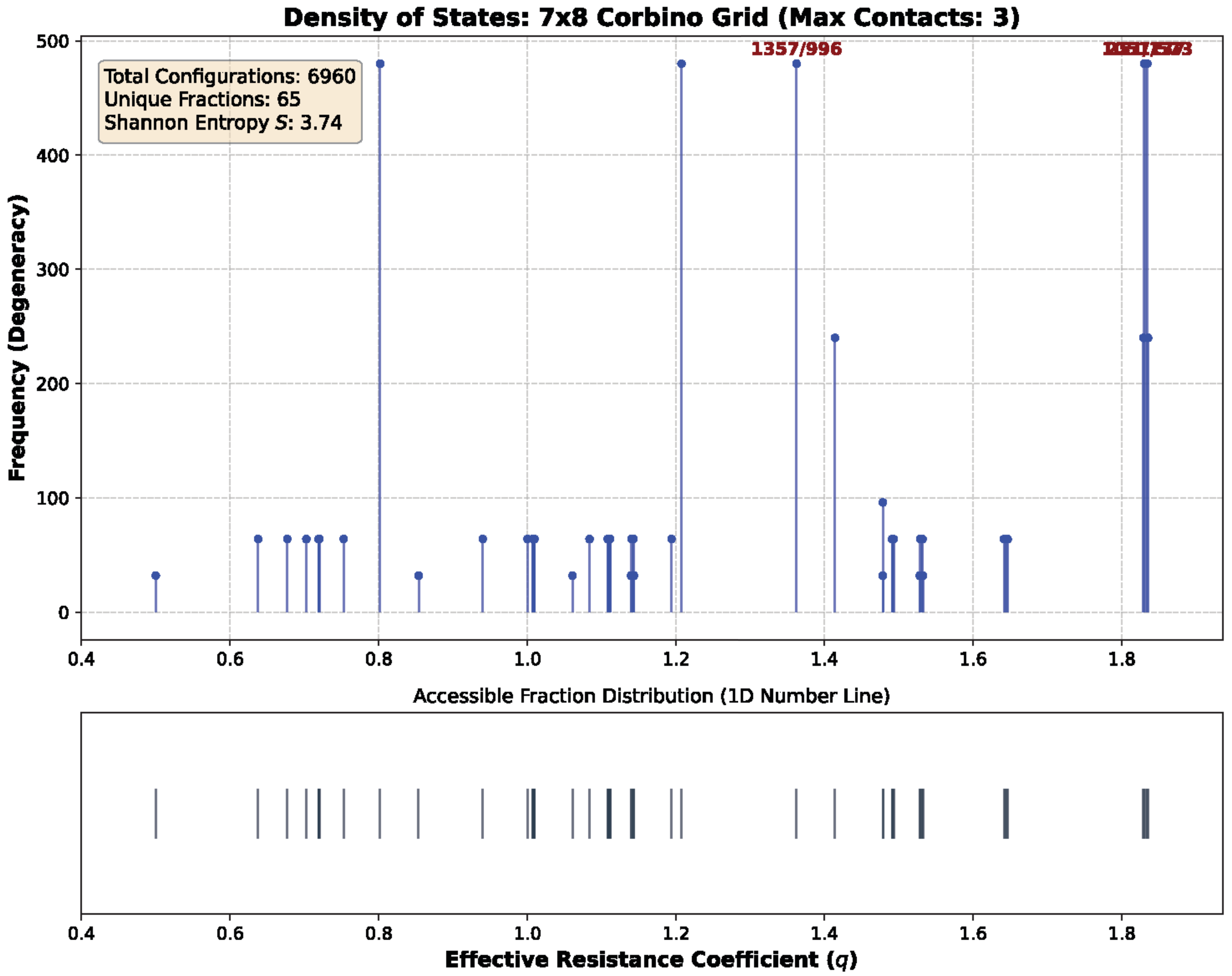
Density of States: 7x8 Corbino Grid (Max Contacts: 3)
Total Configurations: 6960
Unique Fractions: 65
Shannon Entropy S: 3.74
1357/996
Frequency (Degeneracy)
0
100
200
300
400
500
0.4
0.6
0.8
1.0
1.2
1.4
1.6
1.8
Accessible Fraction Distribution (1D Number Line)
Effective Resistance Coefficient (q)

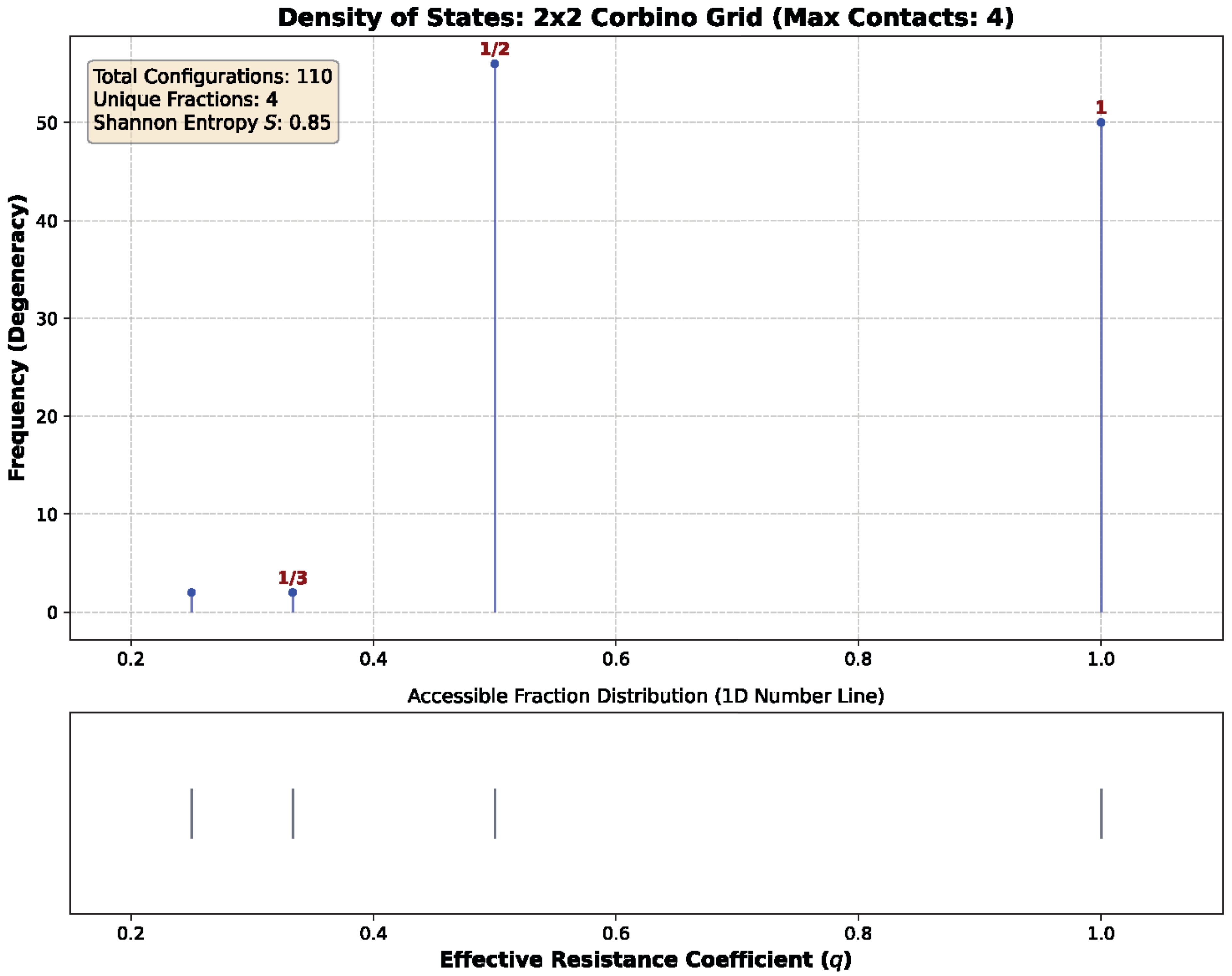
Density of States: 2x2 Corbino Grid (Max Contacts: 4)
Total Configurations: 110
Unique Fractions: 4
Shannon Entropy S: 0.85
1/2
1
1/3
Frequency (Degeneracy)
0
10
20
30
40
50
0.2
0.4
0.6
0.8
1.0
Accessible Fraction Distribution (1D Number Line)
Effective Resistance Coefficient (q)

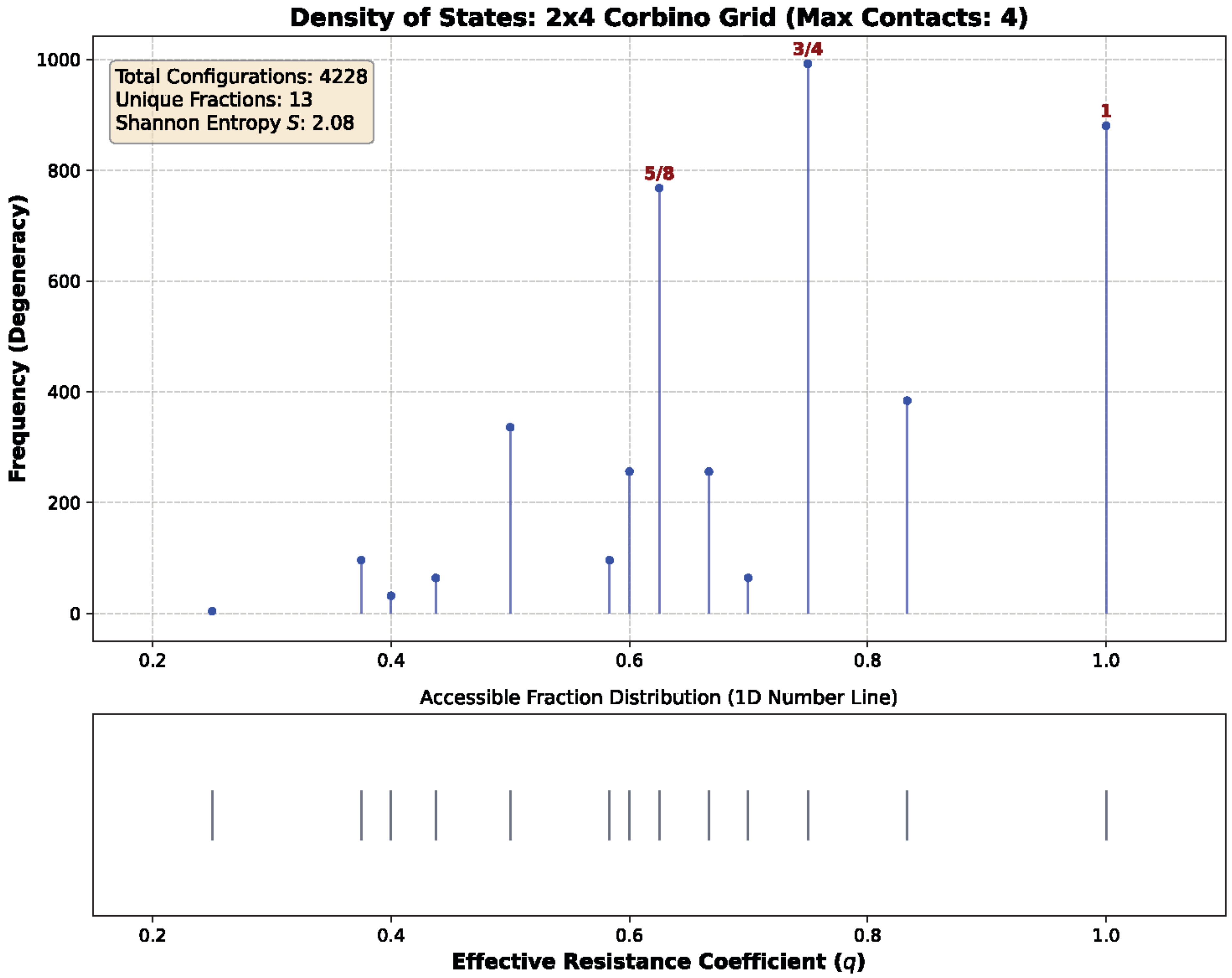
Density of States: 2x4 Corbino Grid (Max Contacts: 4)
Total Configurations: 4228
Unique Fractions: 13
Shannon Entropy S: 2.08
3/4
1
5/8
Frequency (Degeneracy)
1000
800
600
400
200
0
0.2
0.4
0.6
0.8
1.0
Accessible Fraction Distribution (1D Number Line)
0.2
0.4
0.6
0.8
1.0
Effective Resistance Coefficient (q)

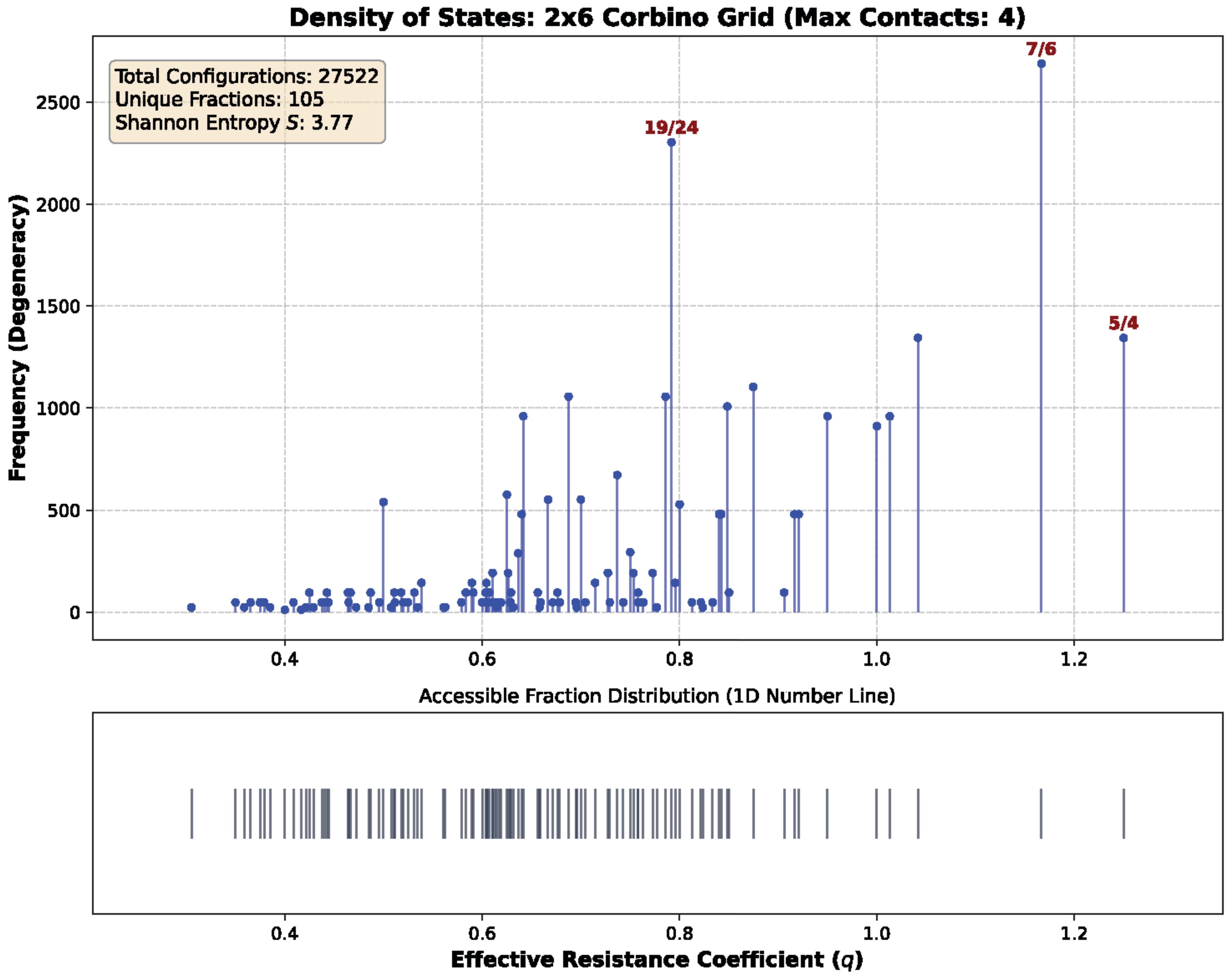

Density of States: 2x6 Corbino Grid (Max Contacts: 4)
Total Configurations: 27522
Unique Fractions: 105
Shannon Entropy S: 3.77
Frequency (Degeneracy)
19/24
7/6
5/4
0
500
1000
1500
2000
2500
0.4
0.6
0.8
1.0
1.2
Accessible Fraction Distribution (1D Number Line)
Effective Resistance Coefficient (q)

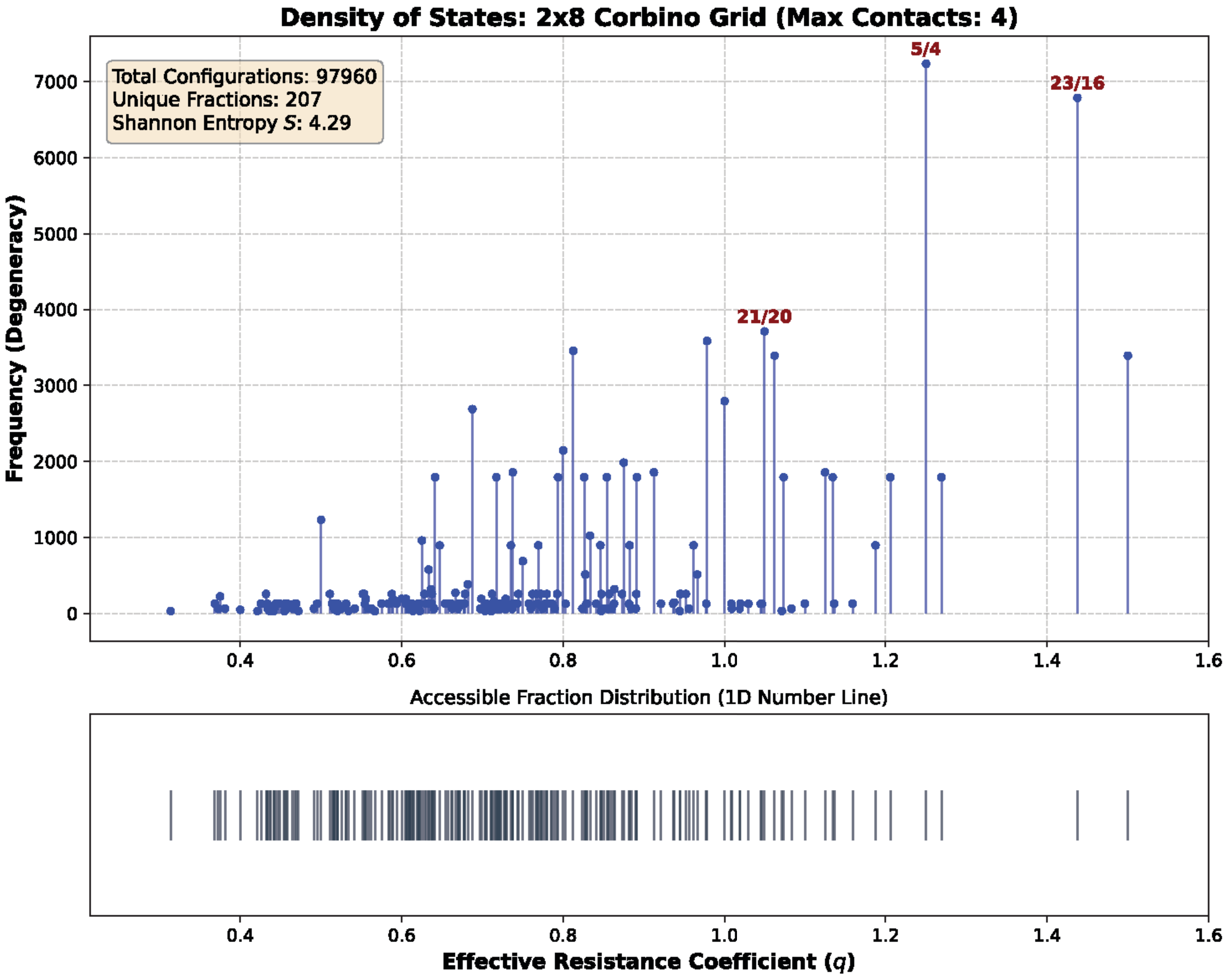

Density of States: 2x8 Corbino Grid (Max Contacts: 4)
Total Configurations: 97960
Unique Fractions: 207
Shannon Entropy S: 4.29
5/4
23/16
21/20
Frequency (Degeneracy)
0
1000
2000
3000
4000
5000
6000
7000
0.4
0.6
0.8
1.0
1.2
1.4
1.6
Accessible Fraction Distribution (1D Number Line)
Effective Resistance Coefficient (q)

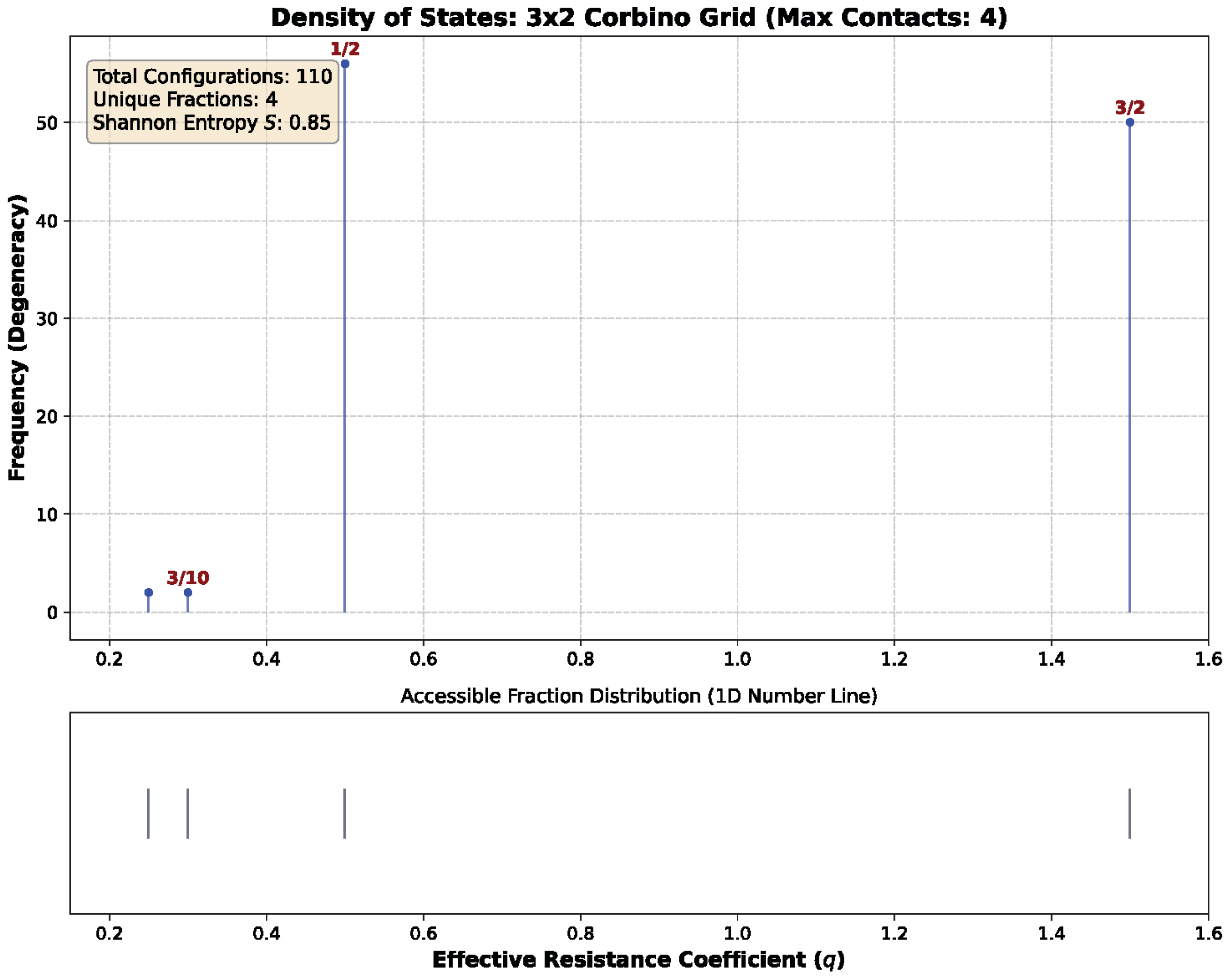

Density of States: 3x2 Corbino Grid (Max Contacts: 4)
Total Configurations: 110
Unique Fractions: 4
Shannon Entropy S: 0.85
1/2
3/2
3/10
Frequency (Degeneracy)
0
10
20
30
40
50
0.2
0.4
0.6
0.8
1.0
1.2
1.4
1.6
Accessible Fraction Distribution (1D Number Line)
Effective Resistance Coefficient (q)

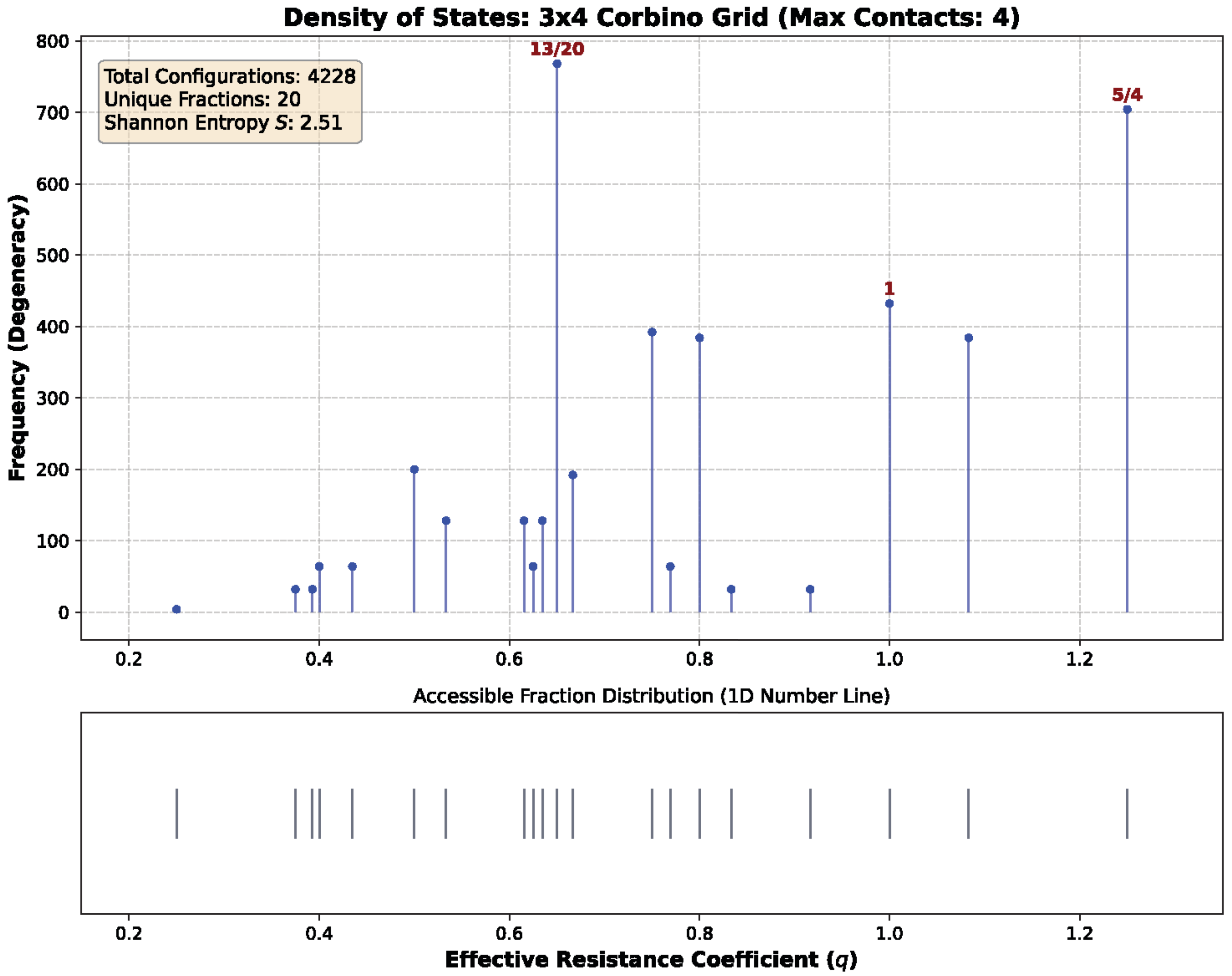
Density of States: 3x4 Corbino Grid (Max Contacts: 4)
Total Configurations: 4228
Unique Fractions: 20
Shannon Entropy S: 2.51
13/20
5/4
1
Frequency (Degeneracy)
0
100
200
300
400
500
600
700
800
0.2
0.4
0.6
0.8
1.0
1.2
Accessible Fraction Distribution (1D Number Line)
Effective Resistance Coefficient (q)

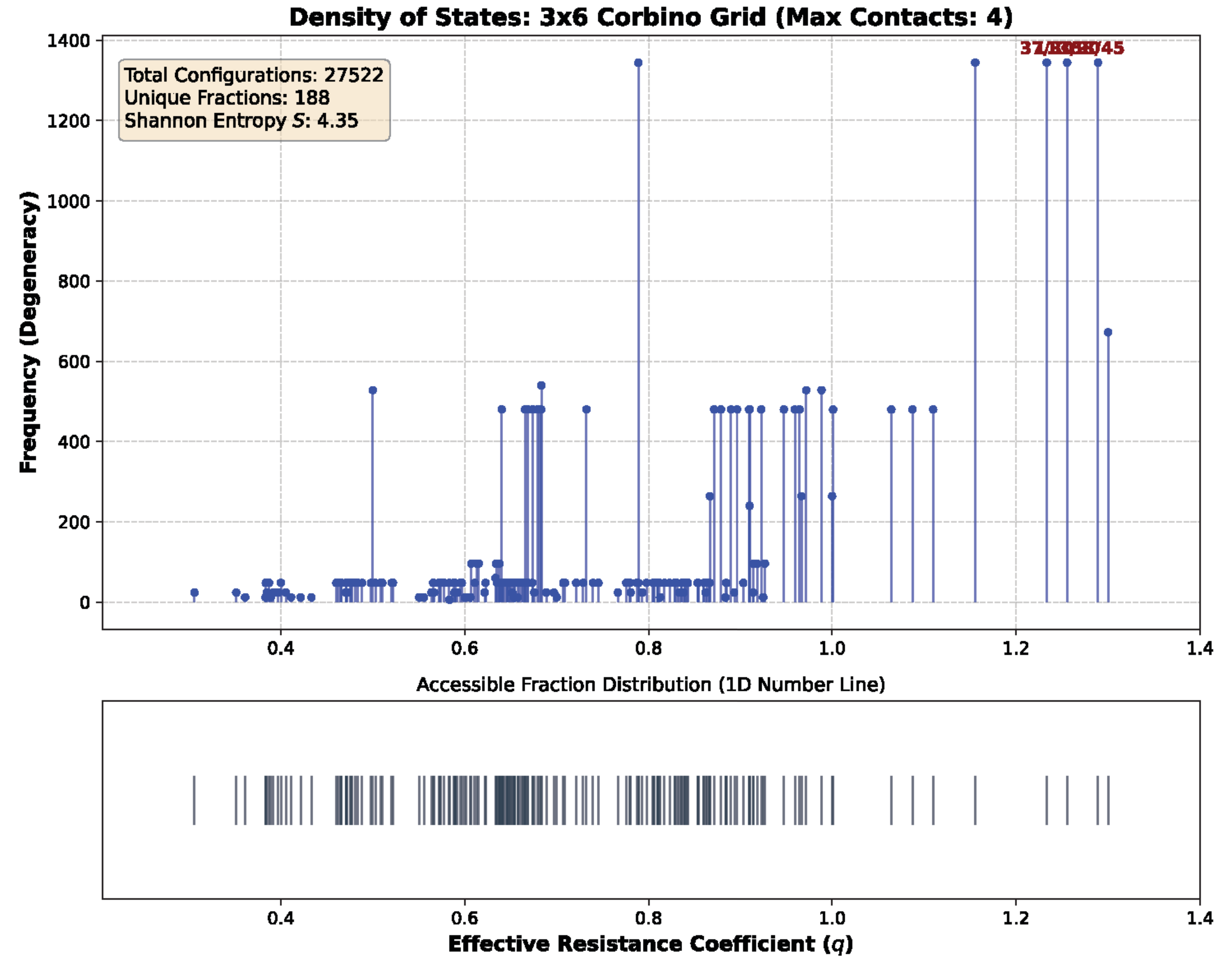
Density of States: 3x6 Corbino Grid (Max Contacts: 4)
Total Configurations: 27522
Unique Fractions: 188
Shannon Entropy S: 4.35
Frequency (Degeneracy)
0
200
400
600
800
1000
1200
1400
0.4
0.6
0.8
1.0
1.2
1.4
Accessible Fraction Distribution (1D Number Line)
Effective Resistance Coefficient (q)

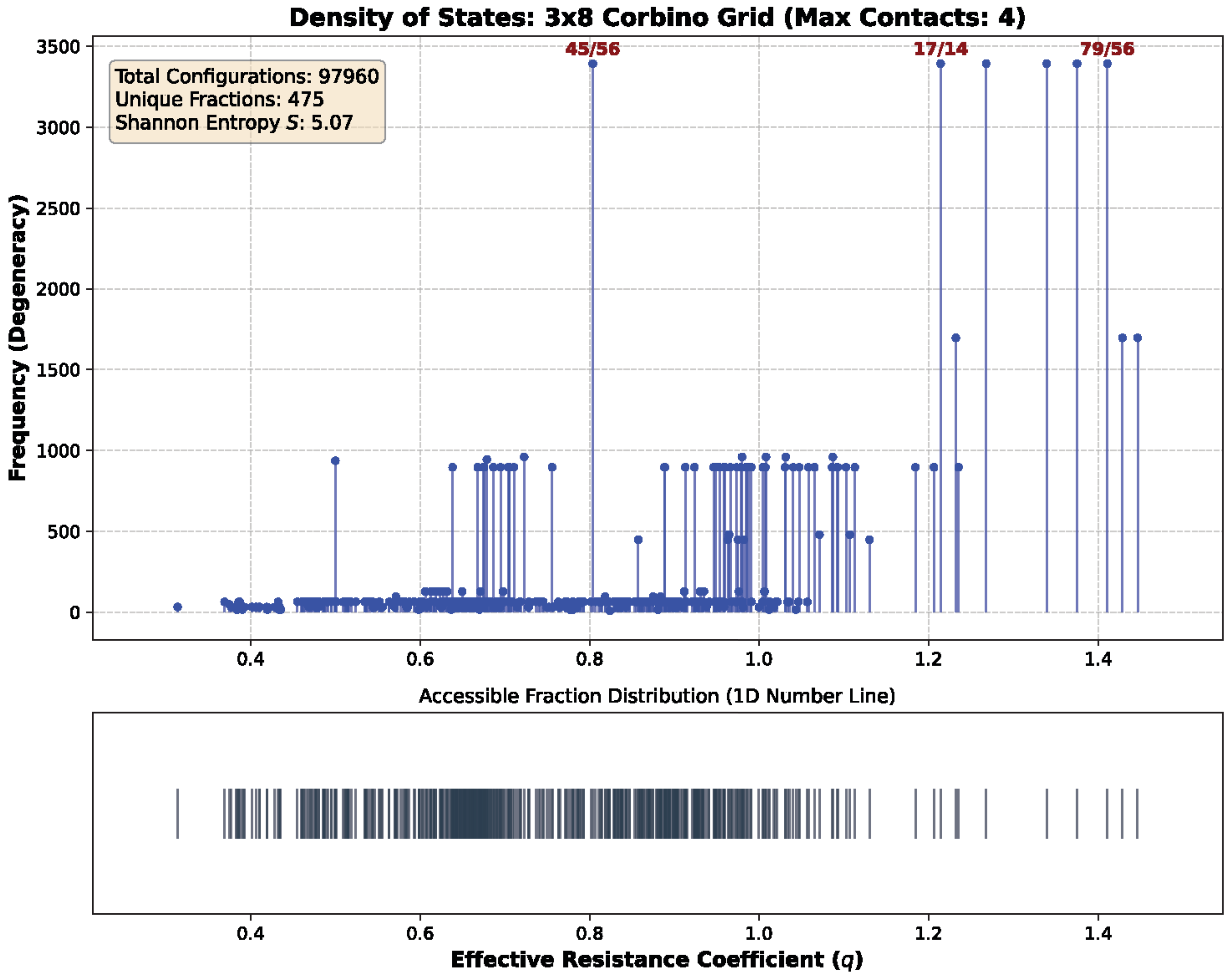
Density of States: 3x8 Corbino Grid (Max Contacts: 4)
Total Configurations: 97960
Unique Fractions: 475
Shannon Entropy S: 5.07
45/56
17/14
79/56
Frequency (Degeneracy)
3500
3000
2500
2000
1500
1000
500
0
0.4
0.6
0.8
1.0
1.2
1.4
Accessible Fraction Distribution (1D Number Line)
Effective Resistance Coefficient (q)

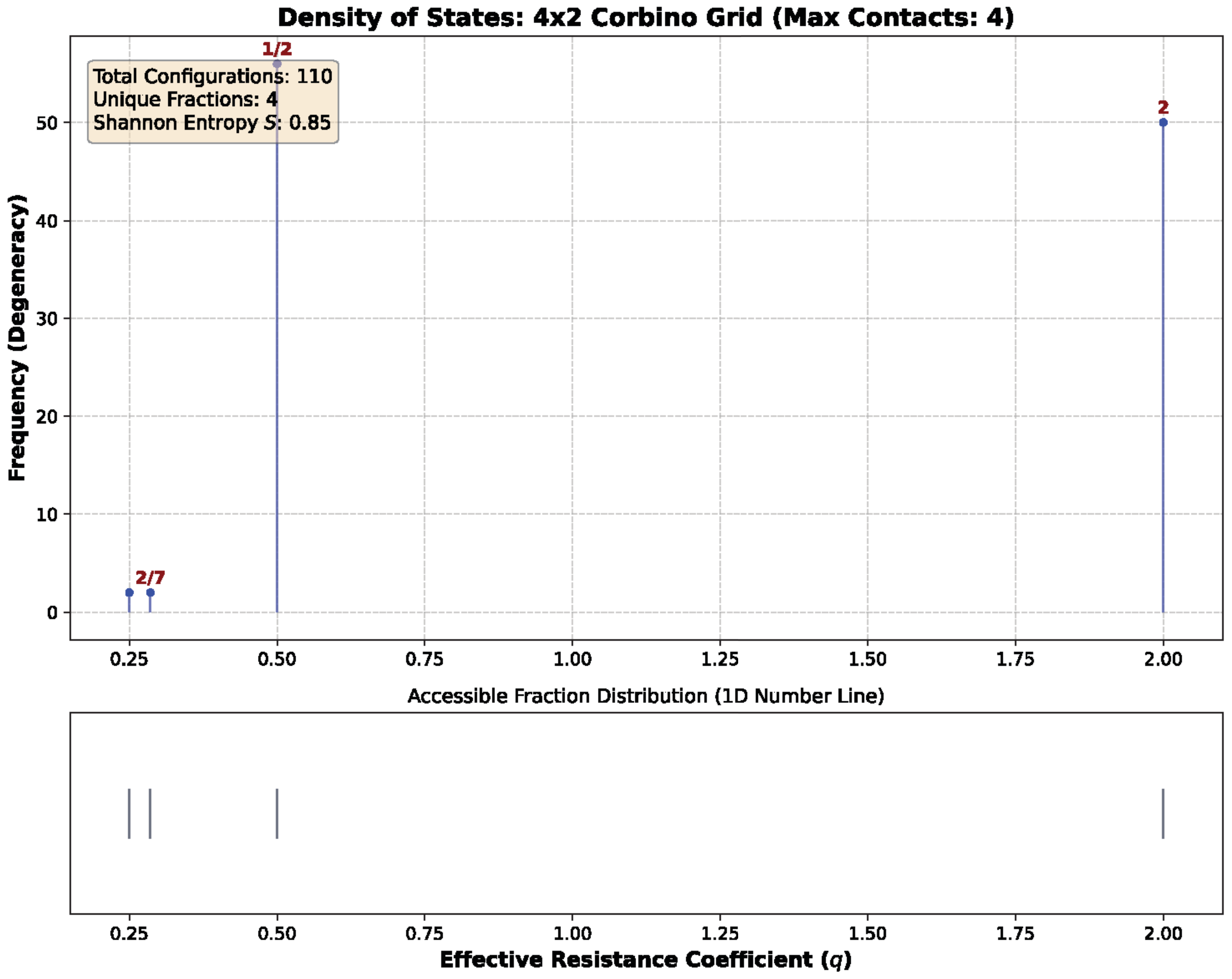
Density of States: 4x2 Corbino Grid (Max Contacts: 4)
Total Configurations: 110
Unique Fractions: 4
Shannon Entropy S: 0.85
1/2
2
2/7
Frequency (Degeneracy)
0
10
20
30
40
50
0.25
0.50
0.75
1.00
1.25
1.50
1.75
2.00
Accessible Fraction Distribution (1D Number Line)
Effective Resistance Coefficient (q)

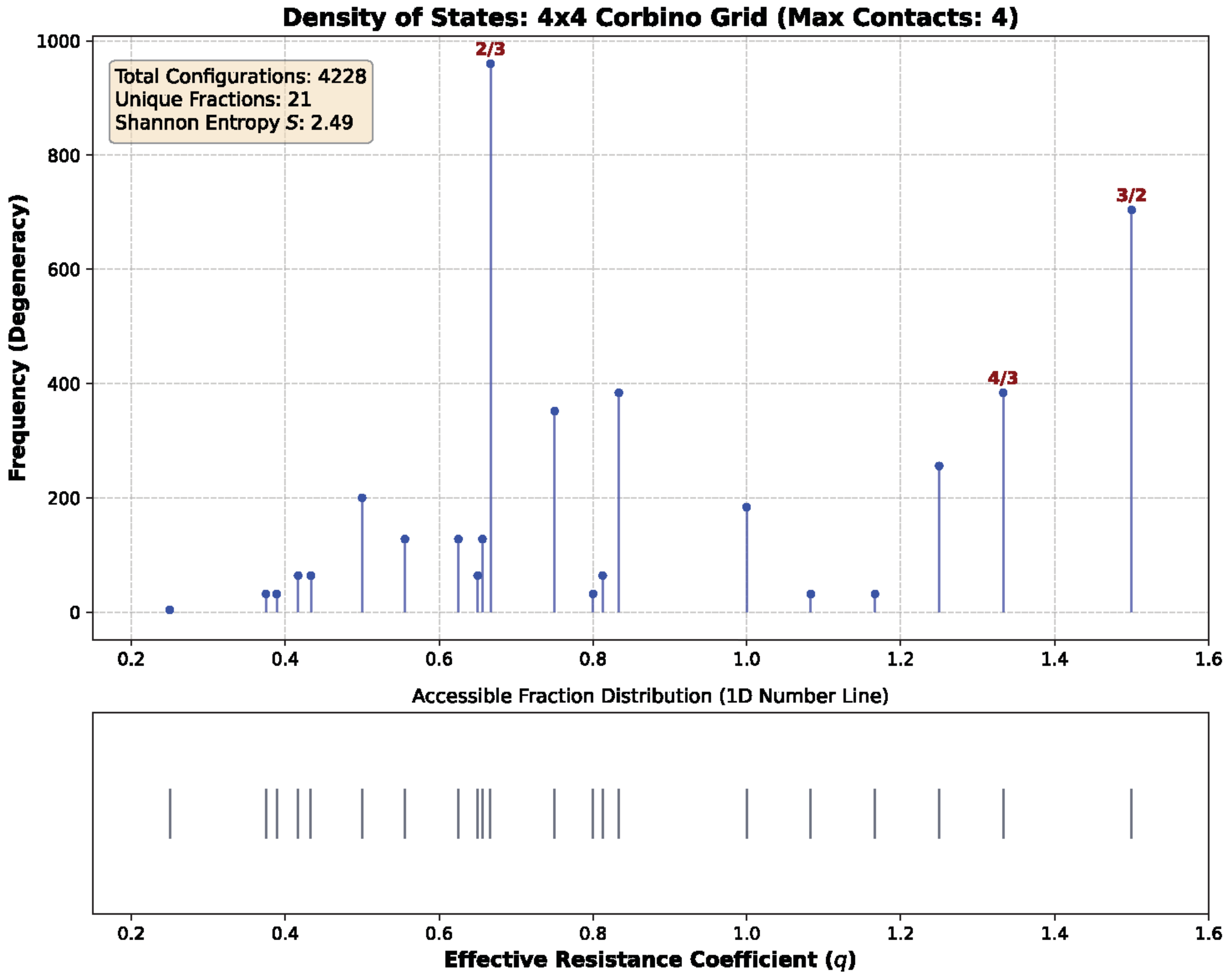

Density of States: 4x4 Corbino Grid (Max Contacts: 4)
Total Configurations: 4228
Unique Fractions: 21
Shannon Entropy S: 2.49
2/3
3/2
4/3
Frequency (Degeneracy)
0
200
400
600
800
1000
0.2
0.4
0.6
0.8
1.0
1.2
1.4
1.6
Accessible Fraction Distribution (1D Number Line)
Effective Resistance Coefficient (q)

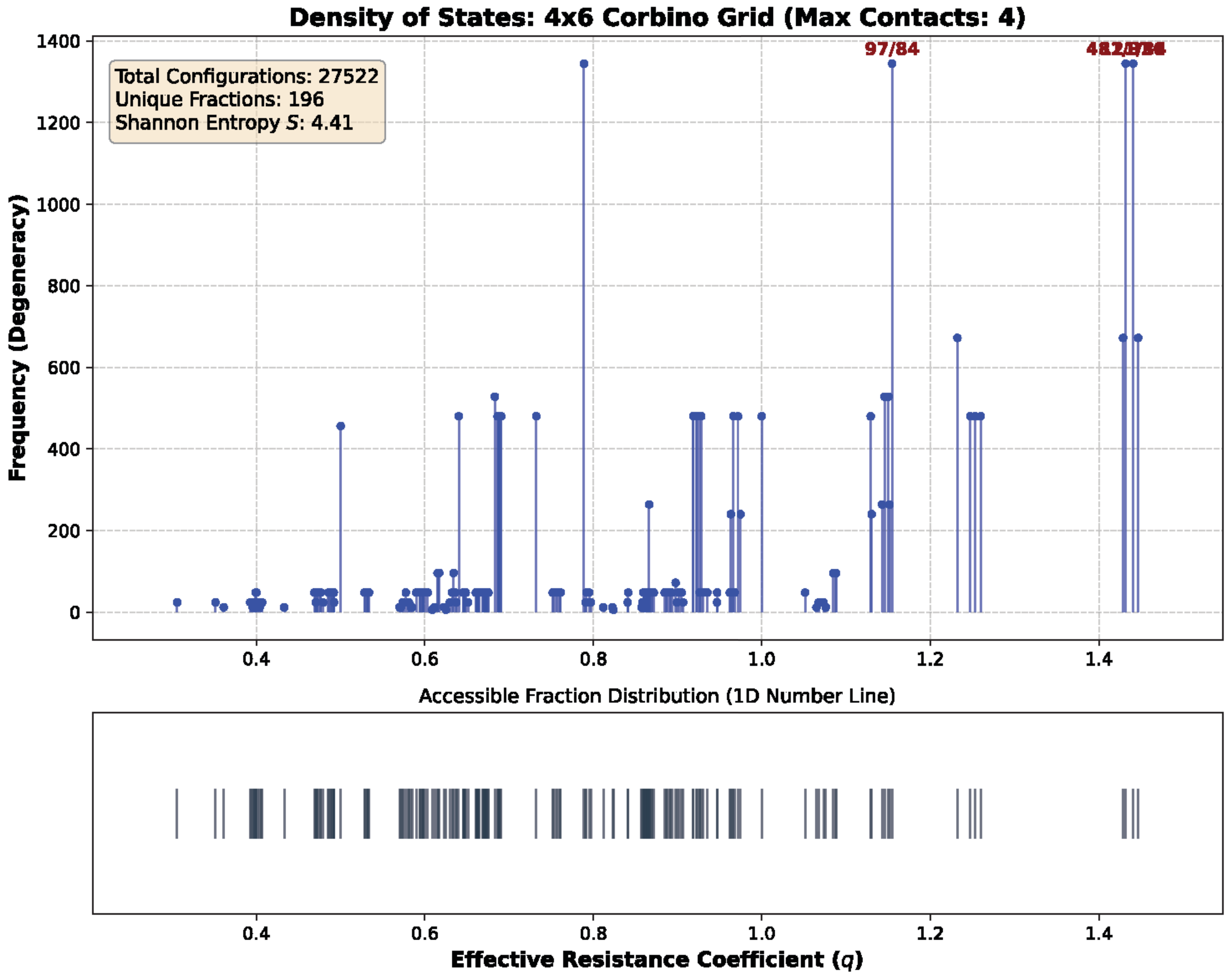

Density of States: 4x6 Corbino Grid (Max Contacts: 4)
Total Configurations: 27522
Unique Fractions: 196
Shannon Entropy S: 4.41
97/84
Frequency (Degeneracy)
0
200
400
600
800
1000
1200
1400
0.4
0.6
0.8
1.0
1.2
1.4
Accessible Fraction Distribution (1D Number Line)
Effective Resistance Coefficient (q)

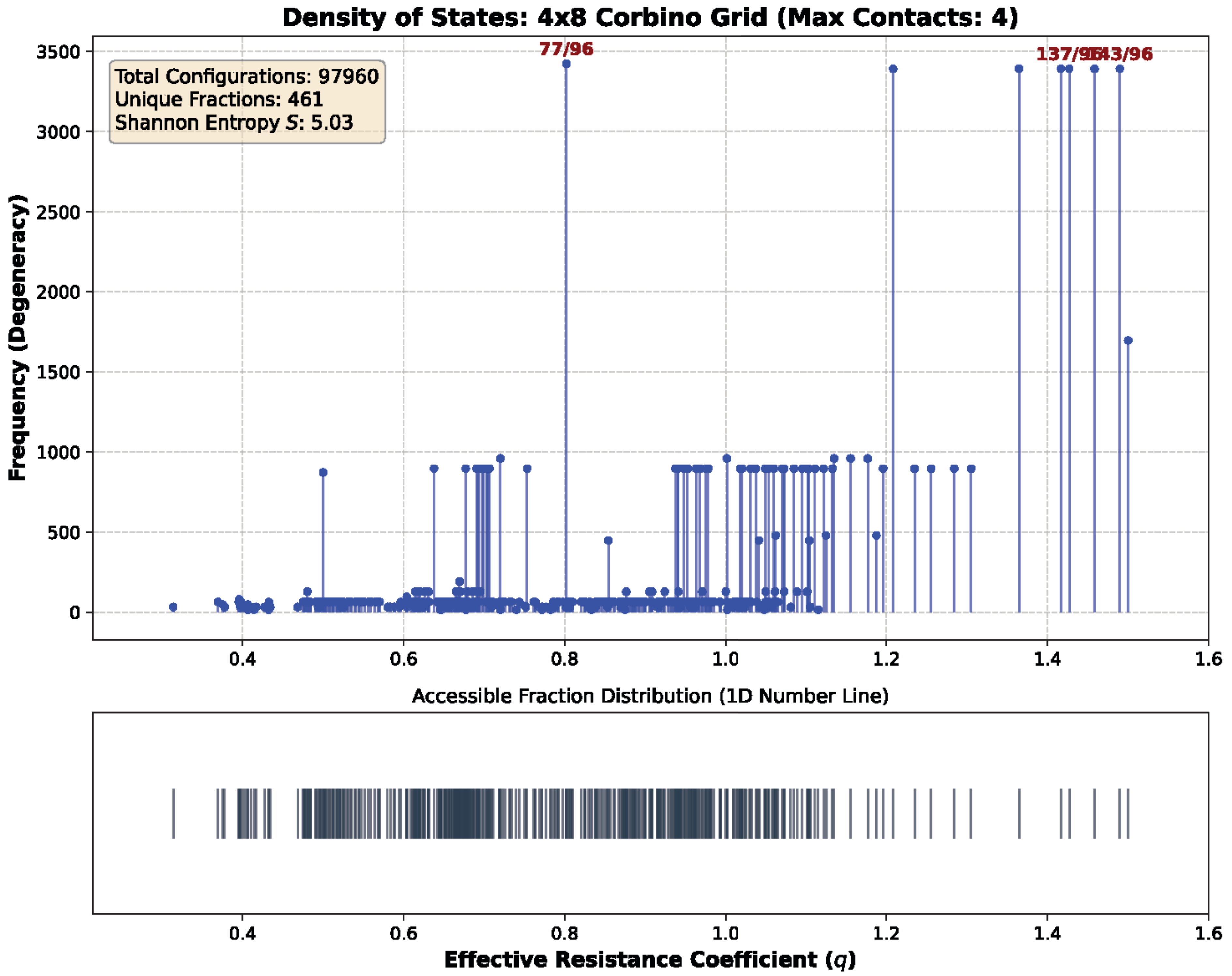

Density of States: 4x8 Corbino Grid (Max Contacts: 4)
Total Configurations: 97960
Unique Fractions: 461
Shannon Entropy S: 5.03
77/96
137/96
143/96
Frequency (Degeneracy)
0
500
1000
1500
2000
2500
3000
3500
0.4
0.6
0.8
1.0
1.2
1.4
1.6
Accessible Fraction Distribution (1D Number Line)
Effective Resistance Coefficient (q)

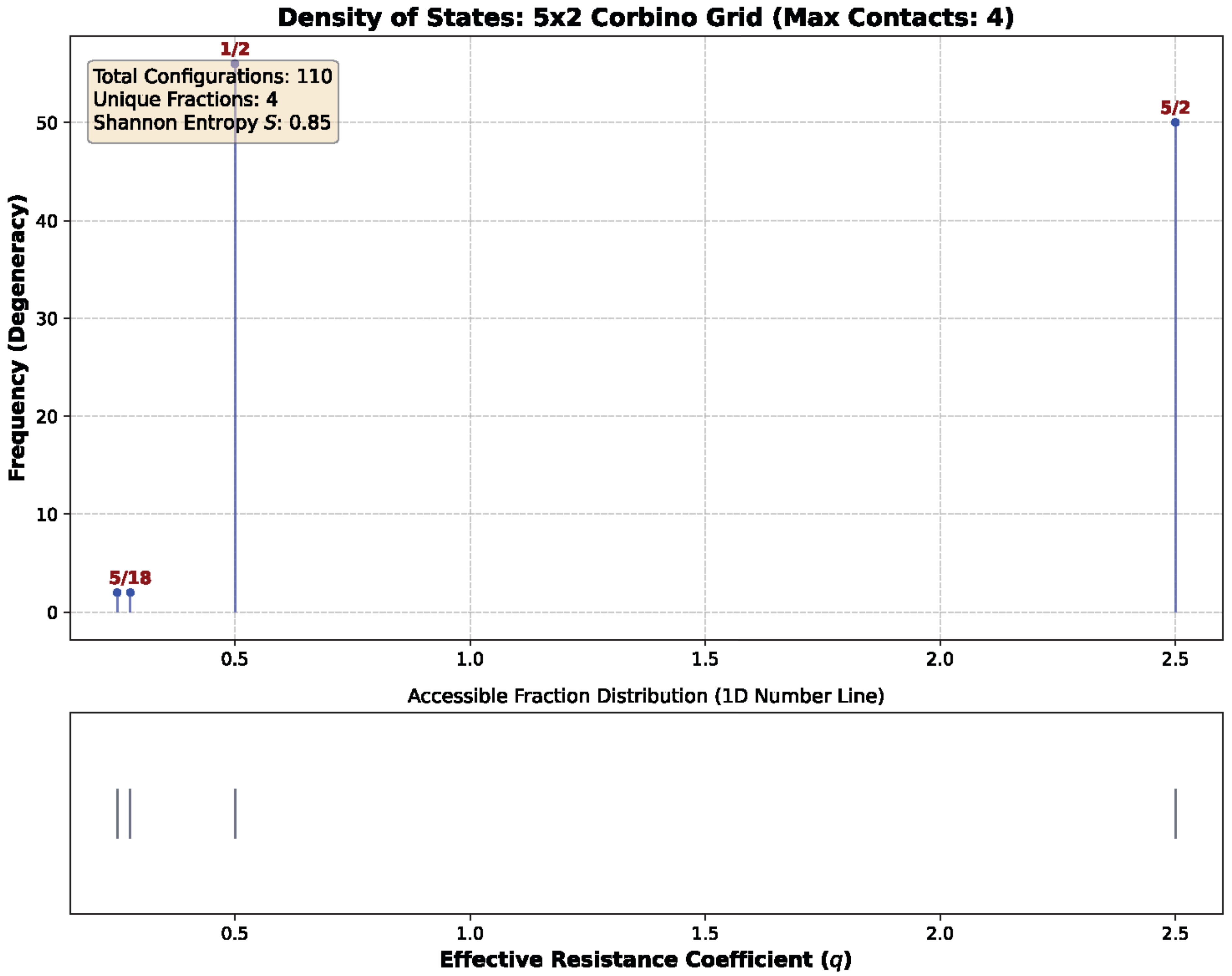

Density of States: 5x2 Corbino Grid (Max Contacts: 4)
Total Configurations: 110
Unique Fractions: 4
Shannon Entropy S: 0.85
1/2
5/2
5/18
Frequency (Degeneracy)
0
10
20
30
40
50
0.5
1.0
1.5
2.0
2.5
Accessible Fraction Distribution (1D Number Line)
0.5
1.0
1.5
2.0
2.5
Effective Resistance Coefficient (q)

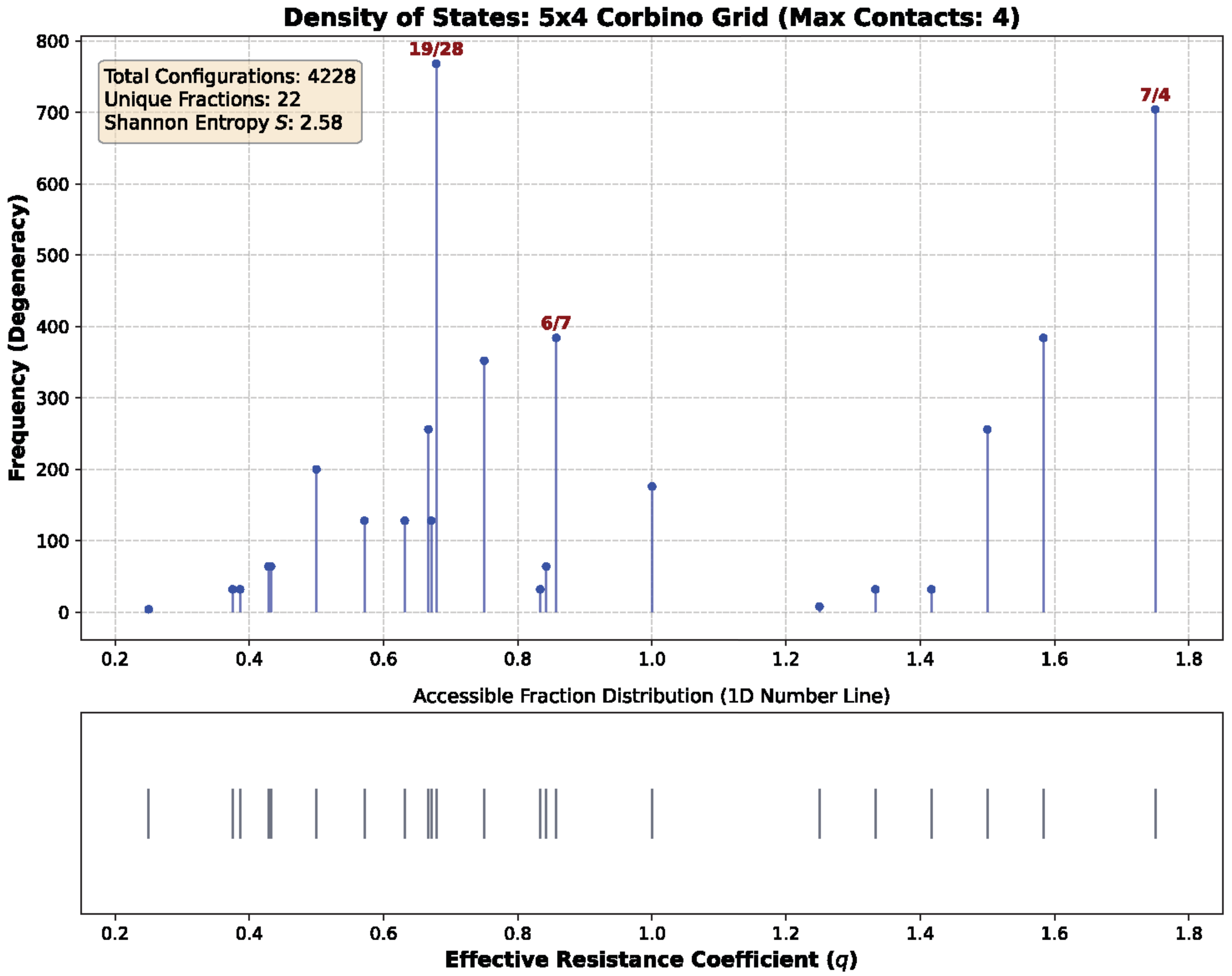
Density of States: 5x4 Corbino Grid (Max Contacts: 4)
Total Configurations: 4228
Unique Fractions: 22
Shannon Entropy S: 2.58
19/28
7/4
6/7
Frequency (Degeneracy)
0
100
200
300
400
500
600
700
800
0.2
0.4
0.6
0.8
1.0
1.2
1.4
1.6
1.8
Accessible Fraction Distribution (1D Number Line)
Effective Resistance Coefficient (q)

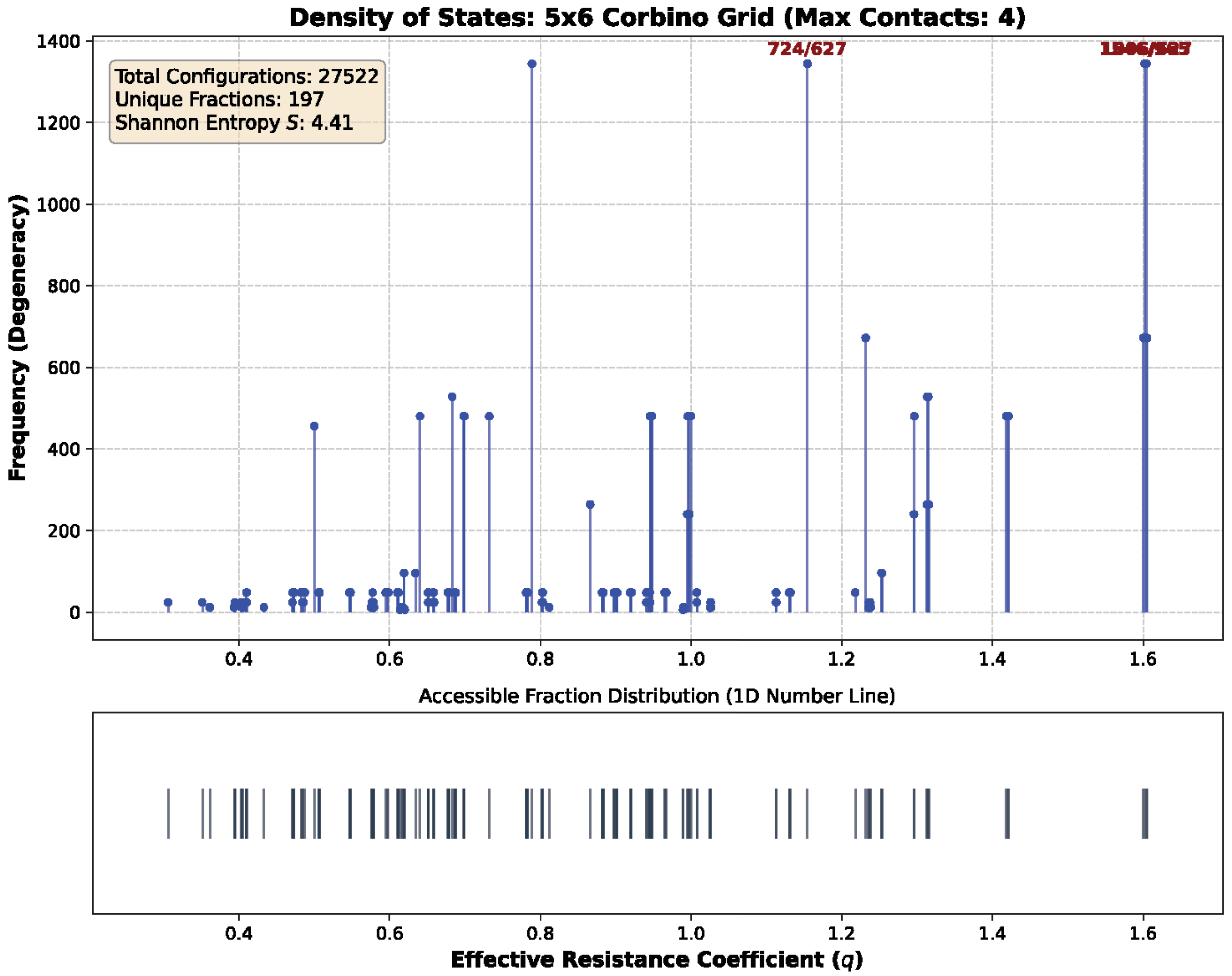
Density of States: 5x6 Corbino Grid (Max Contacts: 4)
Total Configurations: 27522
Unique Fractions: 197
Shannon Entropy S: 4.41
724/627
Frequency (Degeneracy)
1400
1200
1000
800
600
400
200
0
0.4
0.6
0.8
1.0
1.2
1.4
1.6
Accessible Fraction Distribution (1D Number Line)
0.4
0.6
0.8
1.0
1.2
1.4
1.6
Effective Resistance Coefficient (q)

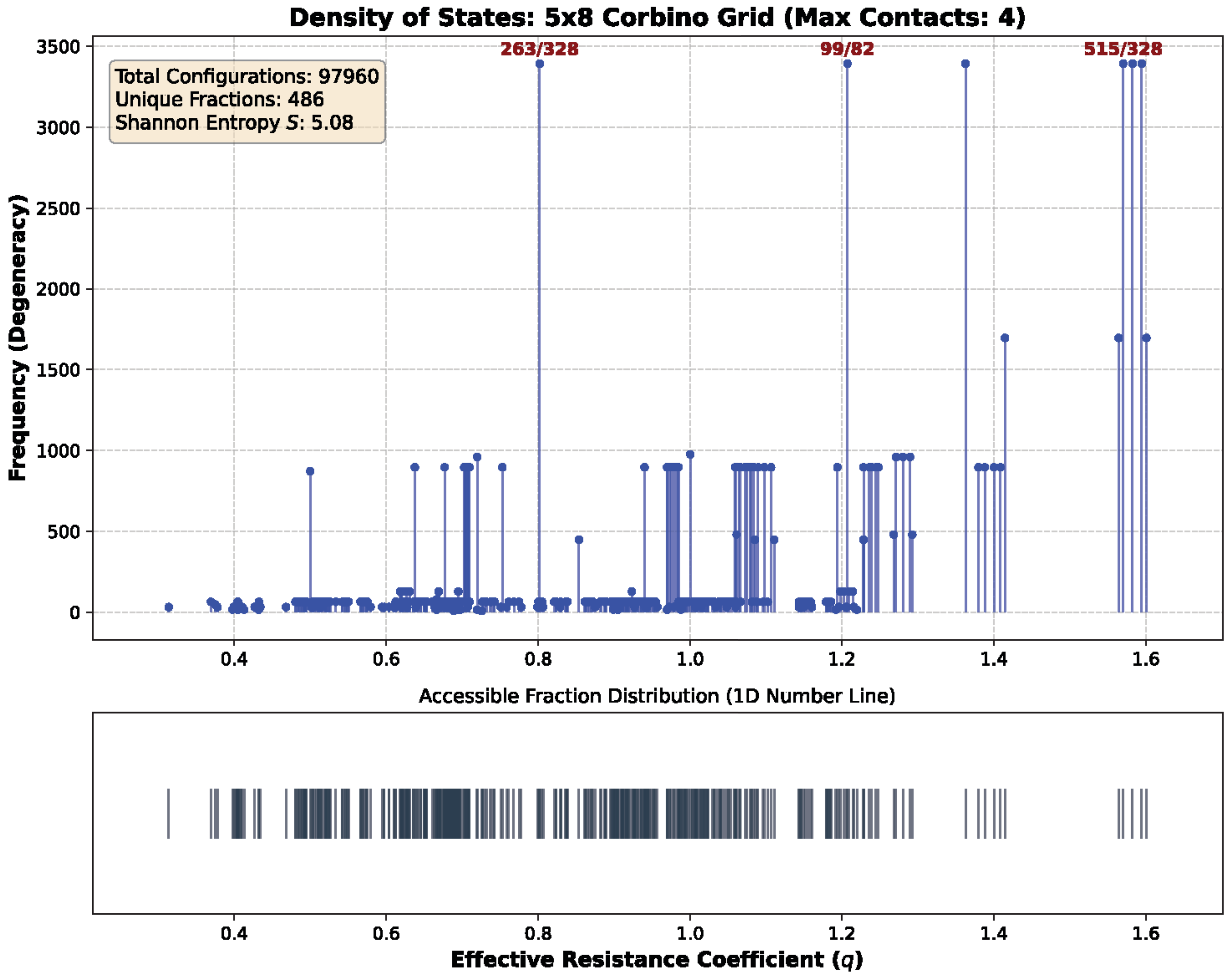
Density of States: 5x8 Corbino Grid (Max Contacts: 4)
Total Configurations: 97960
Unique Fractions: 486
Shannon Entropy S: 5.08
263/328
99/82
515/328
Frequency (Degeneracy)
Accessible Fraction Distribution (1D Number Line)
Effective Resistance Coefficient (q)

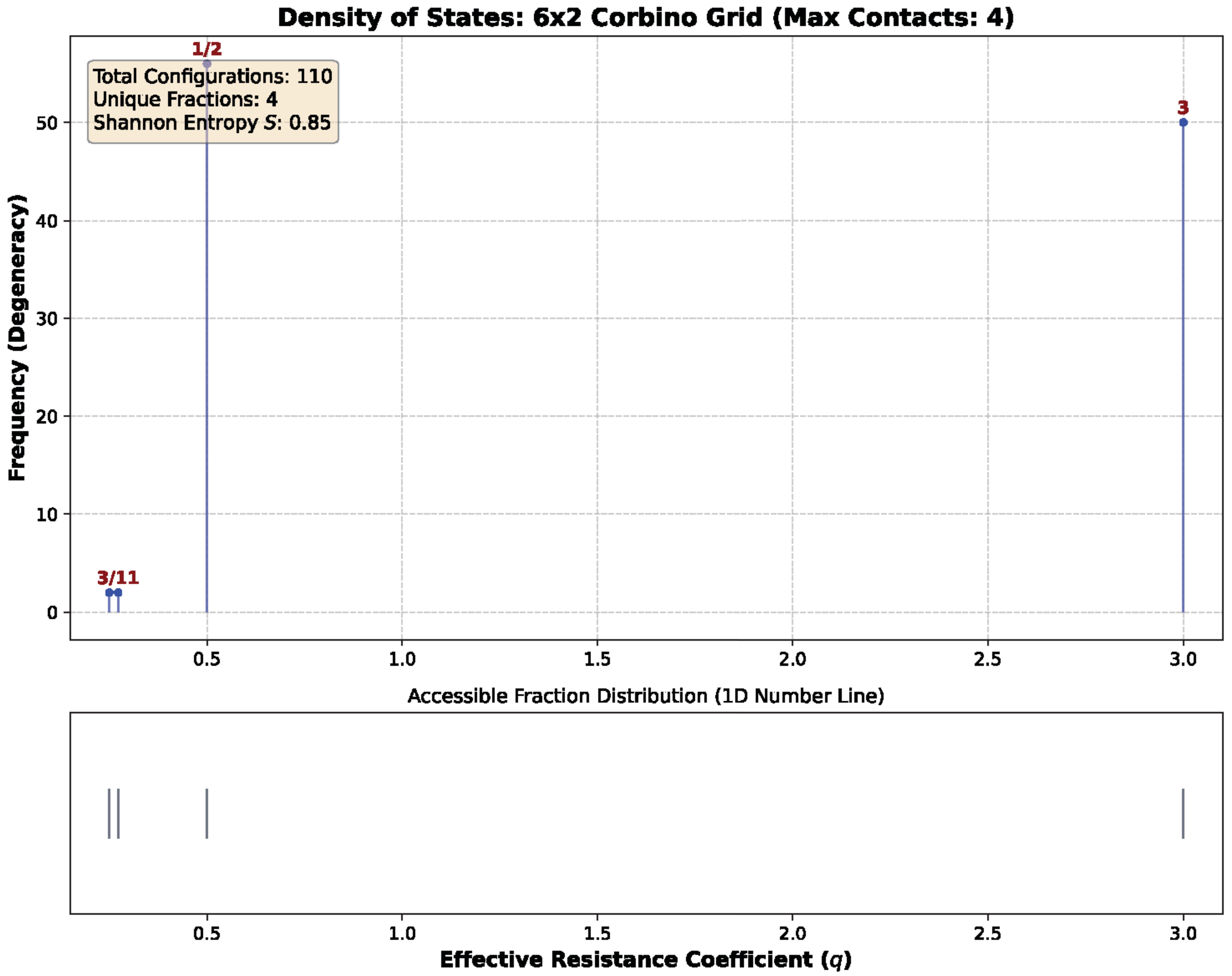
Density of States: 6x2 Corbino Grid (Max Contacts: 4)
Total Configurations: 110
Unique Fractions: 4
Shannon Entropy S: 0.85
1/2
3
3/11
Frequency (Degeneracy)
0
10
20
30
40
50
0.5
1.0
1.5
2.0
2.5
3.0
Accessible Fraction Distribution (1D Number Line)
0.5
1.0
1.5
2.0
2.5
3.0
Effective Resistance Coefficient (q)

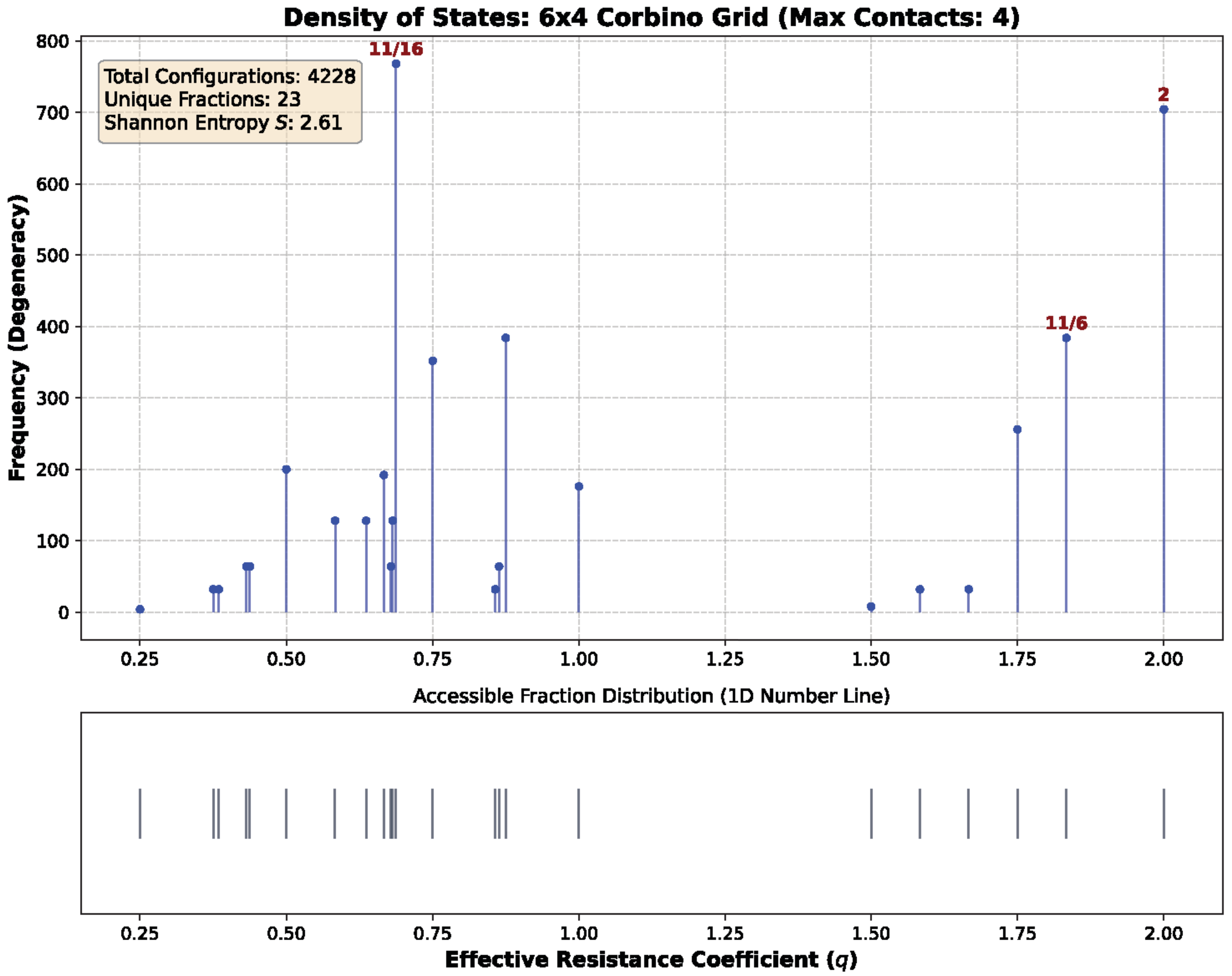
Density of States: 6x4 Corbino Grid (Max Contacts: 4)
Total Configurations: 4228
Unique Fractions: 23
Shannon Entropy S: 2.61
11/16
2
11/6
Frequency (Degeneracy)
0
100
200
300
400
500
600
700
800
0.25
0.50
0.75
1.00
1.25
1.50
1.75
2.00
Accessible Fraction Distribution (1D Number Line)
Effective Resistance Coefficient (q)

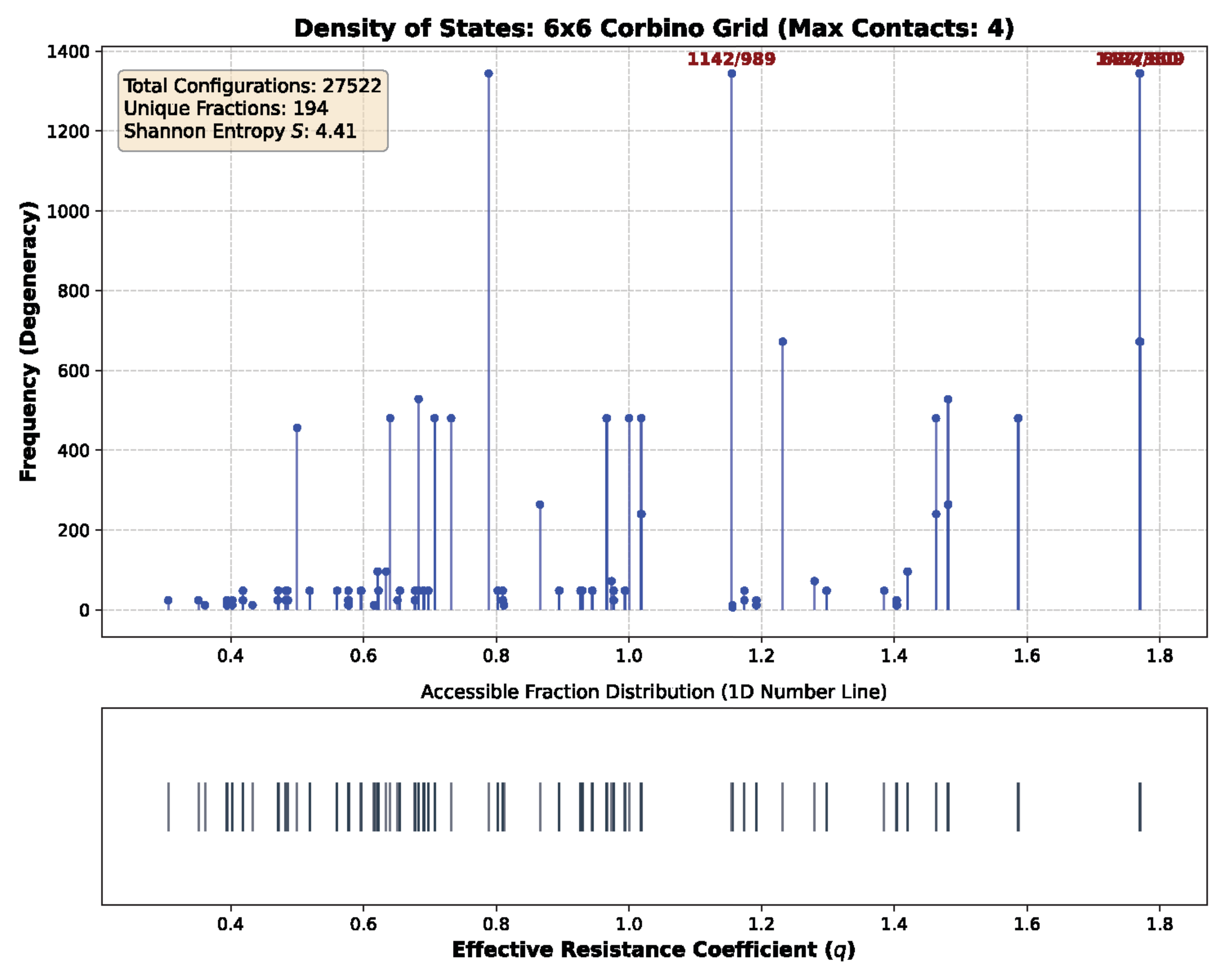
Density of States: 6x6 Corbino Grid (Max Contacts: 4)
Total Configurations: 27522
Unique Fractions: 194
Shannon Entropy S: 4.41
1142/989
Frequency (Degeneracy)
Accessible Fraction Distribution (1D Number Line)
Effective Resistance Coefficient (q)

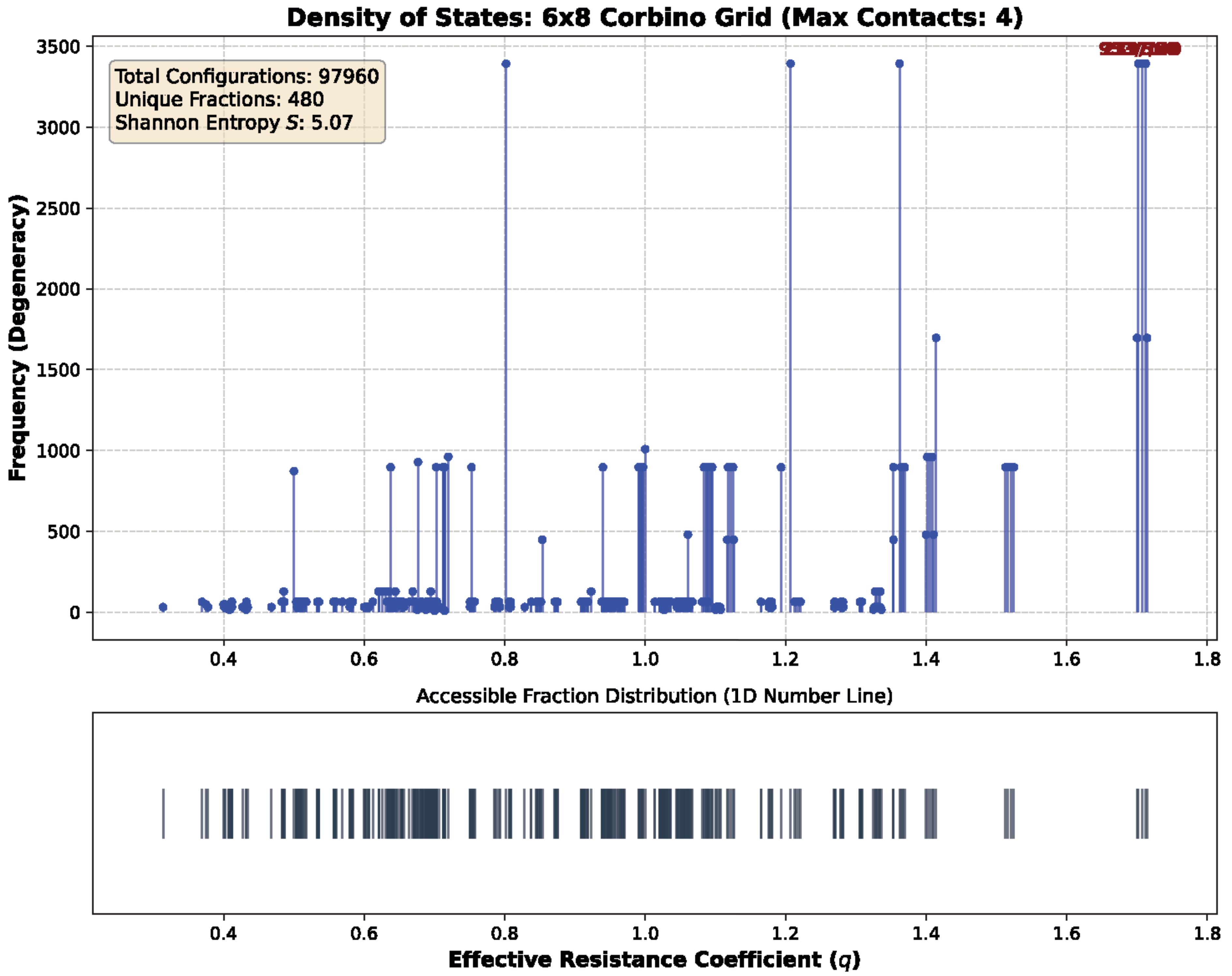

Density of States: 6x8 Corbino Grid (Max Contacts: 4)
Total Configurations: 97960
Unique Fractions: 480
Shannon Entropy S: 5.07
Frequency (Degeneracy)
3500
3000
2500
2000
1500
1000
500
0
0.4
0.6
0.8
1.0
1.2
1.4
1.6
1.8
Accessible Fraction Distribution (1D Number Line)
0.4
0.6
0.8
1.0
1.2
1.4
1.6
1.8
Effective Resistance Coefficient (q)

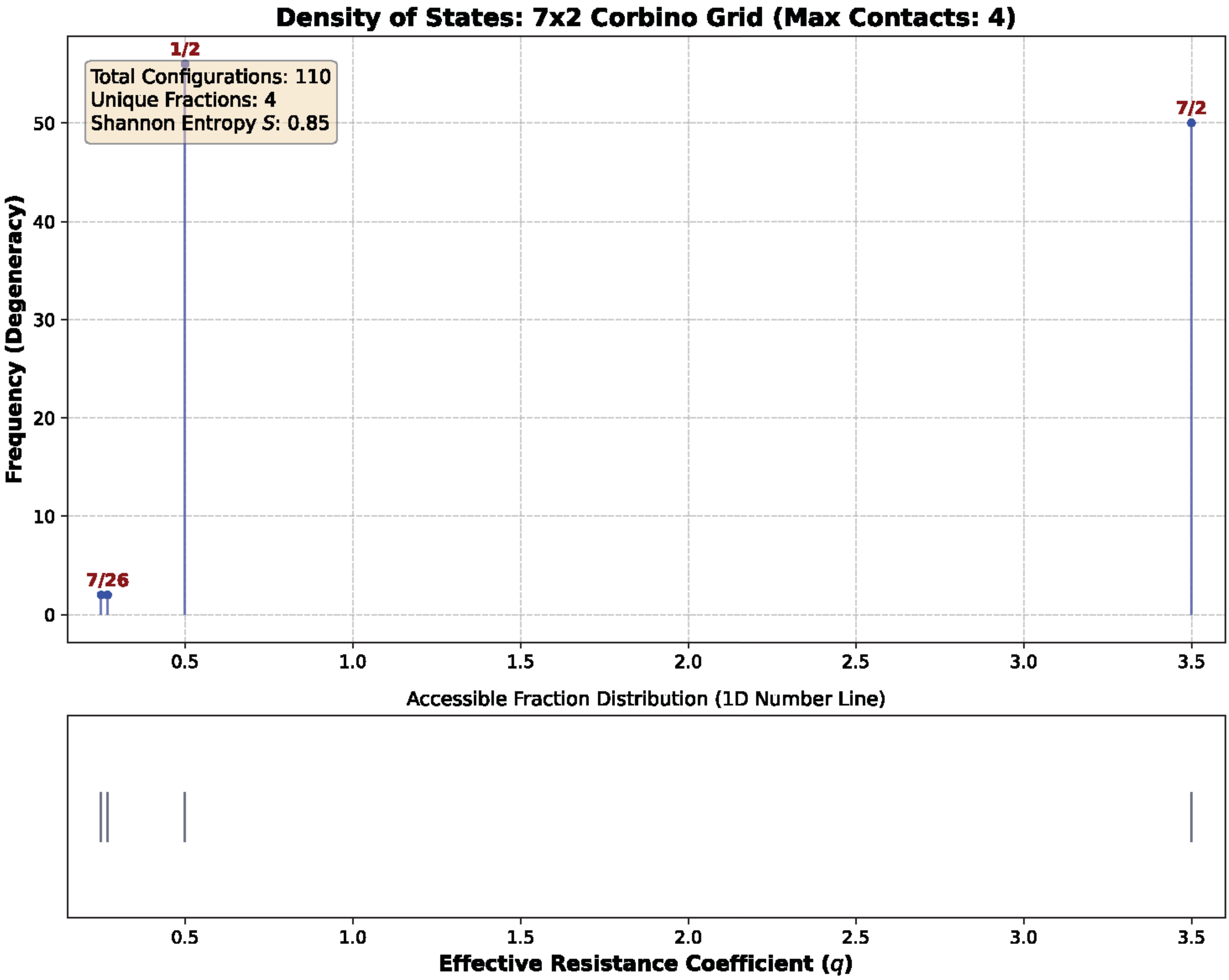

Density of States: 7x2 Corbino Grid (Max Contacts: 4)
Total Configurations: 110
Unique Fractions: 4
Shannon Entropy S: 0.85
1/2
7/2
7/26
Frequency (Degeneracy)
0
10
20
30
40
50
0.5
1.0
1.5
2.0
2.5
3.0
3.5
Accessible Fraction Distribution (1D Number Line)
Effective Resistance Coefficient (q)

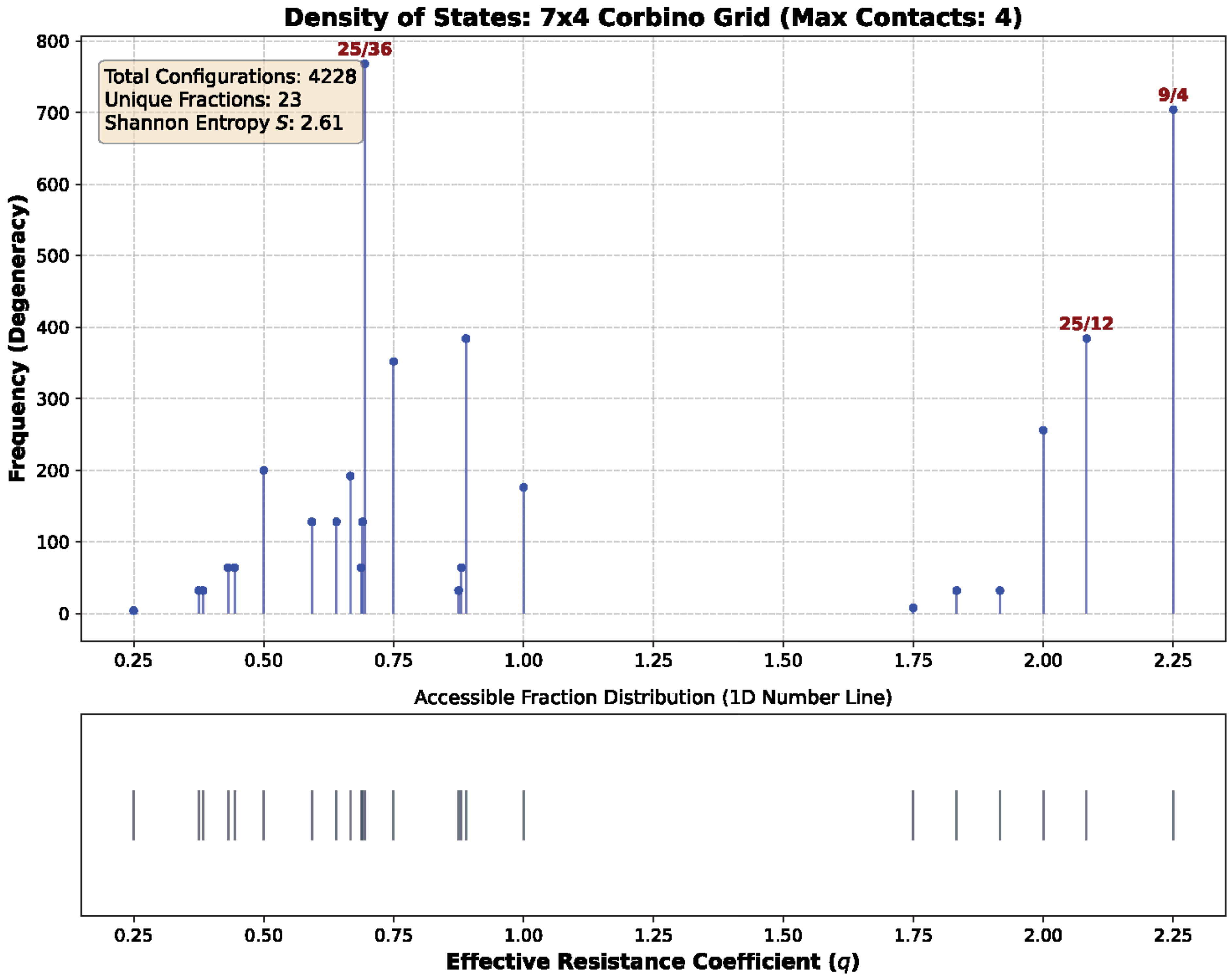

Density of States: 7x4 Corbino Grid (Max Contacts: 4)
Total Configurations: 4228
Unique Fractions: 23
Shannon Entropy S: 2.61
25/36
9/4
25/12
Frequency (Degeneracy)
800
700
600
500
400
300
200
100
0
0.25
0.50
0.75
1.00
1.25
1.50
1.75
2.00
2.25
Accessible Fraction Distribution (1D Number Line)
Effective Resistance Coefficient (q)

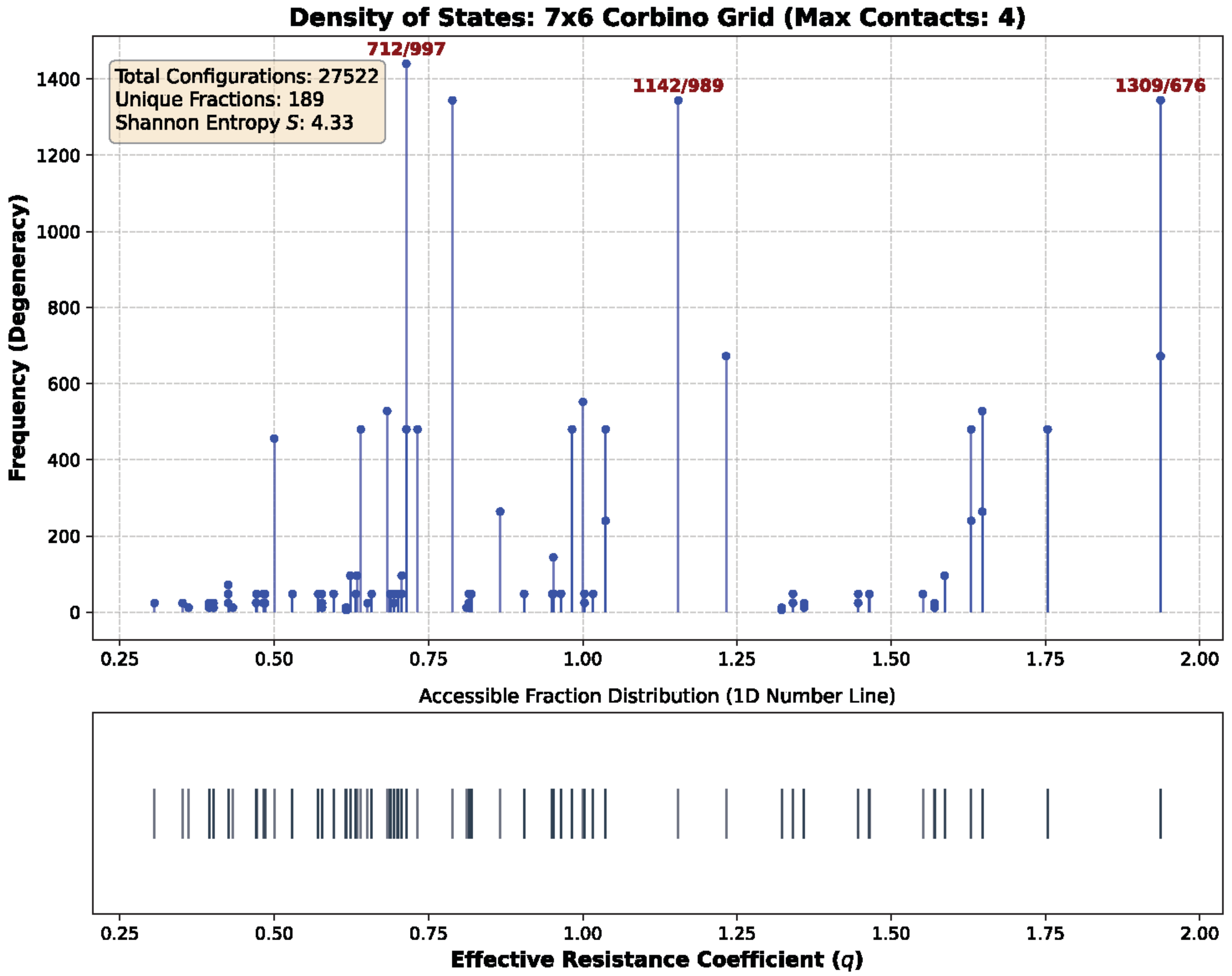

Density of States: 7x6 Corbino Grid (Max Contacts: 4)
Total Configurations: 27522
Unique Fractions: 189
Shannon Entropy S: 4.33
712/997
1142/989
1309/676
Frequency (Degeneracy)
0
200
400
600
800
1000
1200
1400
0.25
0.50
0.75
1.00
1.25
1.50
1.75
2.00
Accessible Fraction Distribution (1D Number Line)
Effective Resistance Coefficient (q)

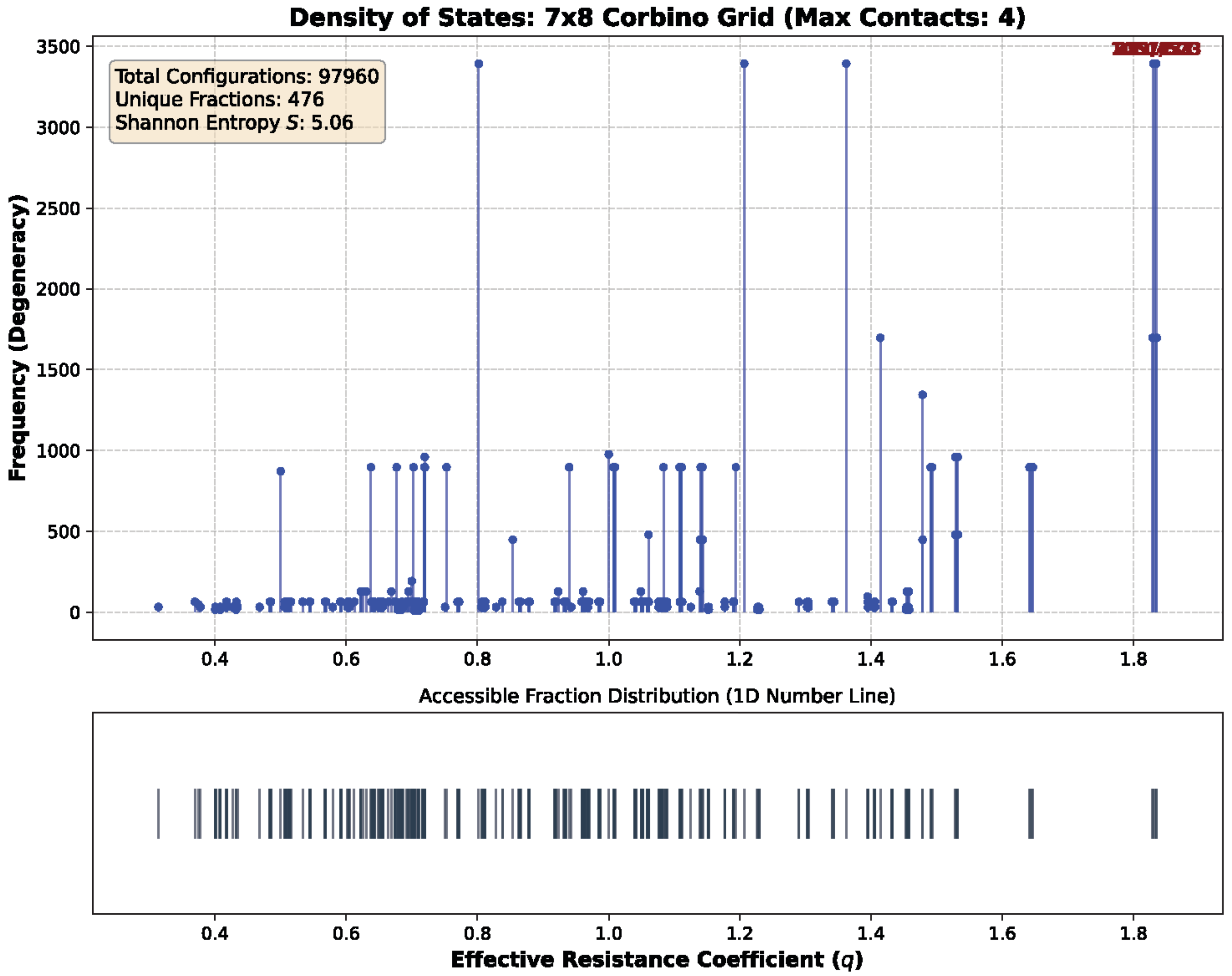
Density of States: 7x8 Corbino Grid (Max Contacts: 4)
Total Configurations: 97960
Unique Fractions: 476
Shannon Entropy S: 5.06
Frequency (Degeneracy)
0
500
1000
1500
2000
2500
3000
3500
0.4
0.6
0.8
1.0
1.2
1.4
1.6
1.8
Accessible Fraction Distribution (1D Number Line)
0.4
0.6
0.8
1.0
1.2
1.4
1.6
1.8
Effective Resistance Coefficient (q)

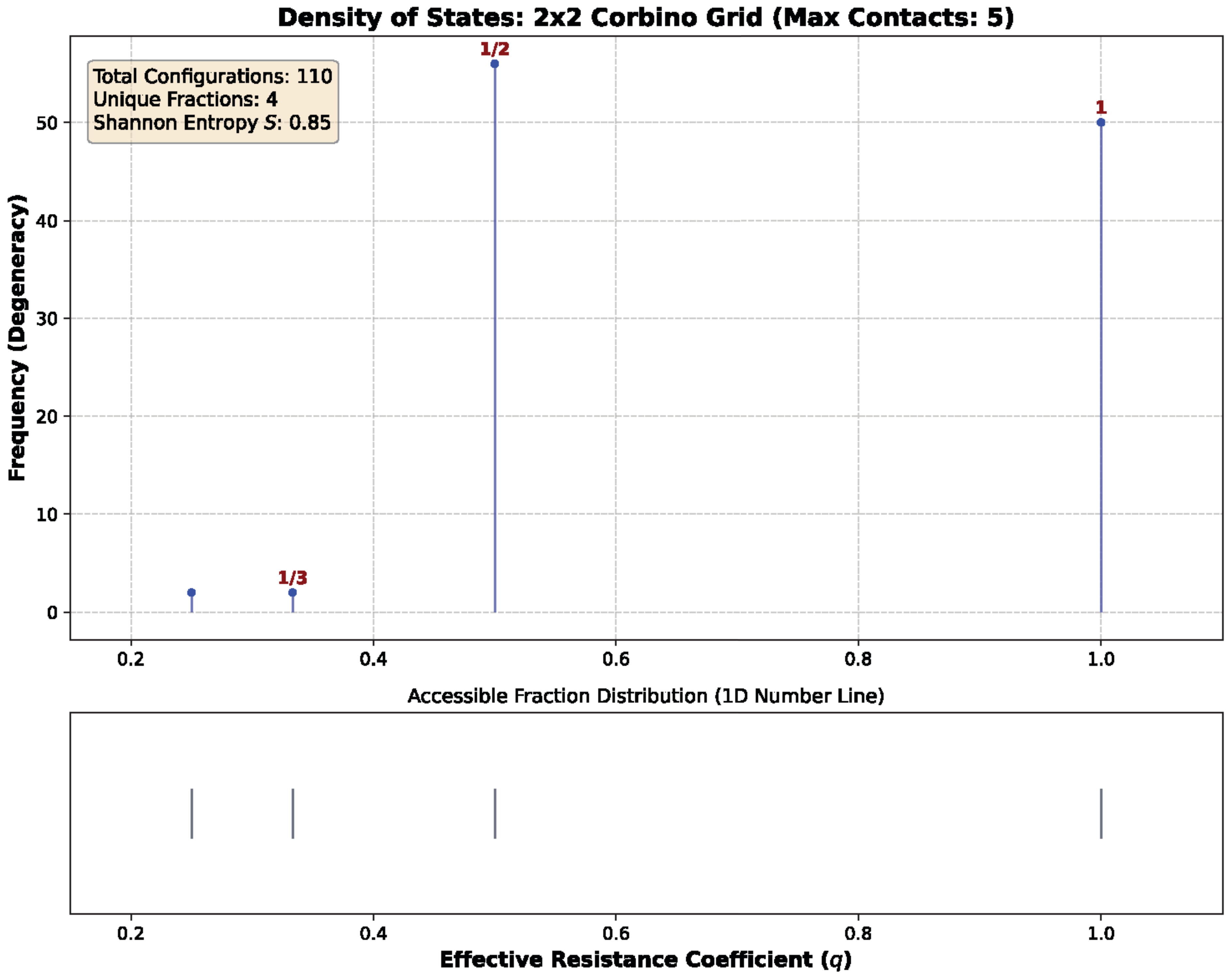

Density of States: 2x2 Corbino Grid (Max Contacts: 5)
Total Configurations: 110
Unique Fractions: 4
Shannon Entropy $S$: 0.85
1/2
1
1/3
Frequency (Degeneracy)
0
10
20
30
40
50
0.2
0.4
0.6
0.8
1.0
Accessible Fraction Distribution (1D Number Line)
Effective Resistance Coefficient ($q$)

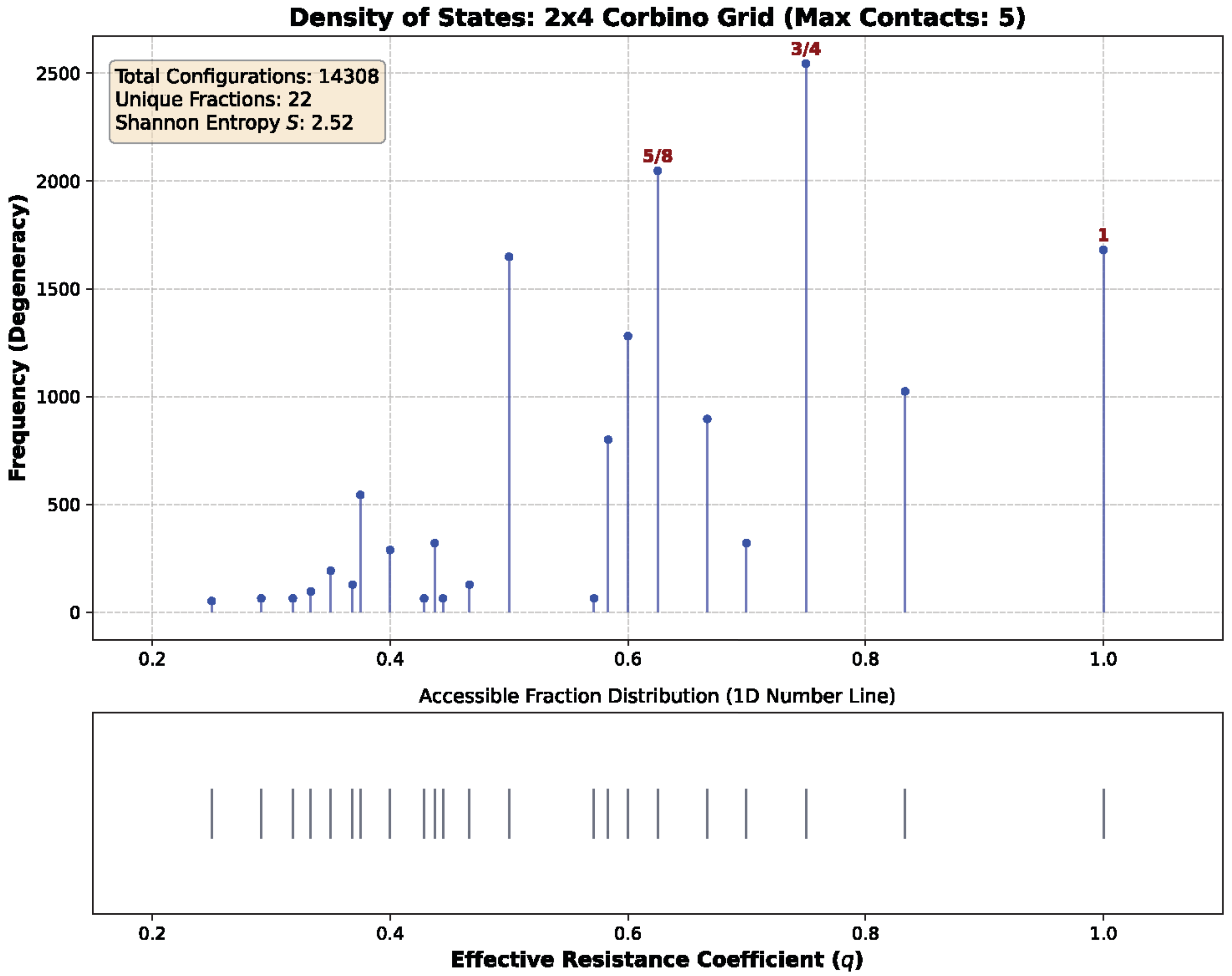
Density of States: 2x4 Corbino Grid (Max Contacts: 5)
Total Configurations: 14308
Unique Fractions: 22
Shannon Entropy S: 2.52
3/4
5/8
1
Frequency (Degeneracy)
2500
2000
1500
1000
500
0
0.2
0.4
0.6
0.8
1.0
Accessible Fraction Distribution (1D Number Line)
0.2
0.4
0.6
0.8
1.0
Effective Resistance Coefficient (q)

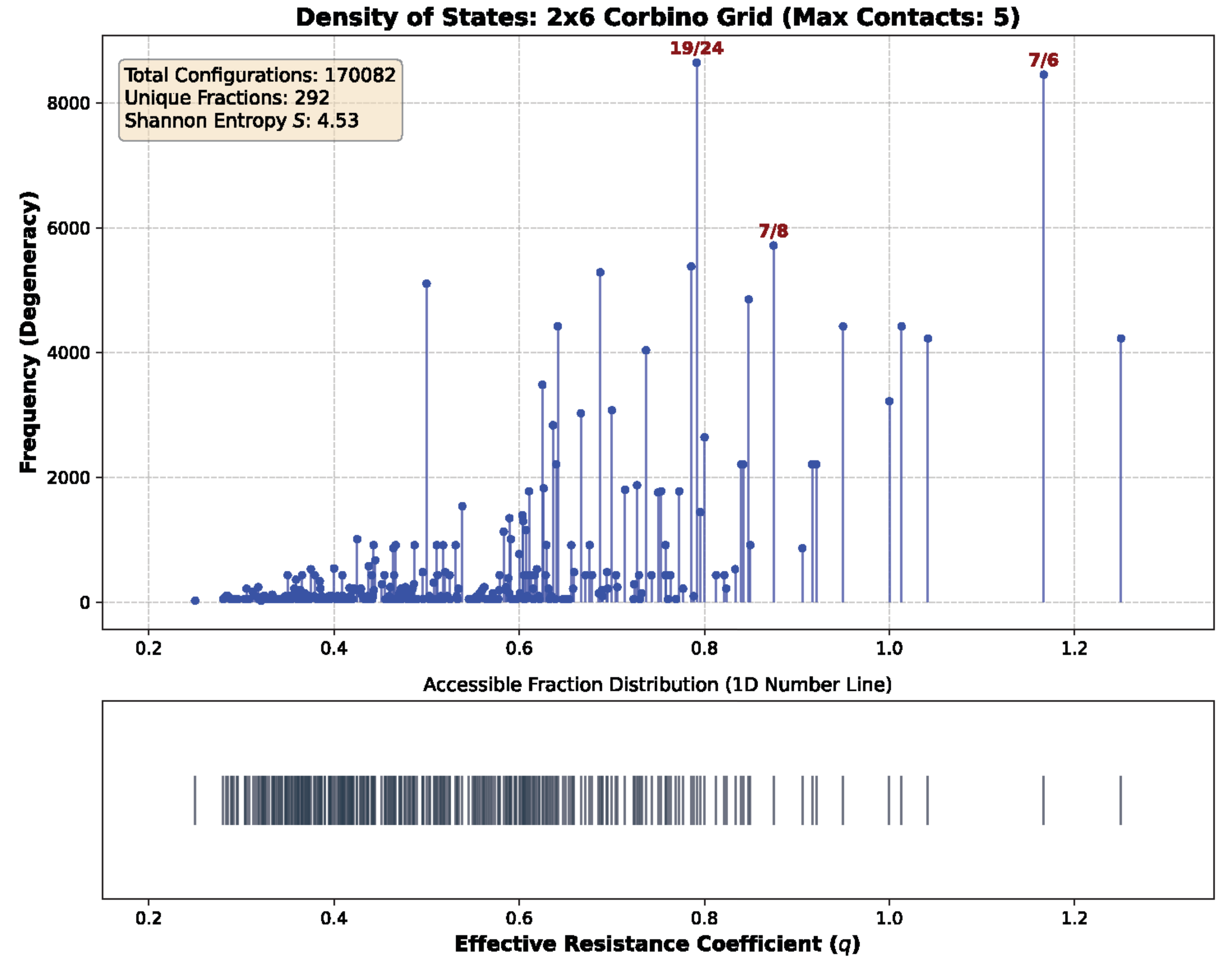
Density of States: 2x6 Corbino Grid (Max Contacts: 5)
Total Configurations: 170082
Unique Fractions: 292
Shannon Entropy S: 4.53
19/24
7/6
7/8
Frequency (Degeneracy)
0
2000
4000
6000
8000
0.2
0.4
0.6
0.8
1.0
1.2
Accessible Fraction Distribution (1D Number Line)
Effective Resistance Coefficient (q)

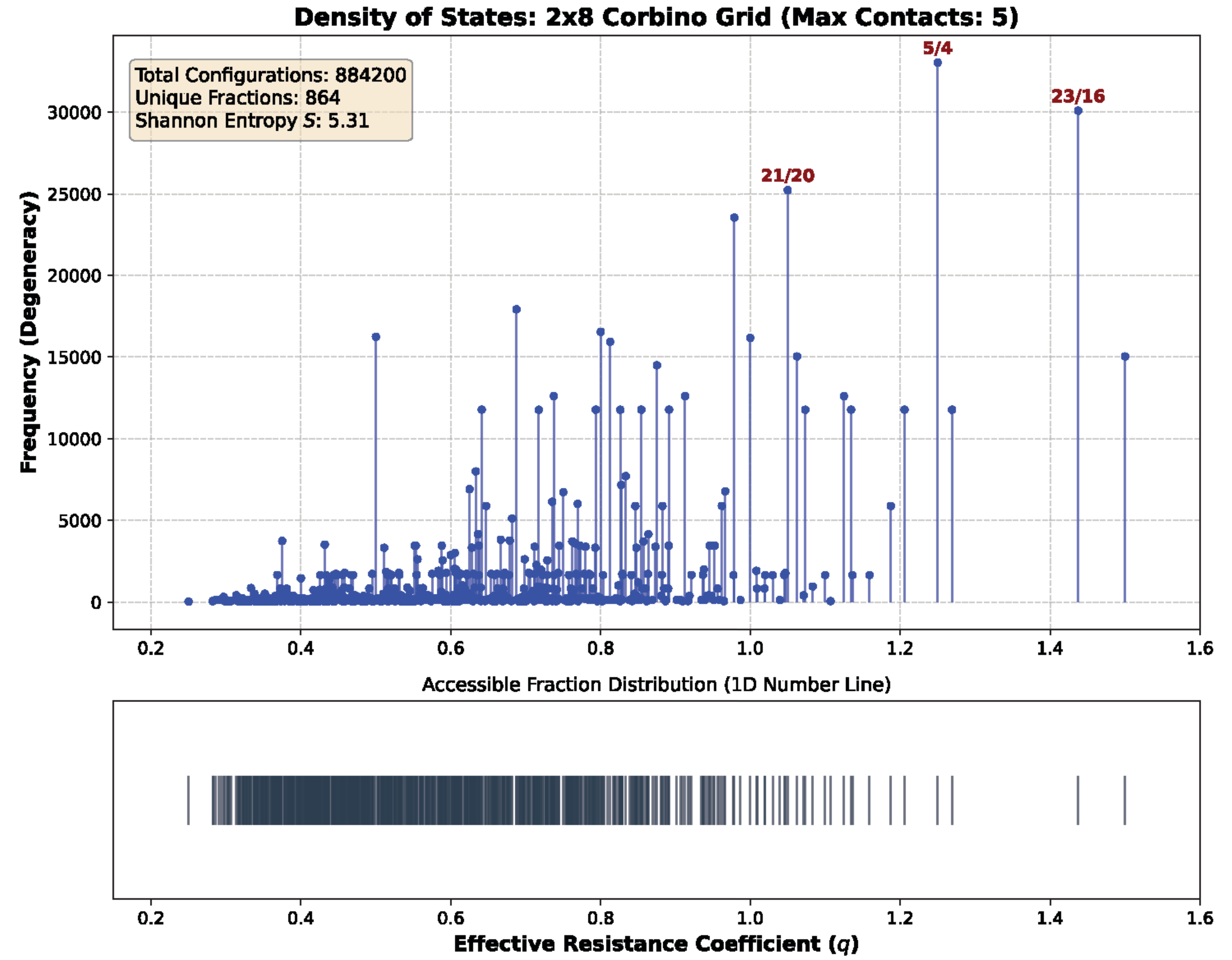
Density of States: 2x8 Corbino Grid (Max Contacts: 5)
Total Configurations: 884200
Unique Fractions: 864
Shannon Entropy S: 5.31
5/4
23/16
21/20
Frequency (Degeneracy)
30000
25000
20000
15000
10000
5000
0
0.2
0.4
0.6
0.8
1.0
1.2
1.4
1.6
Accessible Fraction Distribution (1D Number Line)
Effective Resistance Coefficient (q)

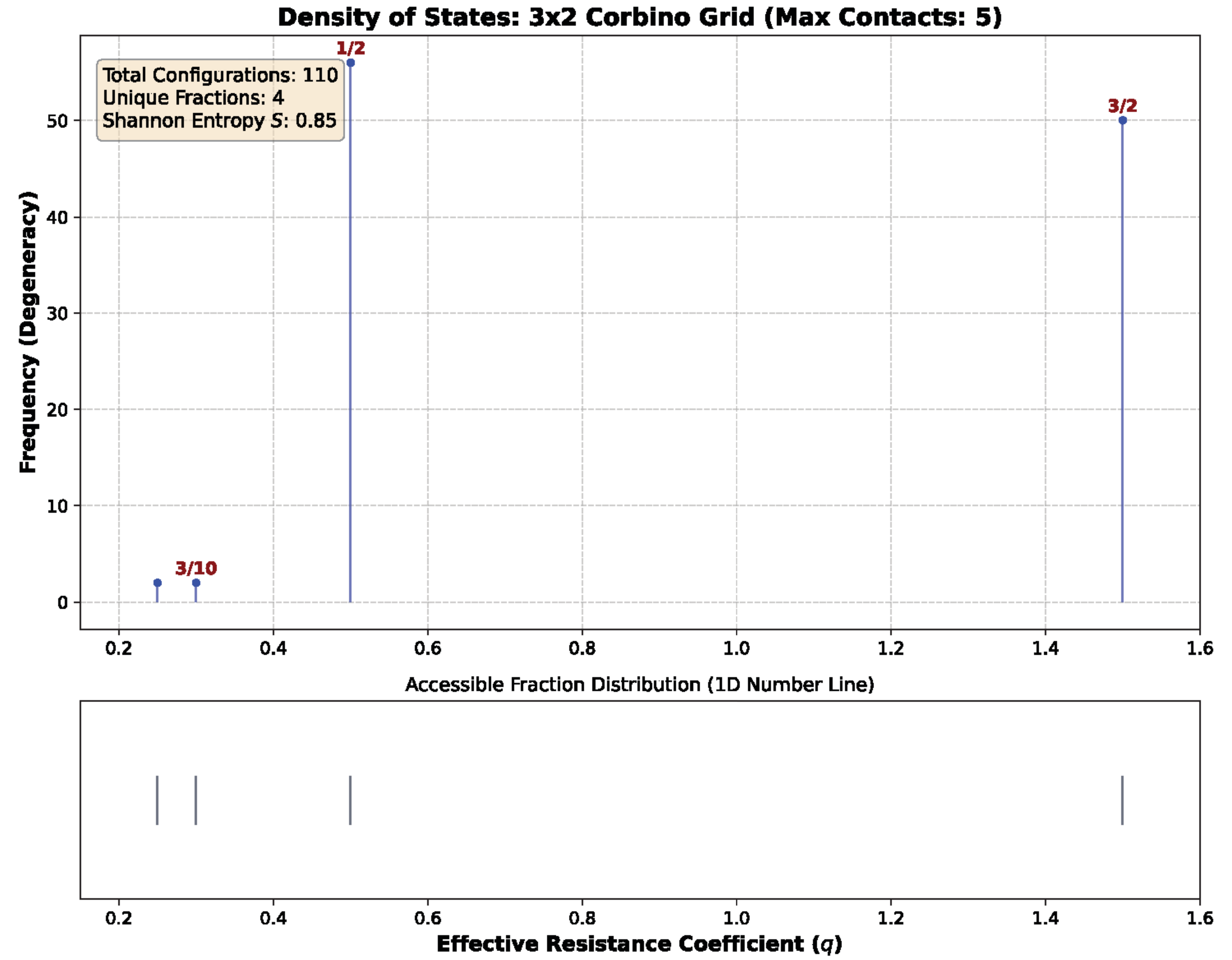

Density of States: 3x2 Corbino Grid (Max Contacts: 5)
Total Configurations: 110
Unique Fractions: 4
Shannon Entropy S: 0.85
1/2
3/2
3/10
Frequency (Degeneracy)
0
10
20
30
40
50
0.2
0.4
0.6
0.8
1.0
1.2
1.4
1.6
Accessible Fraction Distribution (1D Number Line)
Effective Resistance Coefficient (q)

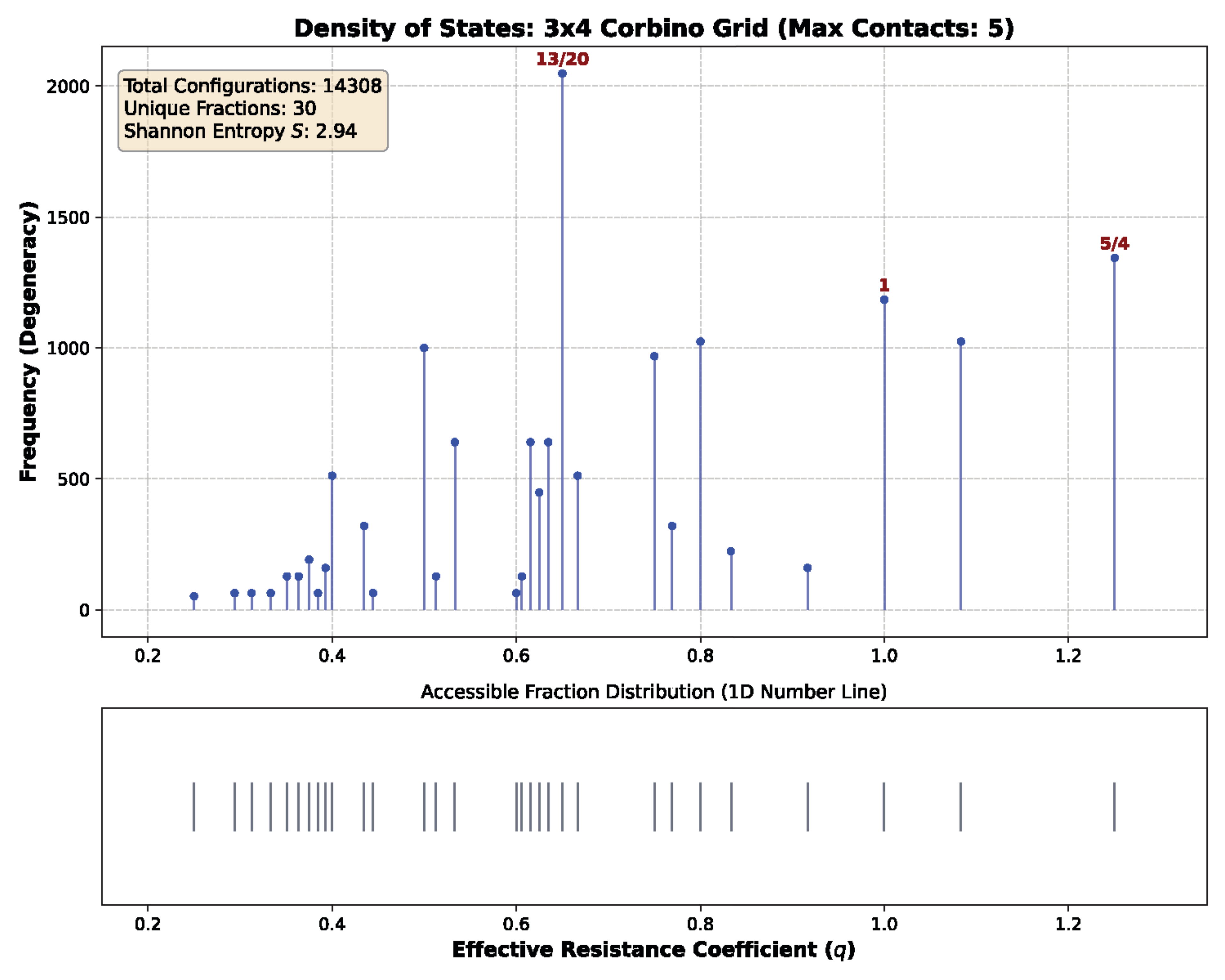

Density of States: 3x4 Corbino Grid (Max Contacts: 5)
Total Configurations: 14308
Unique Fractions: 30
Shannon Entropy S: 2.94
13/20
5/4
1
Frequency (Degeneracy)
0
500
1000
1500
2000
0.2
0.4
0.6
0.8
1.0
1.2
Accessible Fraction Distribution (1D Number Line)
Effective Resistance Coefficient (q)

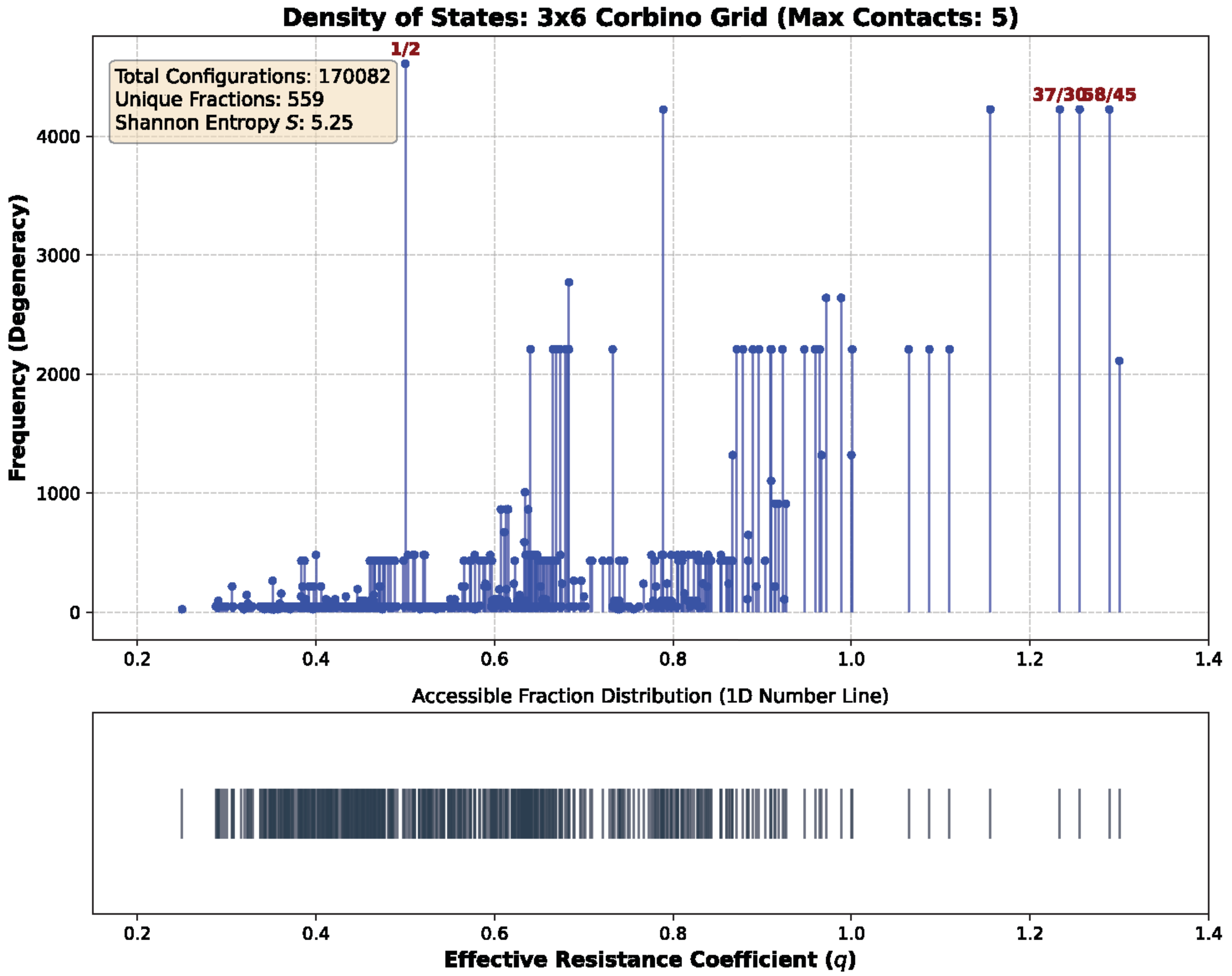

Density of States: 3x6 Corbino Grid (Max Contacts: 5)
Total Configurations: 170082
Unique Fractions: 559
Shannon Entropy S: 5.25
1/2
37/30
58/45
Frequency (Degeneracy)
0
1000
2000
3000
4000
0.2
0.4
0.6
0.8
1.0
1.2
1.4
Accessible Fraction Distribution (1D Number Line)
Effective Resistance Coefficient (q)

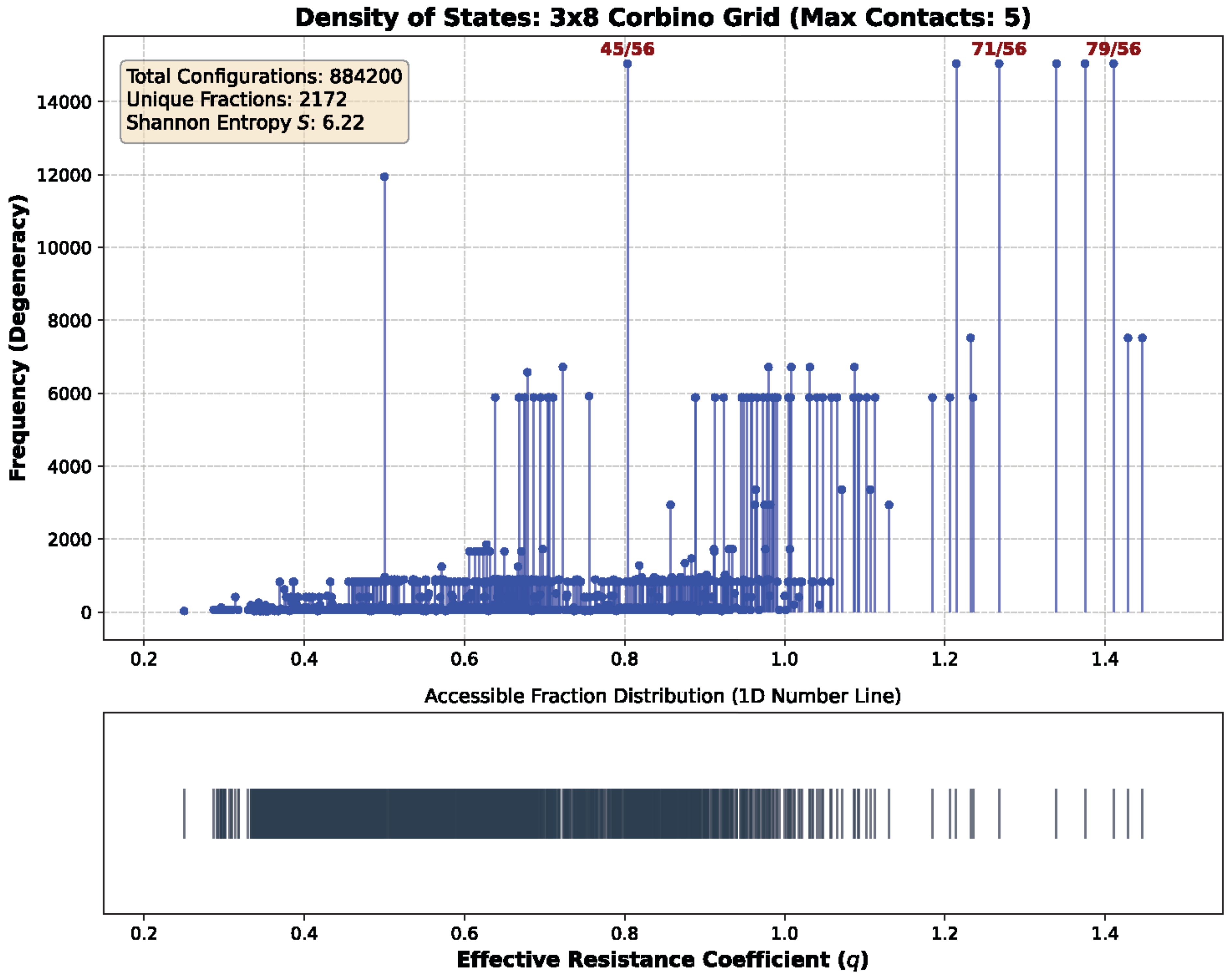

Density of States: 3x8 Corbino Grid (Max Contacts: 5)
Total Configurations: 884200
Unique Fractions: 2172
Shannon Entropy S: 6.22
45/56
71/56
79/56
Frequency (Degeneracy)
0
2000
4000
6000
8000
10000
12000
14000
0.2
0.4
0.6
0.8
1.0
1.2
1.4
Accessible Fraction Distribution (1D Number Line)
Effective Resistance Coefficient (q)

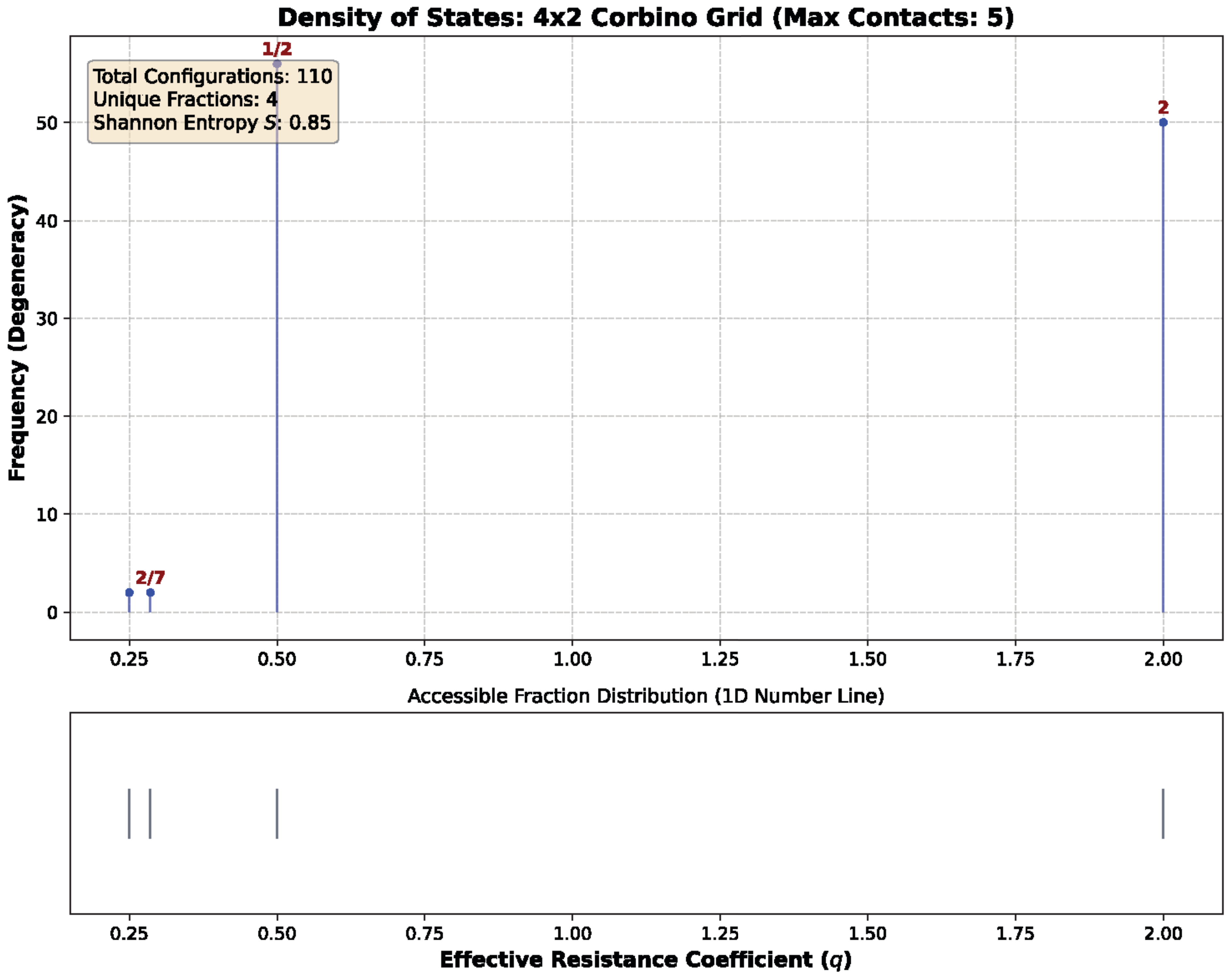

Density of States: 4x2 Corbino Grid (Max Contacts: 5)
Total Configurations: 110
Unique Fractions: 4
Shannon Entropy S: 0.85
1/2
2
2/7
Frequency (Degeneracy)
0
10
20
30
40
50
0.25
0.50
0.75
1.00
1.25
1.50
1.75
2.00
Accessible Fraction Distribution (1D Number Line)
Effective Resistance Coefficient (q)

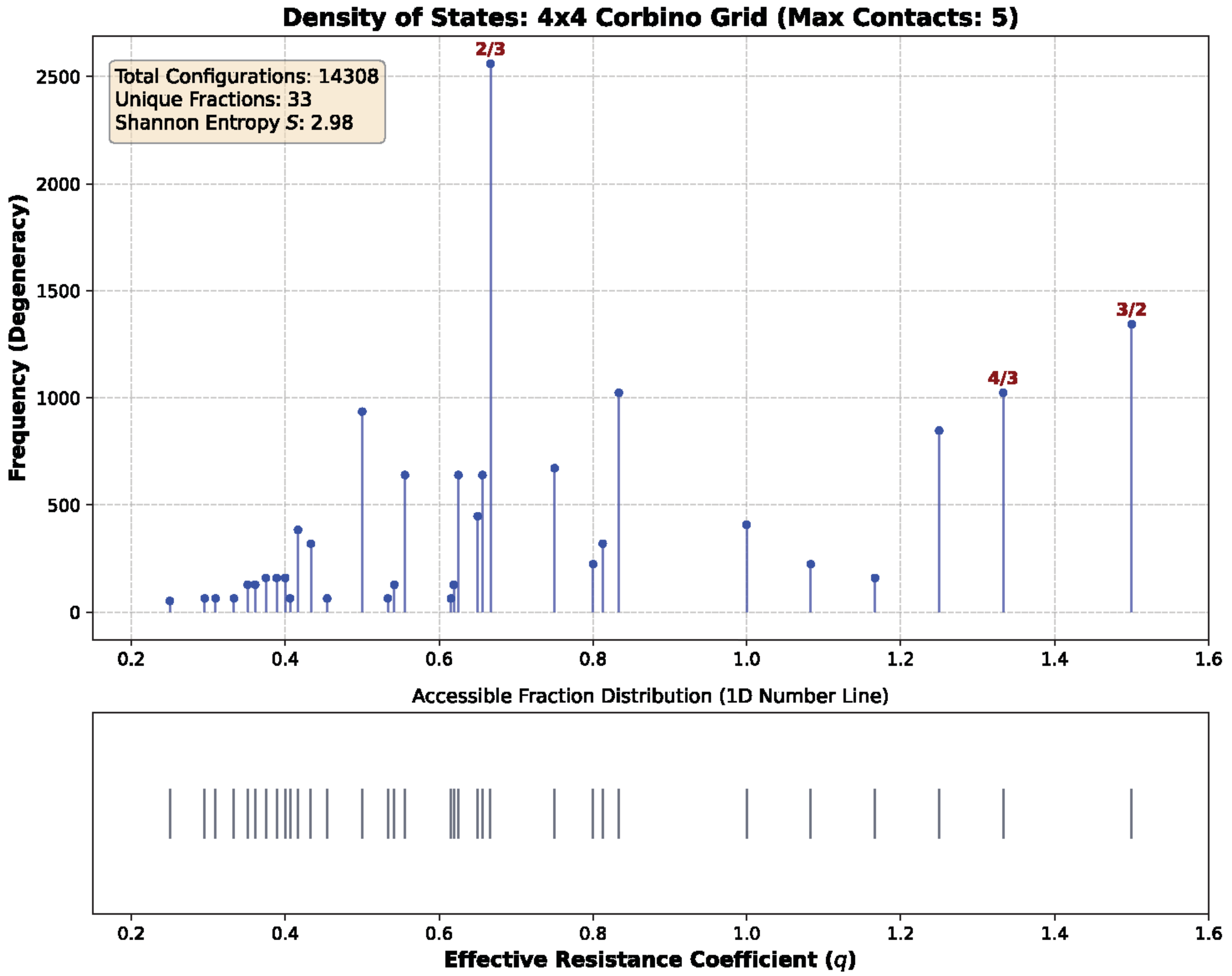
Density of States: 4x4 Corbino Grid (Max Contacts: 5)
Total Configurations: 14308
Unique Fractions: 33
Shannon Entropy S: 2.98
2/3
4/3
3/2
Frequency (Degeneracy)
0
500
1000
1500
2000
2500
0.2
0.4
0.6
0.8
1.0
1.2
1.4
1.6
Accessible Fraction Distribution (1D Number Line)
Effective Resistance Coefficient (q)

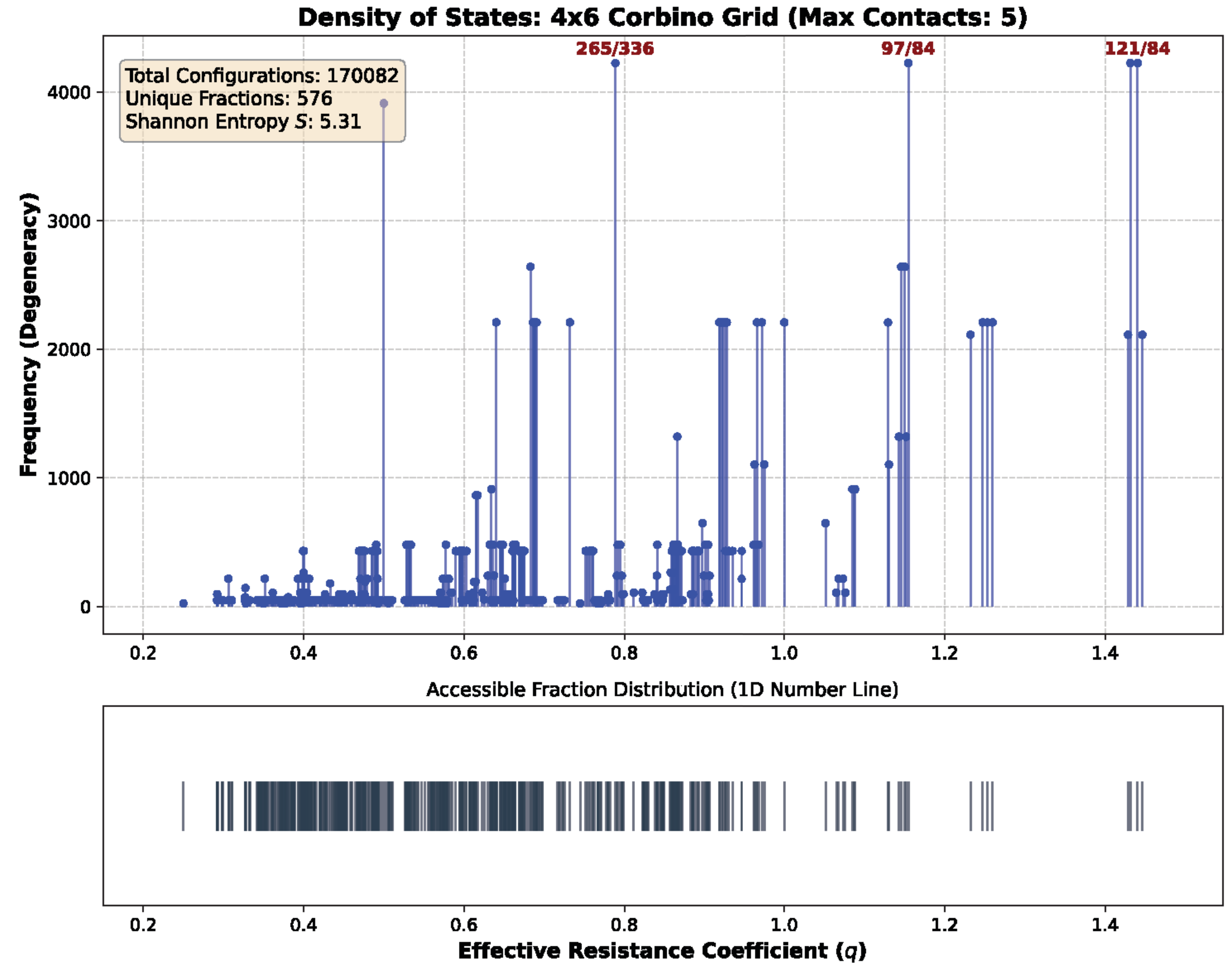
Density of States: 4x6 Corbino Grid (Max Contacts: 5)
Total Configurations: 170082
Unique Fractions: 576
Shannon Entropy S: 5.31
265/336
97/84
121/84
Frequency (Degeneracy)
4000
3000
2000
1000
0
0.2
0.4
0.6
0.8
1.0
1.2
1.4
Accessible Fraction Distribution (1D Number Line)
0.2
0.4
0.6
0.8
1.0
1.2
1.4
Effective Resistance Coefficient (q)

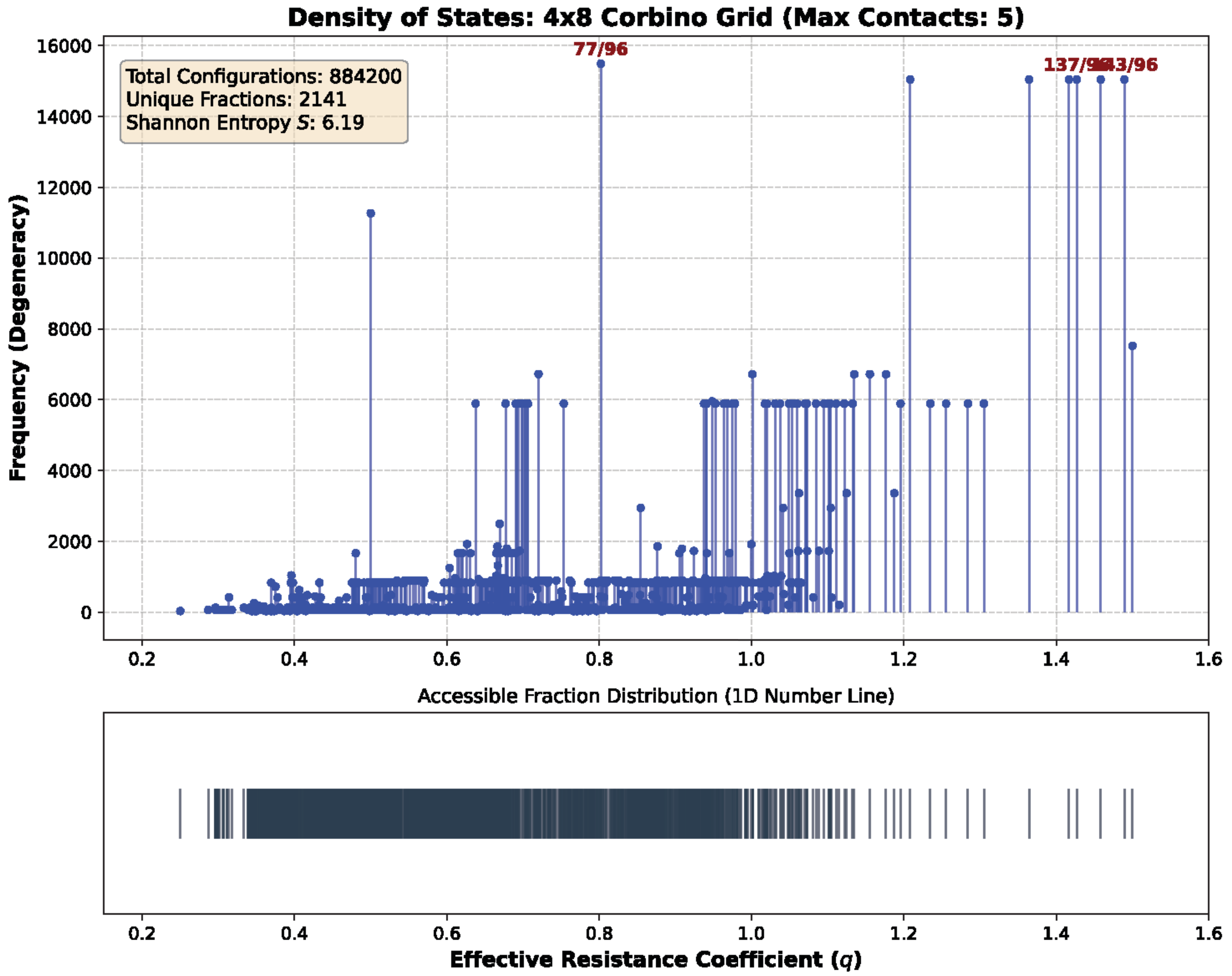

Density of States: 4x8 Corbino Grid (Max Contacts: 5)
Total Configurations: 884200
Unique Fractions: 2141
Shannon Entropy S: 6.19
77/96
137/96
143/96
Frequency (Degeneracy)
16000
14000
12000
10000
8000
6000
4000
2000
0
0.2
0.4
0.6
0.8
1.0
1.2
1.4
1.6
Accessible Fraction Distribution (1D Number Line)
Effective Resistance Coefficient (q)

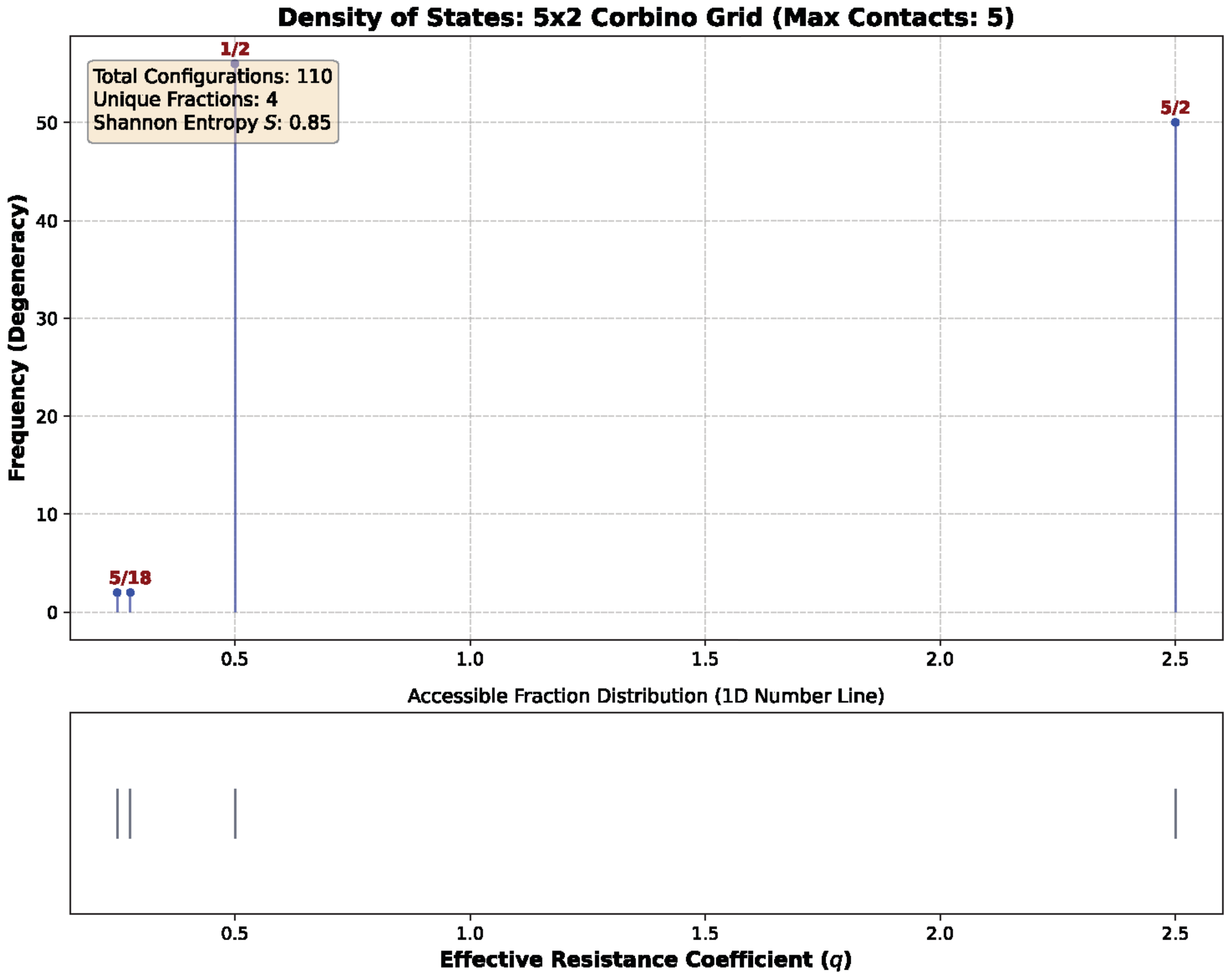
Density of States: 5x2 Corbino Grid (Max Contacts: 5)
Total Configurations: 110
Unique Fractions: 4
Shannon Entropy S: 0.85
1/2
5/2
5/18
Frequency (Degeneracy)
0
10
20
30
40
50
0.5
1.0
1.5
2.0
2.5
Accessible Fraction Distribution (1D Number Line)
Effective Resistance Coefficient (q)

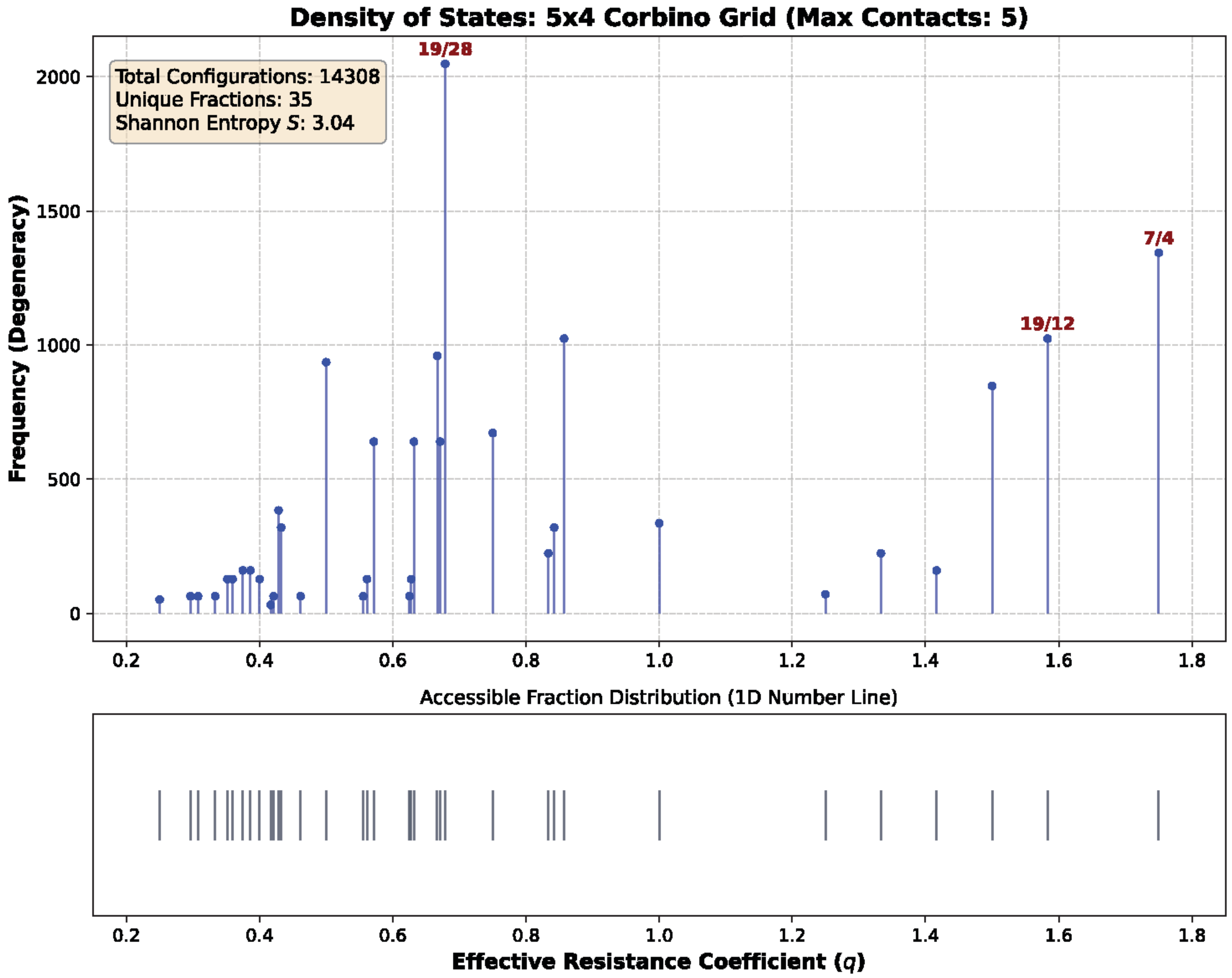
Density of States: 5x4 Corbino Grid (Max Contacts: 5)
Total Configurations: 14308
Unique Fractions: 35
Shannon Entropy S: 3.04
19/28
7/4
19/12
Frequency (Degeneracy)
0
500
1000
1500
2000
0.2
0.4
0.6
0.8
1.0
1.2
1.4
1.6
1.8
Accessible Fraction Distribution (1D Number Line)
Effective Resistance Coefficient (q)

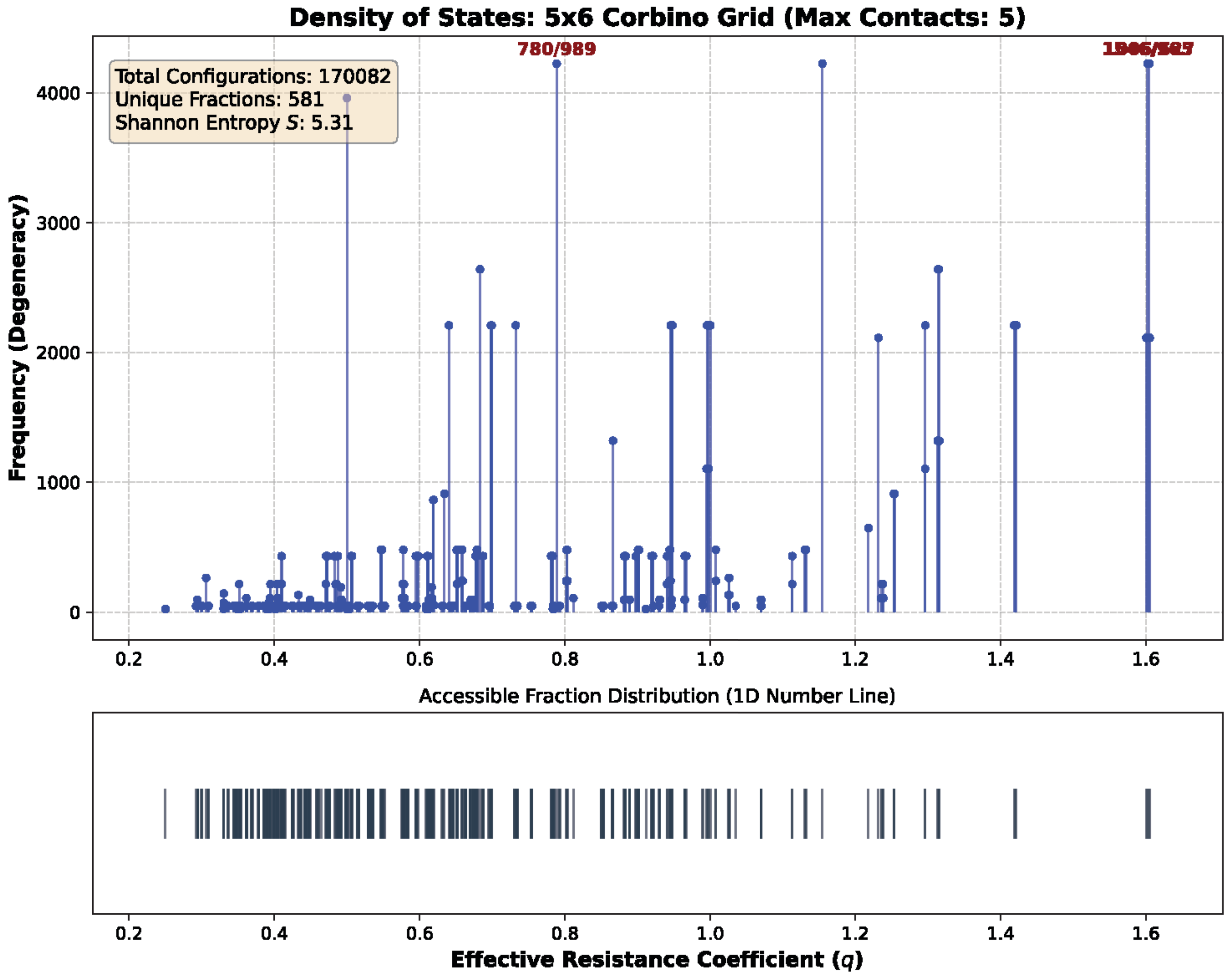

Density of States: 5x6 Corbino Grid (Max Contacts: 5)
780/989
Total Configurations: 170082
Unique Fractions: 581
Shannon Entropy S: 5.31
Frequency (Degeneracy)
4000
3000
2000
1000
0
0.2
0.4
0.6
0.8
1.0
1.2
1.4
1.6
Accessible Fraction Distribution (1D Number Line)
Effective Resistance Coefficient (q)

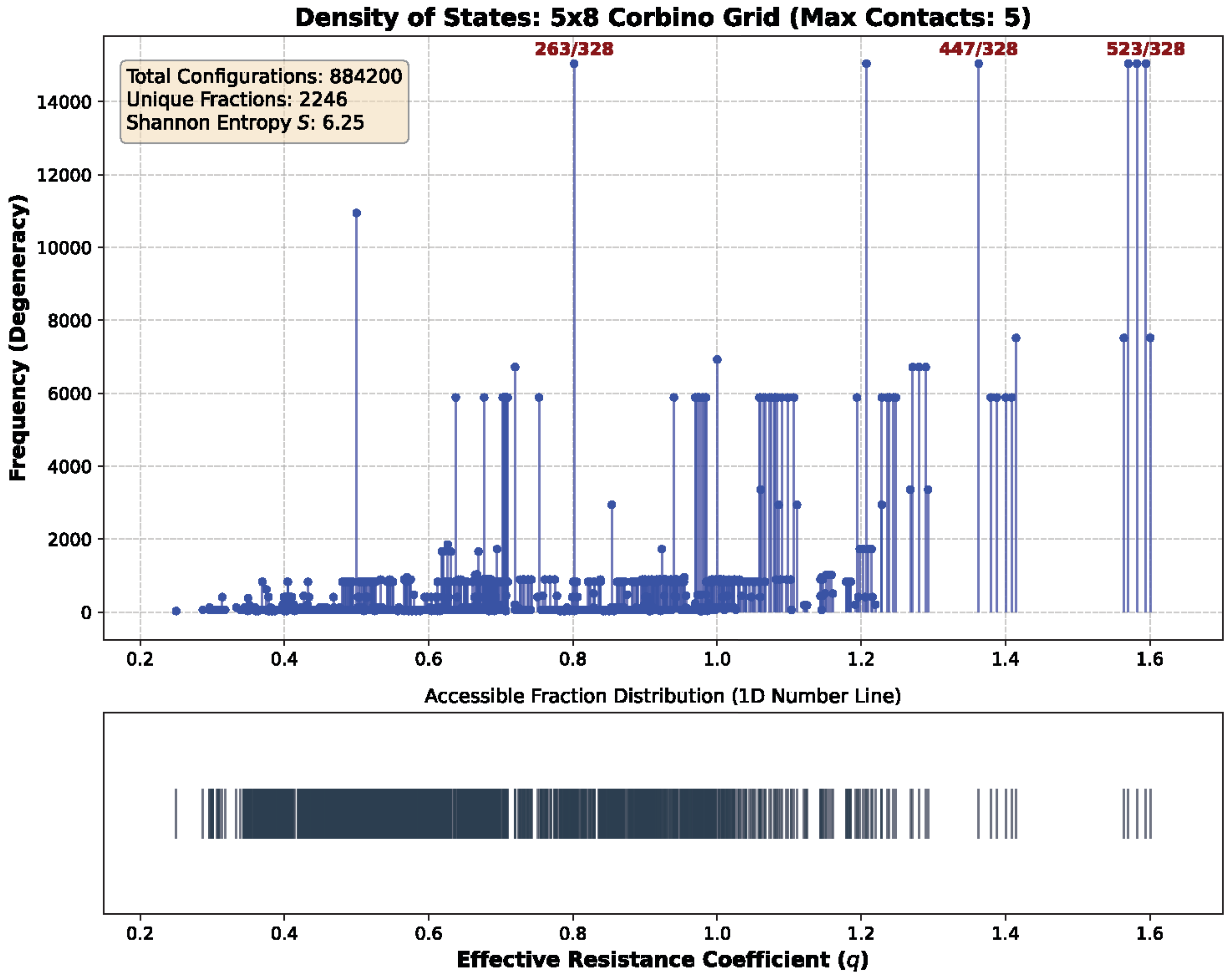

Density of States: 5x8 Corbino Grid (Max Contacts: 5)
Total Configurations: 884200
Unique Fractions: 2246
Shannon Entropy S: 6.25
263/328
447/328
523/328
Frequency (Degeneracy)
0
2000
4000
6000
8000
10000
12000
14000
0.2
0.4
0.6
0.8
1.0
1.2
1.4
1.6
Accessible Fraction Distribution (1D Number Line)
Effective Resistance Coefficient (q)

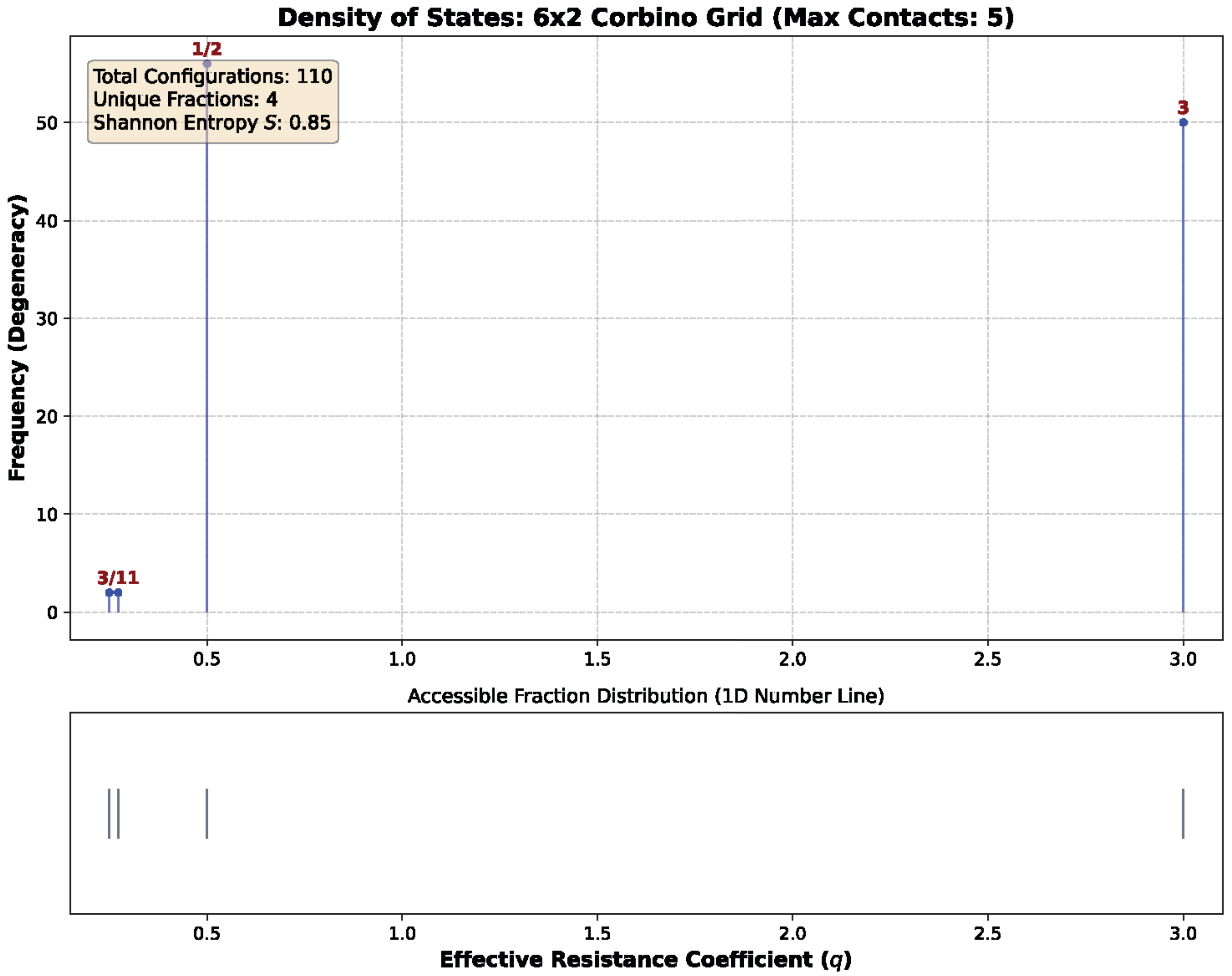

Density of States: 6x2 Corbino Grid (Max Contacts: 5)
Total Configurations: 110
Unique Fractions: 4
Shannon Entropy S: 0.85
1/2
3
3/11
Frequency (Degeneracy)
0
10
20
30
40
50
0.5
1.0
1.5
2.0
2.5
3.0
Accessible Fraction Distribution (1D Number Line)
Effective Resistance Coefficient (q)

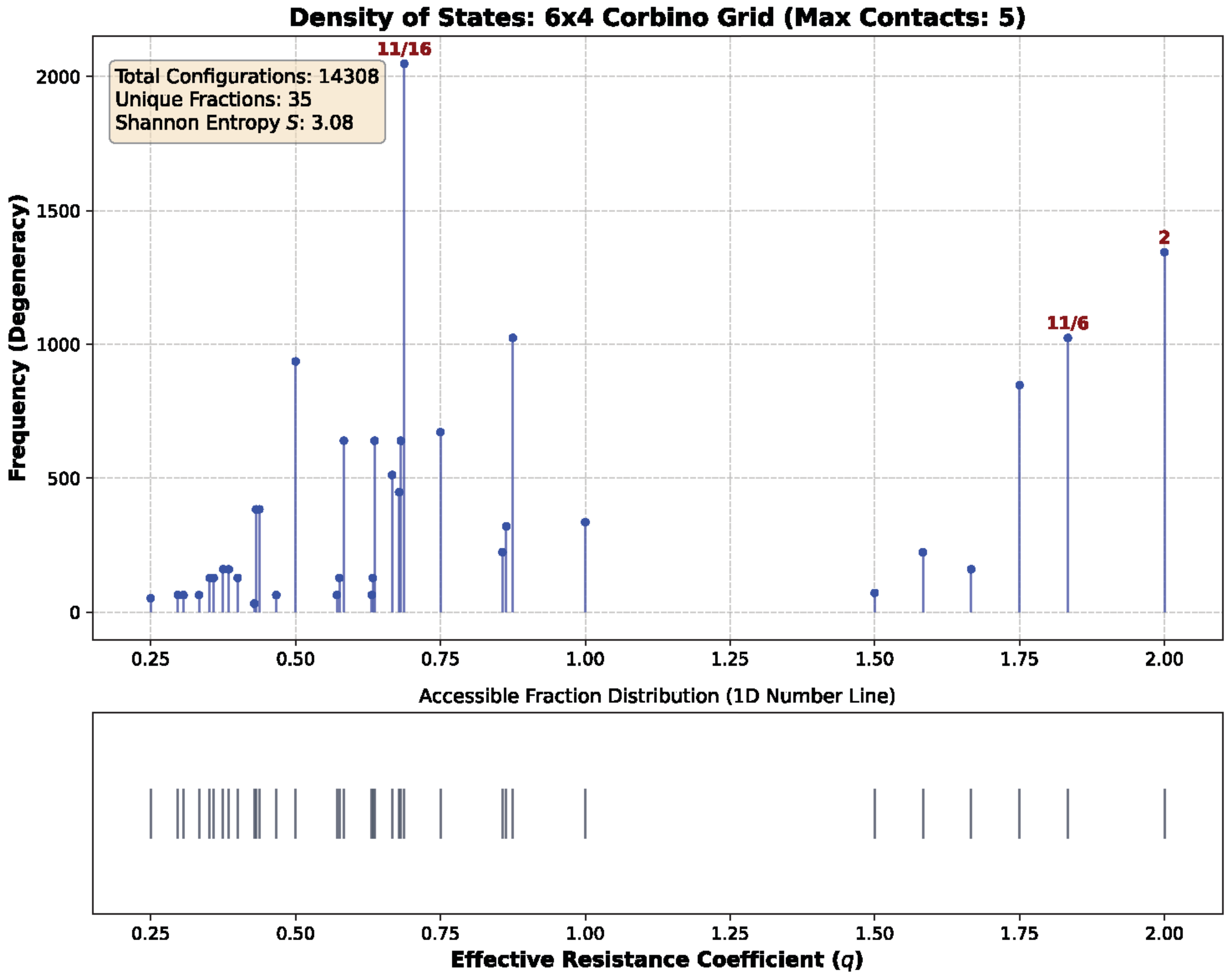

Density of States: 6x4 Corbino Grid (Max Contacts: 5)
Total Configurations: 14308
Unique Fractions: 35
Shannon Entropy S: 3.08
11/16
2
11/6
Frequency (Degeneracy)
0
500
1000
1500
2000
0.25
0.50
0.75
1.00
1.25
1.50
1.75
2.00
Accessible Fraction Distribution (1D Number Line)
Effective Resistance Coefficient (q)

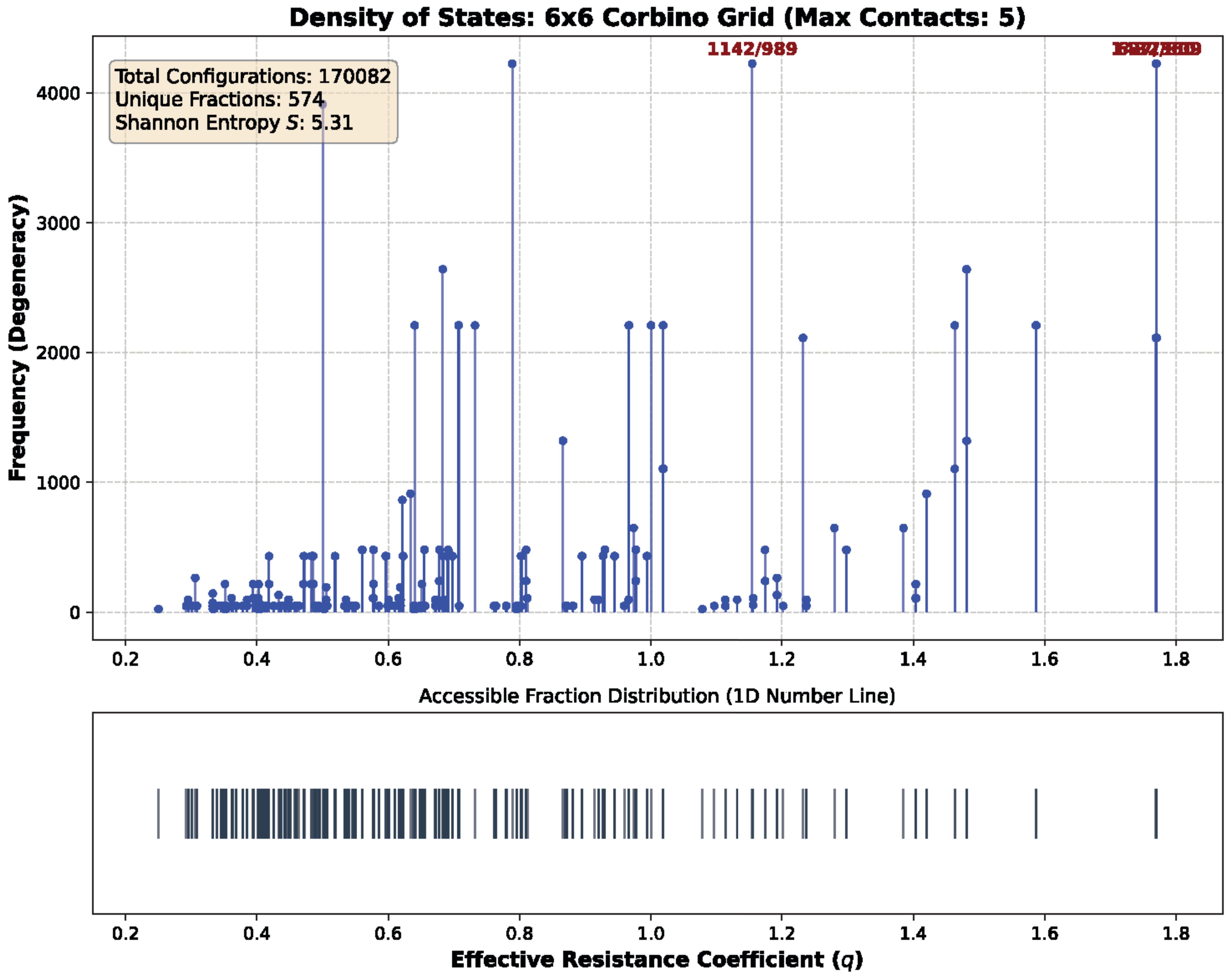
Density of States: 6x6 Corbino Grid (Max Contacts: 5)
1142/989
Total Configurations: 170082
Unique Fractions: 574
Shannon Entropy S: 5.31
Frequency (Degeneracy)
Accessible Fraction Distribution (1D Number Line)
Effective Resistance Coefficient (q)

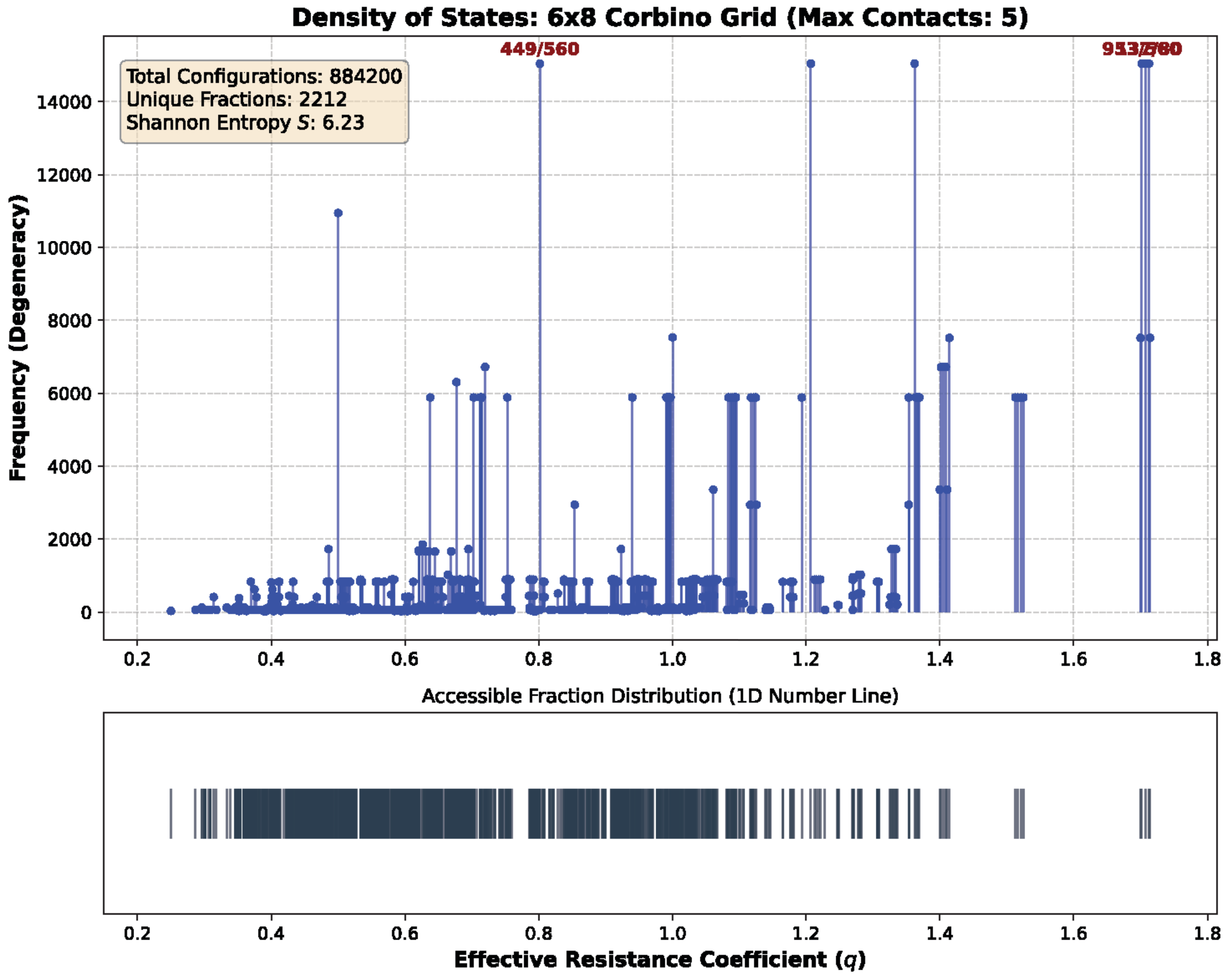
Density of States: 6x8 Corbino Grid (Max Contacts: 5)
449/560
953/560
Total Configurations: 884200
Unique Fractions: 2212
Shannon Entropy S: 6.23
Frequency (Degeneracy)
0
2000
4000
6000
8000
10000
12000
14000
0.2
0.4
0.6
0.8
1.0
1.2
1.4
1.6
1.8
Accessible Fraction Distribution (1D Number Line)
Effective Resistance Coefficient (q)

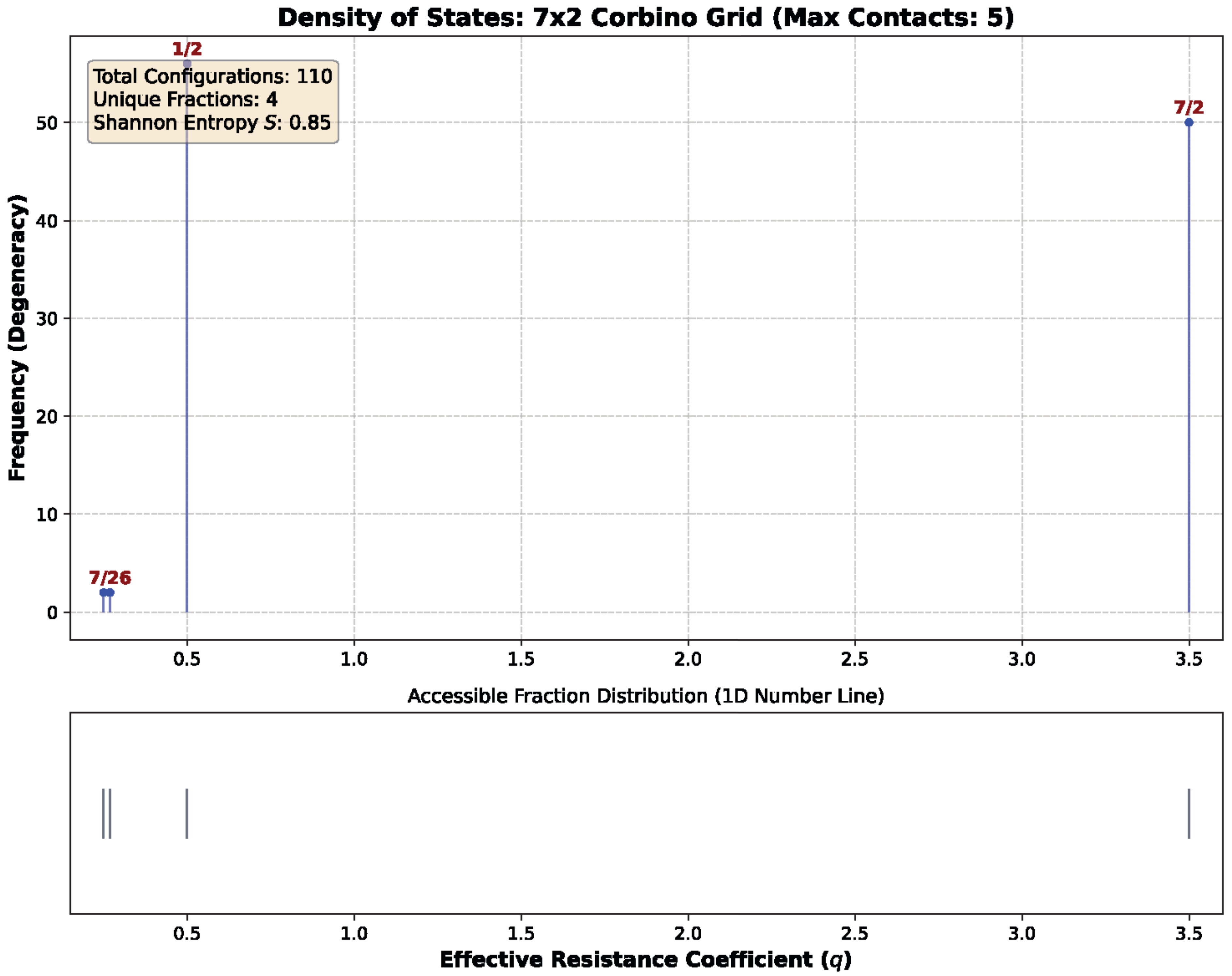

Density of States: 7x2 Corbino Grid (Max Contacts: 5)
Total Configurations: 110
Unique Fractions: 4
Shannon Entropy S: 0.85
1/2
7/2
7/26
Frequency (Degeneracy)
0
10
20
30
40
50
0.5
1.0
1.5
2.0
2.5
3.0
3.5
Accessible Fraction Distribution (1D Number Line)
Effective Resistance Coefficient (q)

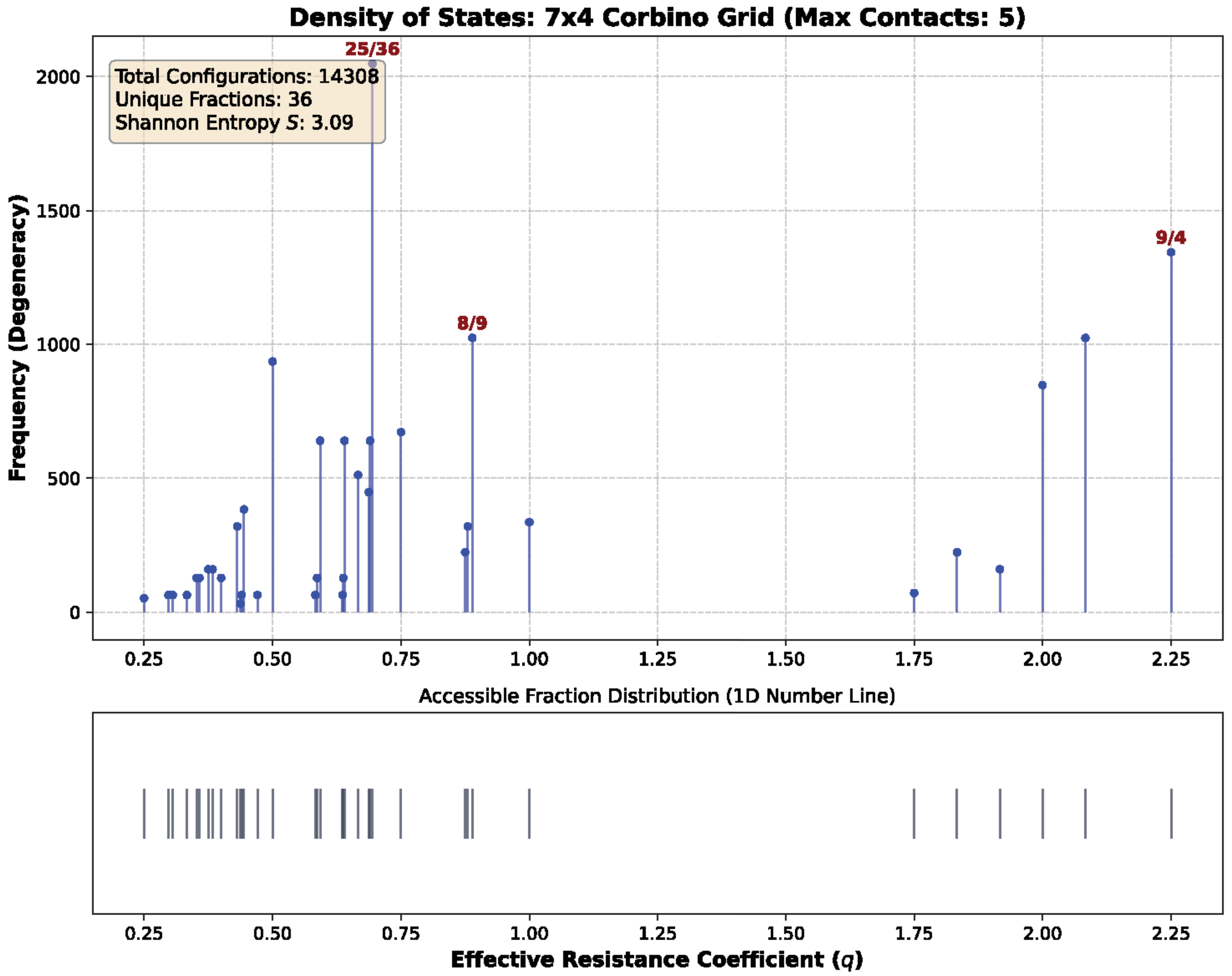
Density of States: 7x4 Corbino Grid (Max Contacts: 5)
Total Configurations: 14308
Unique Fractions: 36
Shannon Entropy S: 3.09
25/36
8/9
9/4
Frequency (Degeneracy)
0
500
1000
1500
2000
0.25
0.50
0.75
1.00
1.25
1.50
1.75
2.00
2.25
Accessible Fraction Distribution (1D Number Line)
Effective Resistance Coefficient (q)

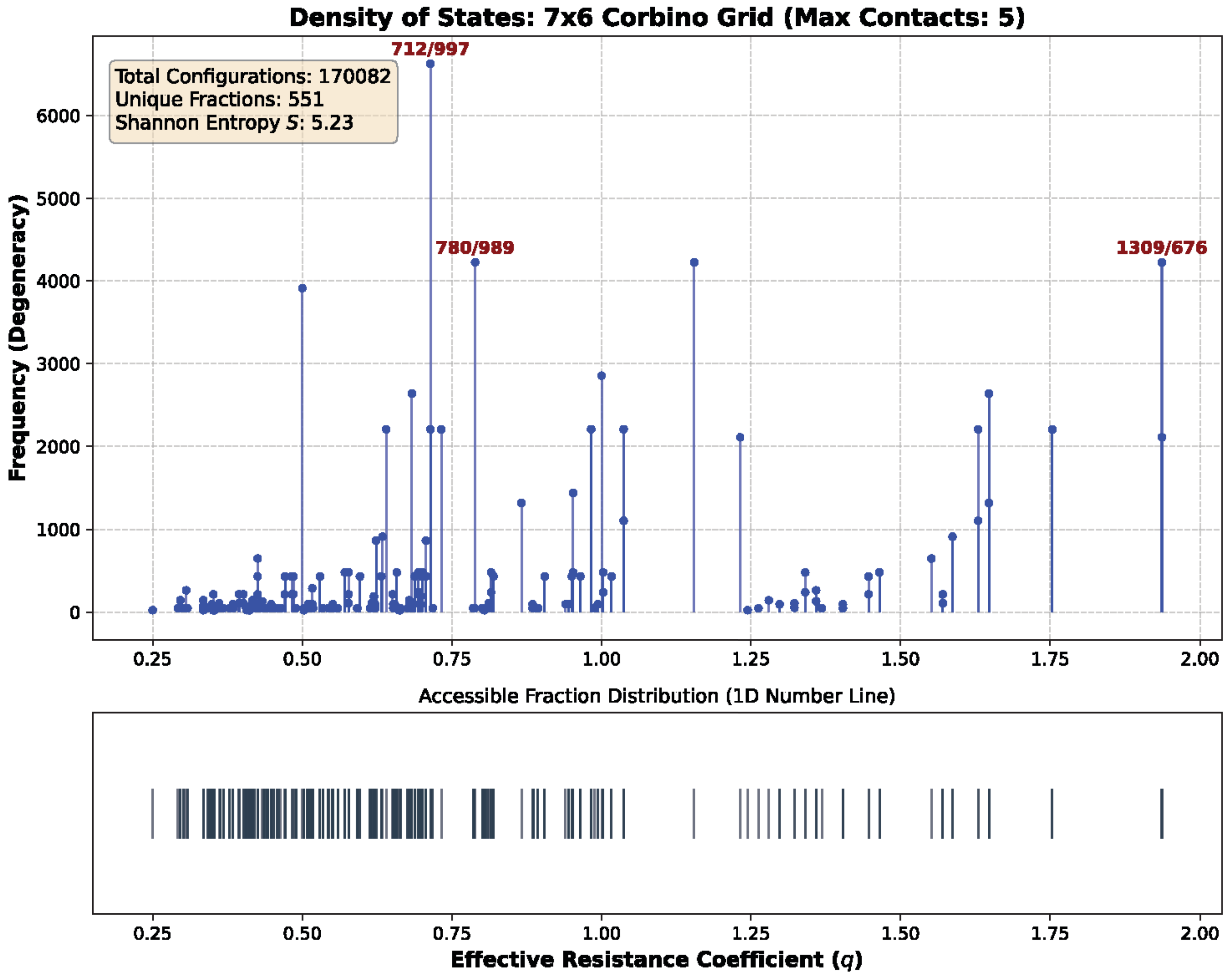

Density of States: 7x6 Corbino Grid (Max Contacts: 5)
Total Configurations: 170082
Unique Fractions: 551
Shannon Entropy S: 5.23
712/997
780/989
1309/676
Frequency (Degeneracy)
0
1000
2000
3000
4000
5000
6000
0.25
0.50
0.75
1.00
1.25
1.50
1.75
2.00
Accessible Fraction Distribution (1D Number Line)
Effective Resistance Coefficient (q)

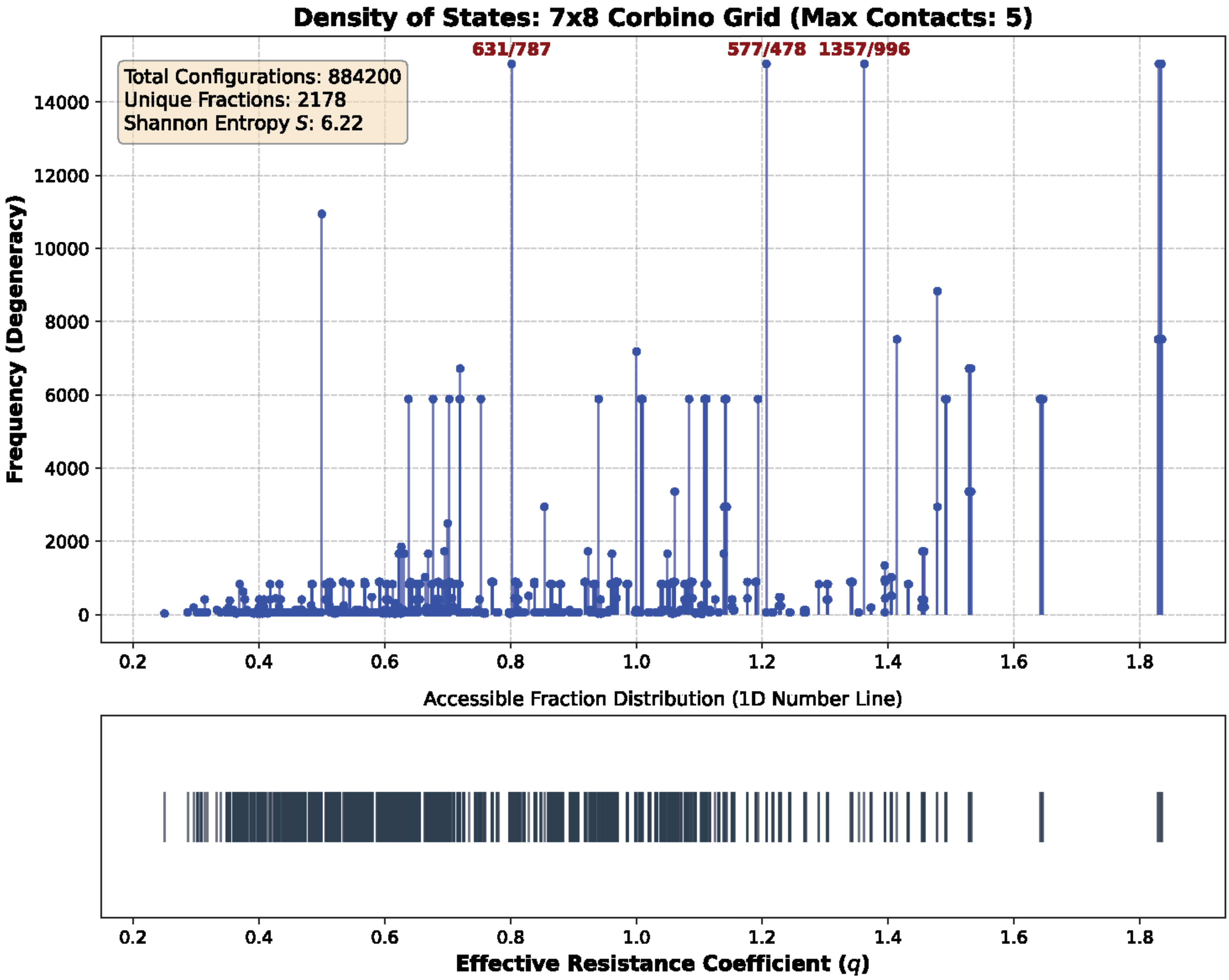

Density of States: 7x8 Corbino Grid (Max Contacts: 5)
Total Configurations: 884200
Unique Fractions: 2178
Shannon Entropy S: 6.22
631/787
577/478
1357/996
Frequency (Degeneracy)
Accessible Fraction Distribution (1D Number Line)
Effective Resistance Coefficient (q)

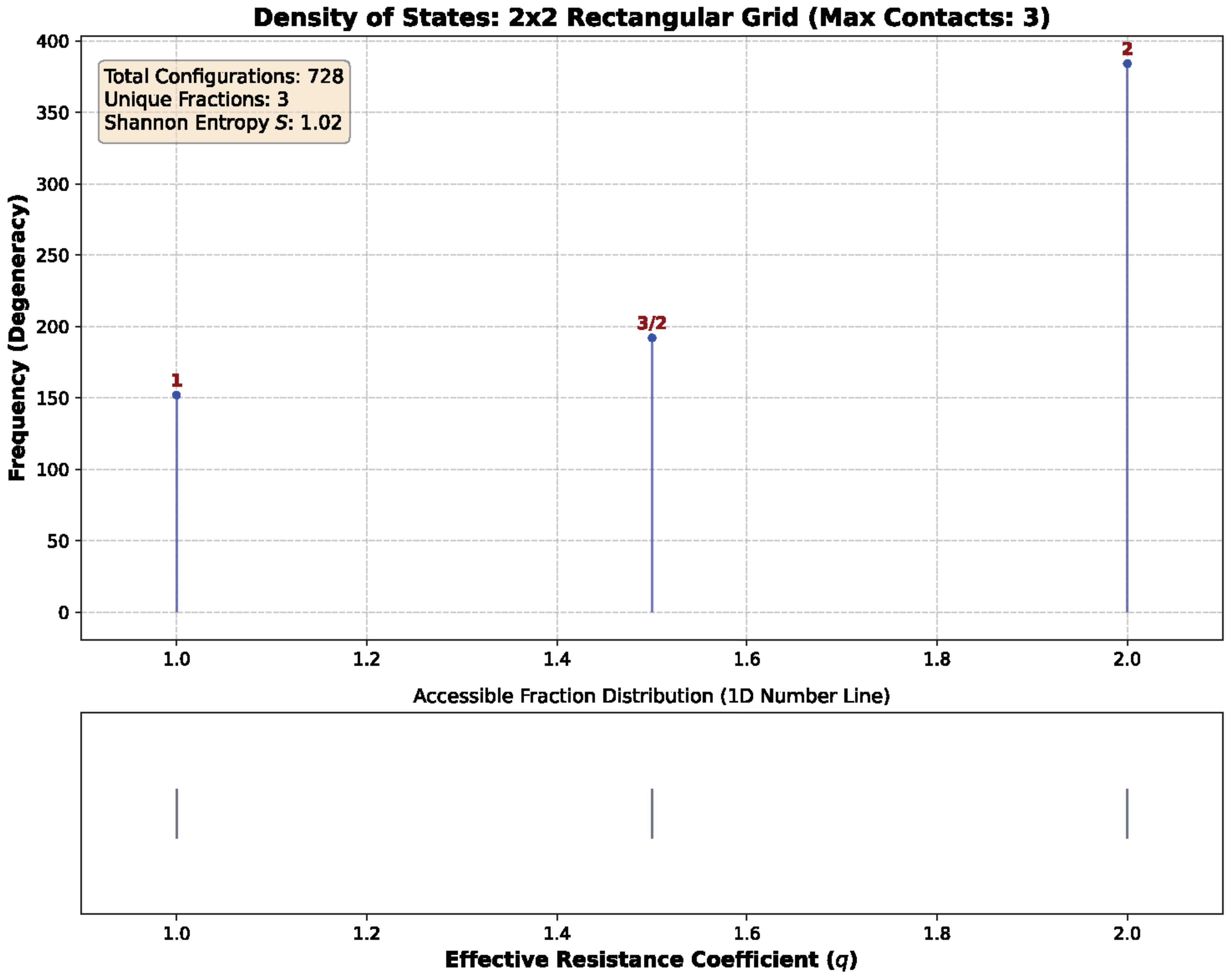
Density of States: 2x2 Rectangular Grid (Max Contacts: 3)
Total Configurations: 728
Unique Fractions: 3
Shannon Entropy S: 1.02
Frequency (Degeneracy)
1
3/2
2
0
50
100
150
200
250
300
350
400
1.0
1.2
1.4
1.6
1.8
2.0
Accessible Fraction Distribution (1D Number Line)
Effective Resistance Coefficient (q)

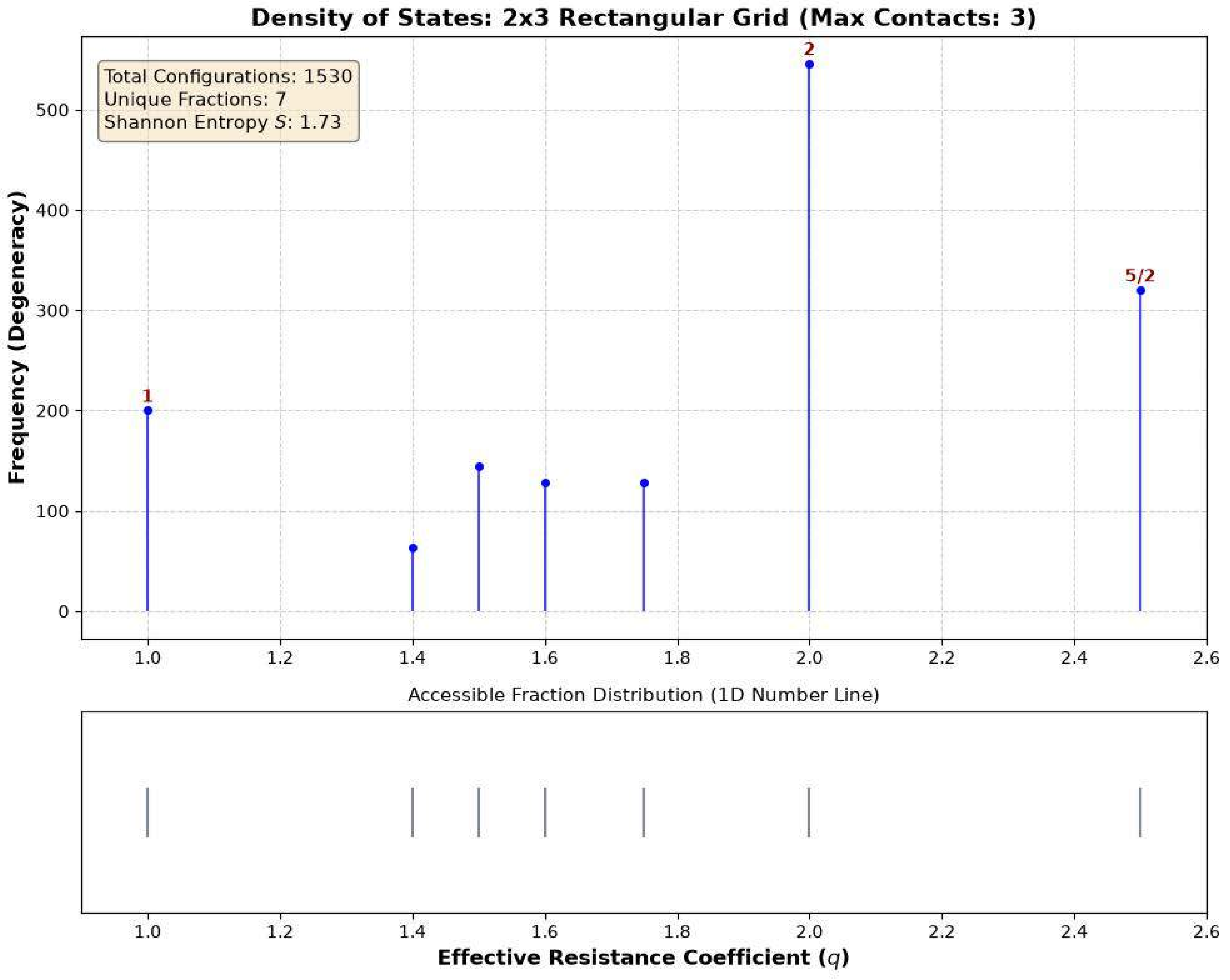
Density of States: 2x3 Rectangular Grid (Max Contacts: 3)
Total Configurations: 1530
Unique Fractions: 7
Shannon Entropy S: 1.73
Frequency (Degeneracy)
1
2
5/2
0
100
200
300
400
500
1.0
1.2
1.4
1.6
1.8
2.0
2.2
2.4
2.6
Accessible Fraction Distribution (1D Number Line)
Effective Resistance Coefficient (q)

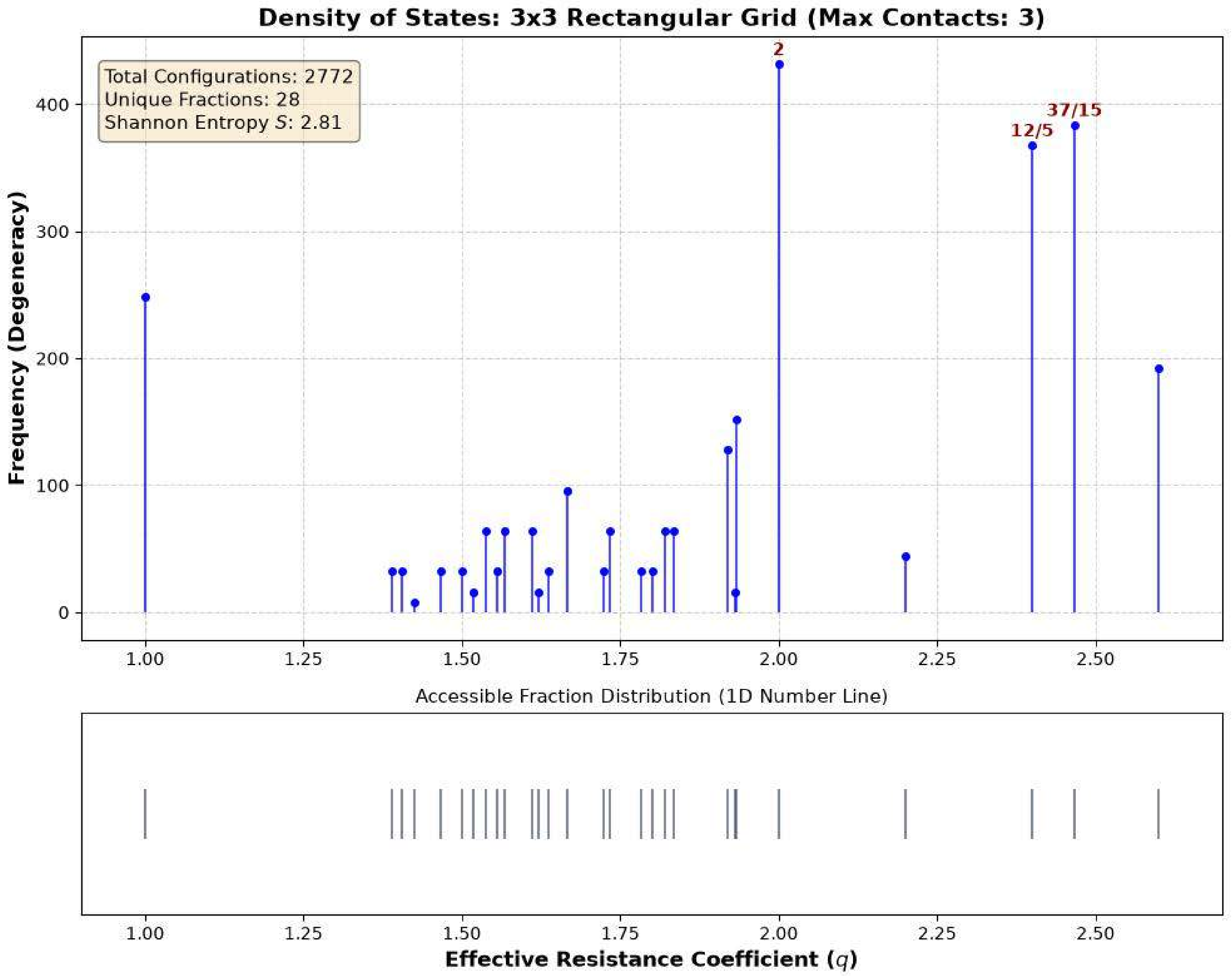

Density of States: 3x3 Rectangular Grid (Max Contacts: 3)
Total Configurations: 2772
Unique Fractions: 28
Shannon Entropy S: 2.81
2
37/15
12/5
Frequency (Degeneracy)
0
100
200
300
400
1.00
1.25
1.50
1.75
2.00
2.25
2.50
Accessible Fraction Distribution (1D Number Line)
Effective Resistance Coefficient (q)

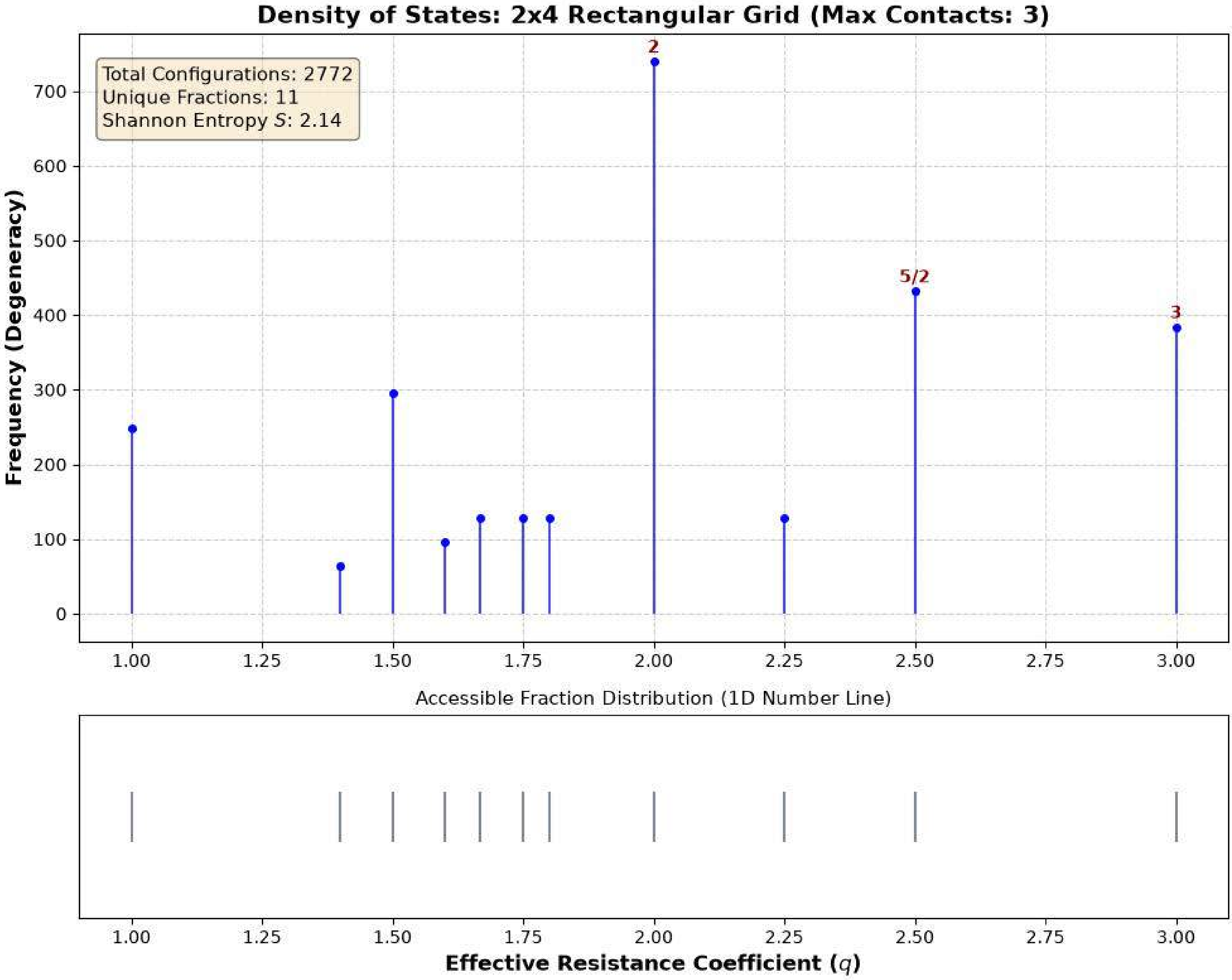
Density of States: 2x4 Rectangular Grid (Max Contacts: 3)
Total Configurations: 2772
Unique Fractions: 11
Shannon Entropy S: 2.14
2
5/2
3
Frequency (Degeneracy)
0
100
200
300
400
500
600
700
1.00
1.25
1.50
1.75
2.00
2.25
2.50
2.75
3.00
Accessible Fraction Distribution (1D Number Line)
Effective Resistance Coefficient (q)

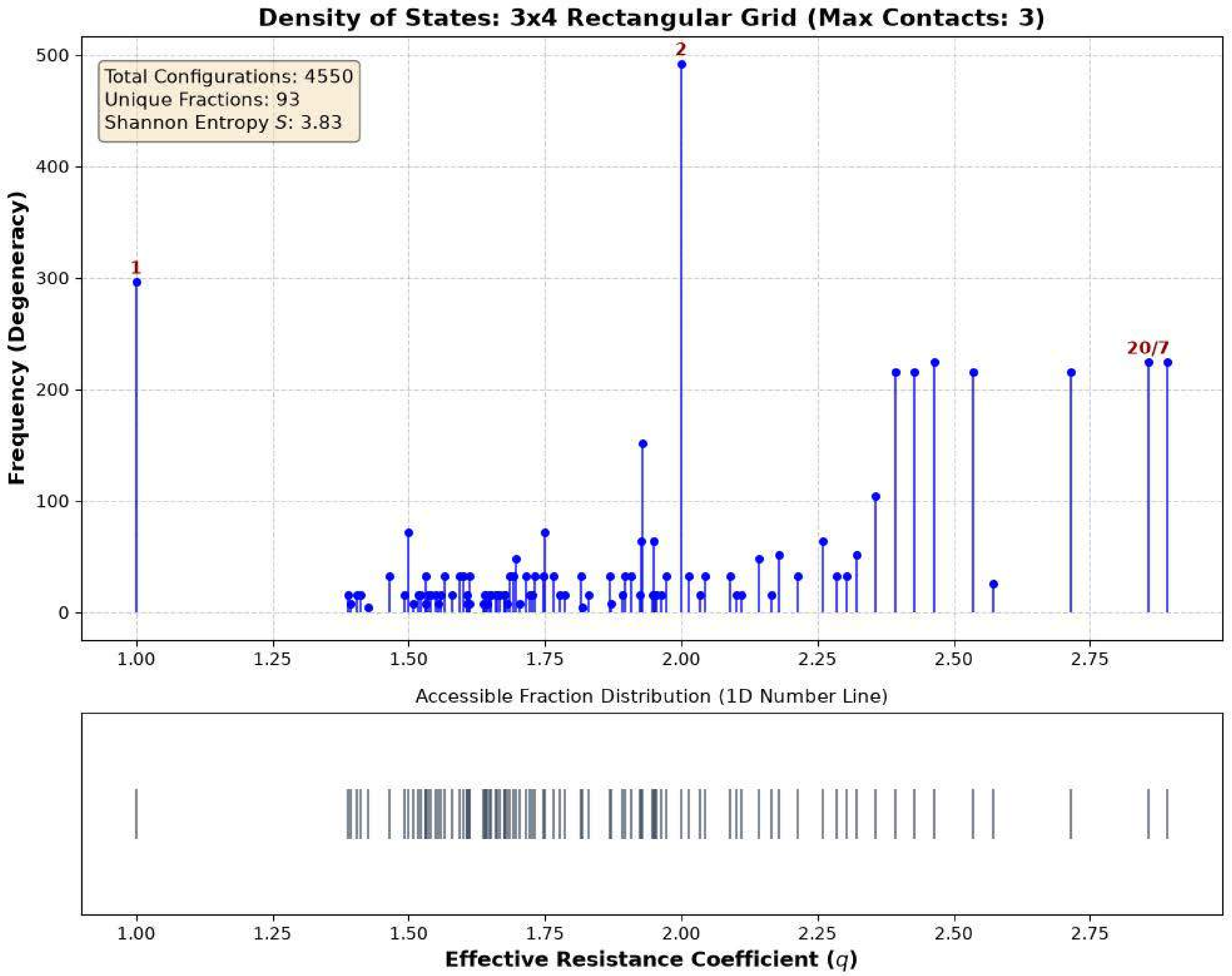

Density of States: 3x4 Rectangular Grid (Max Contacts: 3)
Total Configurations: 4550
Unique Fractions: 93
Shannon Entropy S: 3.83
Frequency (Degeneracy)
1
2
20/7
Accessible Fraction Distribution (1D Number Line)
Effective Resistance Coefficient (q)

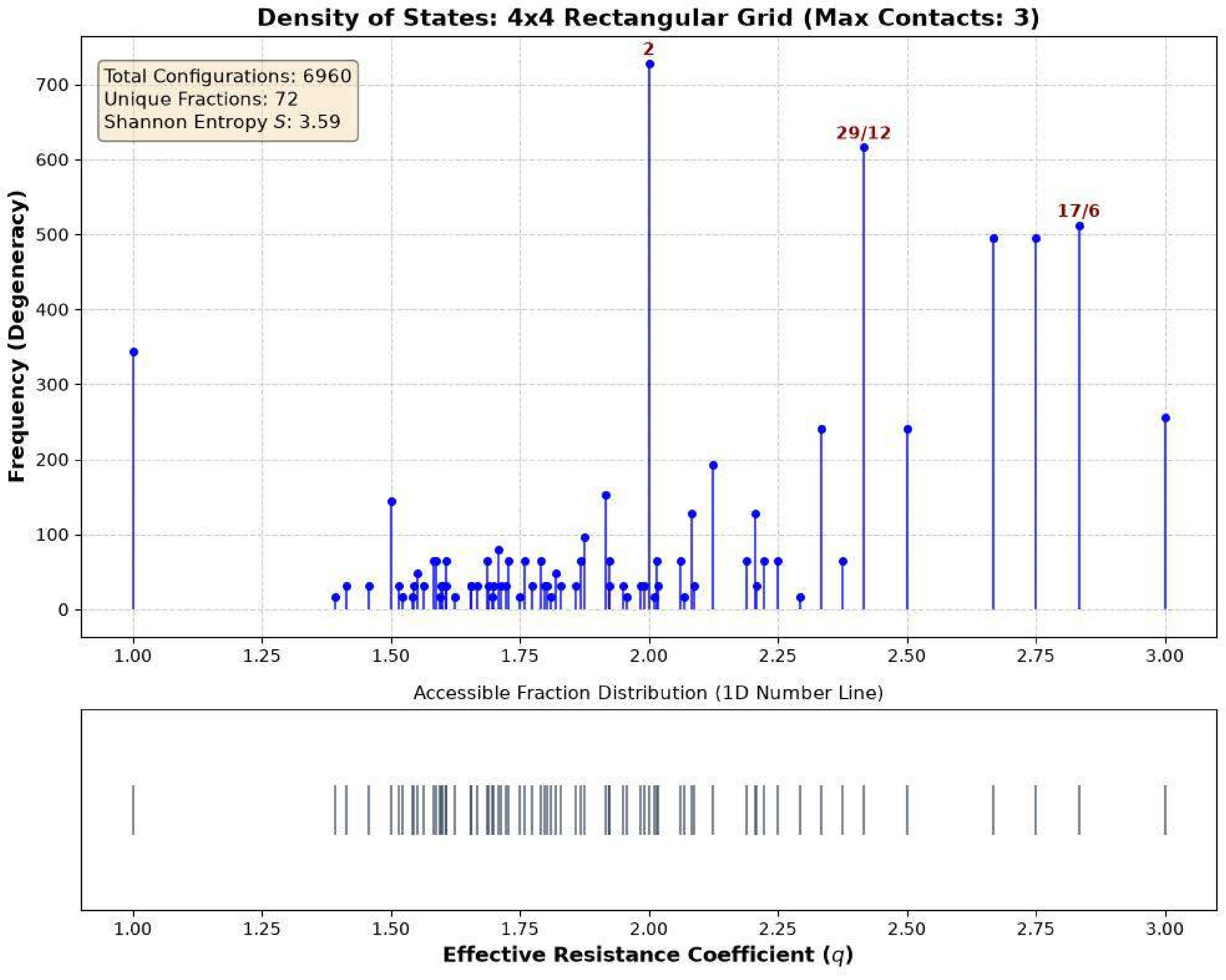

Density of States: 4x4 Rectangular Grid (Max Contacts: 3)
Total Configurations: 6960
Unique Fractions: 72
Shannon Entropy S: 3.59
2
29/12
17/6
Frequency (Degeneracy)
0
100
200
300
400
500
600
700
1.00
1.25
1.50
1.75
2.00
2.25
2.50
2.75
3.00
Accessible Fraction Distribution (1D Number Line)
Effective Resistance Coefficient (q)

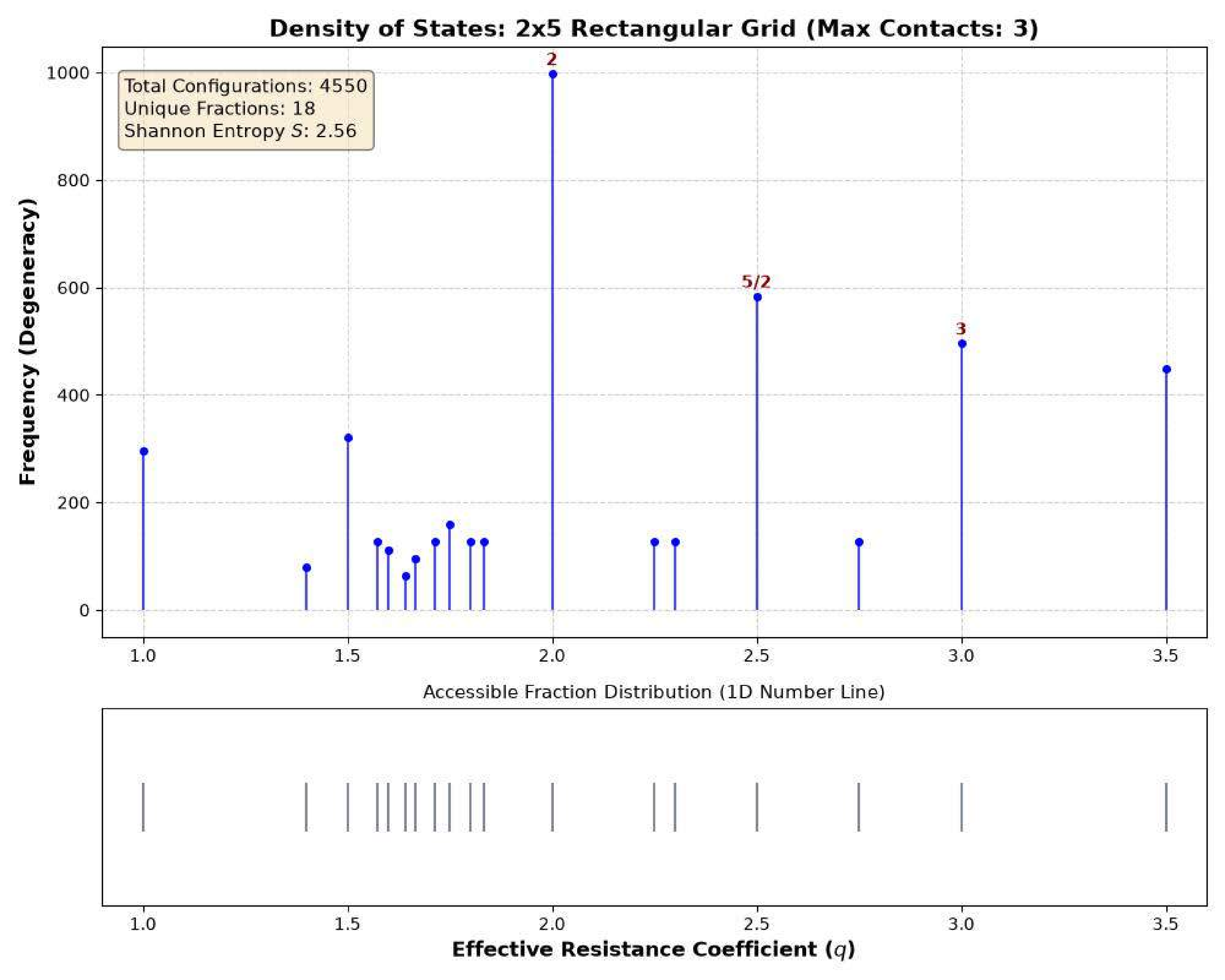

Density of States: 2x5 Rectangular Grid (Max Contacts: 3)
Total Configurations: 4550
Unique Fractions: 18
Shannon Entropy S: 2.56
2
5/2
3
Frequency (Degeneracy)
0
200
400
600
800
1000
1.0
1.5
2.0
2.5
3.0
3.5
Accessible Fraction Distribution (1D Number Line)
Effective Resistance Coefficient (q)

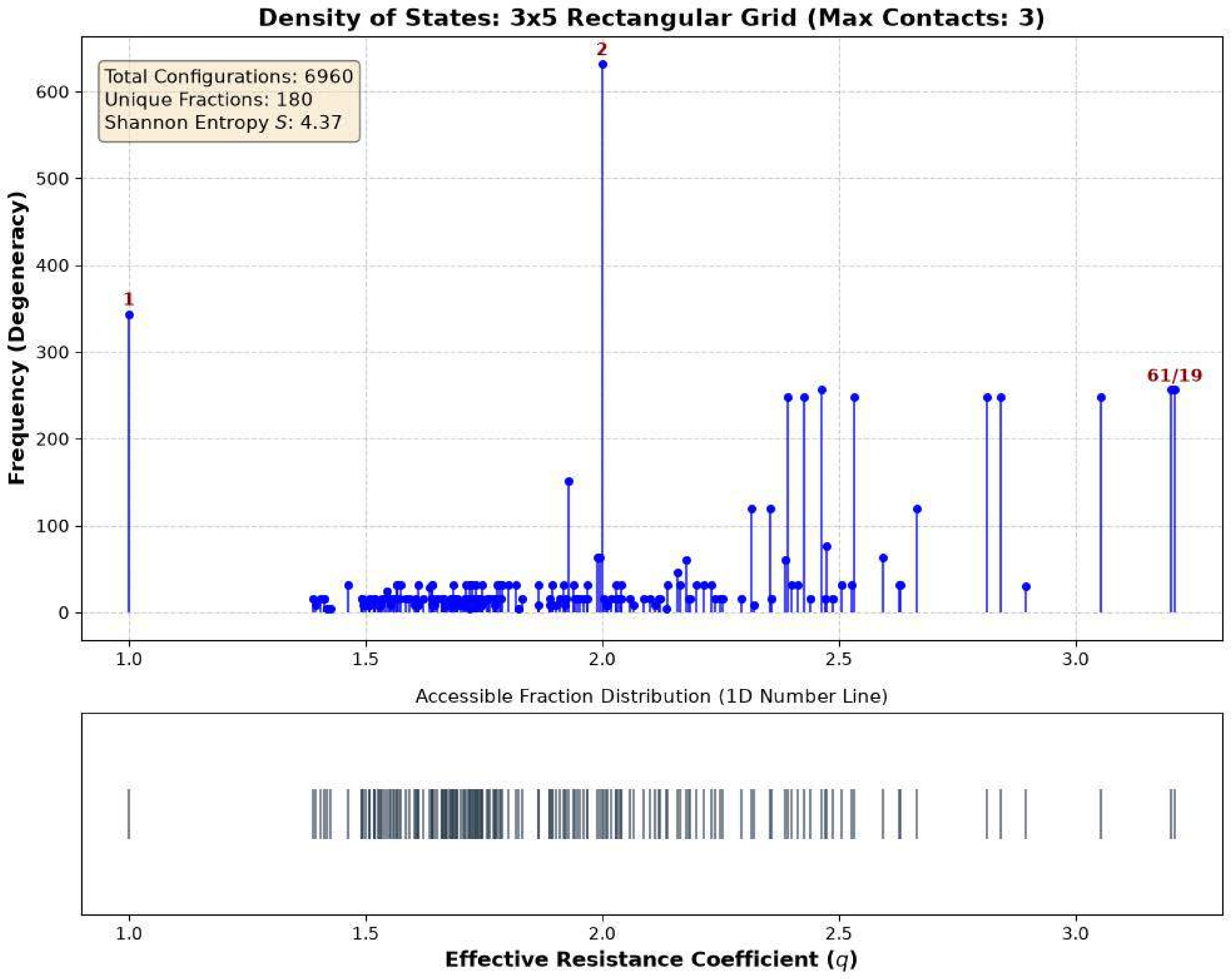

Density of States: 3x5 Rectangular Grid (Max Contacts: 3)
Total Configurations: 6960
Unique Fractions: 180
Shannon Entropy S: 4.37
Frequency (Degeneracy)
1
2
61/19
Accessible Fraction Distribution (1D Number Line)
Effective Resistance Coefficient (q)

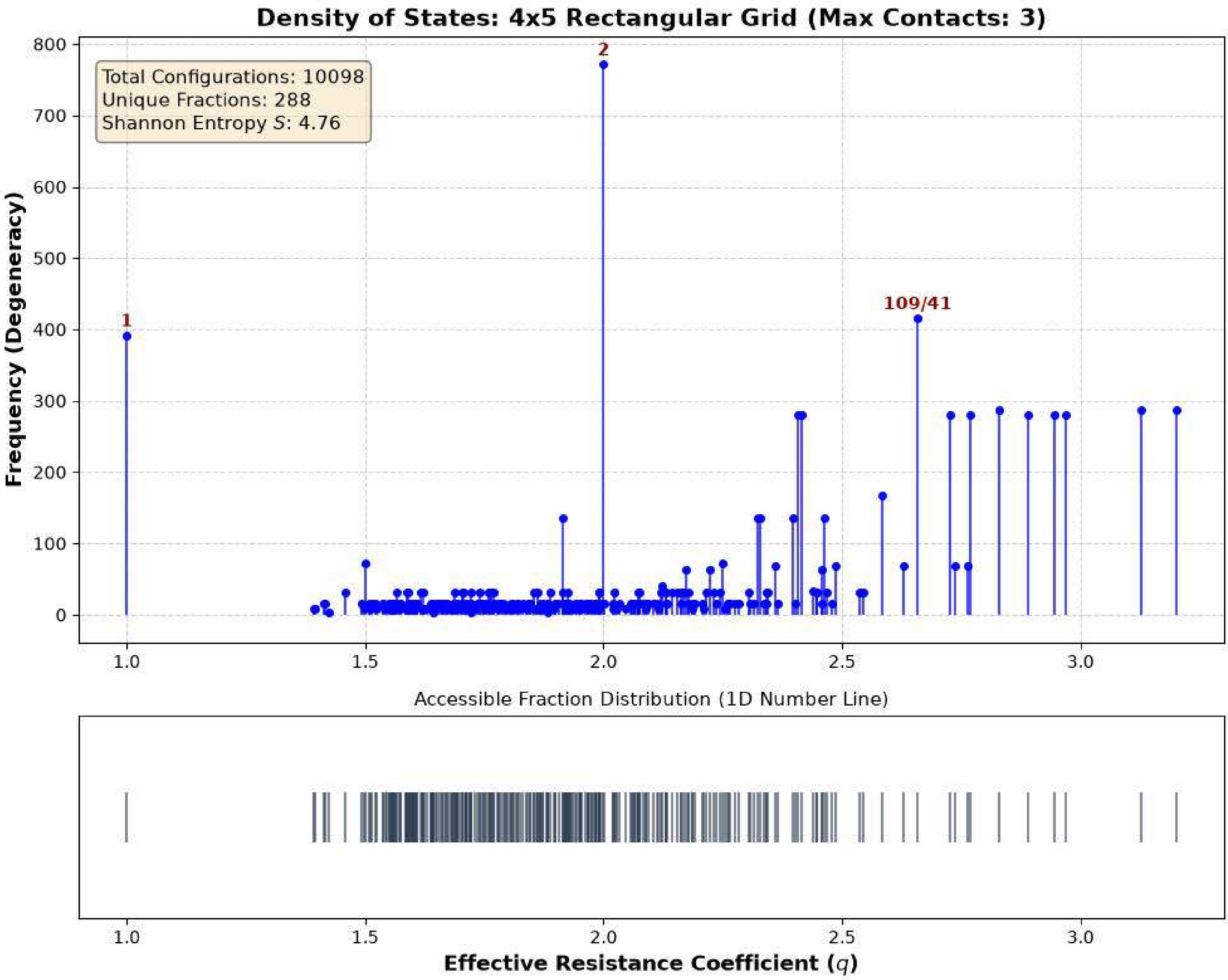

Density of States: 4x5 Rectangular Grid (Max Contacts: 3)
Total Configurations: 10098
Unique Fractions: 288
Shannon Entropy S: 4.76
Frequency (Degeneracy)
1
2
109/41
0
100
200
300
400
500
600
700
800
1.0
1.5
2.0
2.5
3.0
Accessible Fraction Distribution (1D Number Line)
Effective Resistance Coefficient (q)

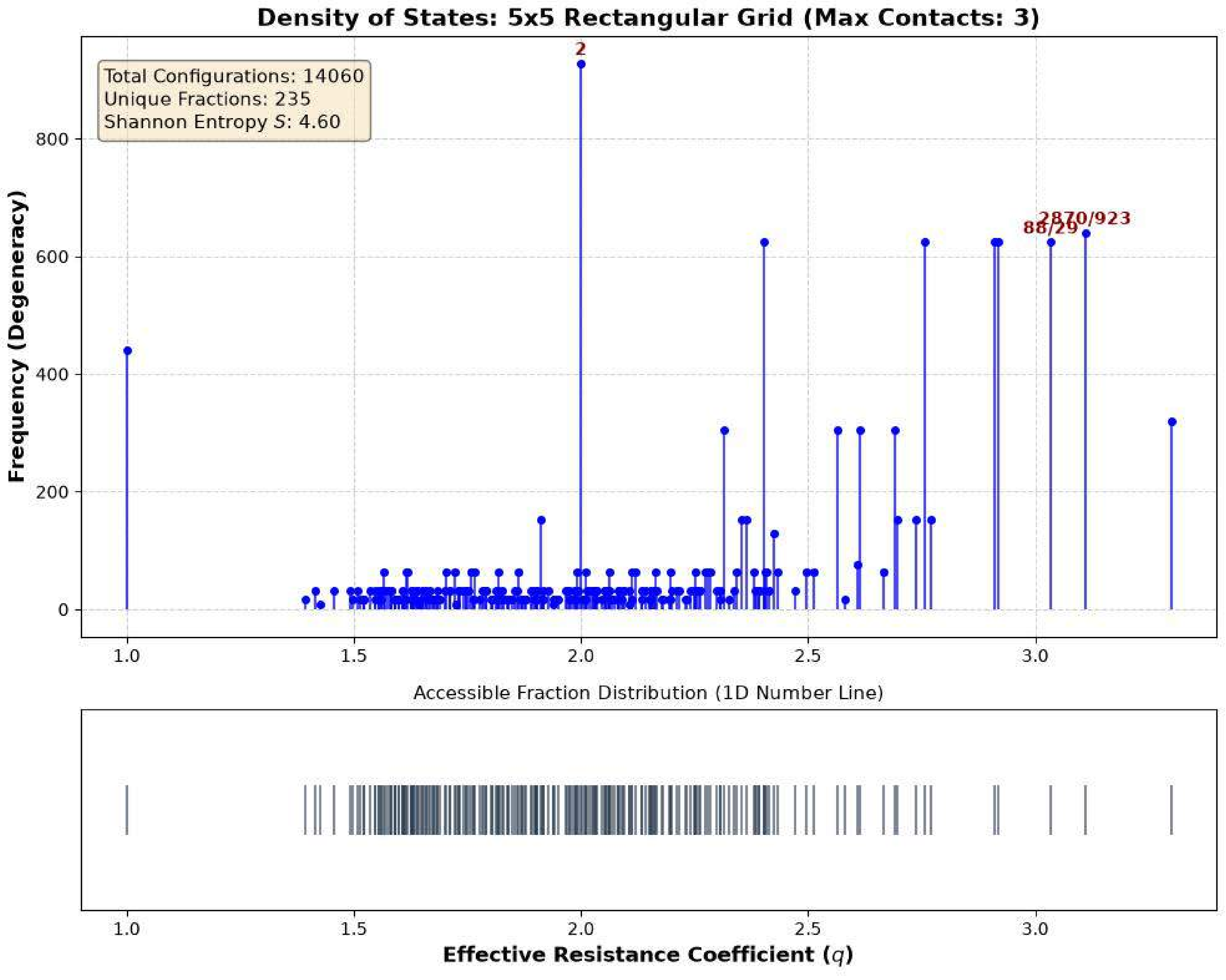
Density of States: 5x5 Rectangular Grid (Max Contacts: 3)
Total Configurations: 14060
Unique Fractions: 235
Shannon Entropy S: 4.60
2
88/29
2870/923
Frequency (Degeneracy)
0
200
400
600
800
1.0
1.5
2.0
2.5
3.0
Accessible Fraction Distribution (1D Number Line)
Effective Resistance Coefficient (q)

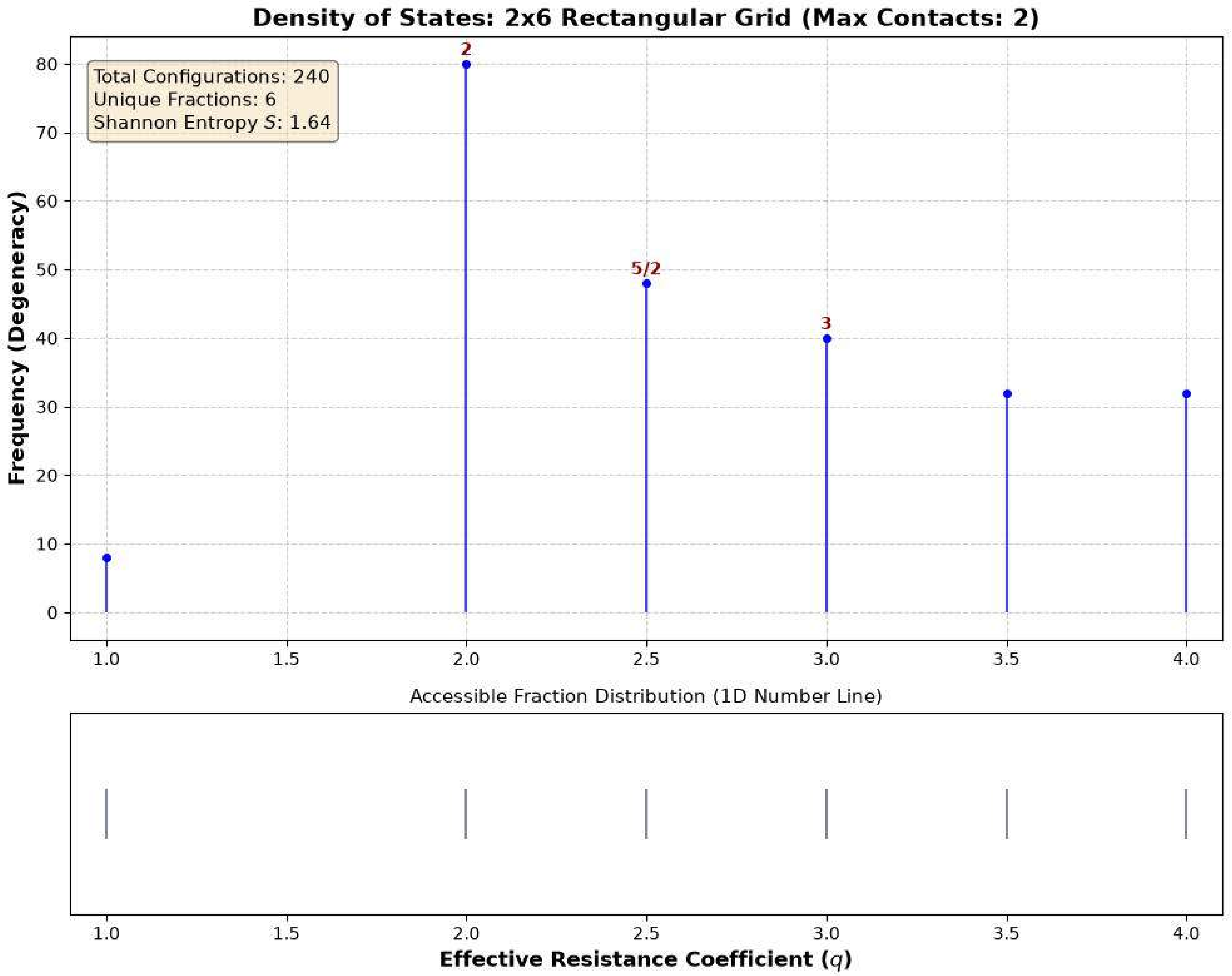

Density of States: 2x6 Rectangular Grid (Max Contacts: 2)
Total Configurations: 240
Unique Fractions: 6
Shannon Entropy S: 1.64
2
5/2
3
Frequency (Degeneracy)
0
10
20
30
40
50
60
70
80
1.0
1.5
2.0
2.5
3.0
3.5
4.0
Accessible Fraction Distribution (1D Number Line)
Effective Resistance Coefficient (q)

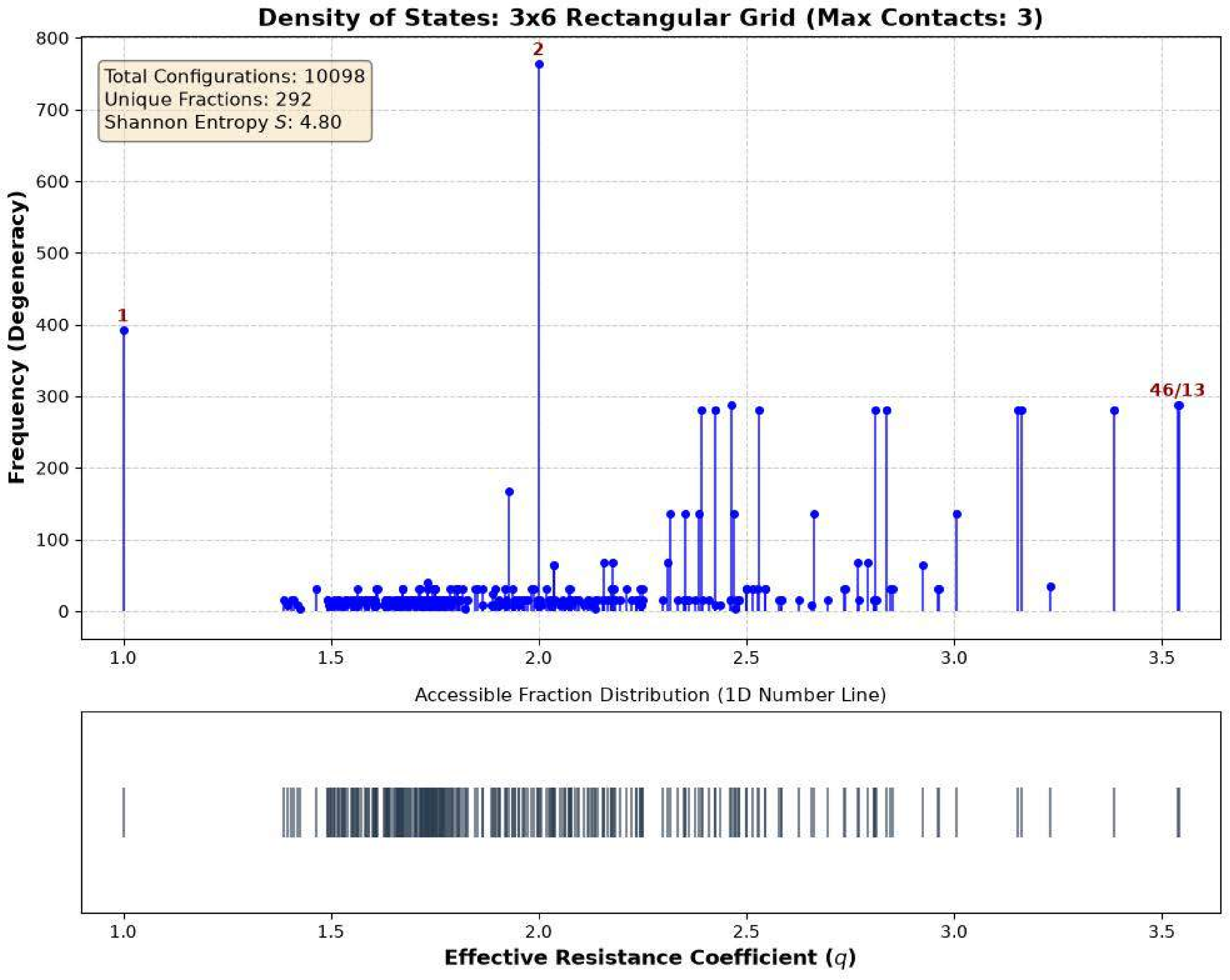

Density of States: 3x6 Rectangular Grid (Max Contacts: 3)
Total Configurations: 10098
Unique Fractions: 292
Shannon Entropy $S$: 4.80
Frequency (Degeneracy)
1
2
46/13
Accessible Fraction Distribution (1D Number Line)
Effective Resistance Coefficient ($q$)

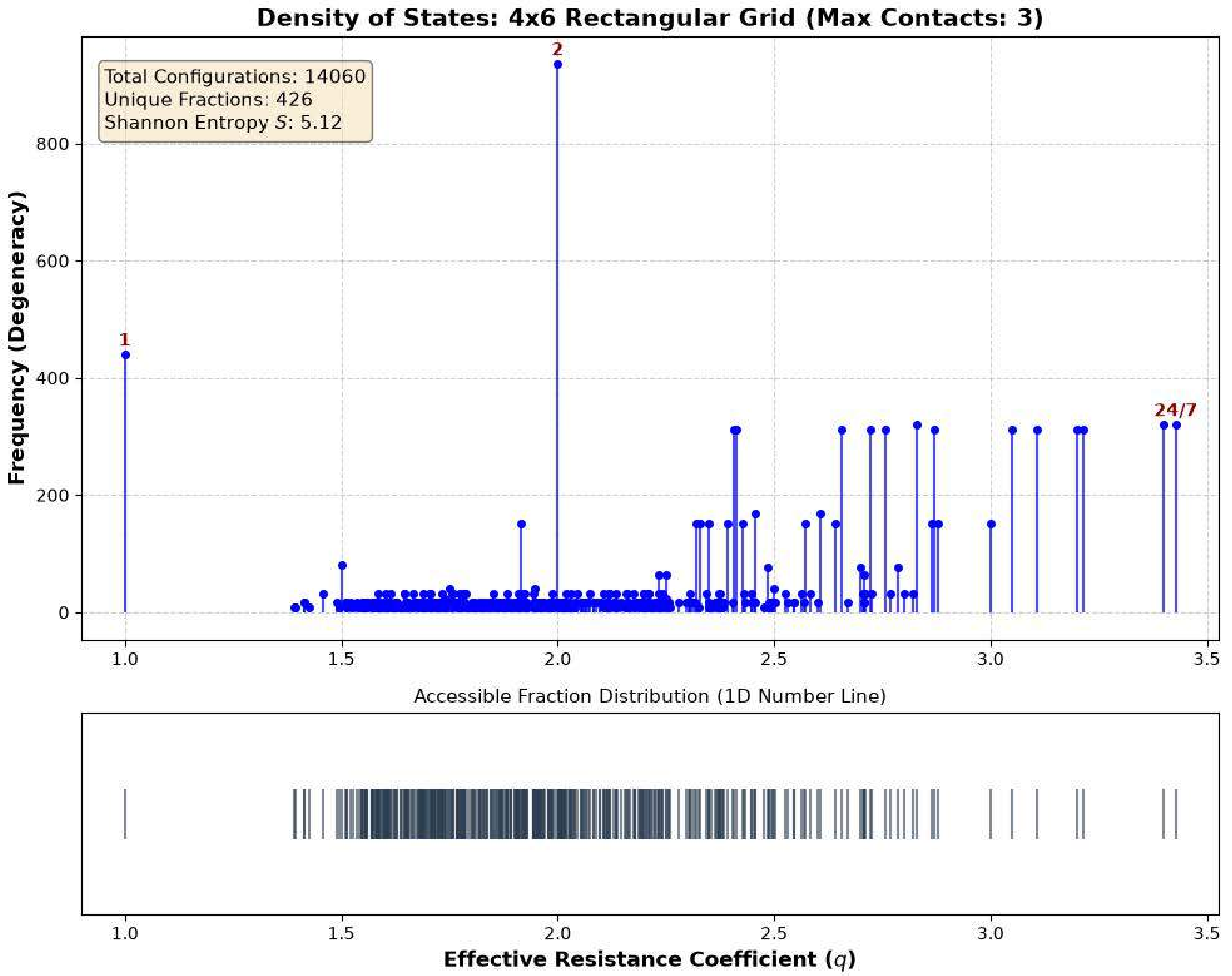

Density of States: 4x6 Rectangular Grid (Max Contacts: 3)
Total Configurations: 14060
Unique Fractions: 426
Shannon Entropy S: 5.12
Frequency (Degeneracy)
0
200
400
600
800
1
2
24/7
1.0
1.5
2.0
2.5
3.0
3.5
Accessible Fraction Distribution (1D Number Line)
Effective Resistance Coefficient (q)

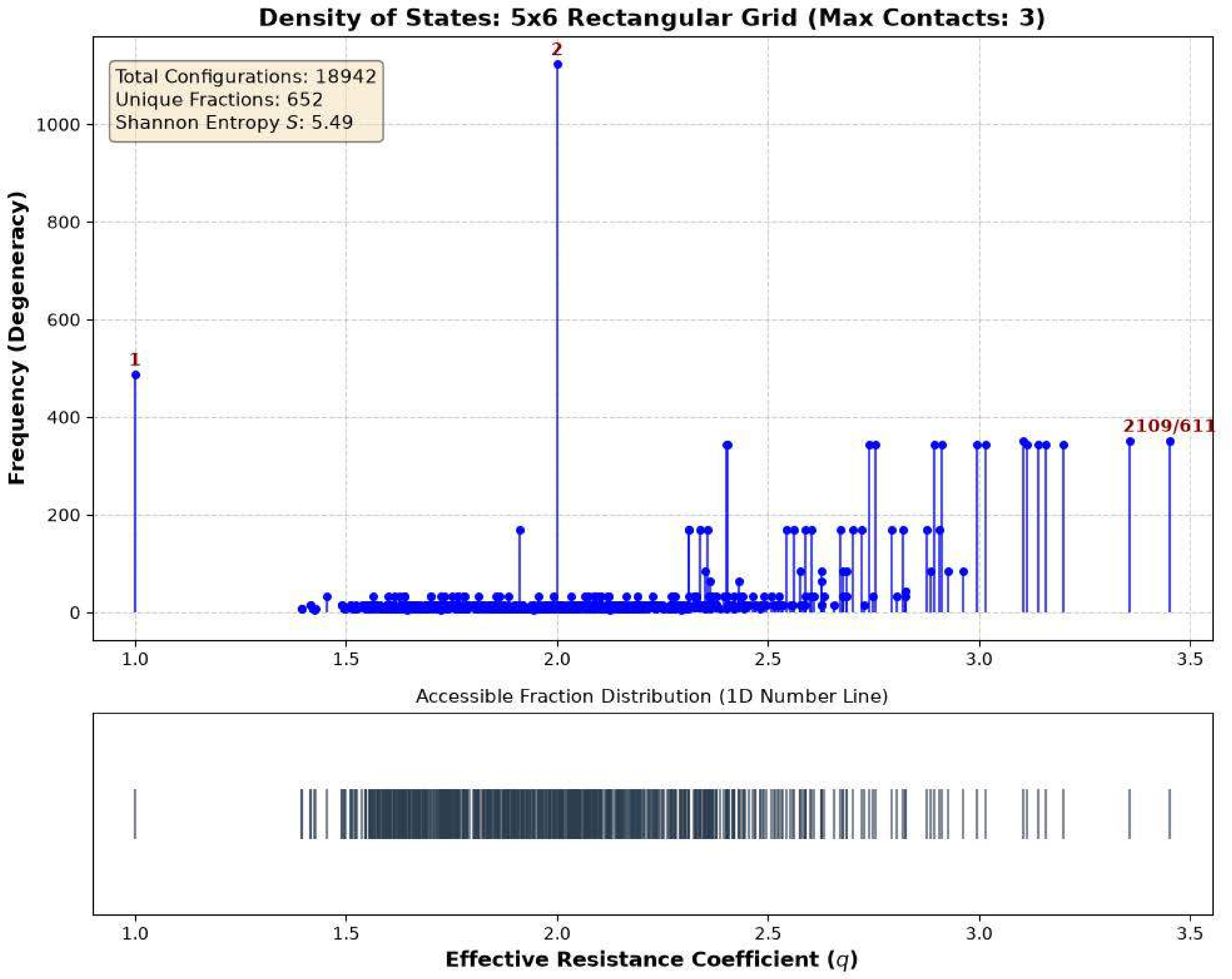

Density of States: 5x6 Rectangular Grid (Max Contacts: 3)
Total Configurations: 18942
Unique Fractions: 652
Shannon Entropy S: 5.49
2
1
2109/611
Frequency (Degeneracy)
0
200
400
600
800
1000
1.0
1.5
2.0
2.5
3.0
3.5
Accessible Fraction Distribution (1D Number Line)
Effective Resistance Coefficient (q)

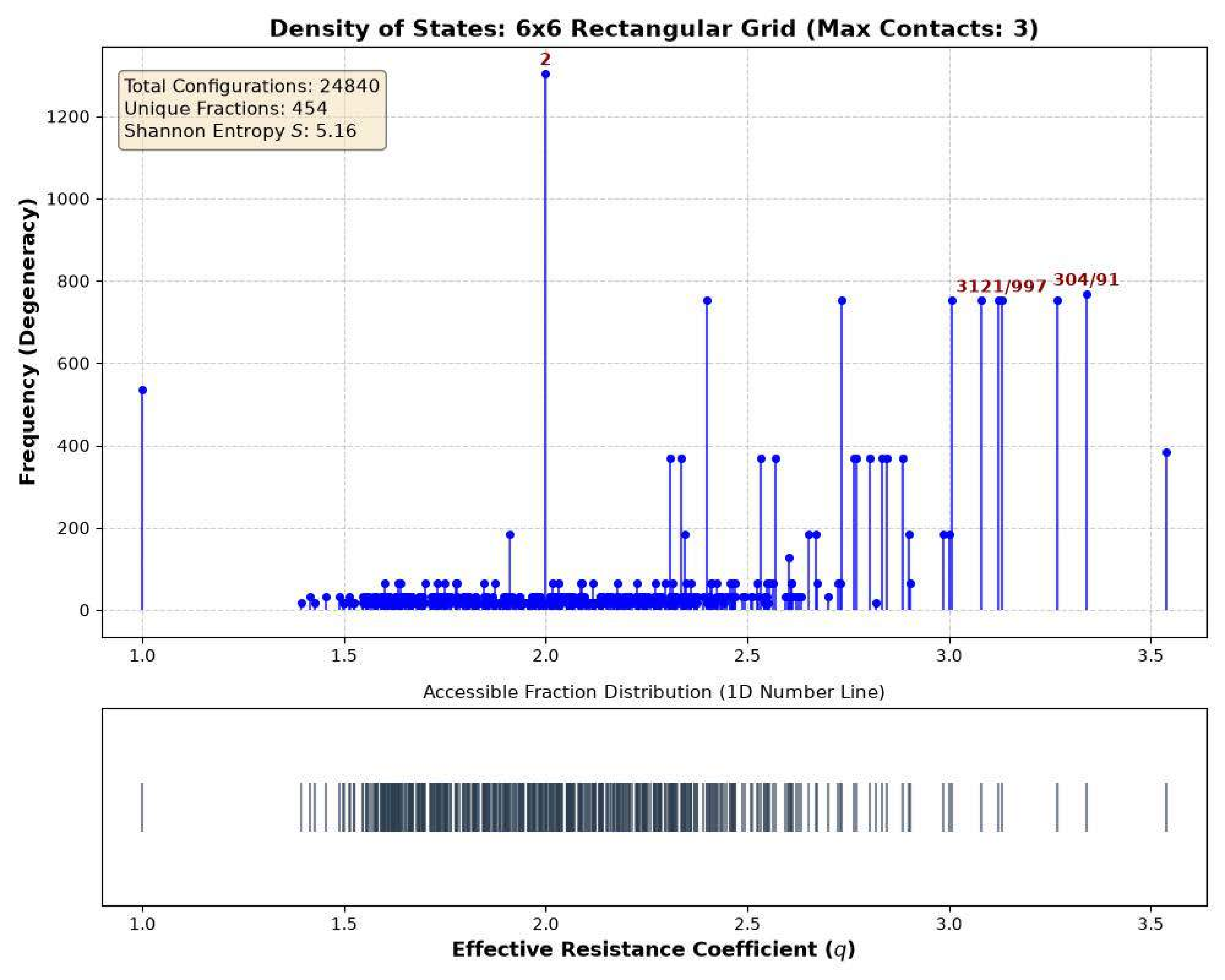

Density of States: 6x6 Rectangular Grid (Max Contacts: 3)
Total Configurations: 24840
Unique Fractions: 454
Shannon Entropy S: 5.16
2
3121/997
304/91
Frequency (Degeneracy)
0
200
400
600
800
1000
1200
1.0
1.5
2.0
2.5
3.0
3.5
Accessible Fraction Distribution (1D Number Line)
Effective Resistance Coefficient (q)

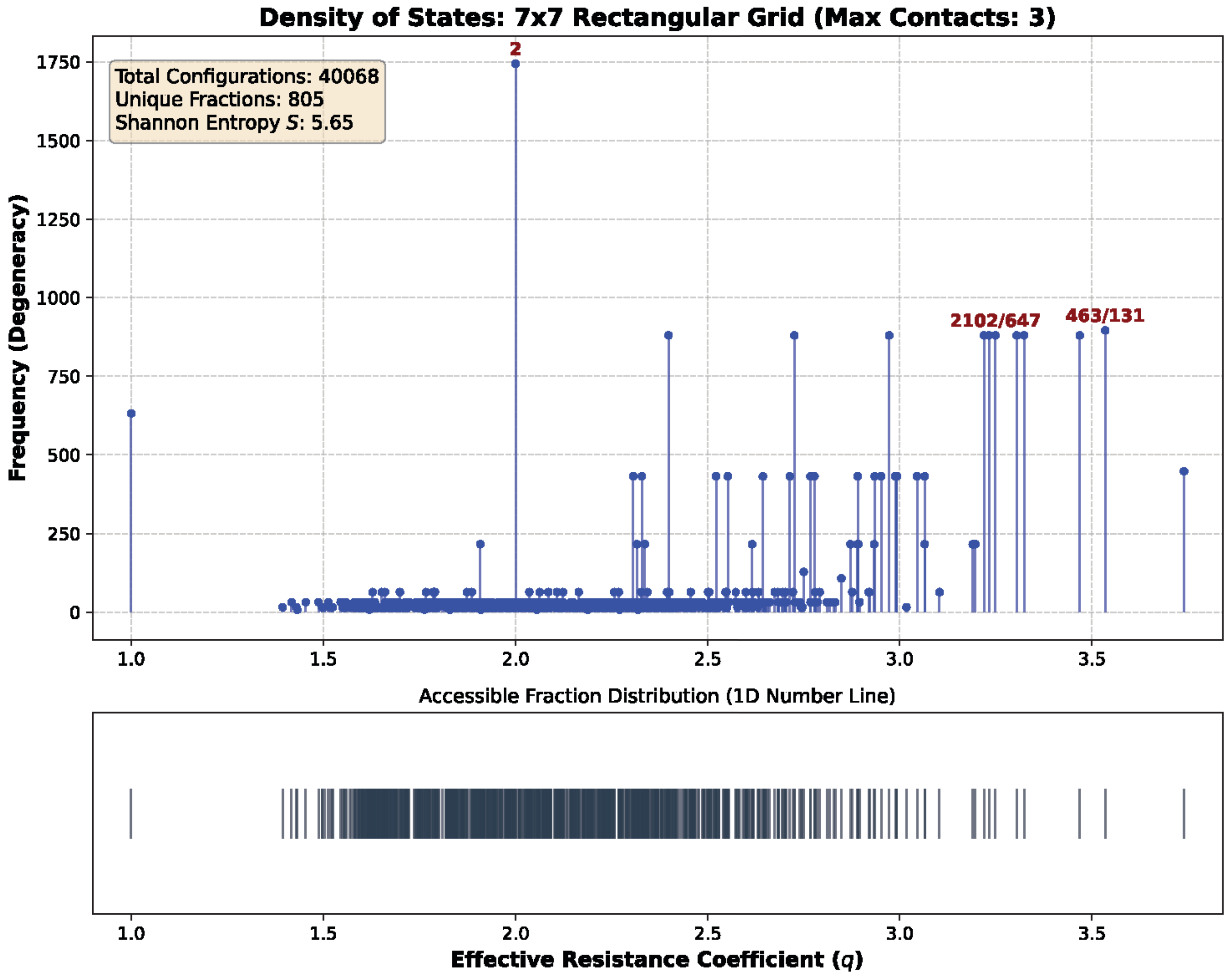

Density of States: 7x7 Rectangular Grid (Max Contacts: 3)
Total Configurations: 40068
Unique Fractions: 805
Shannon Entropy S: 5.65
2
2102/647
463/131
Frequency (Degeneracy)
0
250
500
750
1000
1250
1500
1750
1.0
1.5
2.0
2.5
3.0
3.5
Accessible Fraction Distribution (1D Number Line)
Effective Resistance Coefficient (q)

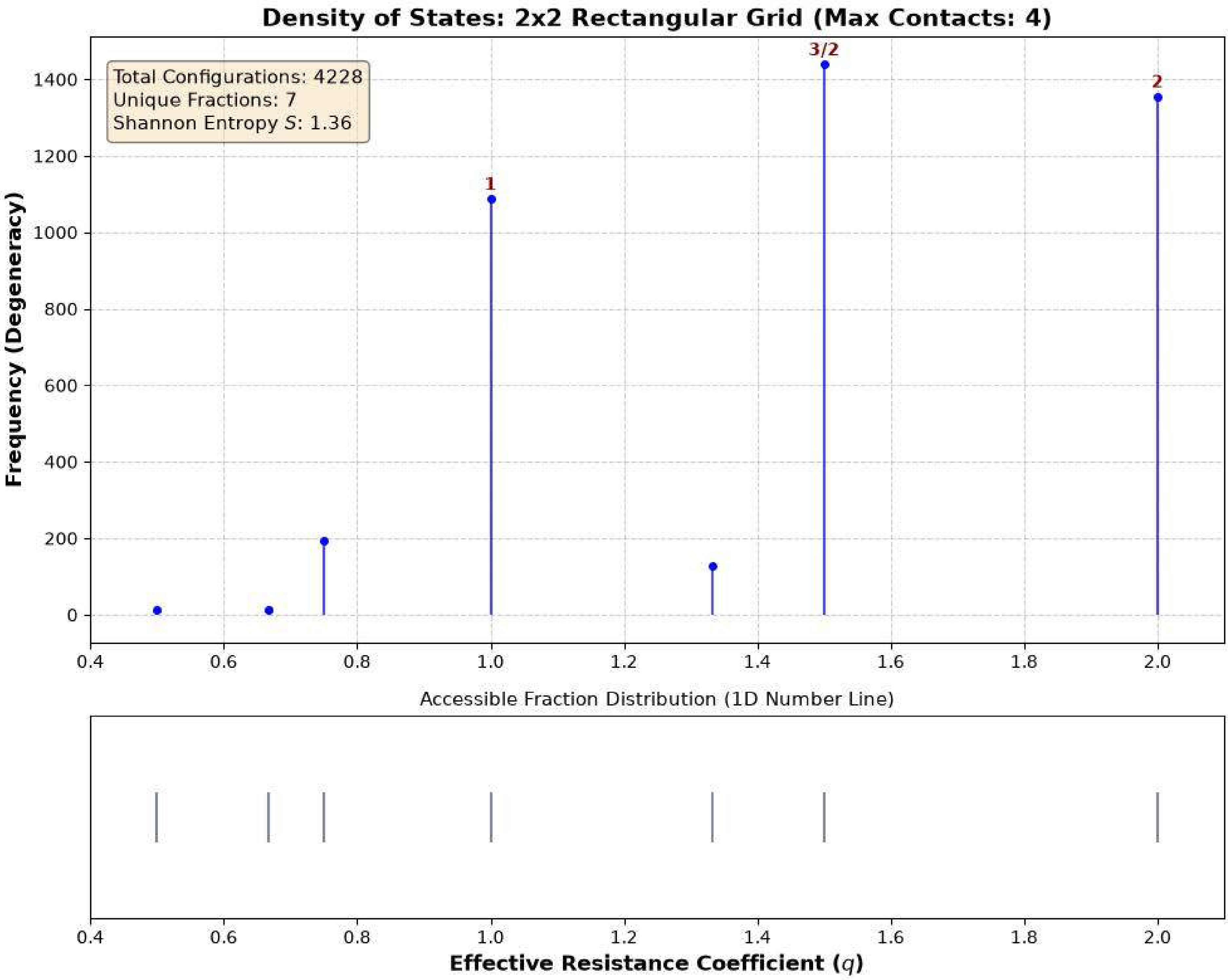

Density of States: 2x2 Rectangular Grid (Max Contacts: 4)
Total Configurations: 4228
Unique Fractions: 7
Shannon Entropy S: 1.36
1
3/2
2
Frequency (Degeneracy)
0
200
400
600
800
1000
1200
1400
0.4
0.6
0.8
1.0
1.2
1.4
1.6
1.8
2.0
Accessible Fraction Distribution (1D Number Line)
Effective Resistance Coefficient (q)

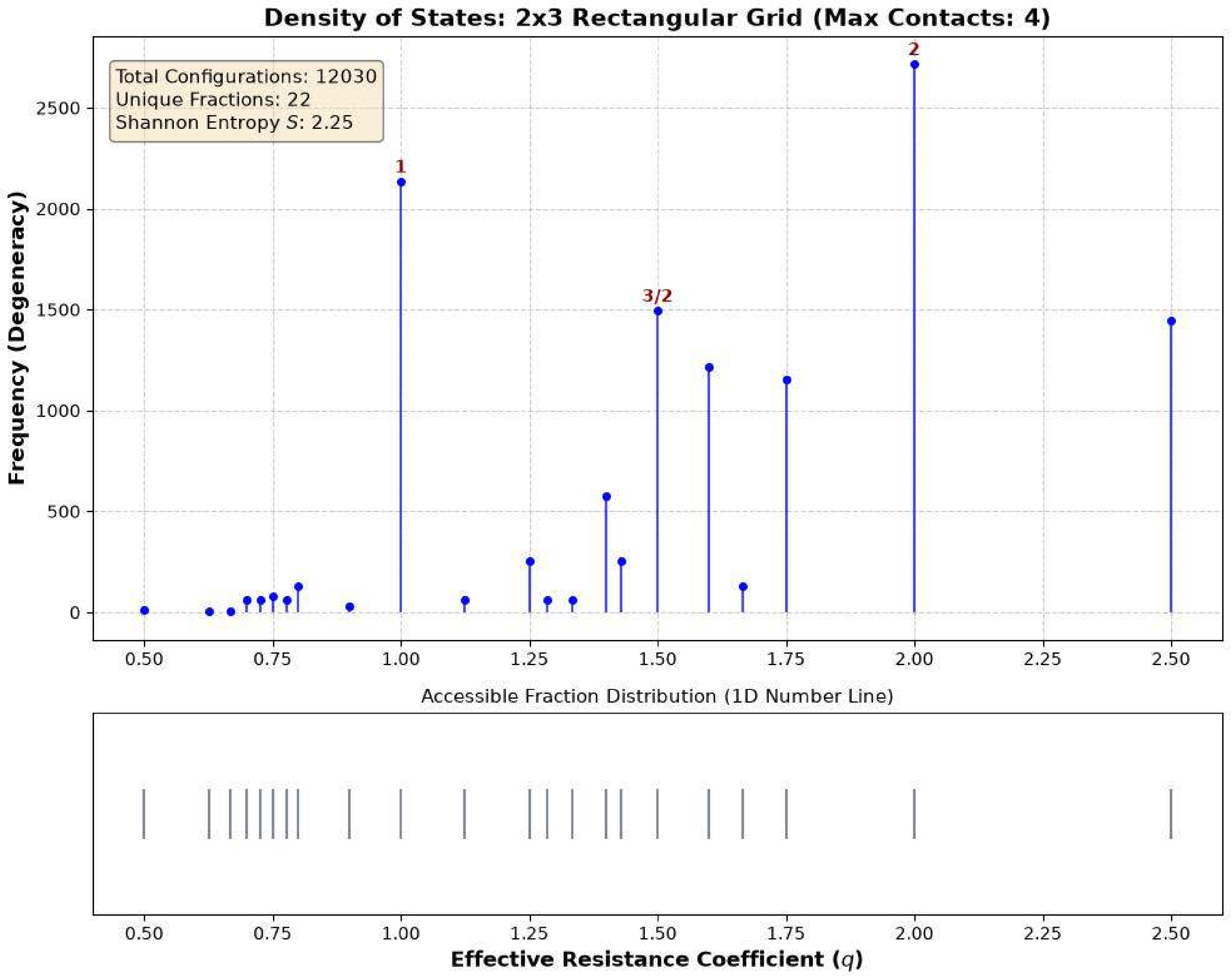

Density of States: 2x3 Rectangular Grid (Max Contacts: 4)
Total Configurations: 12030
Unique Fractions: 22
Shannon Entropy S: 2.25
1
3/2
2
Frequency (Degeneracy)
0
500
1000
1500
2000
2500
0.50
0.75
1.00
1.25
1.50
1.75
2.00
2.25
2.50
Accessible Fraction Distribution (1D Number Line)
Effective Resistance Coefficient (q)

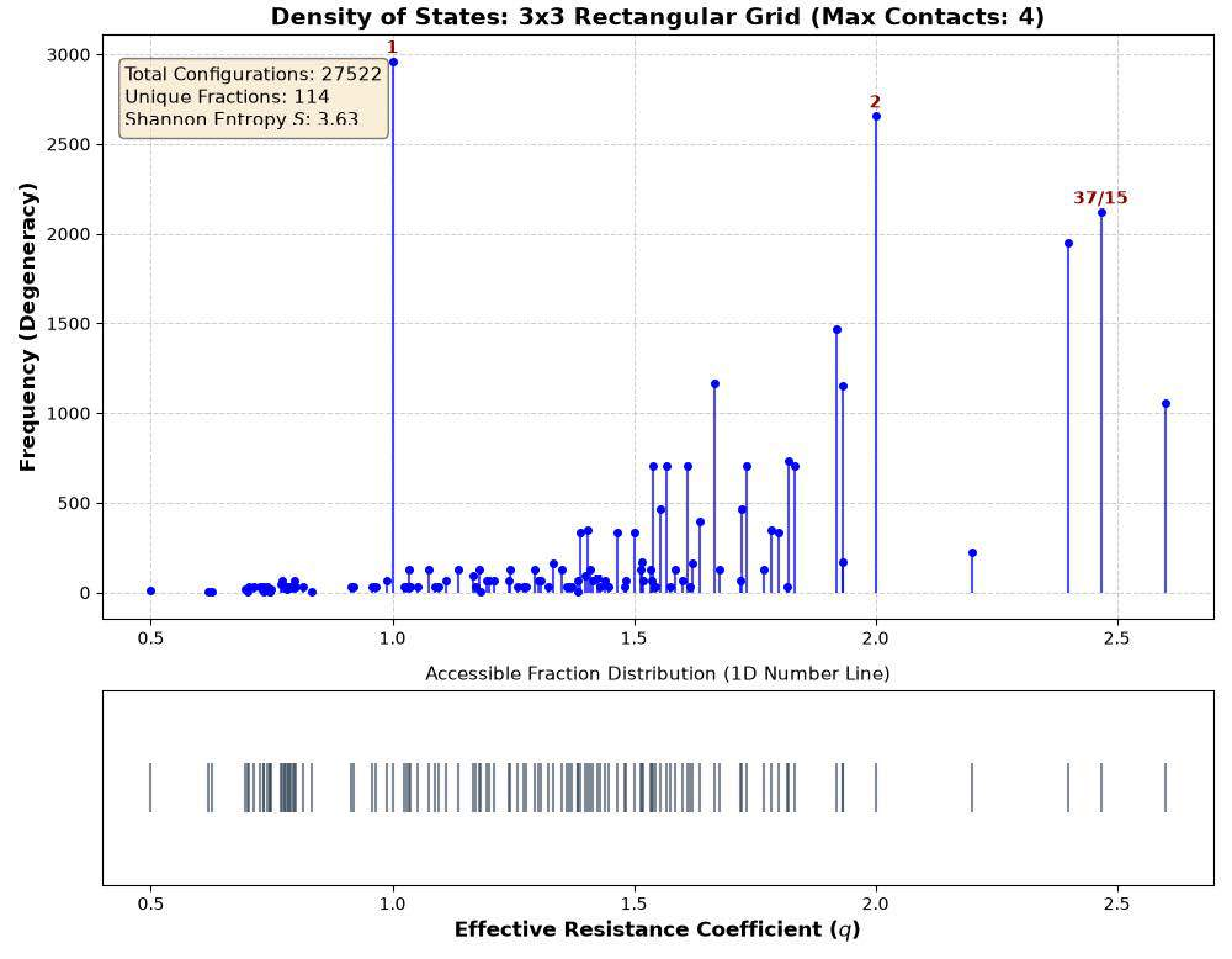

Density of States: 3x3 Rectangular Grid (Max Contacts: 4)
Total Configurations: 27522
Unique Fractions: 114
Shannon Entropy S: 3.63
1
2
37/15
Frequency (Degeneracy)
3000
2500
2000
1500
1000
500
0
0.5
1.0
1.5
2.0
2.5
Accessible Fraction Distribution (1D Number Line)
0.5
1.0
1.5
2.0
2.5
Effective Resistance Coefficient (q)

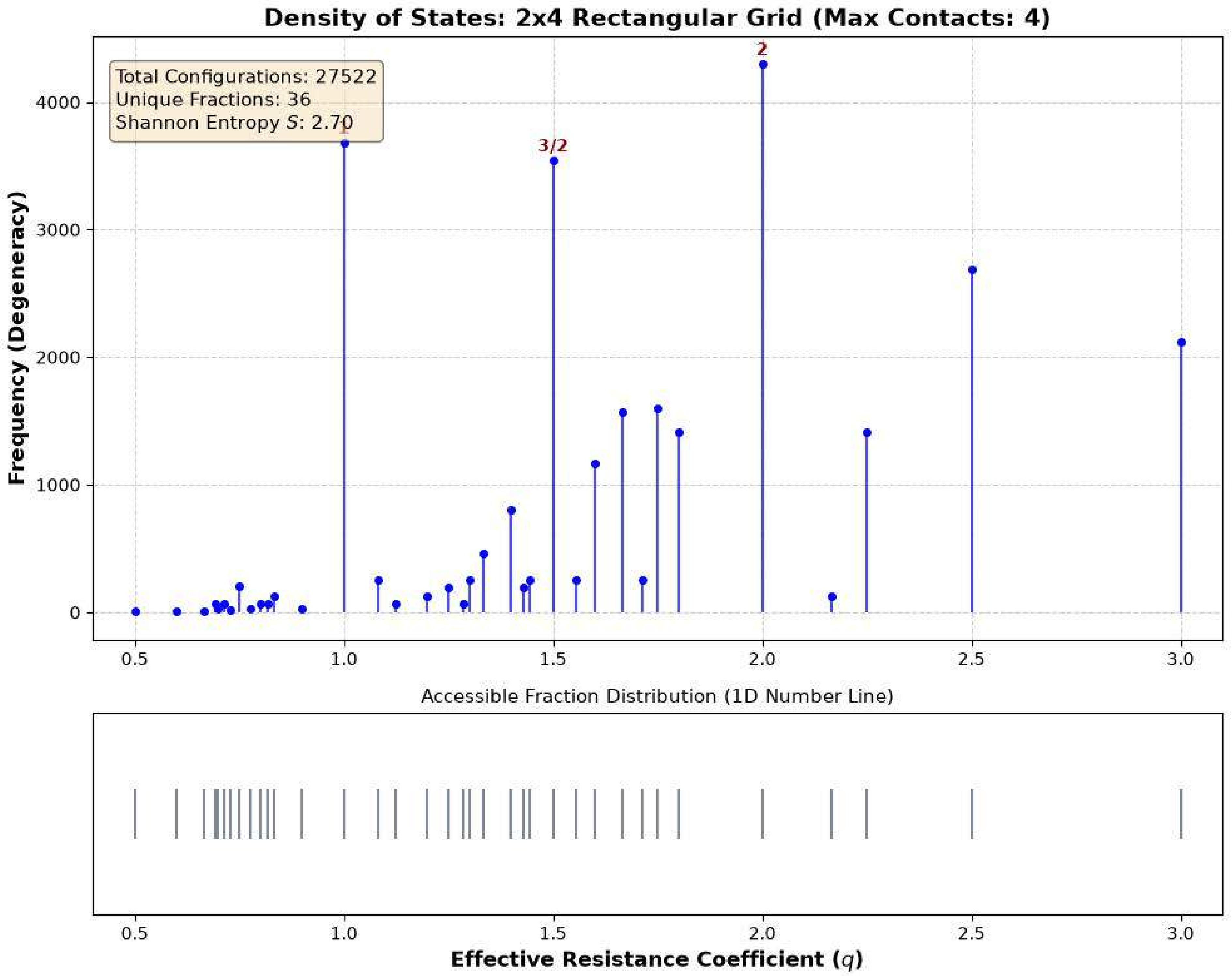
Density of States: 2x4 Rectangular Grid (Max Contacts: 4)
Total Configurations: 27522
Unique Fractions: 36
Shannon Entropy S: 2.70
1
3/2
2
Frequency (Degeneracy)
0
1000
2000
3000
4000
0.5
1.0
1.5
2.0
2.5
3.0
Accessible Fraction Distribution (1D Number Line)
Effective Resistance Coefficient (q)

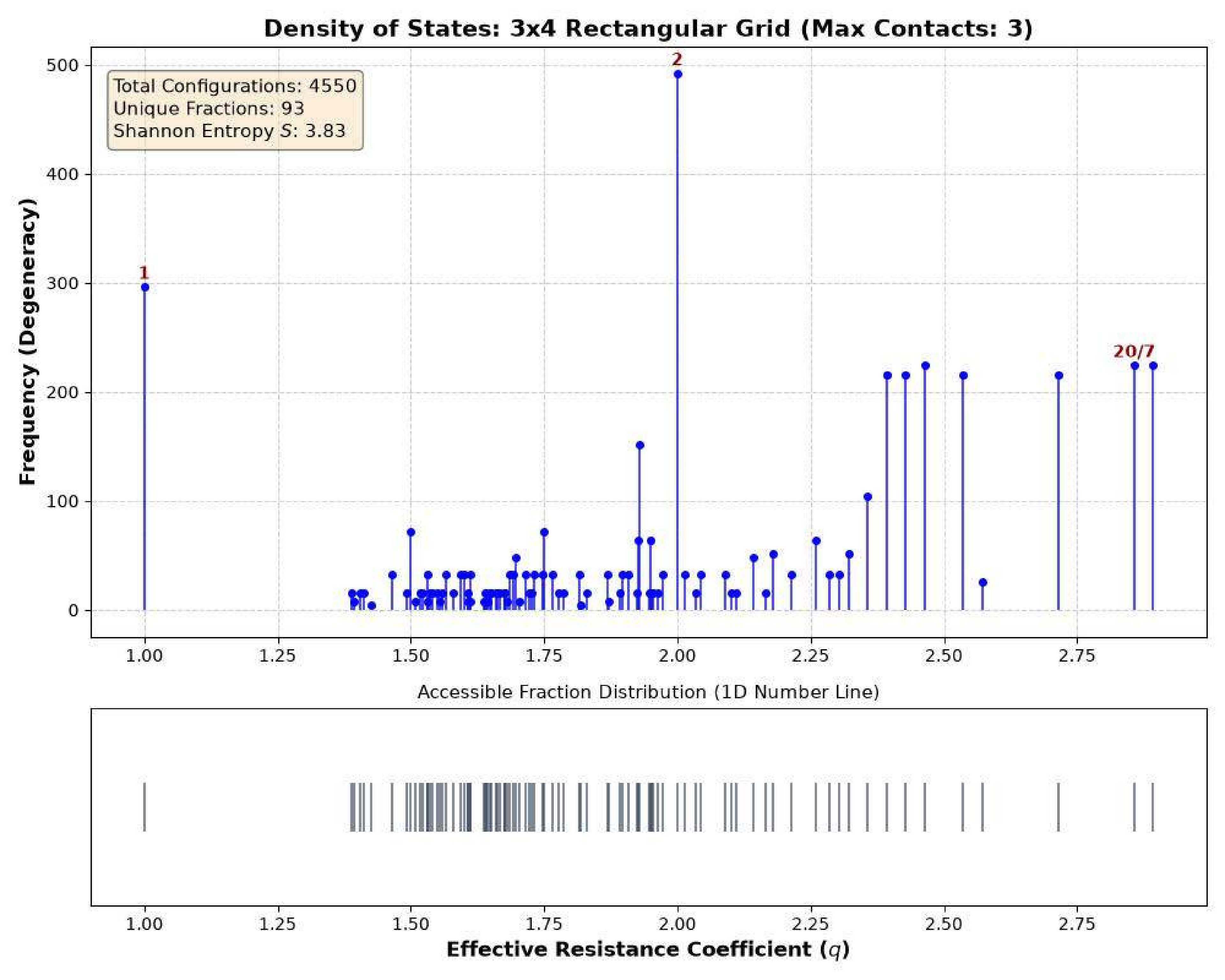

Density of States: 3x4 Rectangular Grid (Max Contacts: 3)
Total Configurations: 4550
Unique Fractions: 93
Shannon Entropy S: 3.83
Frequency (Degeneracy)
1
2
20/7
0
100
200
300
400
500
1.00
1.25
1.50
1.75
2.00
2.25
2.50
2.75
Accessible Fraction Distribution (1D Number Line)
Effective Resistance Coefficient (q)

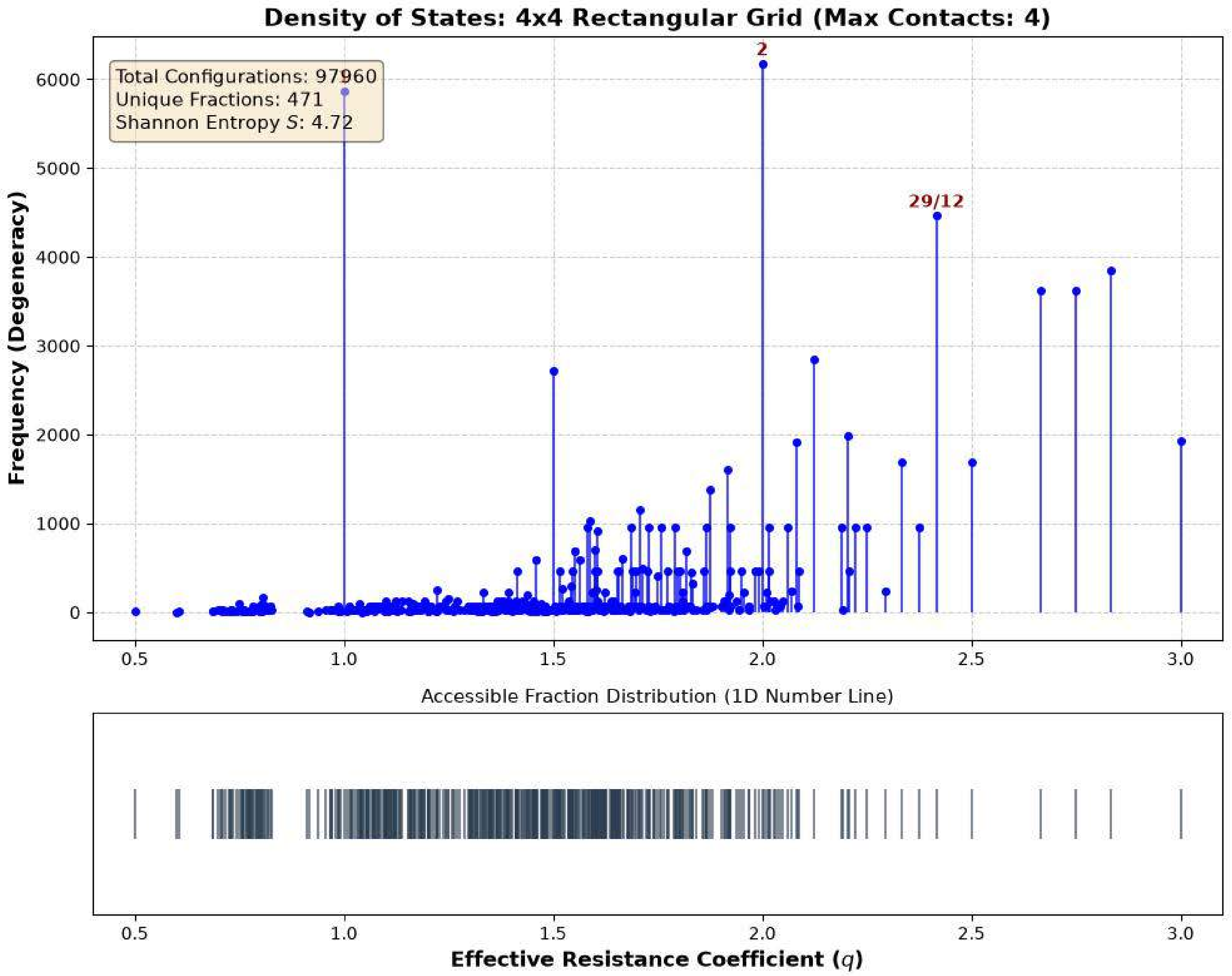
Density of States: 4x4 Rectangular Grid (Max Contacts: 4)
Total Configurations: 97960
Unique Fractions: 471
Shannon Entropy S: 4.72
2
29/12
Frequency (Degeneracy)
0
1000
2000
3000
4000
5000
6000
0.5
1.0
1.5
2.0
2.5
3.0
Accessible Fraction Distribution (1D Number Line)
Effective Resistance Coefficient (q)

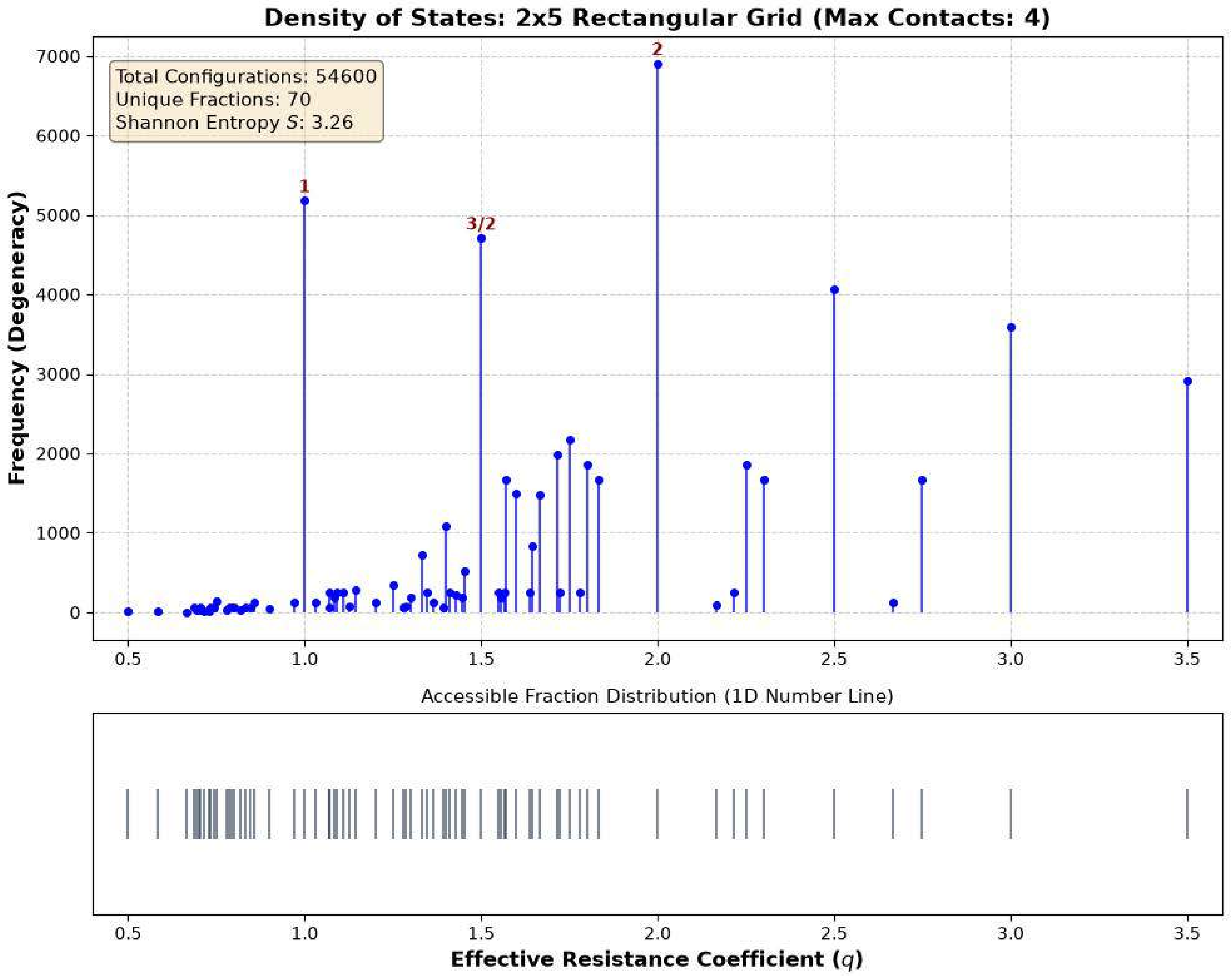
Density of States: 2x5 Rectangular Grid (Max Contacts: 4)
Total Configurations: 54600
Unique Fractions: 70
Shannon Entropy S: 3.26
1
3/2
2
Frequency (Degeneracy)
0
1000
2000
3000
4000
5000
6000
7000
0.5
1.0
1.5
2.0
2.5
3.0
3.5
Accessible Fraction Distribution (1D Number Line)
Effective Resistance Coefficient (q)

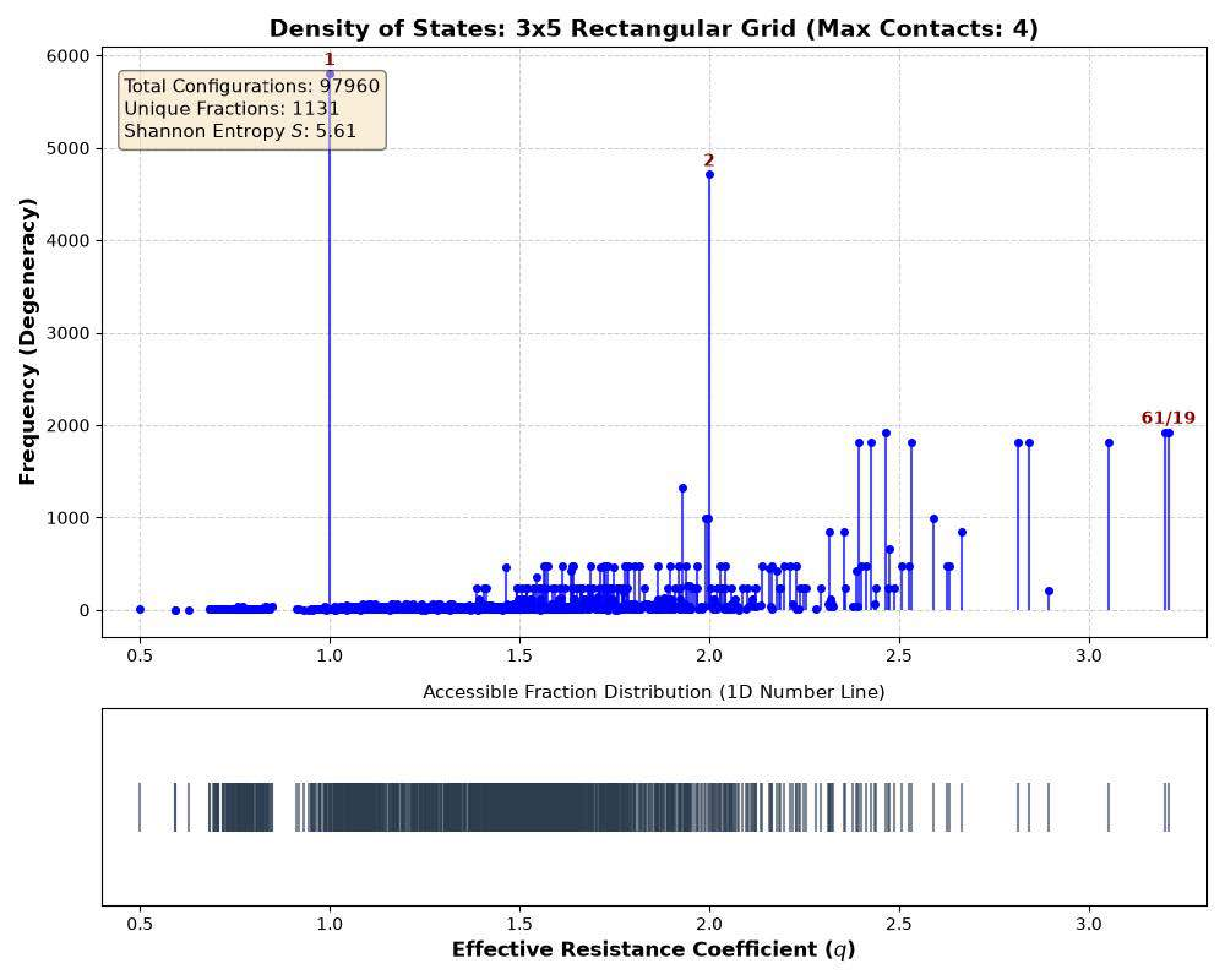
Density of States: 3x5 Rectangular Grid (Max Contacts: 4)
Total Configurations: 97960
Unique Fractions: 1131
Shannon Entropy S: 5.61
1
2
61/19
Frequency (Degeneracy)
0
1000
2000
3000
4000
5000
6000
0.5
1.0
1.5
2.0
2.5
3.0
Accessible Fraction Distribution (1D Number Line)
Effective Resistance Coefficient (q)

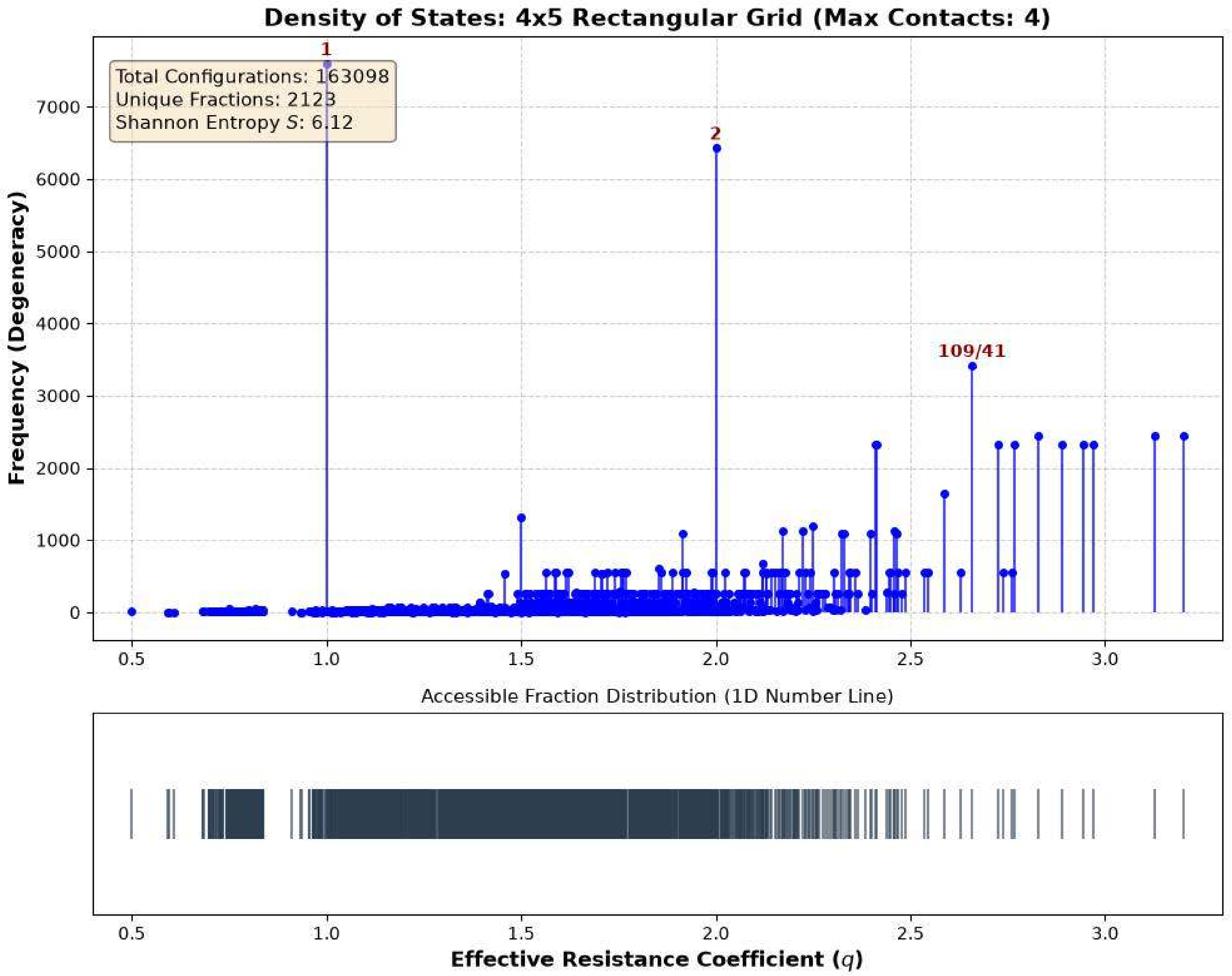

Density of States: 4x5 Rectangular Grid (Max Contacts: 4)
Total Configurations: 163098
Unique Fractions: 2123
Shannon Entropy $S$: 6.12
1
2
109/41
Frequency (Degeneracy)
0
1000
2000
3000
4000
5000
6000
7000
0.5
1.0
1.5
2.0
2.5
3.0
Accessible Fraction Distribution (1D Number Line)
Effective Resistance Coefficient ($q$)

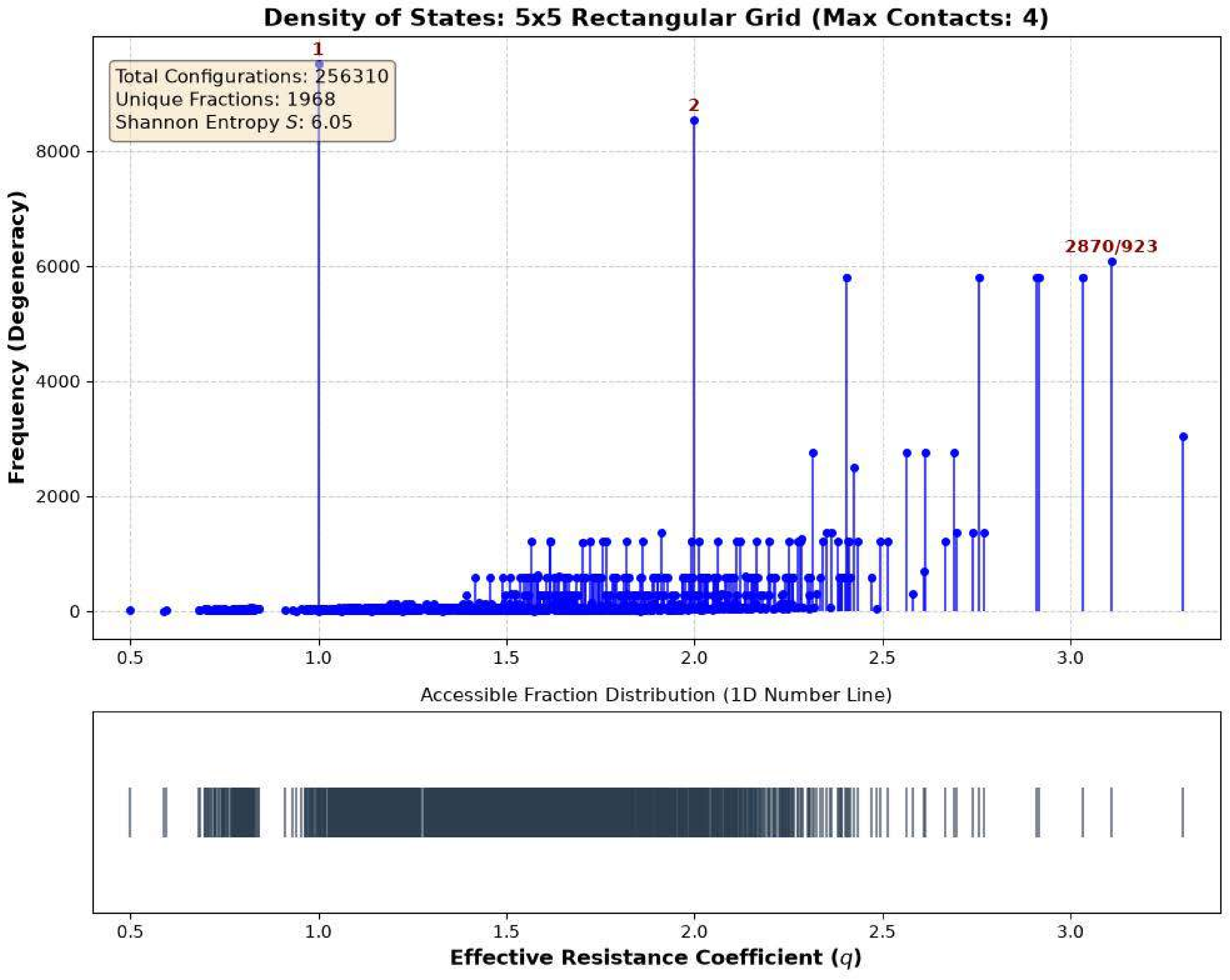

Density of States: 5x5 Rectangular Grid (Max Contacts: 4)
Total Configurations: 256310
Unique Fractions: 1968
Shannon Entropy S: 6.05
1
2
2870/923
Frequency (Degeneracy)
0
2000
4000
6000
8000
0.5
1.0
1.5
2.0
2.5
3.0
Accessible Fraction Distribution (1D Number Line)
0.5
1.0
1.5
2.0
2.5
3.0
Effective Resistance Coefficient (q)

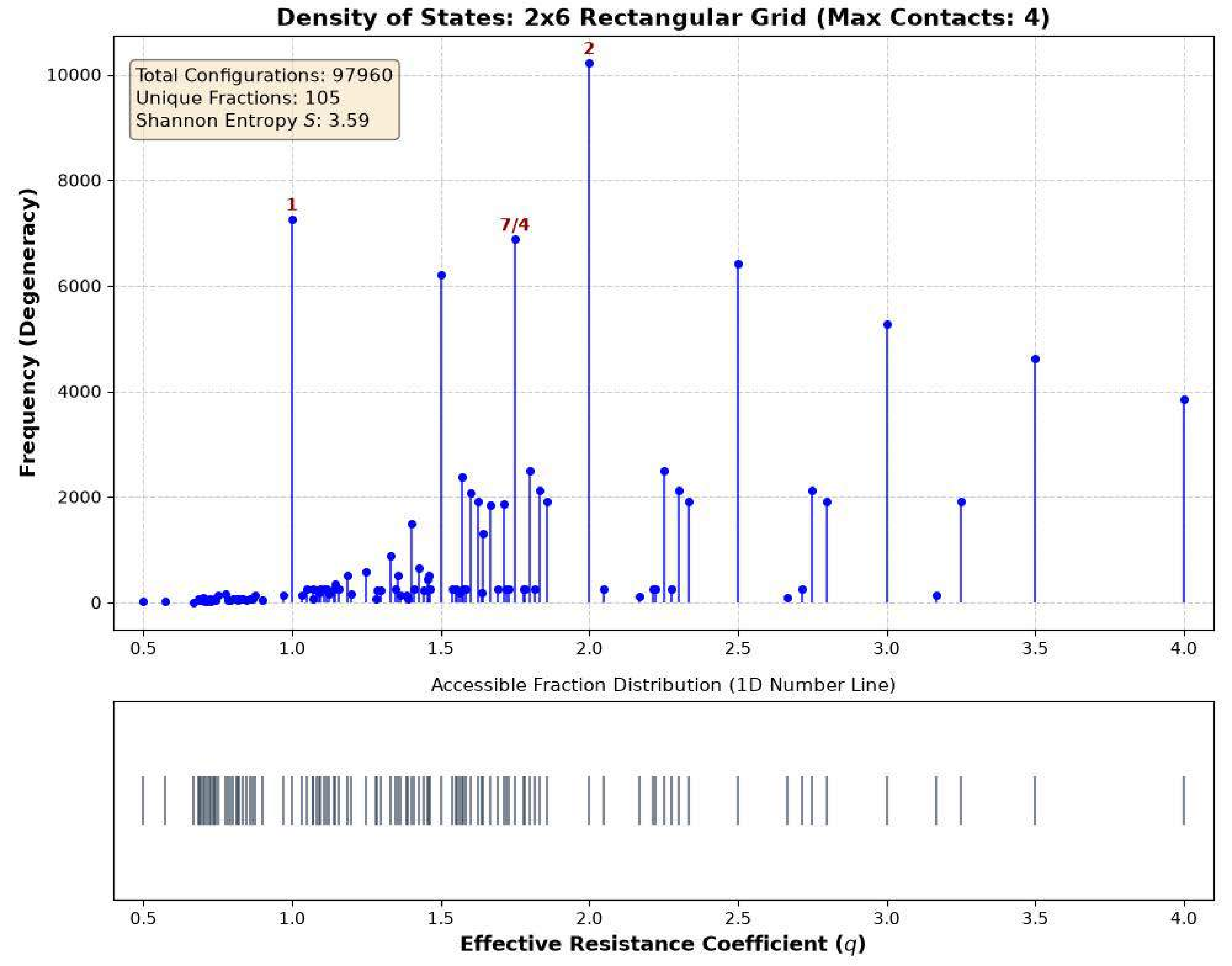

Density of States: 2x6 Rectangular Grid (Max Contacts: 4)
Total Configurations: 97960
Unique Fractions: 105
Shannon Entropy S: 3.59
Frequency (Degeneracy)
1
7/4
2
0
2000
4000
6000
8000
10000
0.5
1.0
1.5
2.0
2.5
3.0
3.5
4.0
Accessible Fraction Distribution (1D Number Line)
Effective Resistance Coefficient (q)

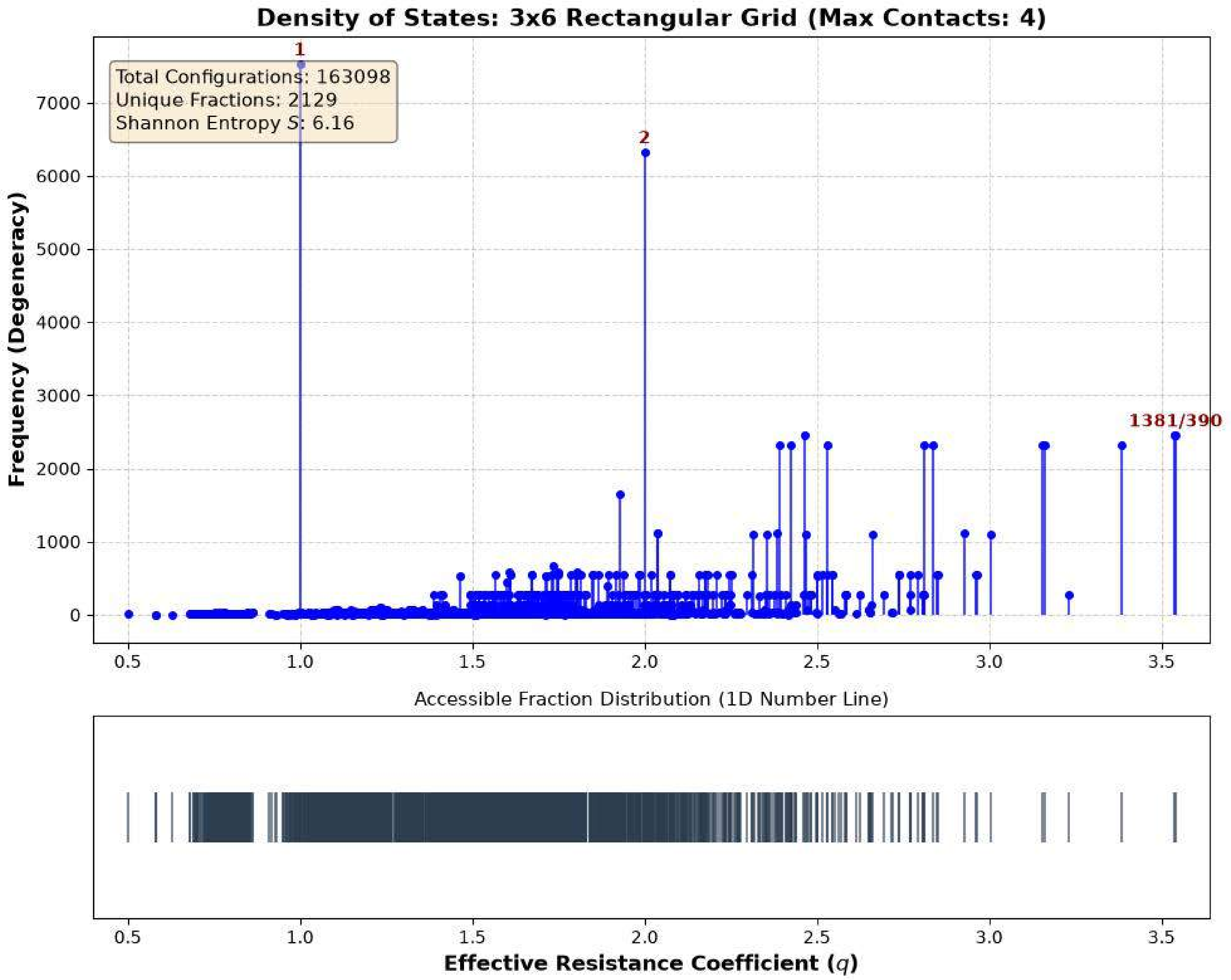
Density of States: 3x6 Rectangular Grid (Max Contacts: 4)
Total Configurations: 163098
Unique Fractions: 2129
Shannon Entropy S: 6.16
1
2
1381/390
Frequency (Degeneracy)
0
1000
2000
3000
4000
5000
6000
7000
0.5
1.0
1.5
2.0
2.5
3.0
3.5
Accessible Fraction Distribution (1D Number Line)
Effective Resistance Coefficient (q)

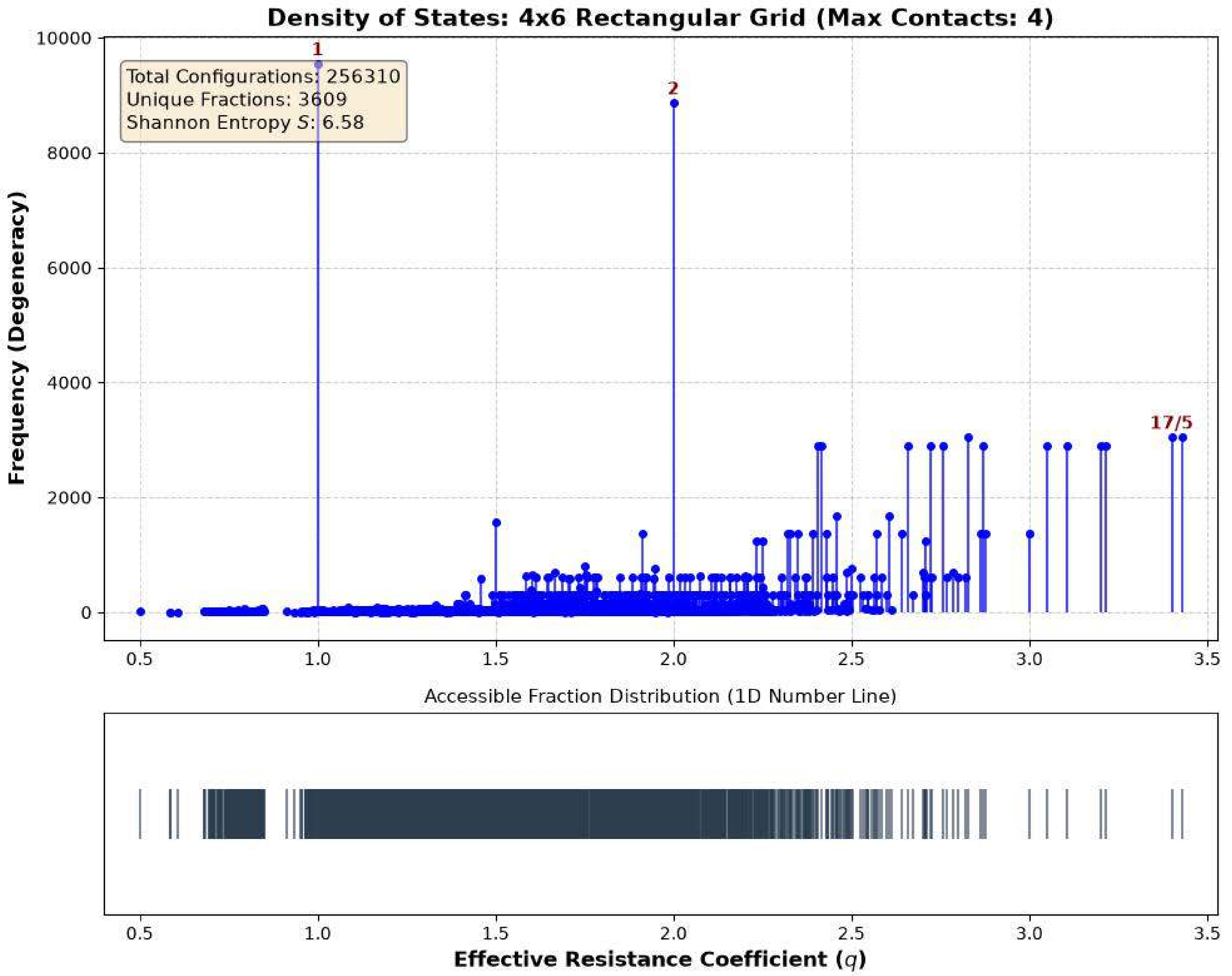

Density of States: 4x6 Rectangular Grid (Max Contacts: 4)
Total Configurations: 256310
Unique Fractions: 3609
Shannon Entropy S: 6.58
1
2
17/5
Frequency (Degeneracy)
10000
8000
6000
4000
2000
0
0.5
1.0
1.5
2.0
2.5
3.0
3.5
Accessible Fraction Distribution (1D Number Line)
Effective Resistance Coefficient (q)

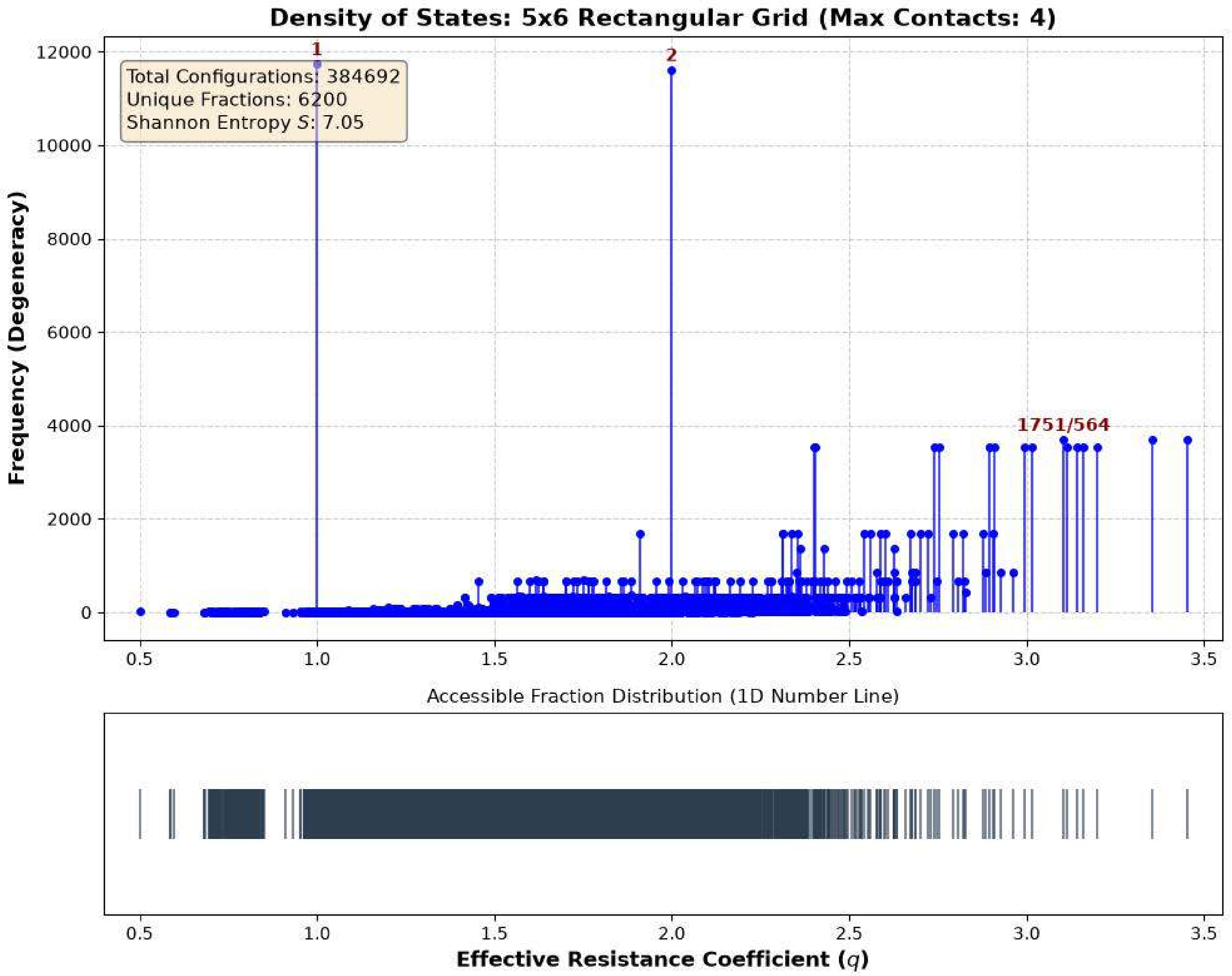

Density of States: 5x6 Rectangular Grid (Max Contacts: 4)
Total Configurations: 384692
Unique Fractions: 6200
Shannon Entropy S: 7.05
1
2
1751/564
Frequency (Degeneracy)
12000
10000
8000
6000
4000
2000
0
0.5
1.0
1.5
2.0
2.5
3.0
3.5
Accessible Fraction Distribution (1D Number Line)
Effective Resistance Coefficient (q)

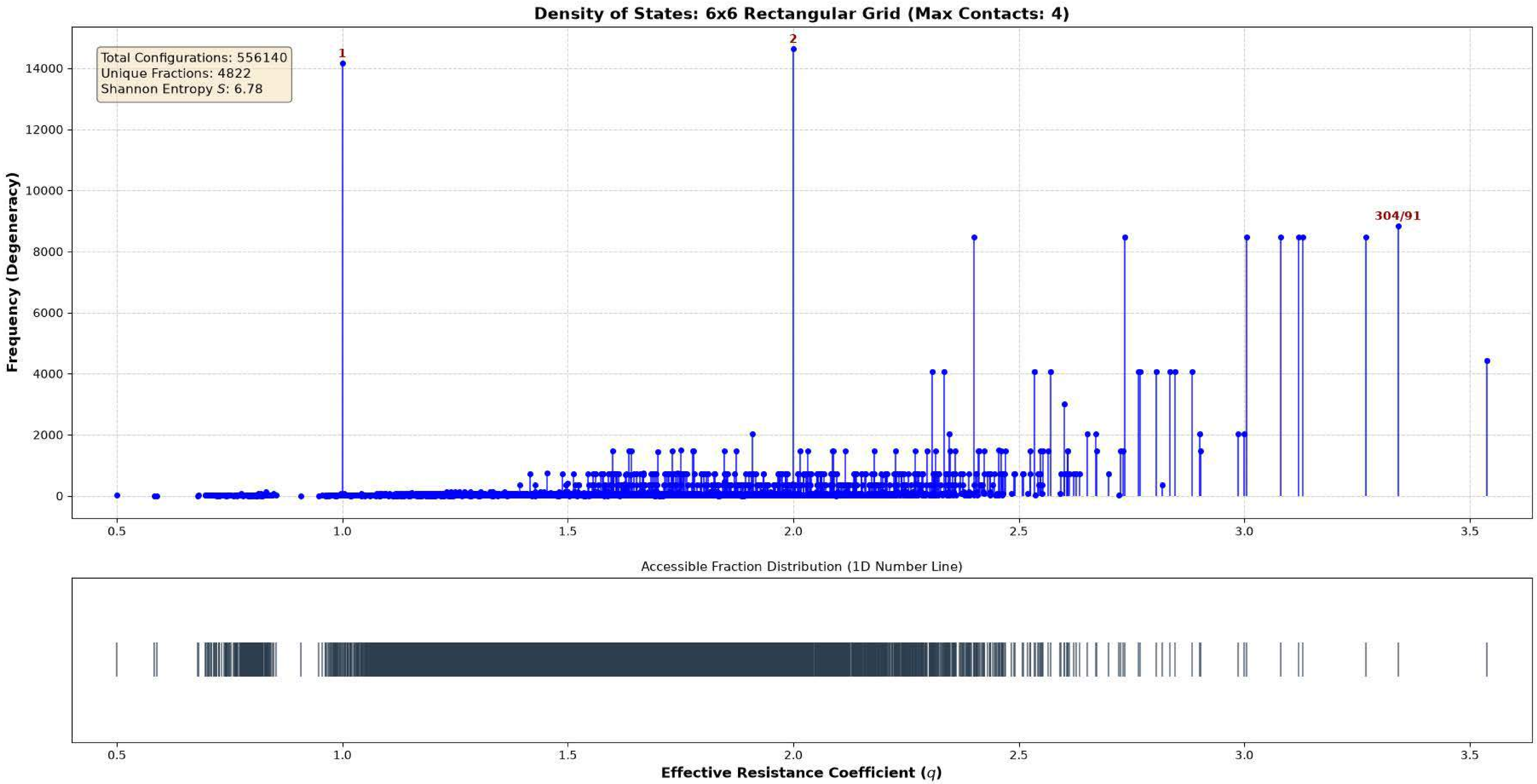

Density of States: 6x6 Rectangular Grid (Max Contacts: 4)
Total Configurations: 556140
Unique Fractions: 4822
Shannon Entropy S: 6.78
1
2
304/91
Frequency (Degeneracy)
0
2000
4000
6000
8000
10000
12000
14000
0.5
1.0
1.5
2.0
2.5
3.0
3.5
Accessible Fraction Distribution (1D Number Line)
Effective Resistance Coefficient (q)

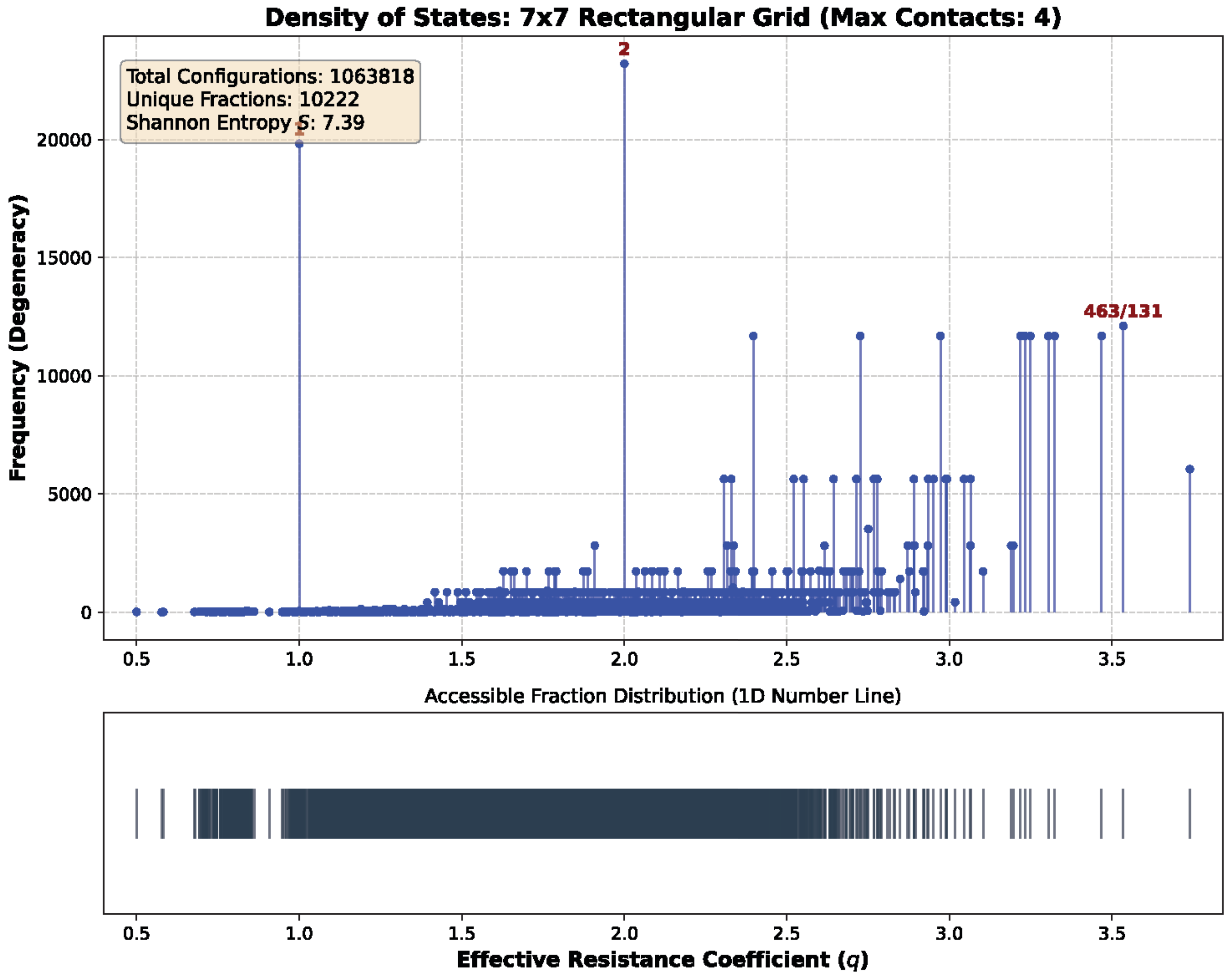

Density of States: 7x7 Rectangular Grid (Max Contacts: 4)
Total Configurations: 1063818
Unique Fractions: 10222
Shannon Entropy S: 7.39
1
2
463/131
Frequency (Degeneracy)
0
5000
10000
15000
20000
0.5
1.0
1.5
2.0
2.5
3.0
3.5
Accessible Fraction Distribution (1D Number Line)
Effective Resistance Coefficient (q)

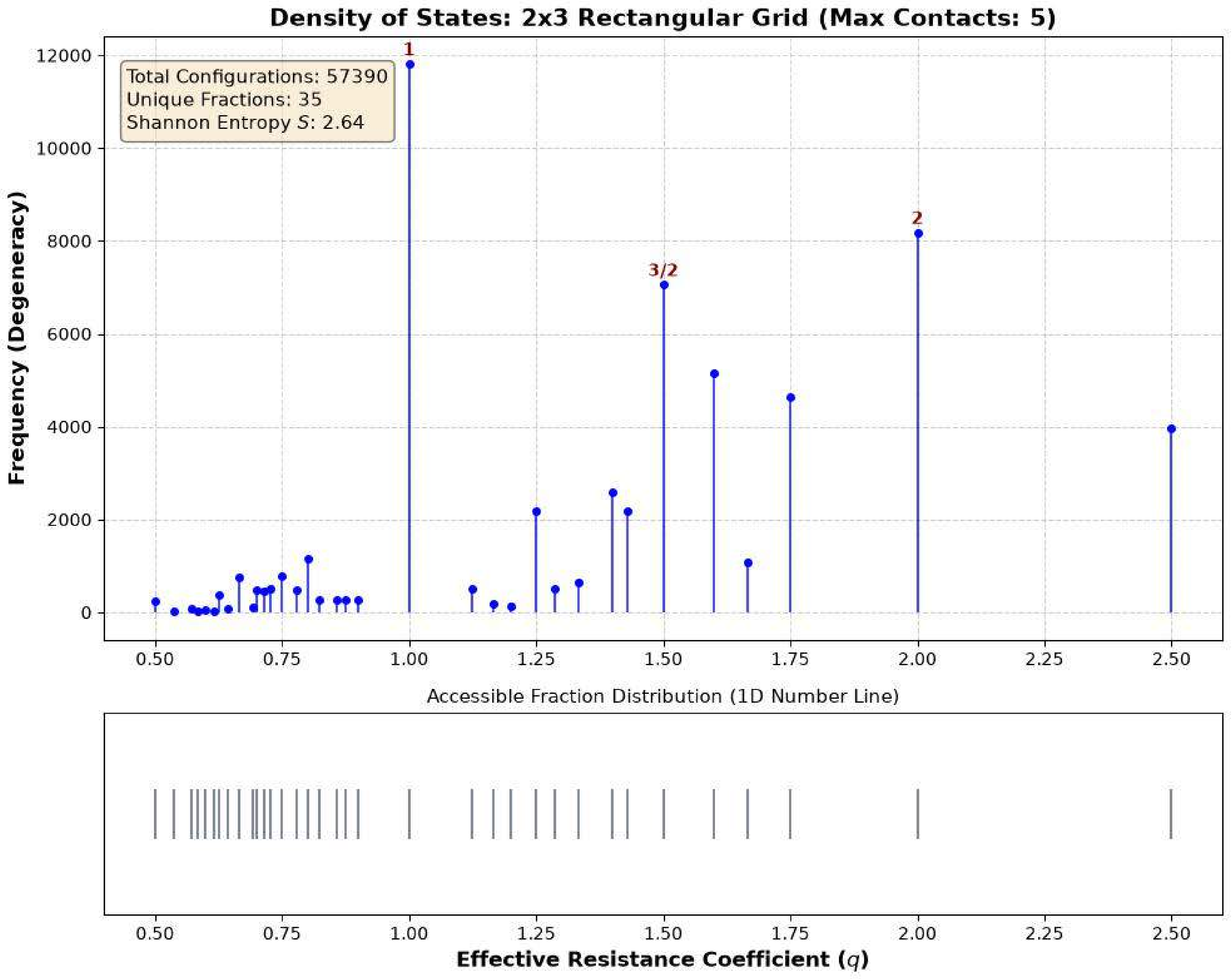

Density of States: 2x3 Rectangular Grid (Max Contacts: 5)
Total Configurations: 57390
Unique Fractions: 35
Shannon Entropy S: 2.64
1
3/2
2
Frequency (Degeneracy)
0
2000
4000
6000
8000
10000
12000
0.50
0.75
1.00
1.25
1.50
1.75
2.00
2.25
2.50
Accessible Fraction Distribution (1D Number Line)
Effective Resistance Coefficient (q)

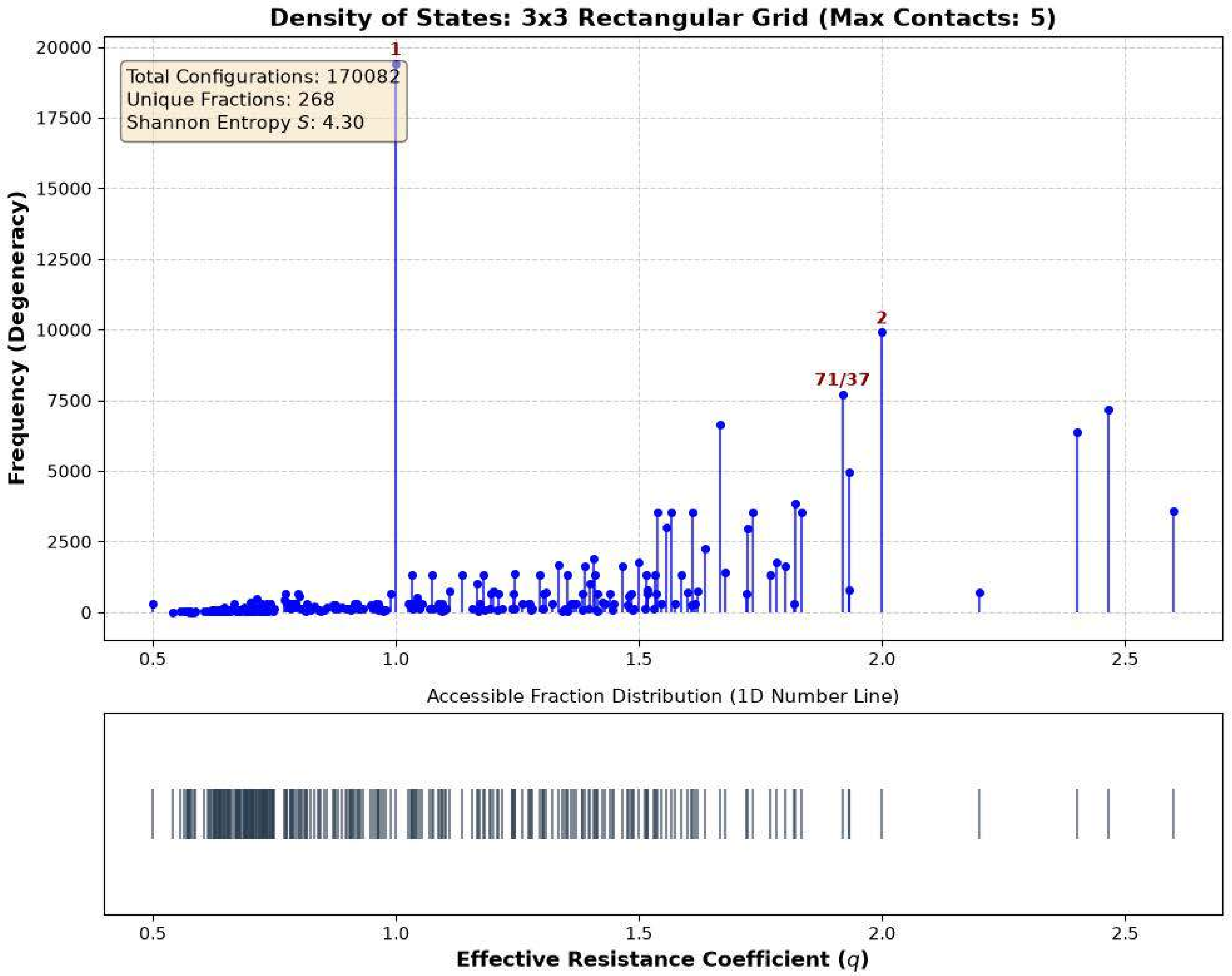
Density of States: 3x3 Rectangular Grid (Max Contacts: 5)
Total Configurations: 170082
Unique Fractions: 268
Shannon Entropy S: 4.30
1
2
71/37
Frequency (Degeneracy)
0
2500
5000
7500
10000
12500
15000
17500
20000
0.5
1.0
1.5
2.0
2.5
Accessible Fraction Distribution (1D Number Line)
Effective Resistance Coefficient (q)

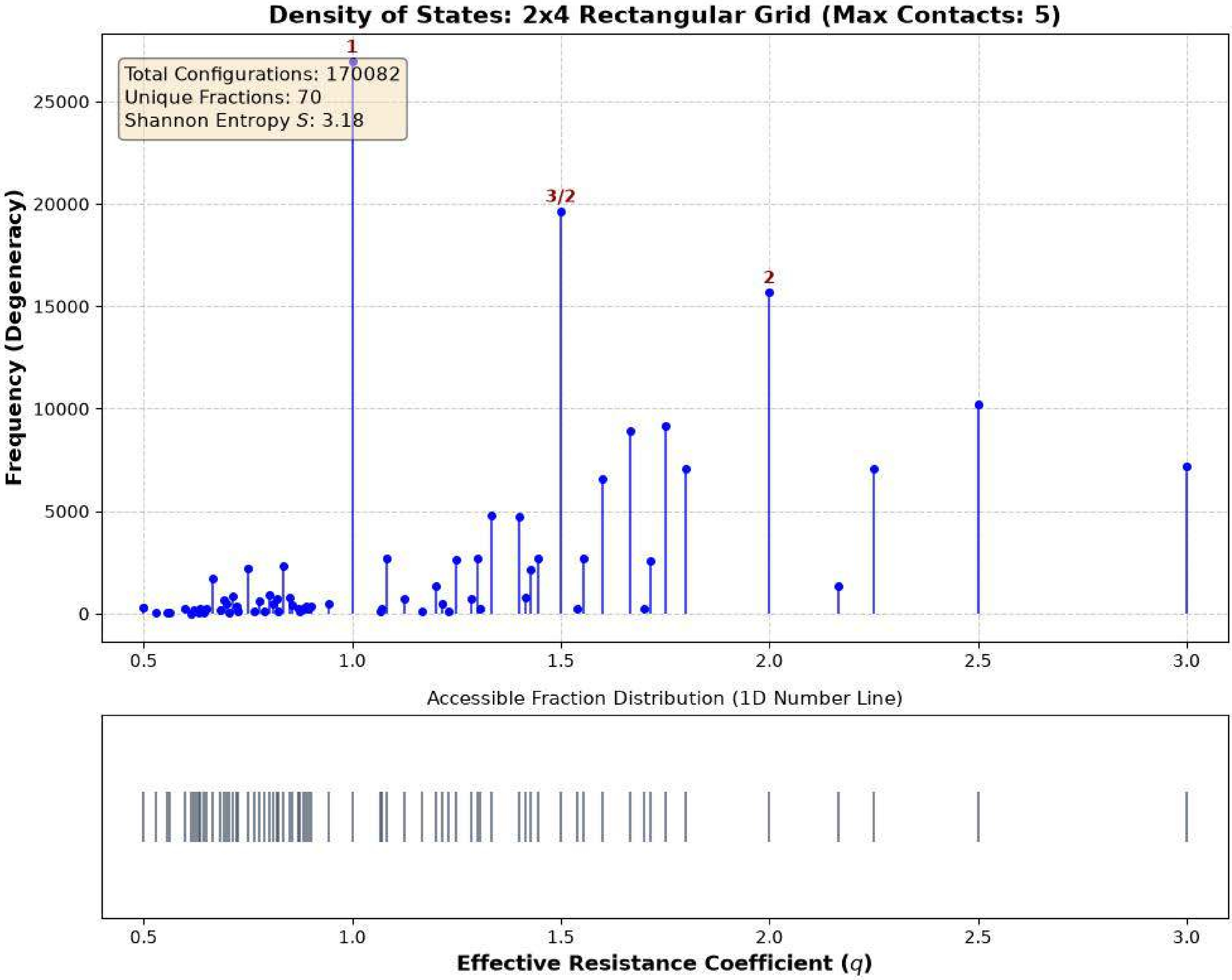
Density of States: 2x4 Rectangular Grid (Max Contacts: 5)
Total Configurations: 170082
Unique Fractions: 70
Shannon Entropy S: 3.18
1
3/2
2
Frequency (Degeneracy)
0
5000
10000
15000
20000
25000
0.5
1.0
1.5
2.0
2.5
3.0
Accessible Fraction Distribution (1D Number Line)
Effective Resistance Coefficient (q)

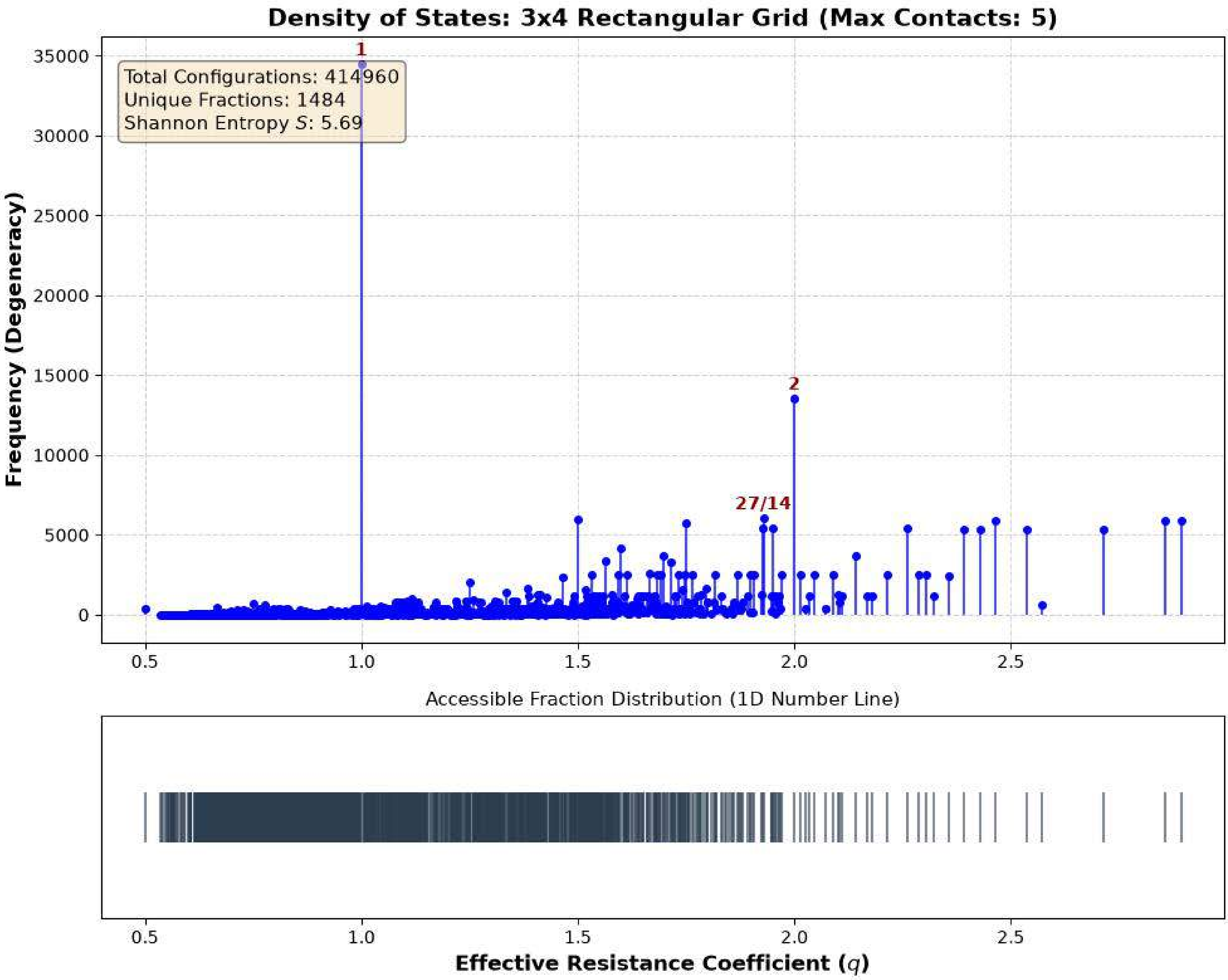

Density of States: 3x4 Rectangular Grid (Max Contacts: 5)
Total Configurations: 414960
Unique Fractions: 1484
Shannon Entropy S: 5.69
Frequency (Degeneracy)
1
2
27/14
35000
30000
25000
20000
15000
10000
5000
0
0.5
1.0
1.5
2.0
2.5
Accessible Fraction Distribution (1D Number Line)
Effective Resistance Coefficient (q)

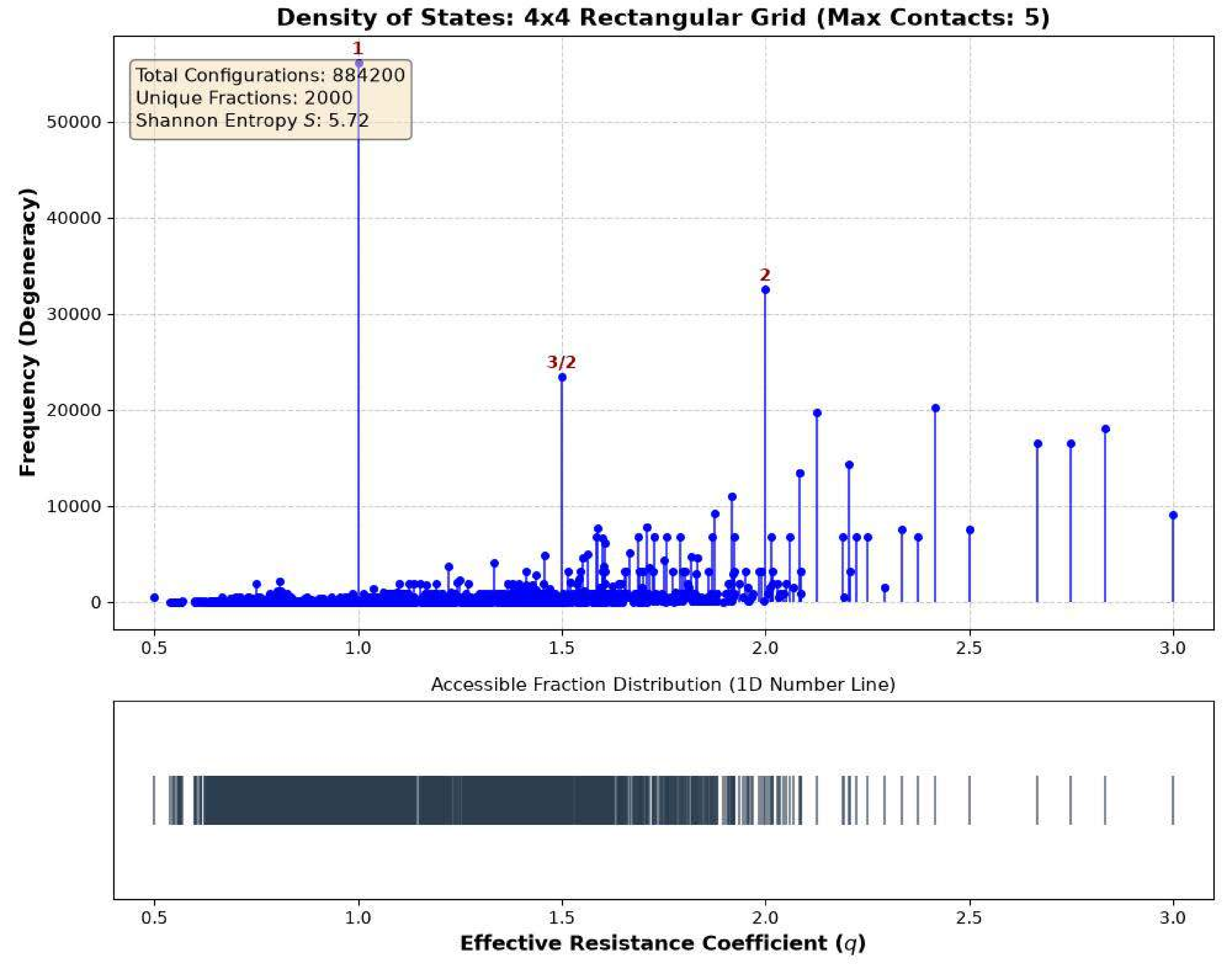

Density of States: 4x4 Rectangular Grid (Max Contacts: 5)
Total Configurations: 884200
Unique Fractions: 2000
Shannon Entropy S: 5.72
1
3/2
2
Frequency (Degeneracy)
0
10000
20000
30000
40000
50000
0.5
1.0
1.5
2.0
2.5
3.0
Accessible Fraction Distribution (1D Number Line)
Effective Resistance Coefficient (q)

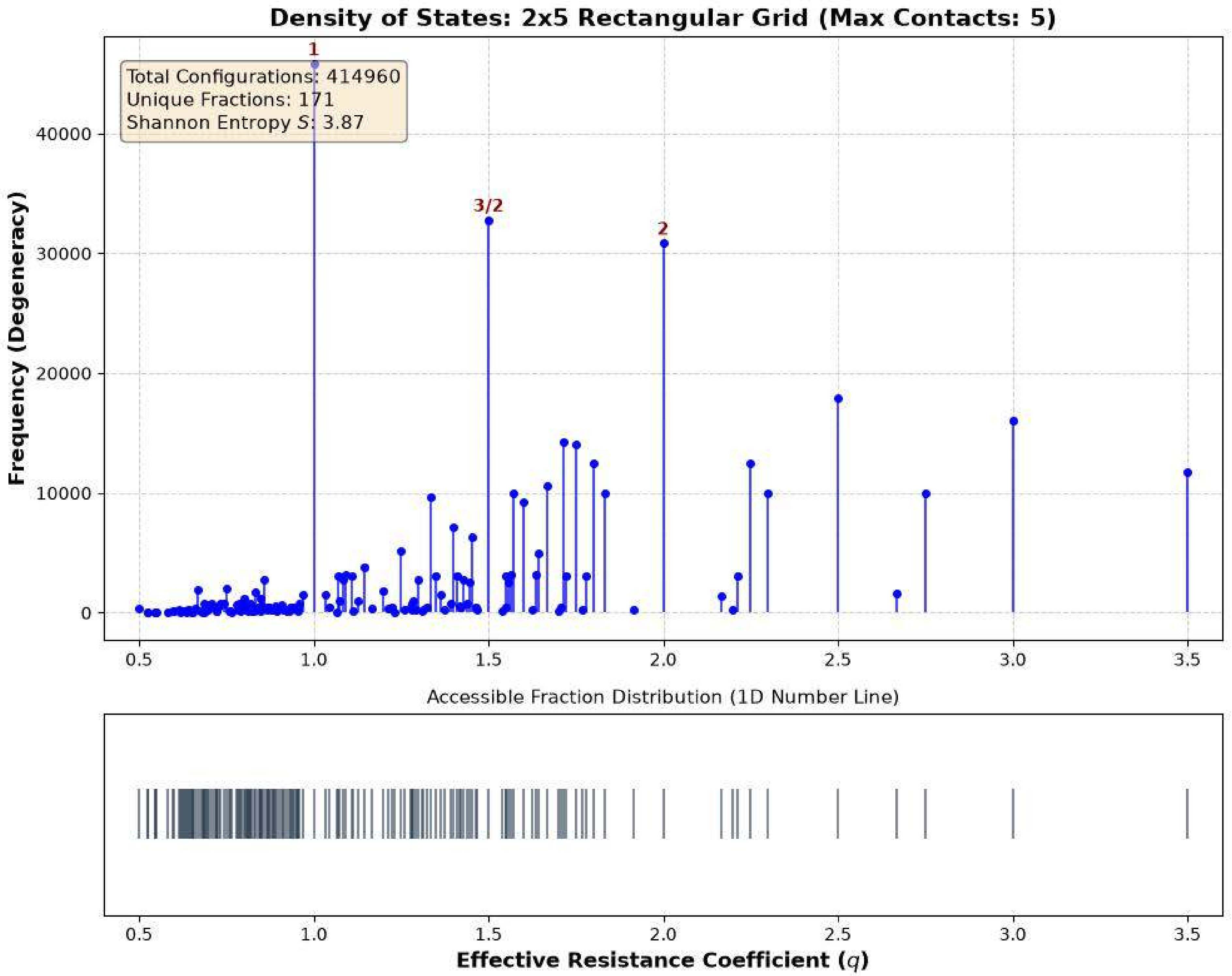

Density of States: 2x5 Rectangular Grid (Max Contacts: 5)
Total Configurations: 414960
Unique Fractions: 171
Shannon Entropy S: 3.87
1
3/2
2
Frequency (Degeneracy)
0
10000
20000
30000
40000
0.5
1.0
1.5
2.0
2.5
3.0
3.5
Accessible Fraction Distribution (1D Number Line)
Effective Resistance Coefficient (q)

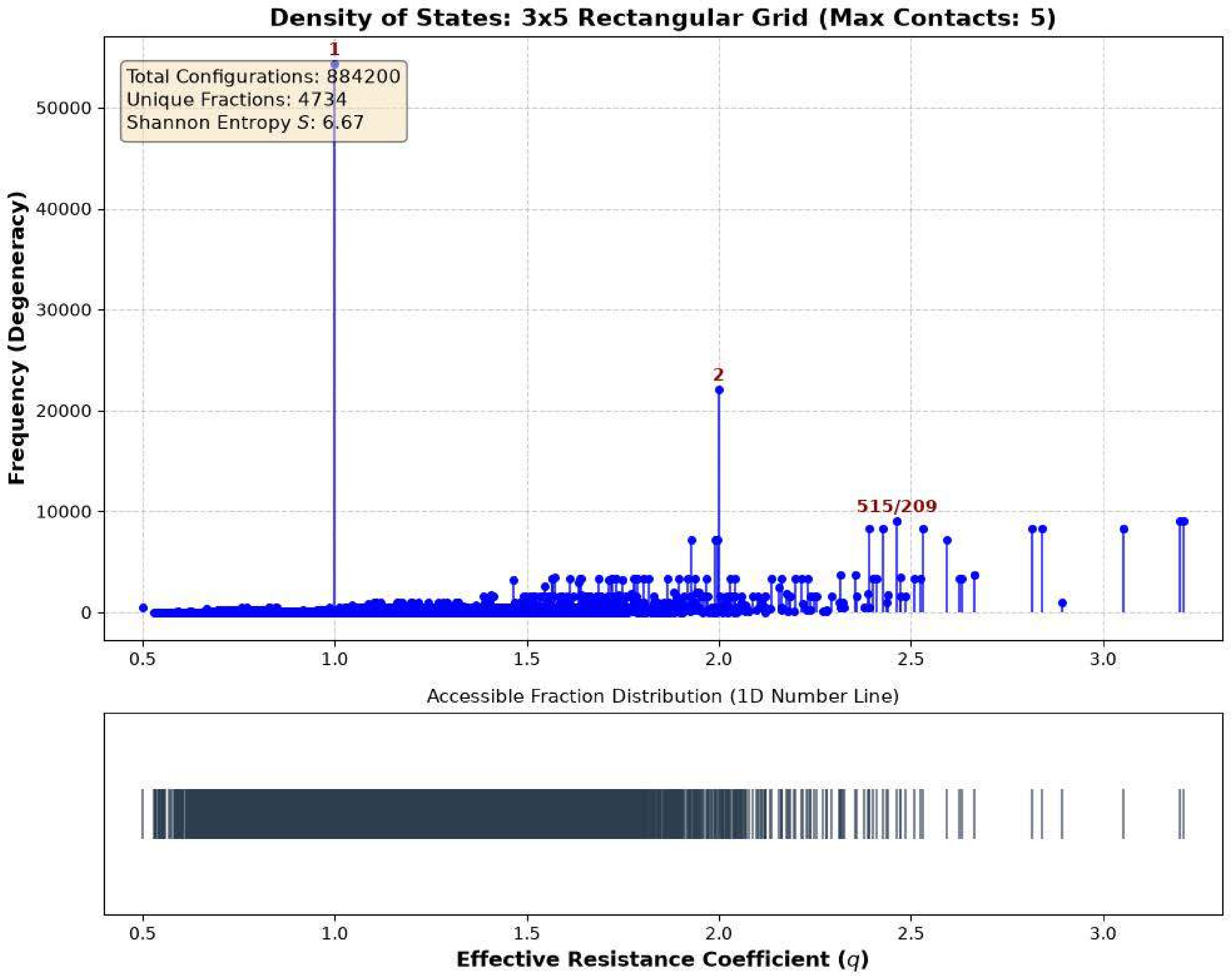

Density of States: 3x5 Rectangular Grid (Max Contacts: 5)
Total Configurations: 884200
Unique Fractions: 4734
Shannon Entropy S: 6.67
1
2
515/209
Frequency (Degeneracy)
0
10000
20000
30000
40000
50000
0.5
1.0
1.5
2.0
2.5
3.0
Accessible Fraction Distribution (1D Number Line)
Effective Resistance Coefficient (q)

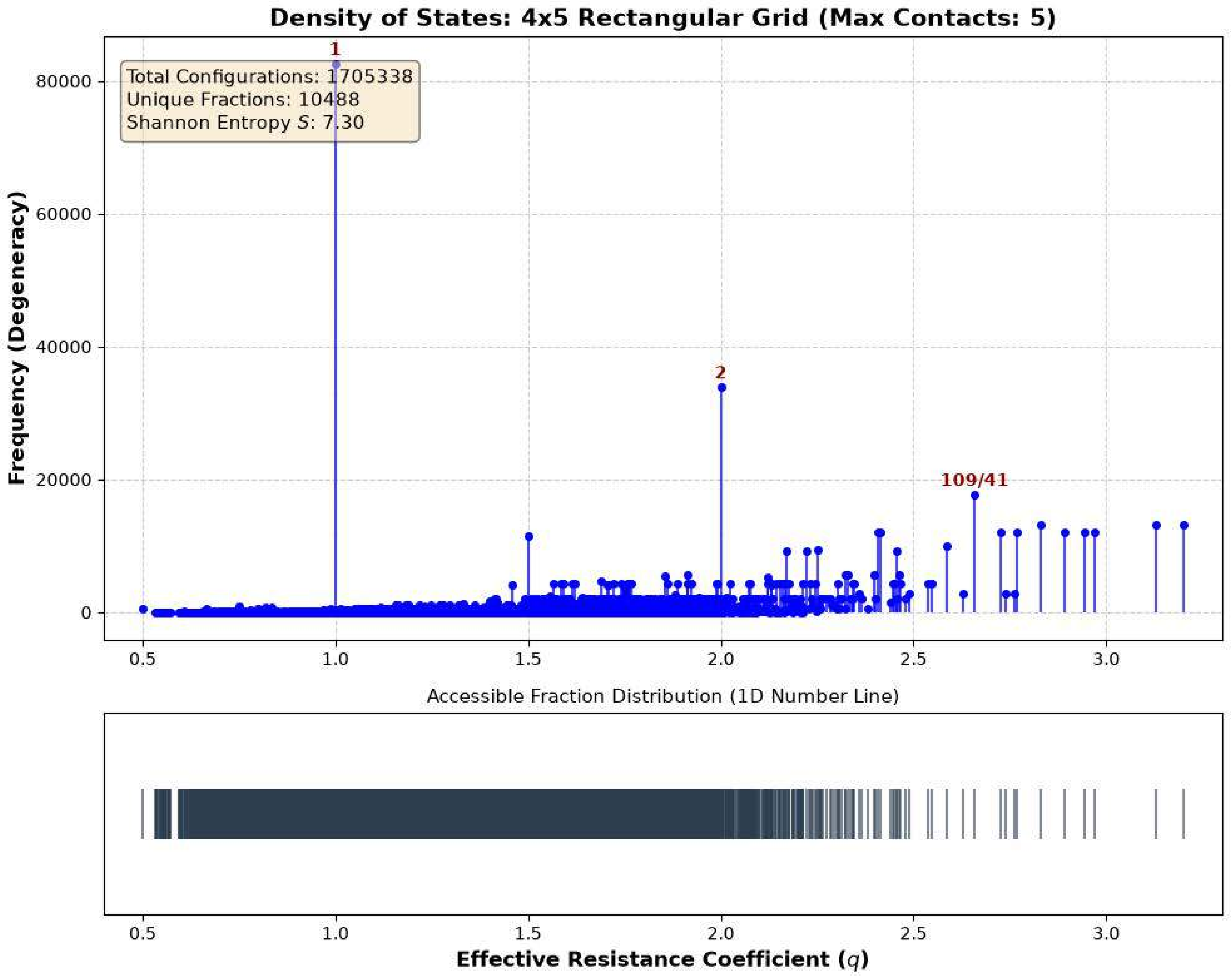
Density of States: 4x5 Rectangular Grid (Max Contacts: 5)
Total Configurations: 1705338
Unique Fractions: 10488
Shannon Entropy S: 7.30
Frequency (Degeneracy)
1
2
109/41
80000
60000
40000
20000
0
0.5
1.0
1.5
2.0
2.5
3.0
Accessible Fraction Distribution (1D Number Line)
Effective Resistance Coefficient (q)

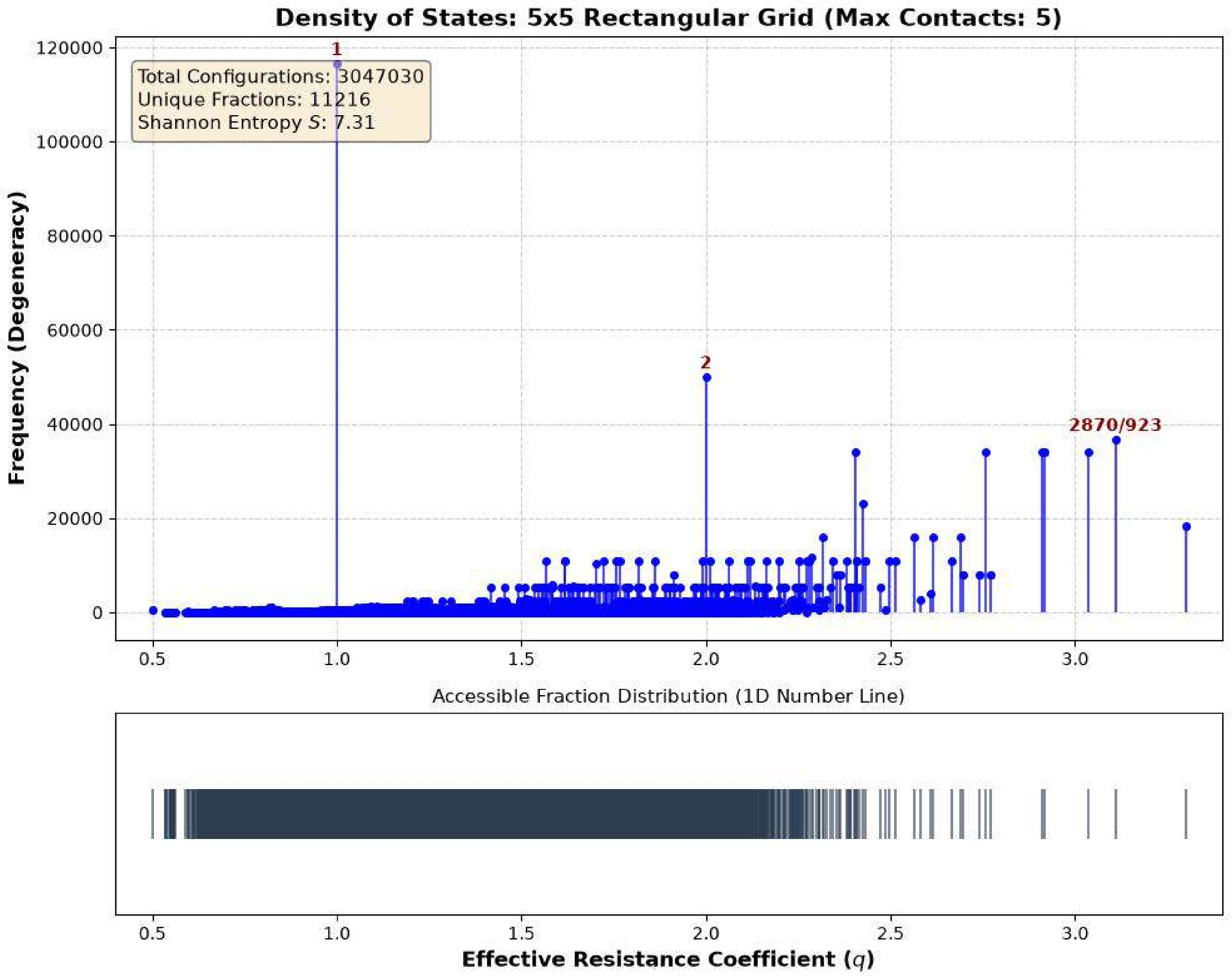
Density of States: 5x5 Rectangular Grid (Max Contacts: 5)
Total Configurations: 3047030
Unique Fractions: 11216
Shannon Entropy $S$: 7.31
1
2
2870/923
Frequency (Degeneracy)
120000
100000
80000
60000
40000
20000
0
0.5
1.0
1.5
2.0
2.5
3.0
Accessible Fraction Distribution (1D Number Line)
Effective Resistance Coefficient ($q$)

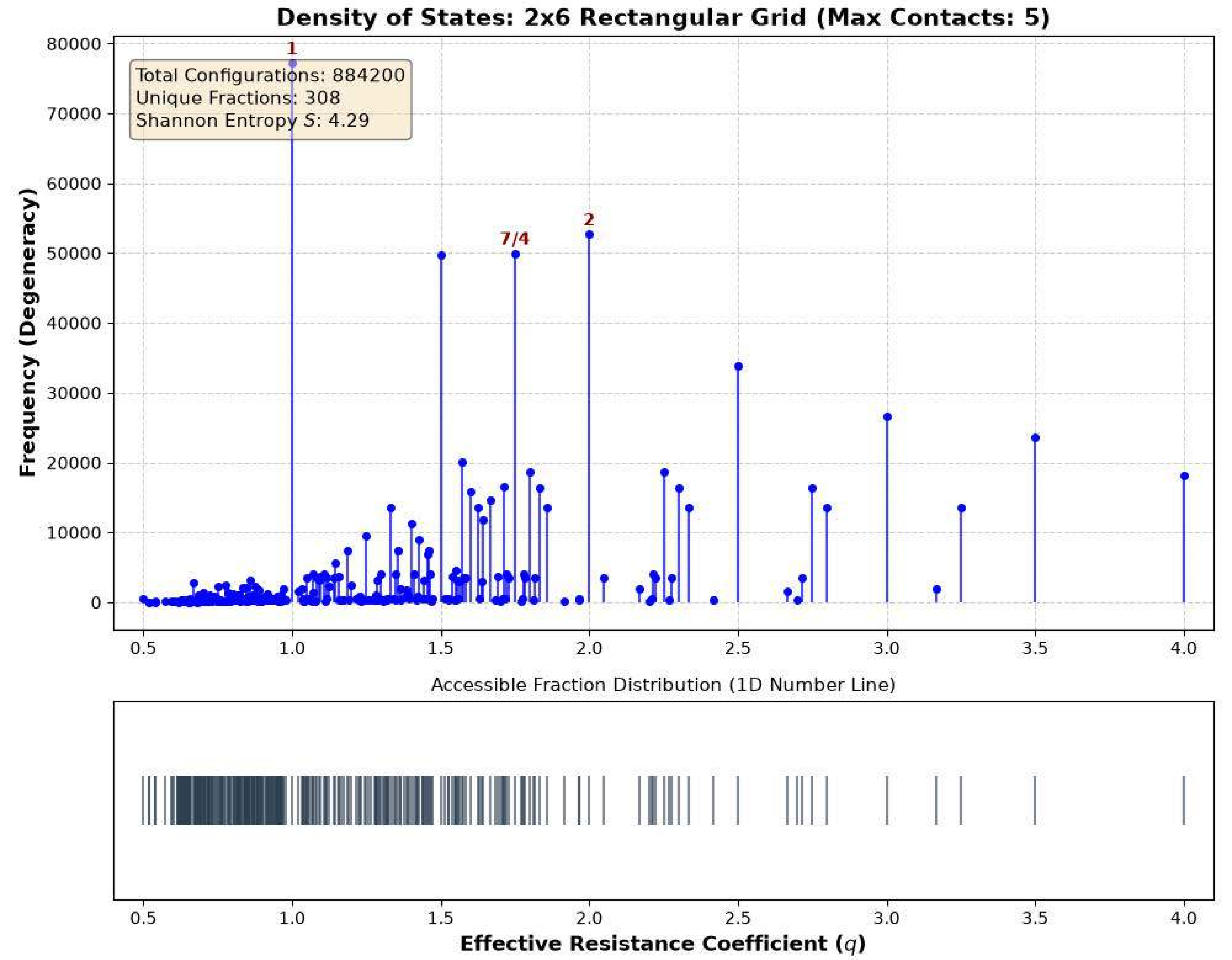
Density of States: 2x6 Rectangular Grid (Max Contacts: 5)
Total Configurations: 884200
Unique Fractions: 308
Shannon Entropy S: 4.29
1
7/4
2
Frequency (Degeneracy)
0
10000
20000
30000
40000
50000
60000
70000
80000
0.5
1.0
1.5
2.0
2.5
3.0
3.5
4.0
Accessible Fraction Distribution (1D Number Line)
Effective Resistance Coefficient (q)

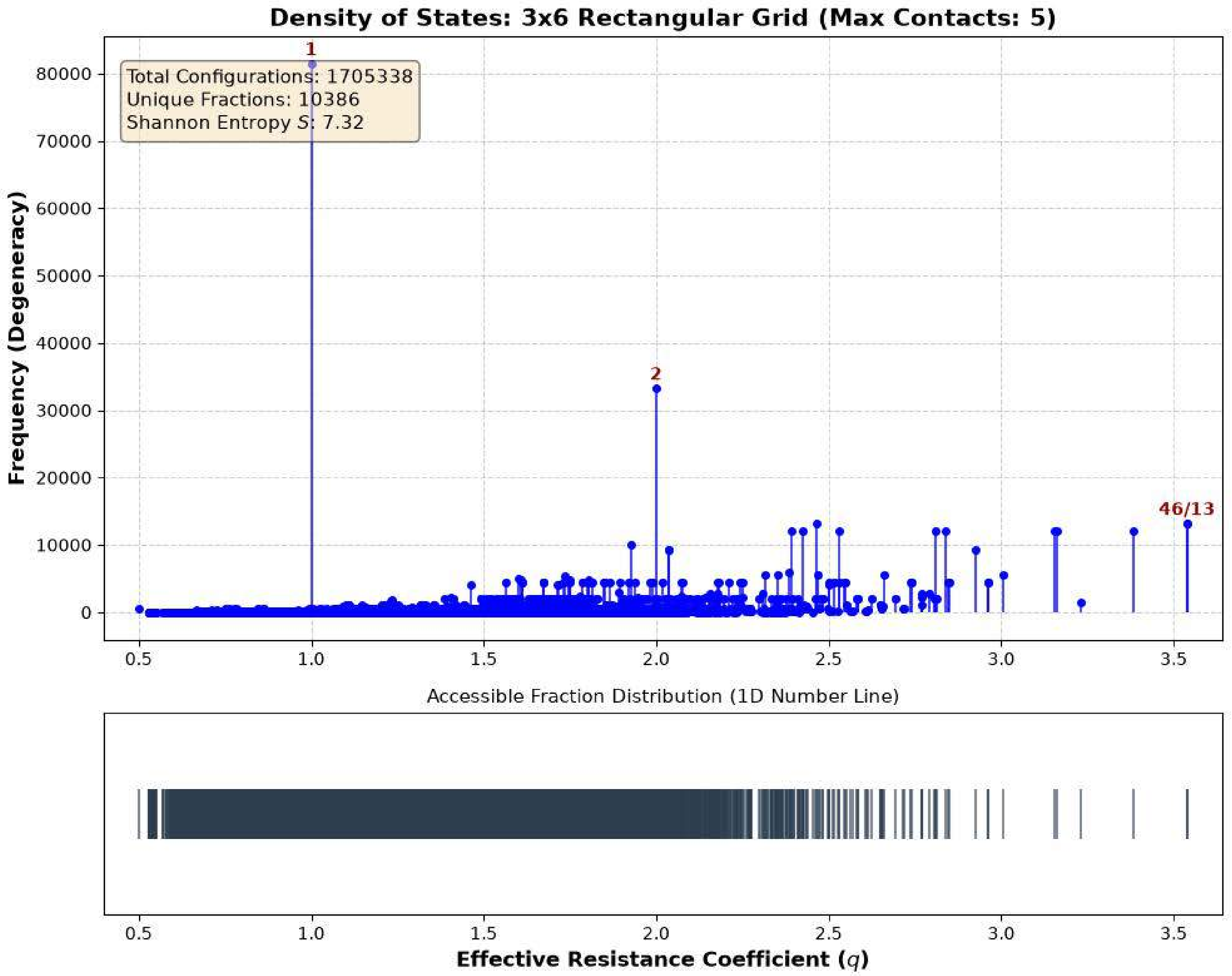
Density of States: 3x6 Rectangular Grid (Max Contacts: 5)
Total Configurations: 1705338
Unique Fractions: 10386
Shannon Entropy S: 7.32
1
2
46/13
Frequency (Degeneracy)
0
10000
20000
30000
40000
50000
60000
70000
80000
0.5
1.0
1.5
2.0
2.5
3.0
3.5
Accessible Fraction Distribution (1D Number Line)
Effective Resistance Coefficient (q)

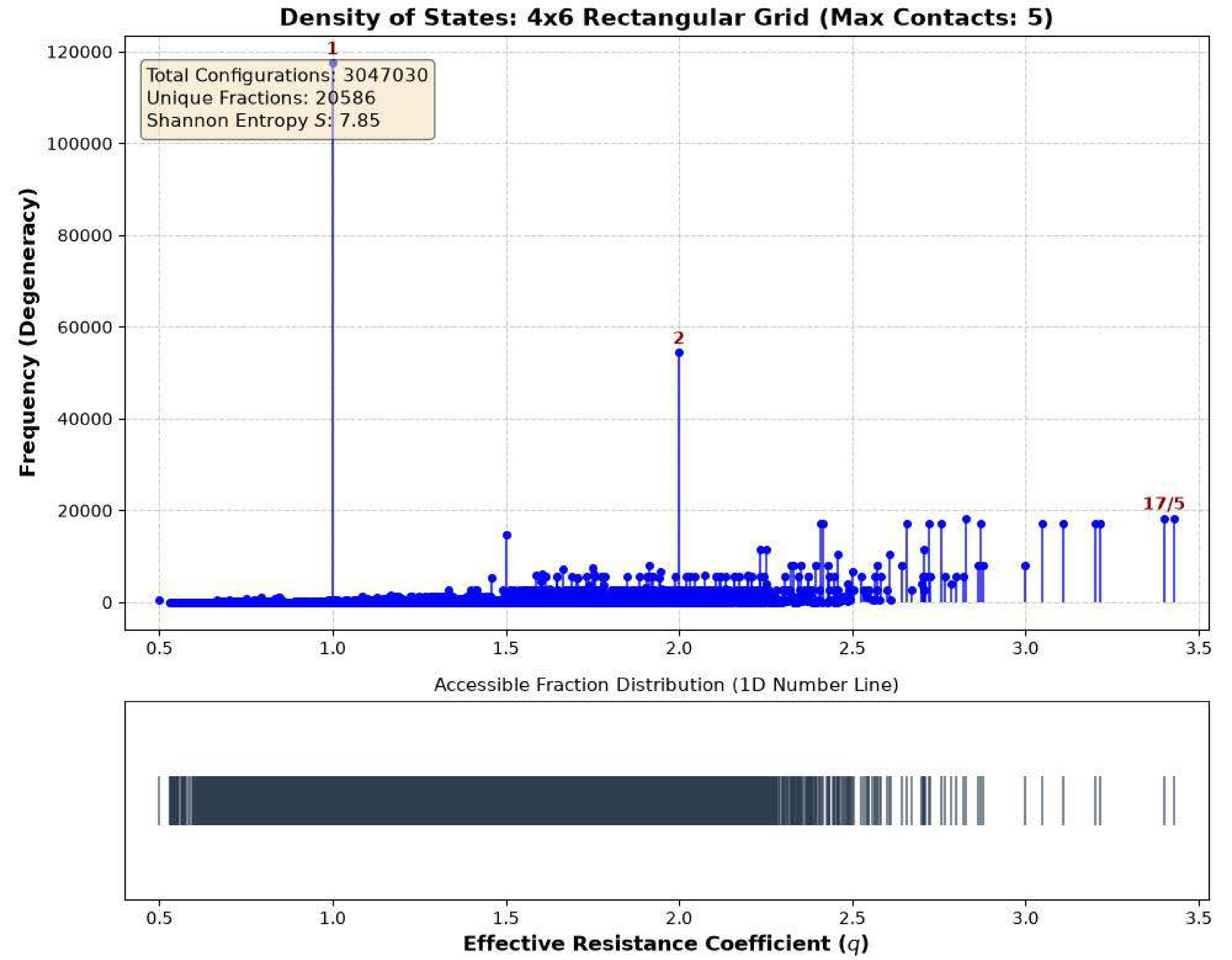

Density of States: 4x6 Rectangular Grid (Max Contacts: 5)
Total Configurations: 3047030
Unique Fractions: 20586
Shannon Entropy S: 7.85
1
2
17/5
Frequency (Degeneracy)
120000
100000
80000
60000
40000
20000
0
0.5
1.0
1.5
2.0
2.5
3.0
3.5
Accessible Fraction Distribution (1D Number Line)
Effective Resistance Coefficient (q)

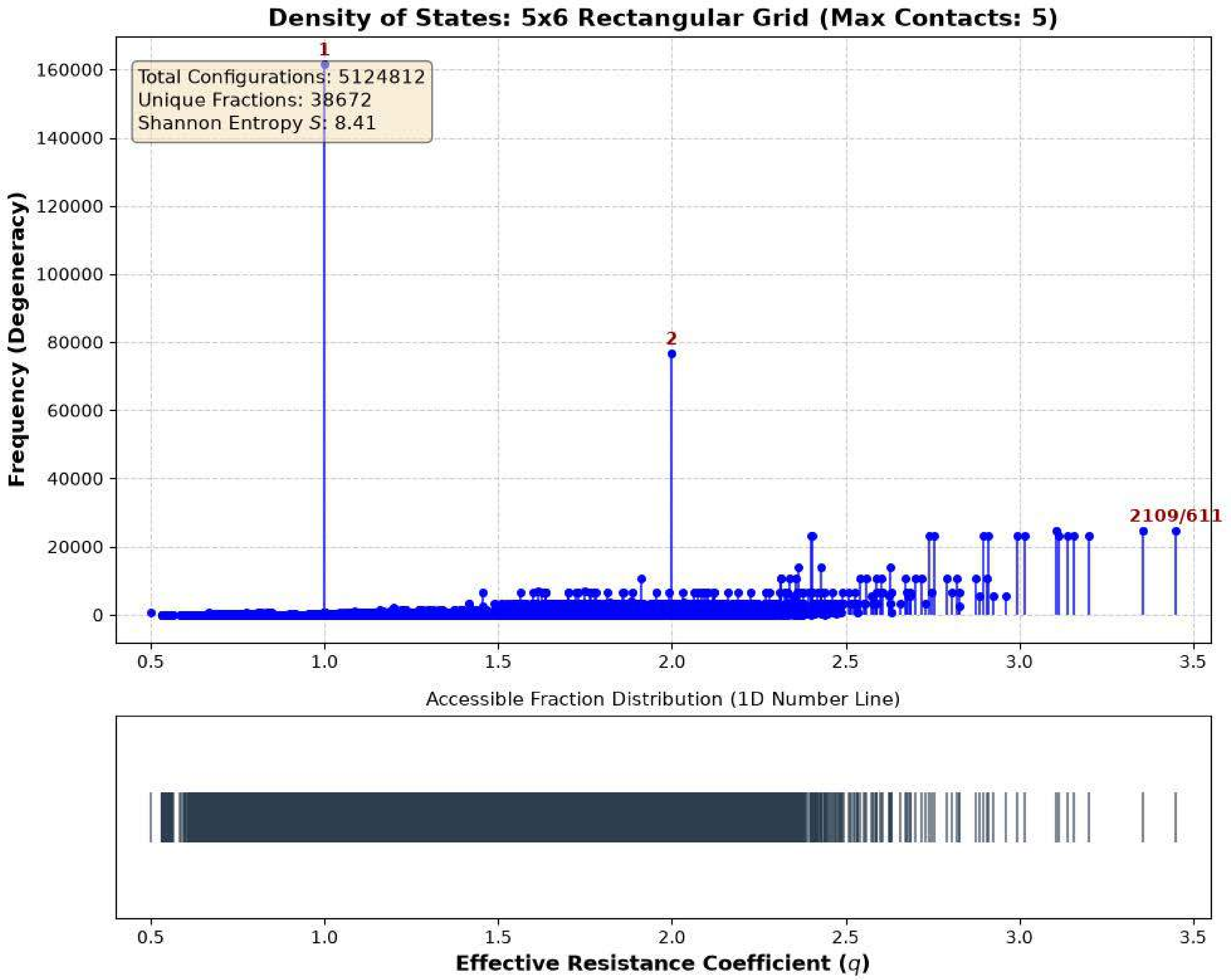

Density of States: 5x6 Rectangular Grid (Max Contacts: 5)
Total Configurations: 5124812
Unique Fractions: 38672
Shannon Entropy S: 8.41
1
2
2109/611
Frequency (Degeneracy)
0
20000
40000
60000
80000
100000
120000
140000
160000
0.5
1.0
1.5
2.0
2.5
3.0
3.5
Accessible Fraction Distribution (1D Number Line)
Effective Resistance Coefficient (q)

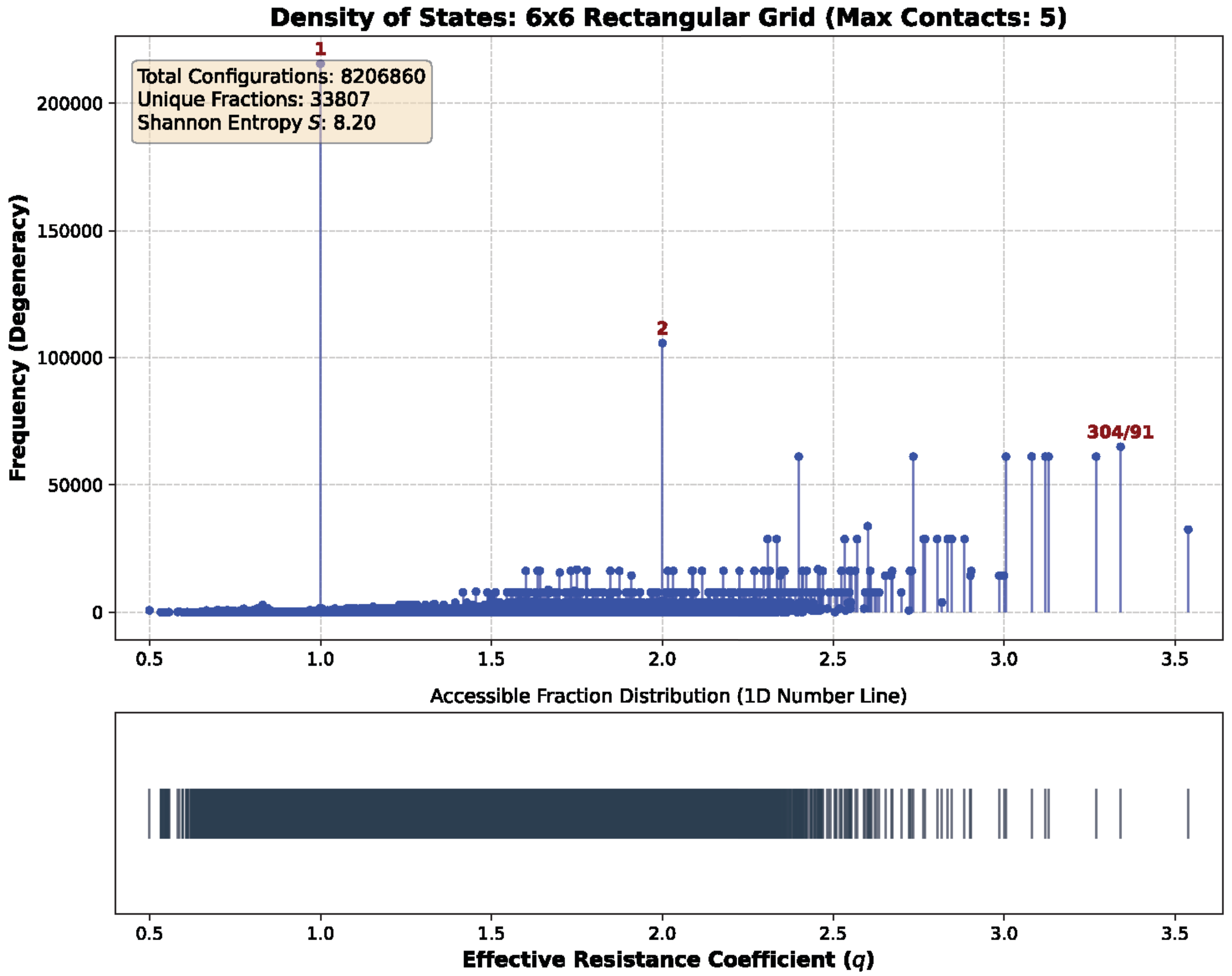

Density of States: 6x6 Rectangular Grid (Max Contacts: 5)
Total Configurations: 8206860
Unique Fractions: 33807
Shannon Entropy S: 8.20
1
2
304/91
Frequency (Degeneracy)
0
50000
100000
150000
200000
0.5
1.0
1.5
2.0
2.5
3.0
3.5
Accessible Fraction Distribution (1D Number Line)
Effective Resistance Coefficient (q)

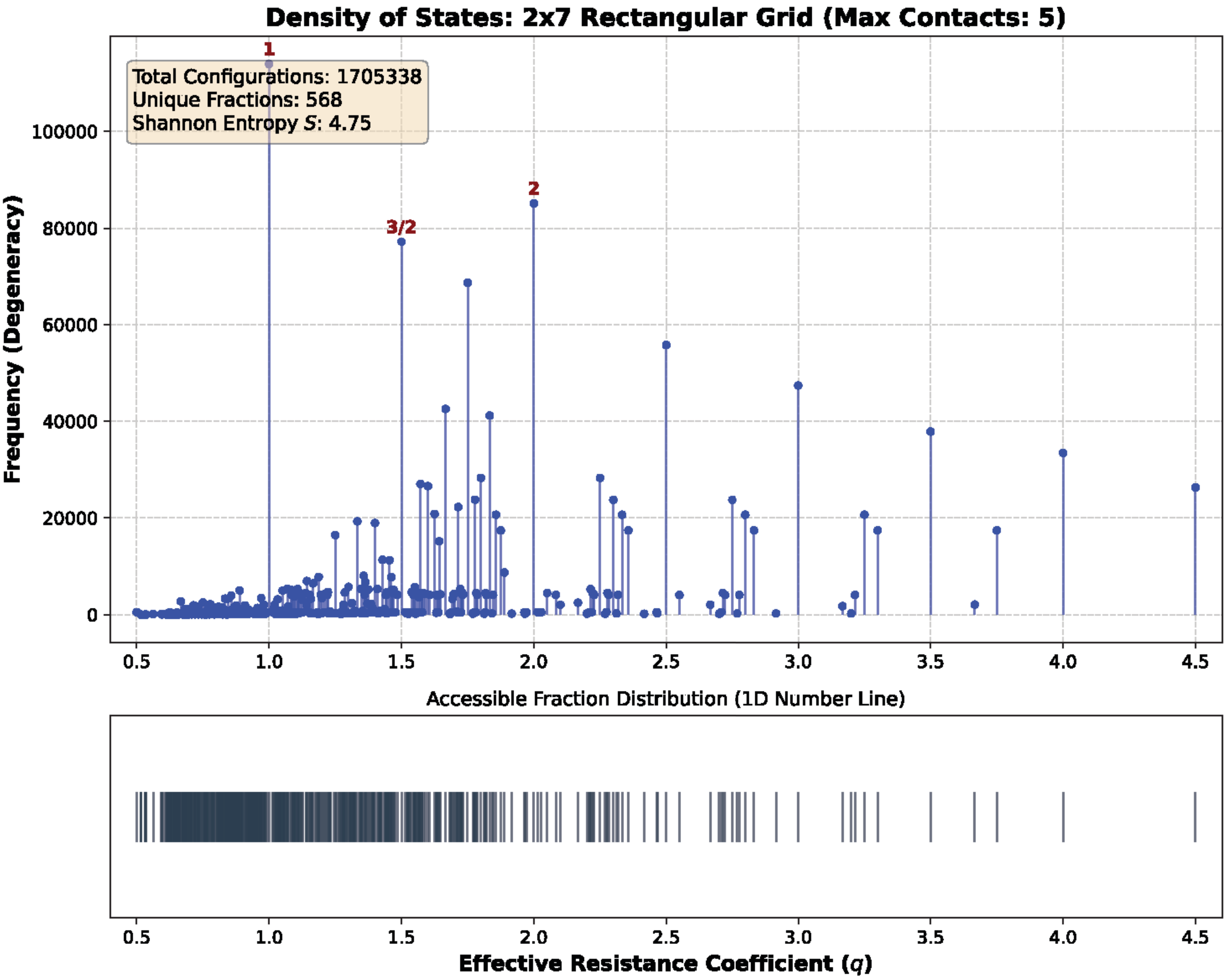

Density of States: 2x7 Rectangular Grid (Max Contacts: 5)
Total Configurations: 1705338
Unique Fractions: 568
Shannon Entropy S: 4.75
1
3/2
2
Frequency (Degeneracy)
0
20000
40000
60000
80000
100000
0.5
1.0
1.5
2.0
2.5
3.0
3.5
4.0
4.5
Accessible Fraction Distribution (1D Number Line)
Effective Resistance Coefficient (q)

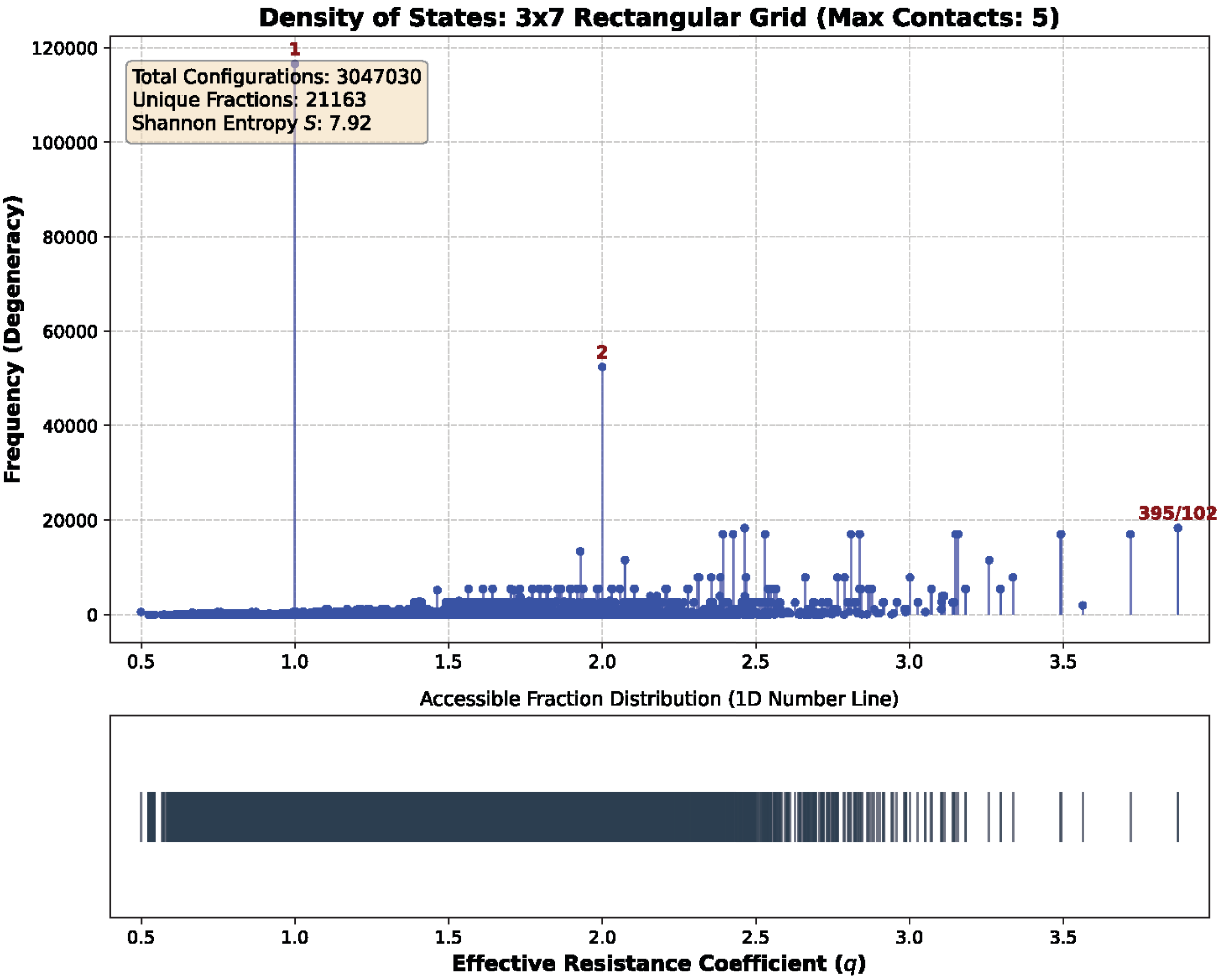
Density of States: 3x7 Rectangular Grid (Max Contacts: 5)
Total Configurations: 3047030
Unique Fractions: 21163
Shannon Entropy S: 7.92
1
2
395/102
Frequency (Degeneracy)
0
20000
40000
60000
80000
100000
120000
0.5
1.0
1.5
2.0
2.5
3.0
3.5
Accessible Fraction Distribution (1D Number Line)
Effective Resistance Coefficient (q)

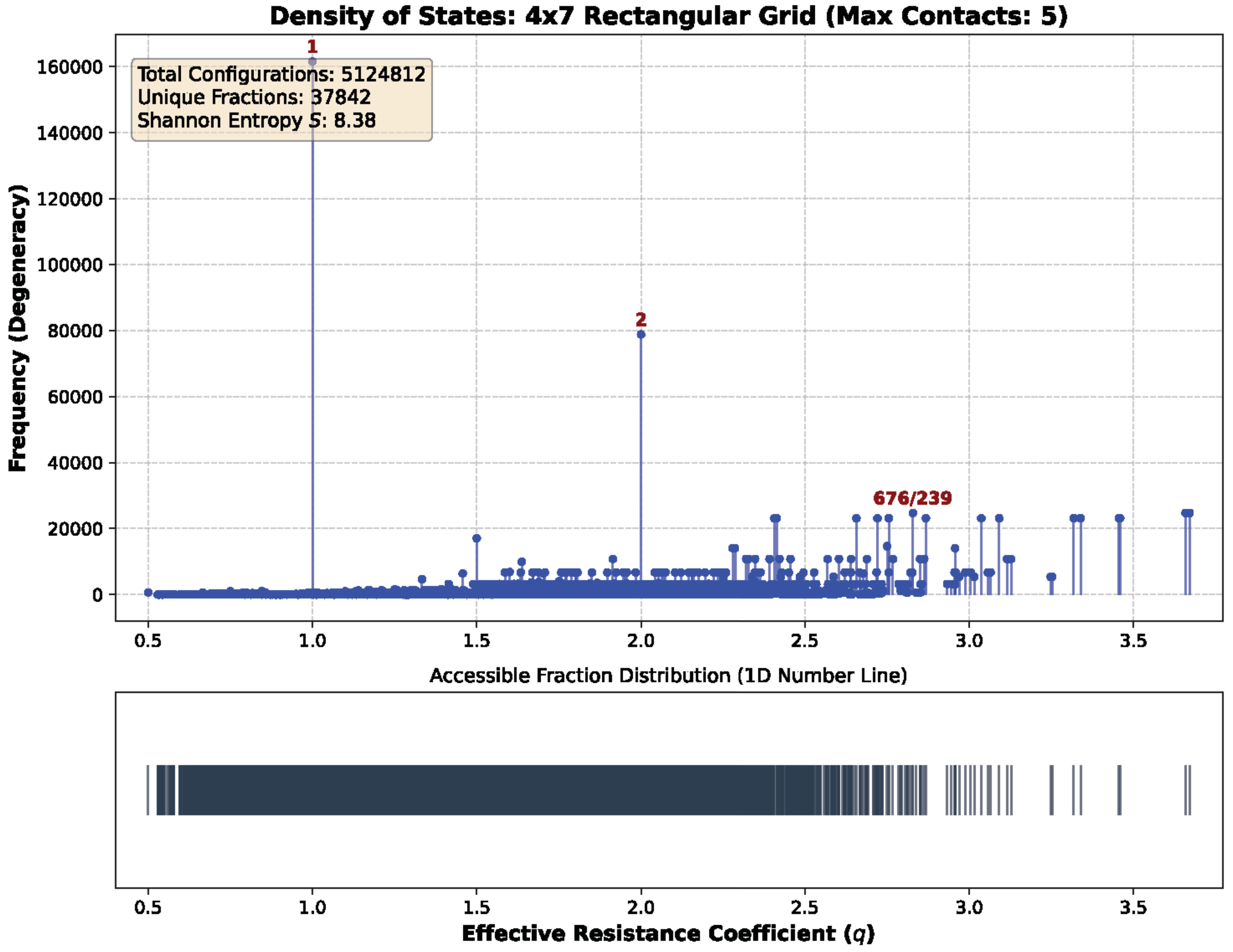
Density of States: 4x7 Rectangular Grid (Max Contacts: 5)
Total Configurations: 5124812
Unique Fractions: 37842
Shannon Entropy S: 8.38
Frequency (Degeneracy)
160000
140000
120000
100000
80000
60000
40000
20000
0
1
2
676/239
0.5
1.0
1.5
2.0
2.5
3.0
3.5
Accessible Fraction Distribution (1D Number Line)
Effective Resistance Coefficient (q)

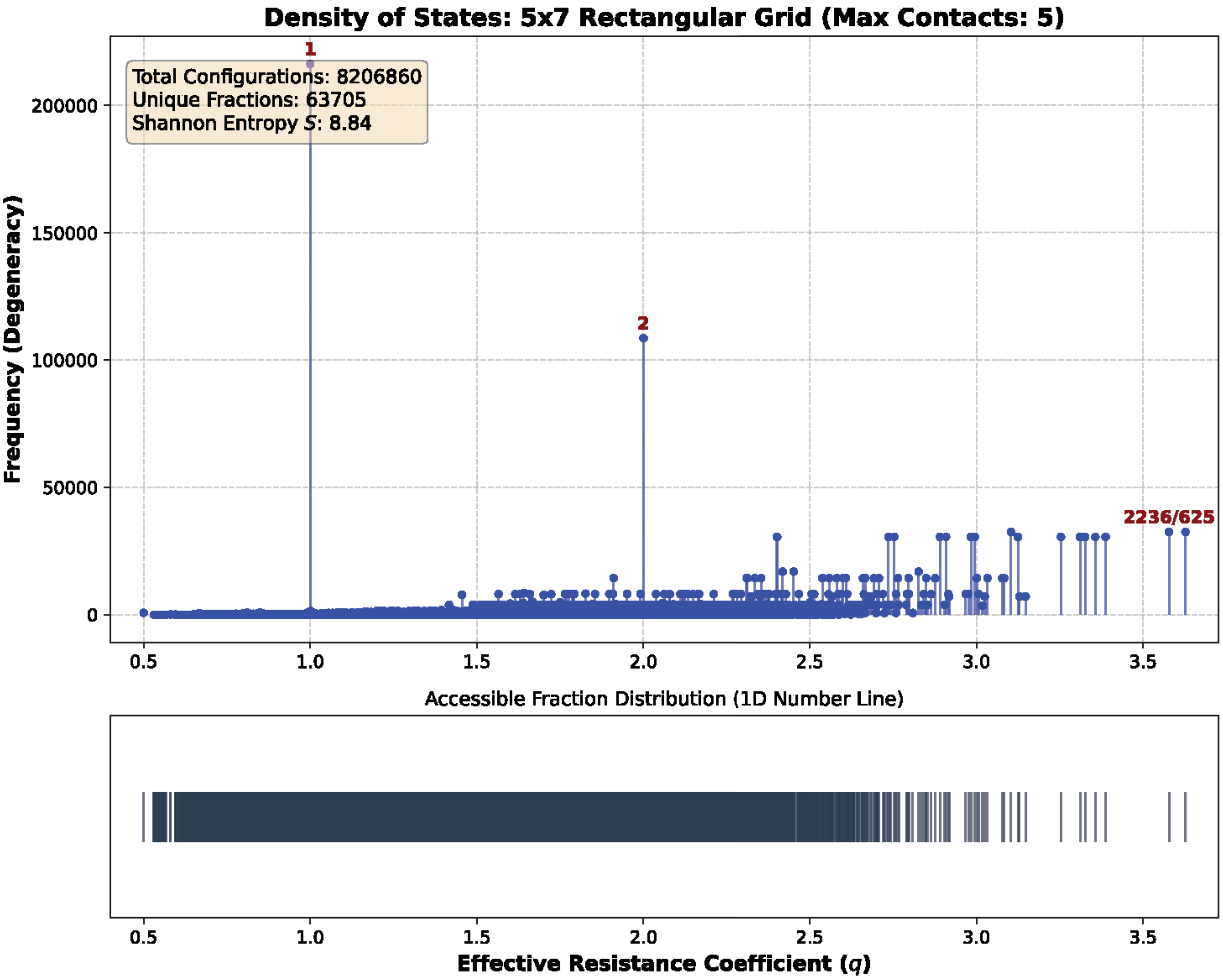

Density of States: 5x7 Rectangular Grid (Max Contacts: 5)
Total Configurations: 8206860
Unique Fractions: 63705
Shannon Entropy S: 8.84
1
2
2236/625
Frequency (Degeneracy)
0
50000
100000
150000
200000
0.5
1.0
1.5
2.0
2.5
3.0
3.5
Accessible Fraction Distribution (1D Number Line)
Effective Resistance Coefficient (q)

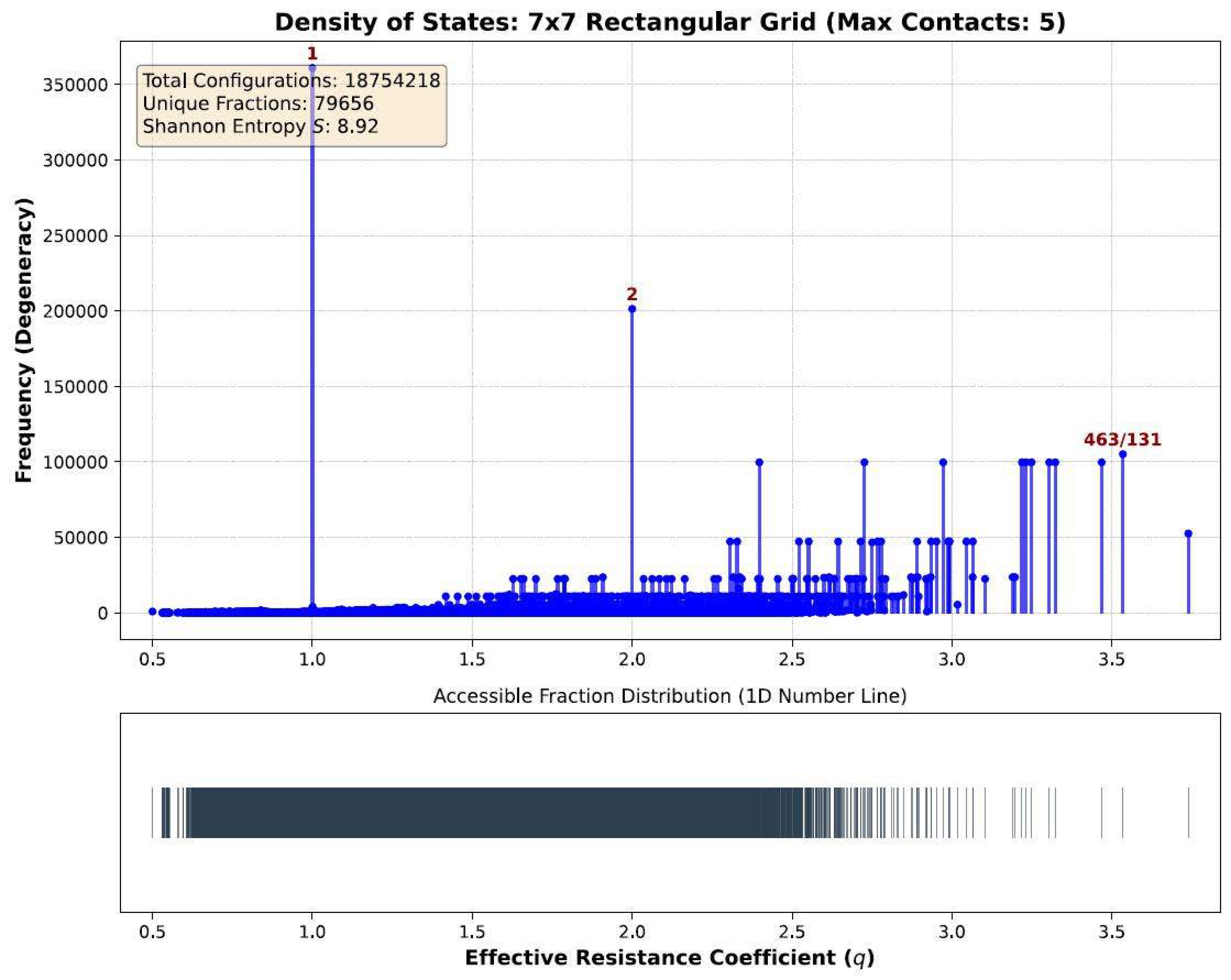

Density of States: 7x7 Rectangular Grid (Max Contacts: 5)
Total Configurations: 18754218
Unique Fractions: 79656
Shannon Entropy S: 8.92
1
2
463/131
Frequency (Degeneracy)
0
50000
100000
150000
200000
250000
300000
350000
0.5
1.0
1.5
2.0
2.5
3.0
3.5
Accessible Fraction Distribution (1D Number Line)
Effective Resistance Coefficient (q)

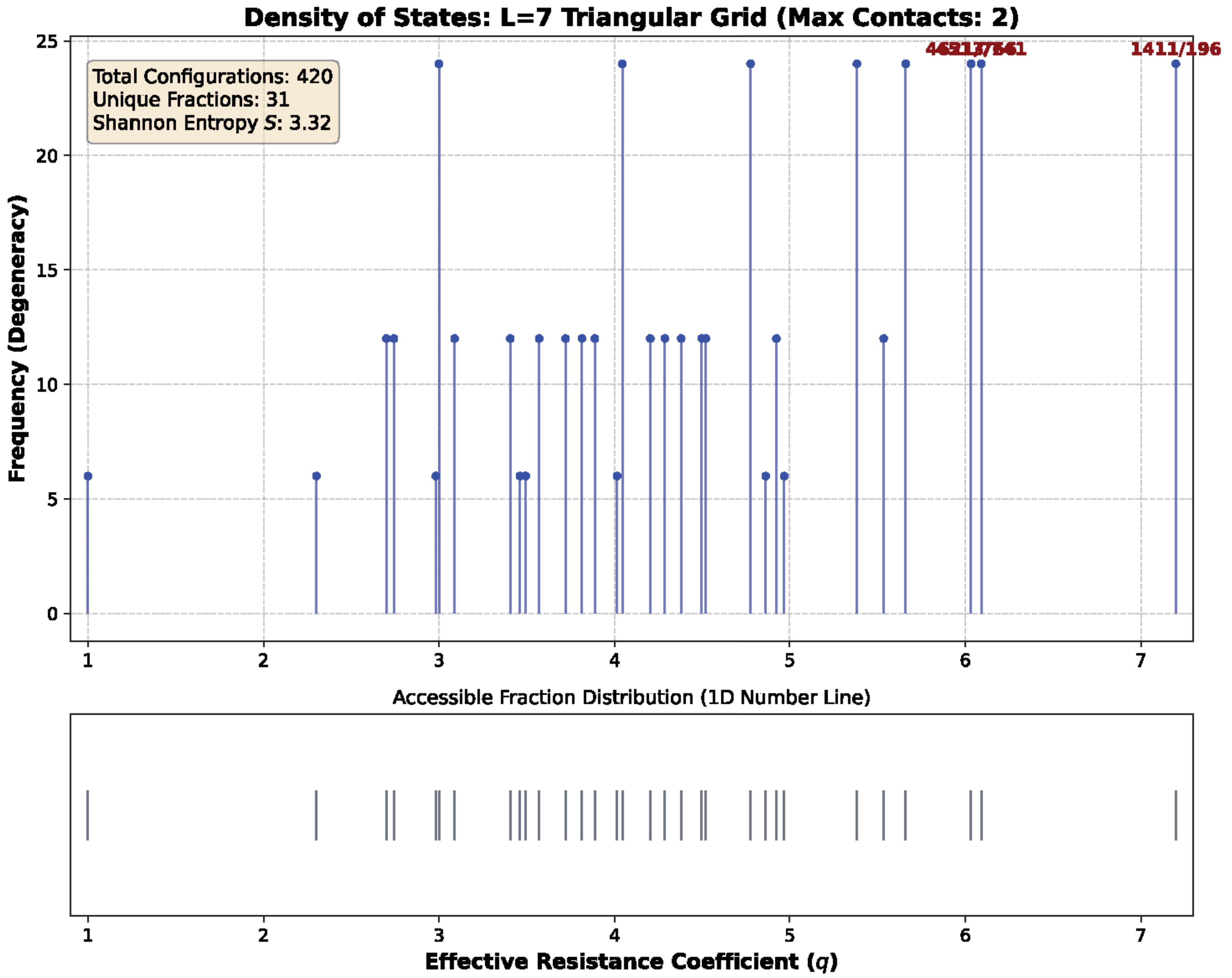
Density of States: L=7 Triangular Grid (Max Contacts: 2)
Total Configurations: 420
Unique Fractions: 31
Shannon Entropy S: 3.32
1411/196
Frequency (Degeneracy)
Accessible Fraction Distribution (1D Number Line)
Effective Resistance Coefficient (q)

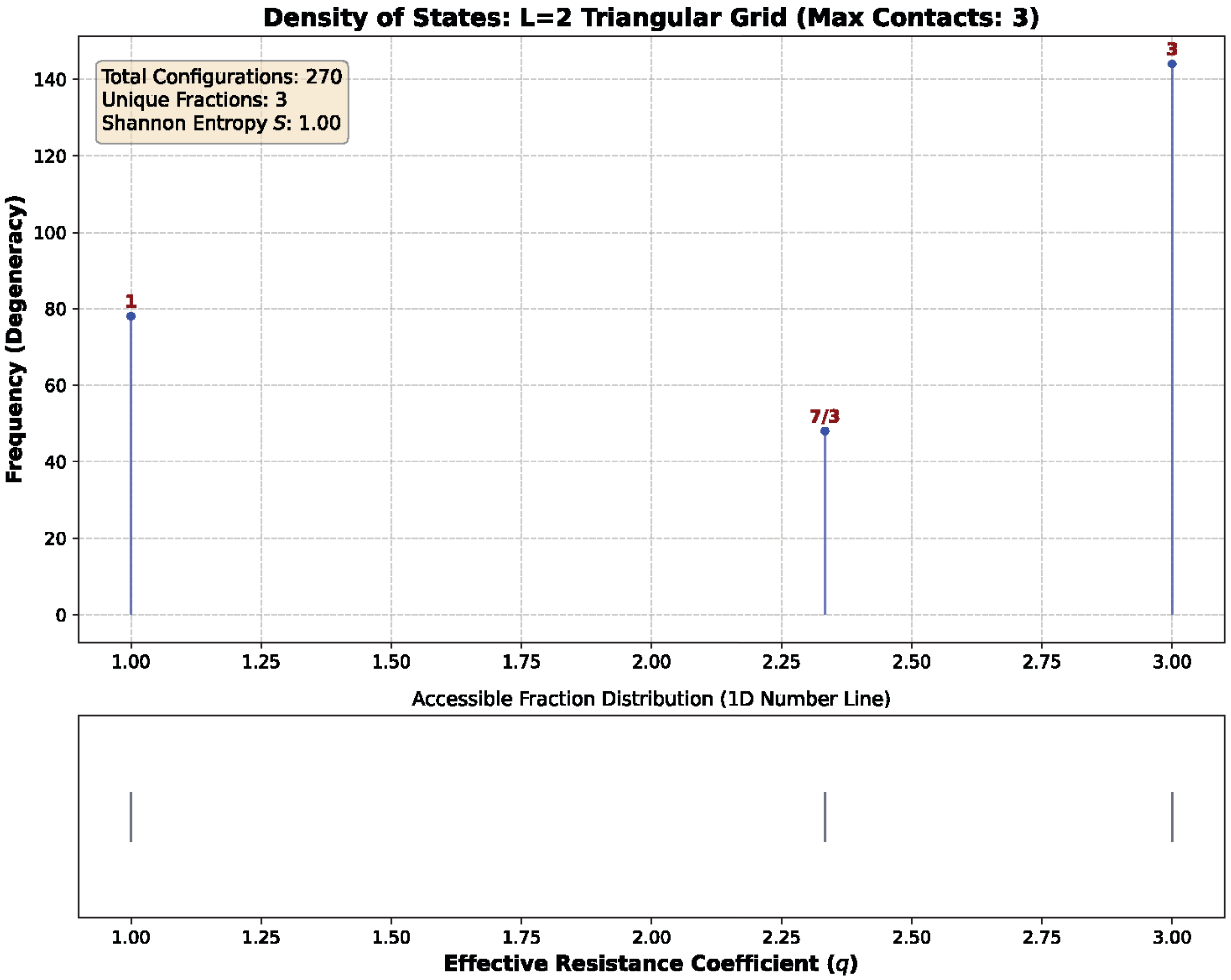

Density of States: L=2 Triangular Grid (Max Contacts: 3)
Total Configurations: 270
Unique Fractions: 3
Shannon Entropy S: 1.00
Frequency (Degeneracy)
1
7/3
3
0
20
40
60
80
100
120
140
1.00
1.25
1.50
1.75
2.00
2.25
2.50
2.75
3.00
Accessible Fraction Distribution (1D Number Line)
Effective Resistance Coefficient (q)

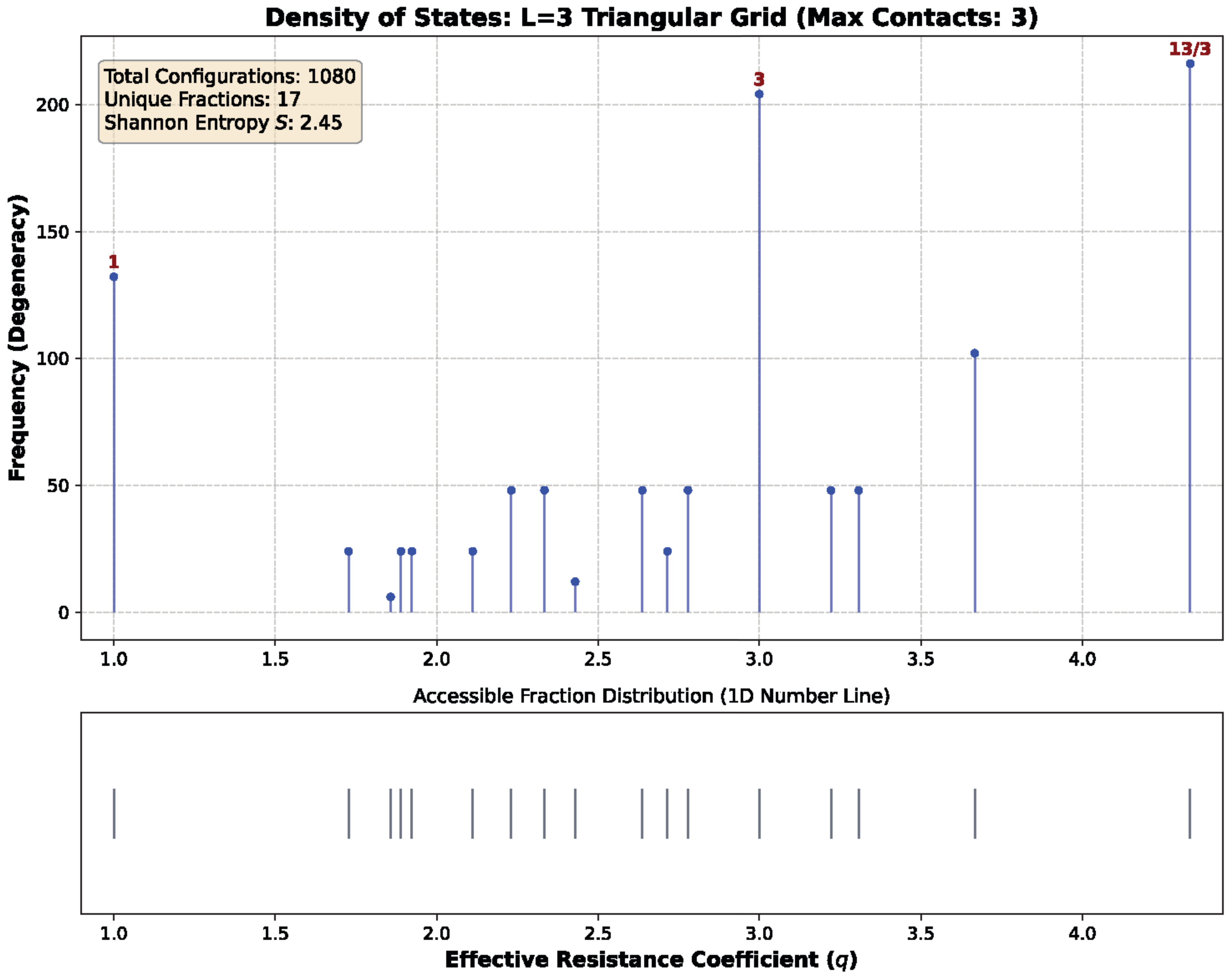
Density of States: L=3 Triangular Grid (Max Contacts: 3)
Total Configurations: 1080
Unique Fractions: 17
Shannon Entropy S: 2.45
1
3
13/3
Frequency (Degeneracy)
0
50
100
150
200
1.0
1.5
2.0
2.5
3.0
3.5
4.0
Accessible Fraction Distribution (1D Number Line)
Effective Resistance Coefficient (q)

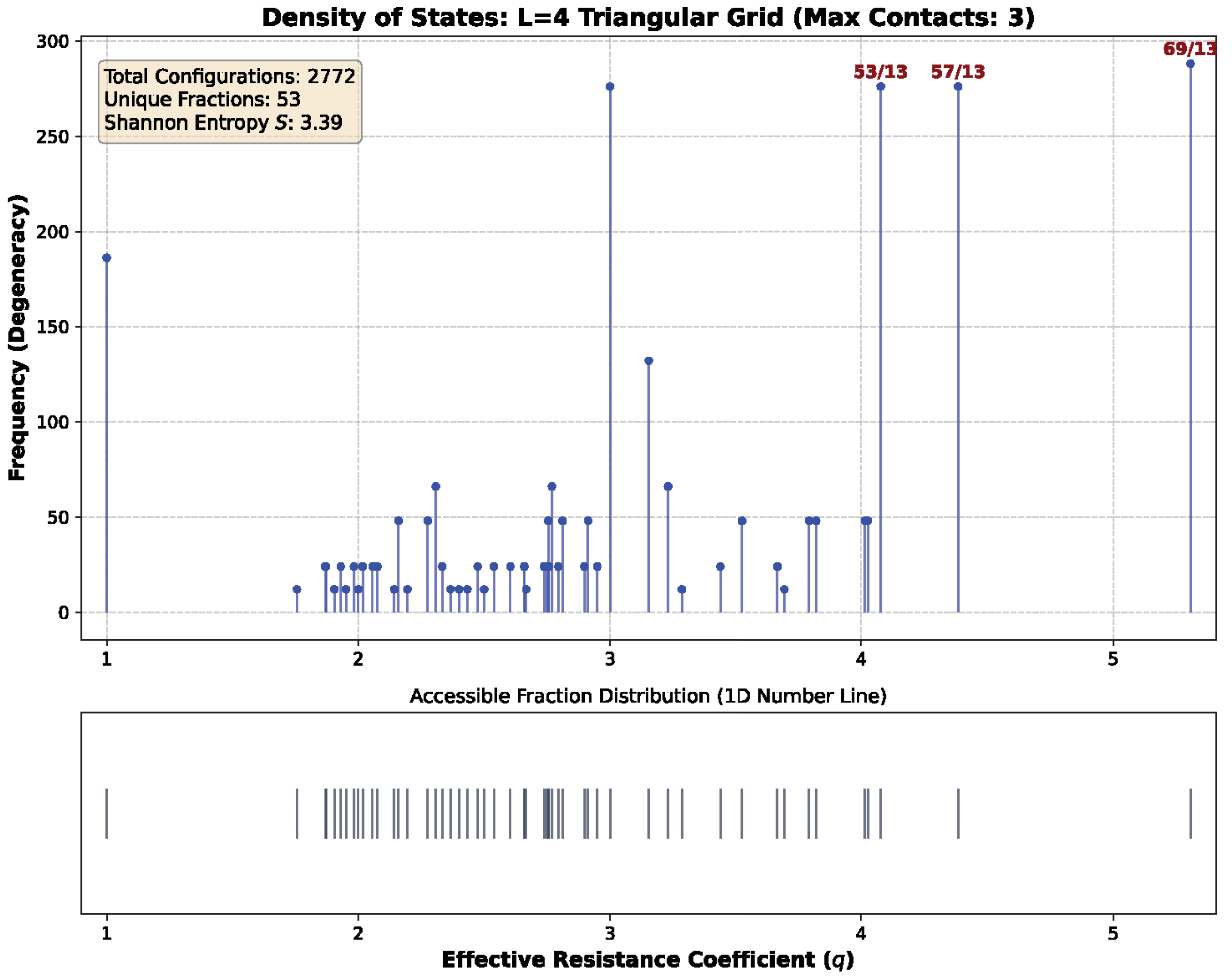

Density of States: L=4 Triangular Grid (Max Contacts: 3)
Total Configurations: 2772
Unique Fractions: 53
Shannon Entropy S: 3.39
53/13
57/13
69/13
Frequency (Degeneracy)
0
50
100
150
200
250
300
1
2
3
4
5
Accessible Fraction Distribution (1D Number Line)
Effective Resistance Coefficient (q)

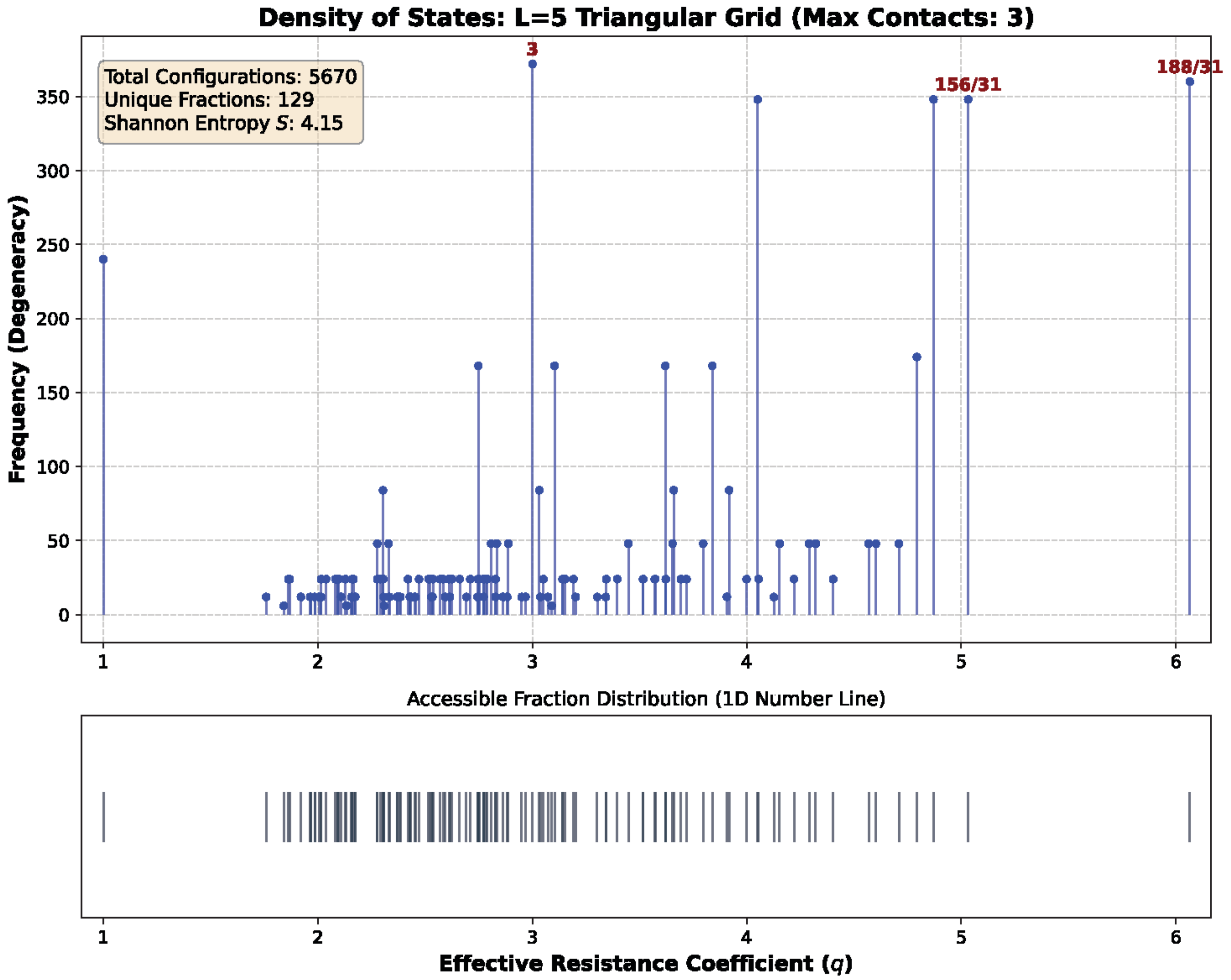
Density of States: L=5 Triangular Grid (Max Contacts: 3)
Total Configurations: 5670
Unique Fractions: 129
Shannon Entropy S: 4.15
3
156/31
188/31
Frequency (Degeneracy)
0
50
100
150
200
250
300
350
1
2
3
4
5
6
Accessible Fraction Distribution (1D Number Line)
Effective Resistance Coefficient (q)

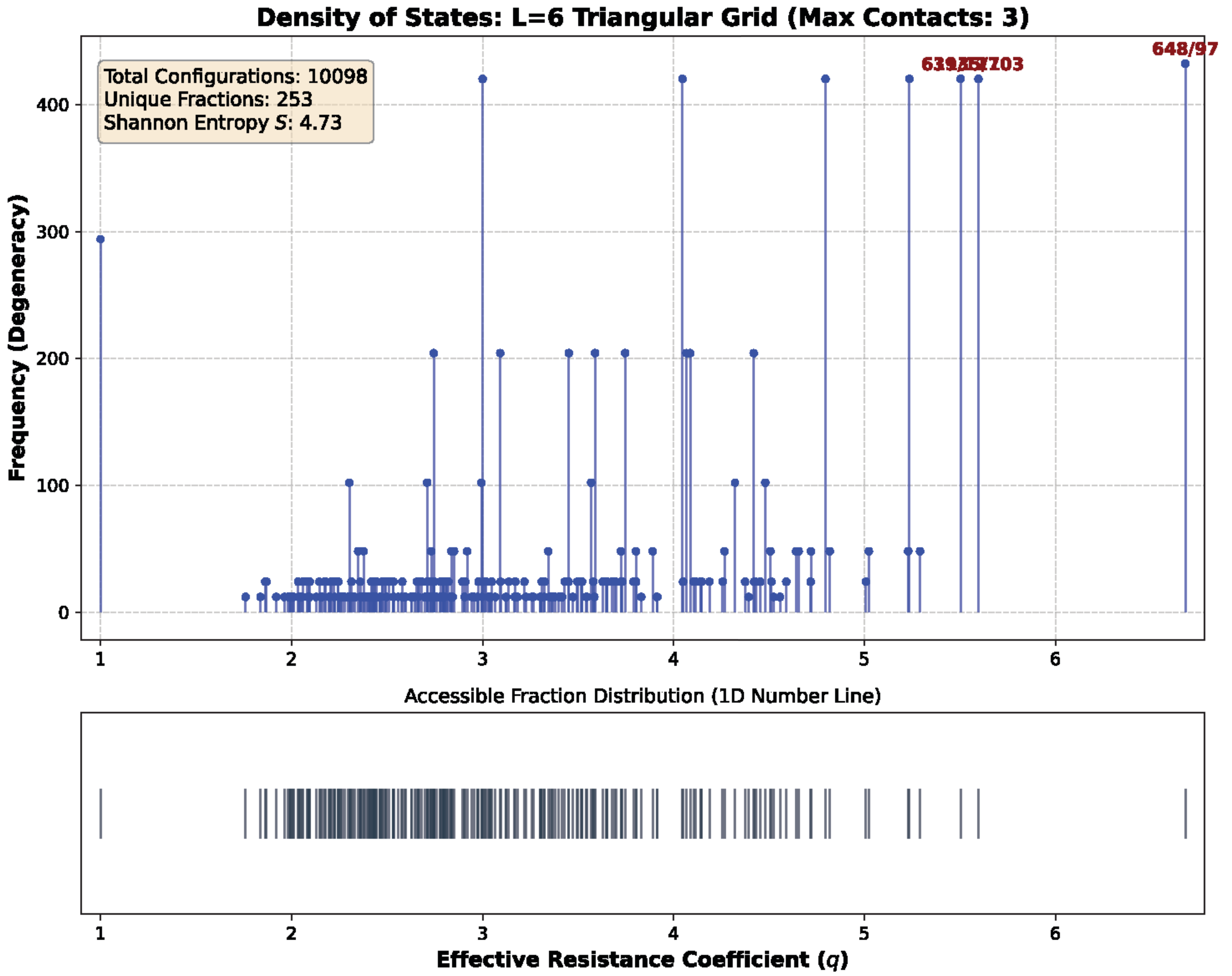
Density of States: L=6 Triangular Grid (Max Contacts: 3)
Total Configurations: 10098
Unique Fractions: 253
Shannon Entropy S: 4.73
648/97
Frequency (Degeneracy)
0
100
200
300
400
1
2
3
4
5
6
Accessible Fraction Distribution (1D Number Line)
Effective Resistance Coefficient (q)

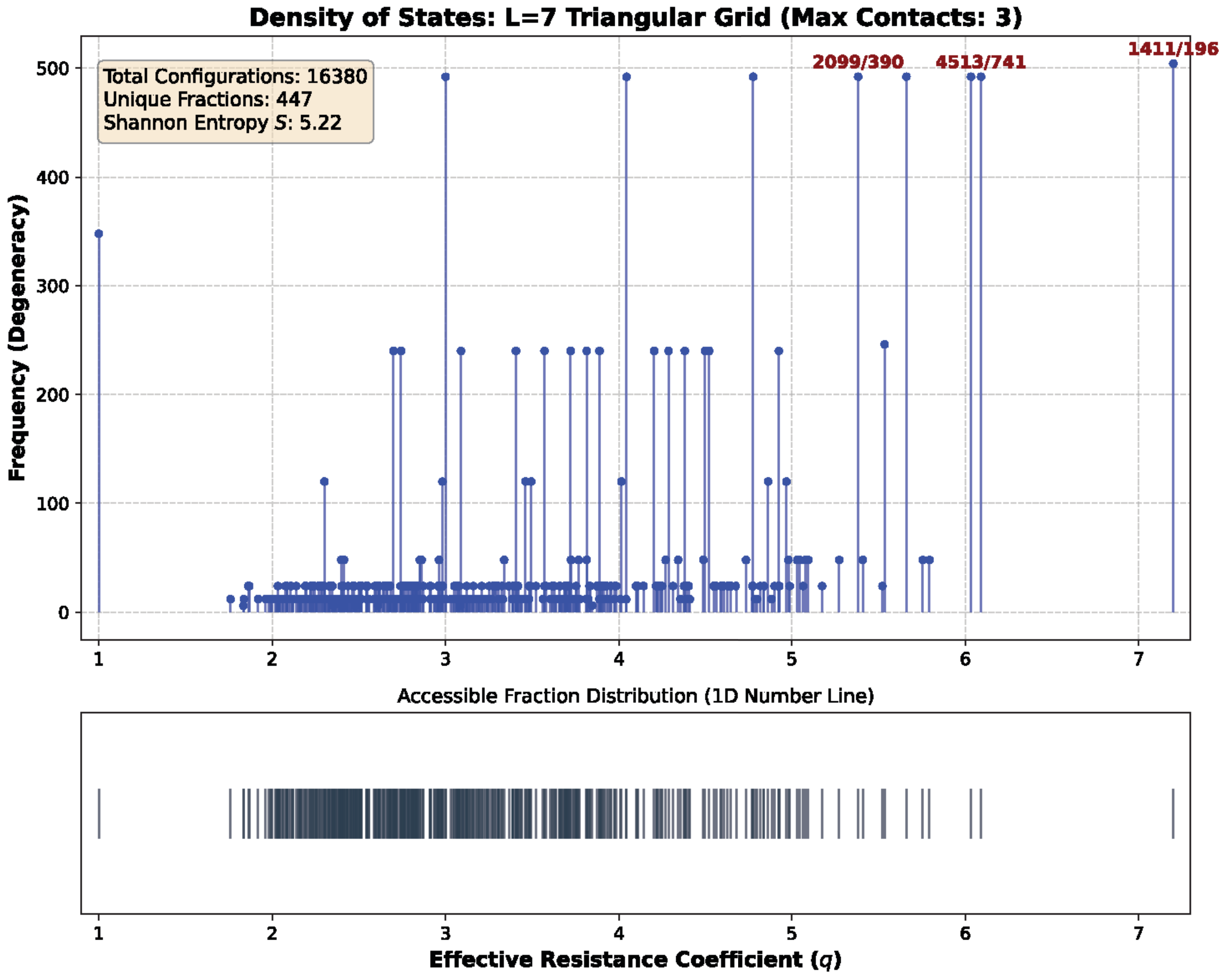

Density of States: L=7 Triangular Grid (Max Contacts: 3)
Total Configurations: 16380
Unique Fractions: 447
Shannon Entropy S: 5.22
2099/390
4513/741
1411/196
Frequency (Degeneracy)
0
100
200
300
400
500
1
2
3
4
5
6
7
Accessible Fraction Distribution (1D Number Line)
Effective Resistance Coefficient (q)

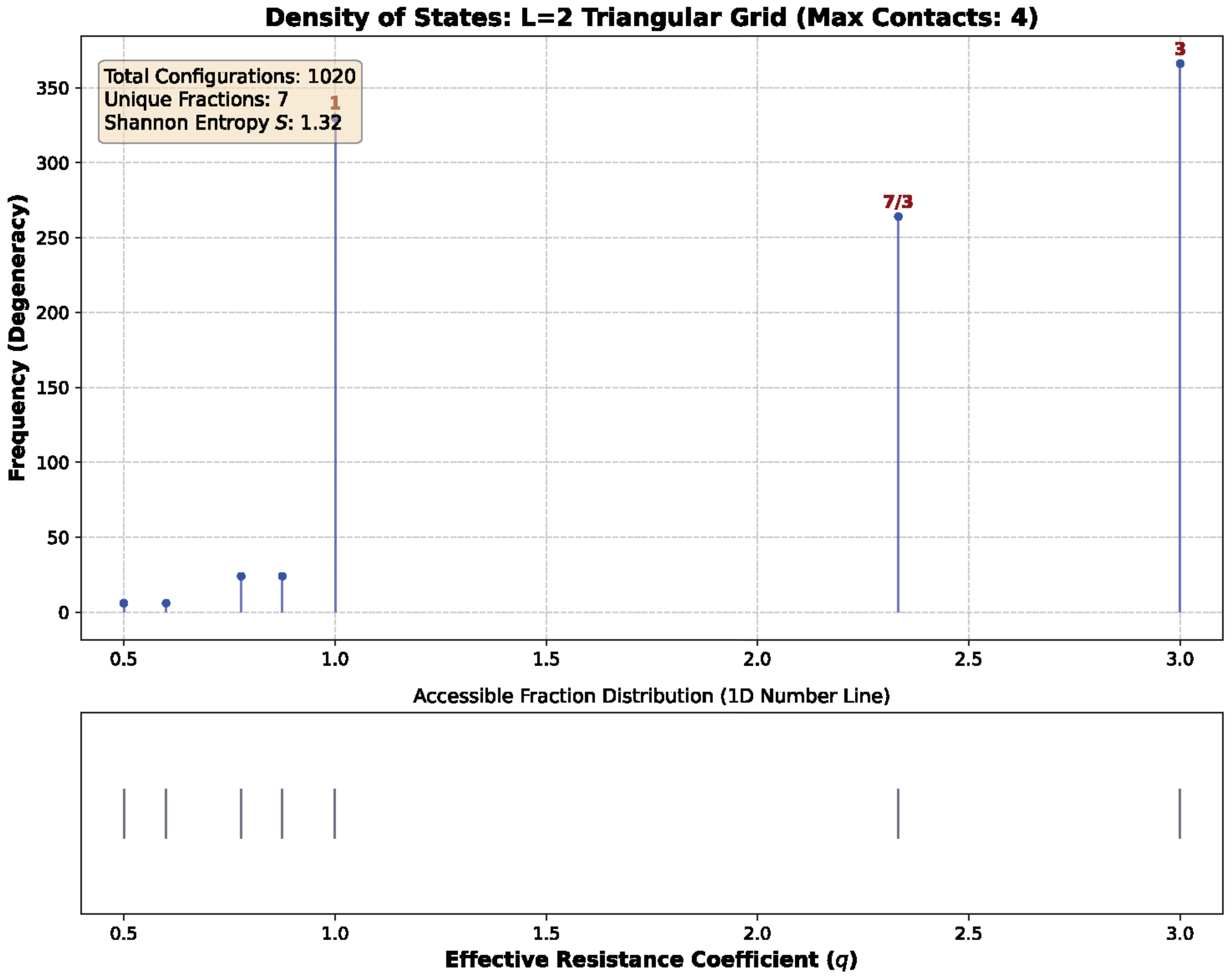
Density of States: L=2 Triangular Grid (Max Contacts: 4)
Total Configurations: 1020
Unique Fractions: 7
Shannon Entropy S: 1.32
1
7/3
3
Frequency (Degeneracy)
0
50
100
150
200
250
300
350
0.5
1.0
1.5
2.0
2.5
3.0
Accessible Fraction Distribution (1D Number Line)
Effective Resistance Coefficient (q)

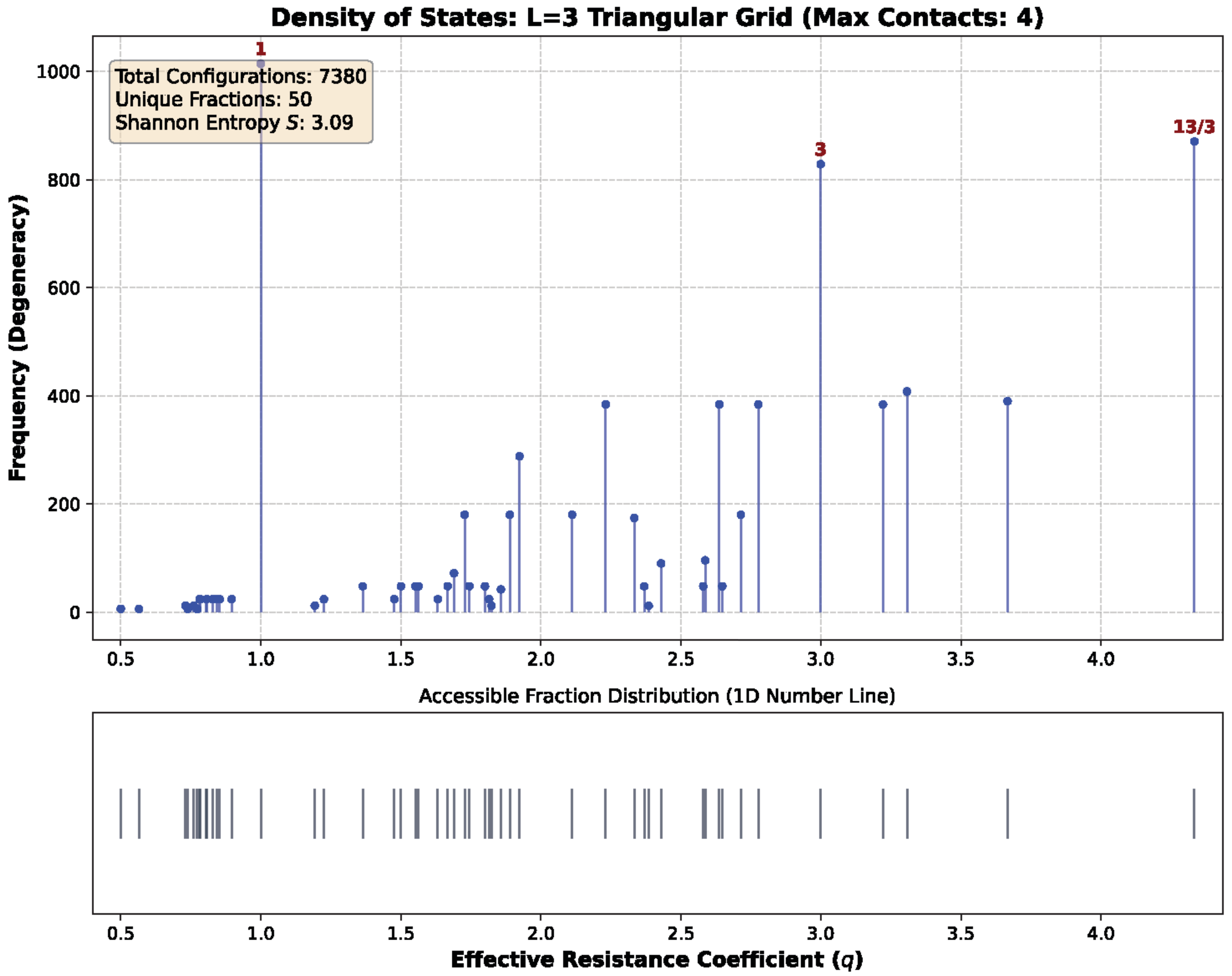
Density of States: L=3 Triangular Grid (Max Contacts: 4)
Total Configurations: 7380
Unique Fractions: 50
Shannon Entropy S: 3.09
1
3
13/3
Frequency (Degeneracy)
0
200
400
600
800
1000
0.5
1.0
1.5
2.0
2.5
3.0
3.5
4.0
Accessible Fraction Distribution (1D Number Line)
Effective Resistance Coefficient (q)

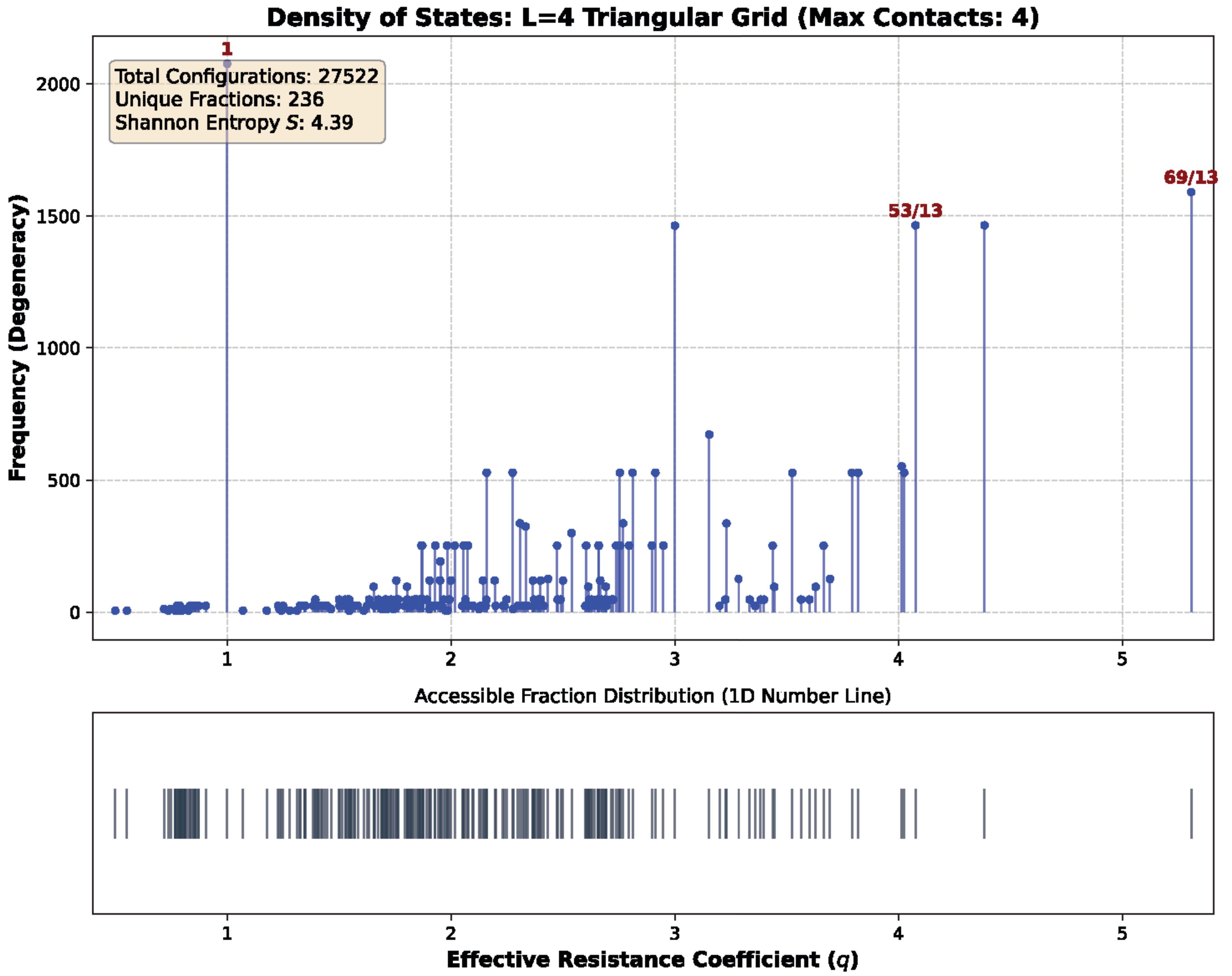

Density of States: L=4 Triangular Grid (Max Contacts: 4)
Total Configurations: 27522
Unique Fractions: 236
Shannon Entropy S: 4.39
1
53/13
69/13
Frequency (Degeneracy)
0
500
1000
1500
2000
1
2
3
4
5
Accessible Fraction Distribution (1D Number Line)
Effective Resistance Coefficient (q)

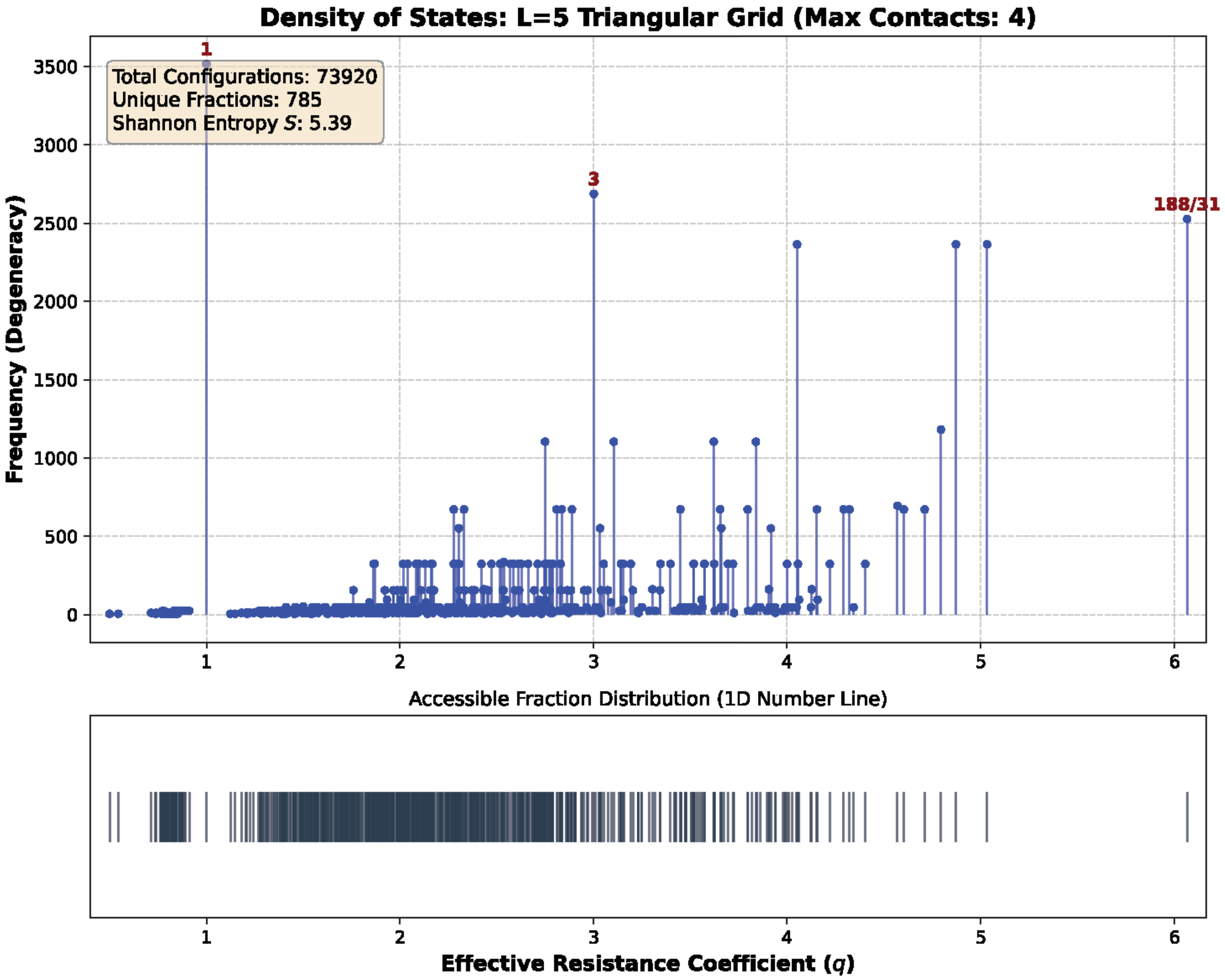

Density of States: L=5 Triangular Grid (Max Contacts: 4)
Total Configurations: 73920
Unique Fractions: 785
Shannon Entropy S: 5.39
1
3
188/31
Frequency (Degeneracy)
0
500
1000
1500
2000
2500
3000
3500
1
2
3
4
5
6
Accessible Fraction Distribution (1D Number Line)
1
2
3
4
5
6
Effective Resistance Coefficient (q)

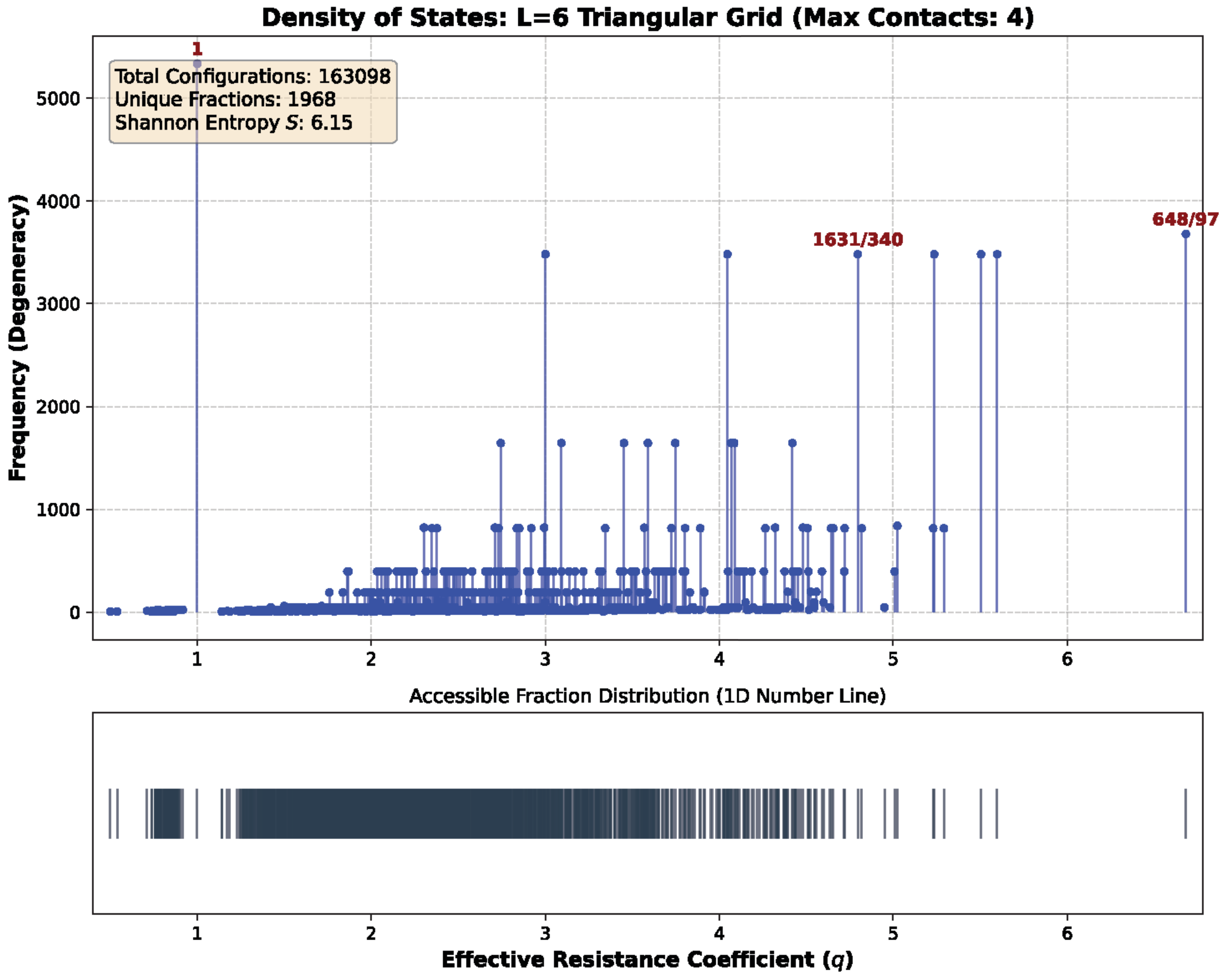

Density of States: L=6 Triangular Grid (Max Contacts: 4)
Total Configurations: 163098
Unique Fractions: 1968
Shannon Entropy $S$: 6.15
1
1631/340
648/97
Frequency (Degeneracy)
0
1000
2000
3000
4000
5000
1
2
3
4
5
6
Accessible Fraction Distribution (1D Number Line)
Effective Resistance Coefficient ($q$)

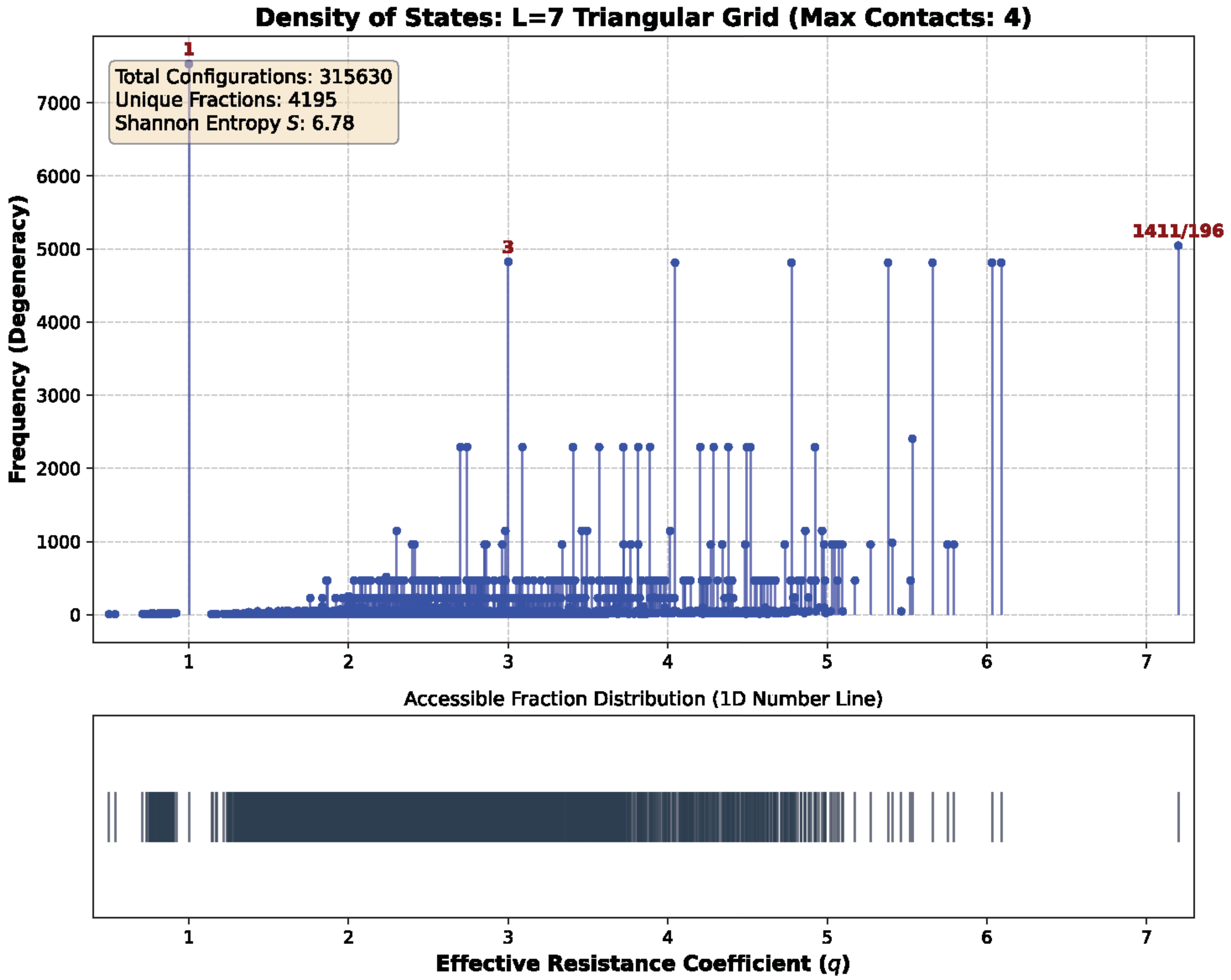

Density of States: L=7 Triangular Grid (Max Contacts: 4)
Total Configurations: 315630
Unique Fractions: 4195
Shannon Entropy S: 6.78
1
3
1411/196
Frequency (Degeneracy)
0
1000
2000
3000
4000
5000
6000
7000
1
2
3
4
5
6
7
Accessible Fraction Distribution (1D Number Line)
Effective Resistance Coefficient (q)

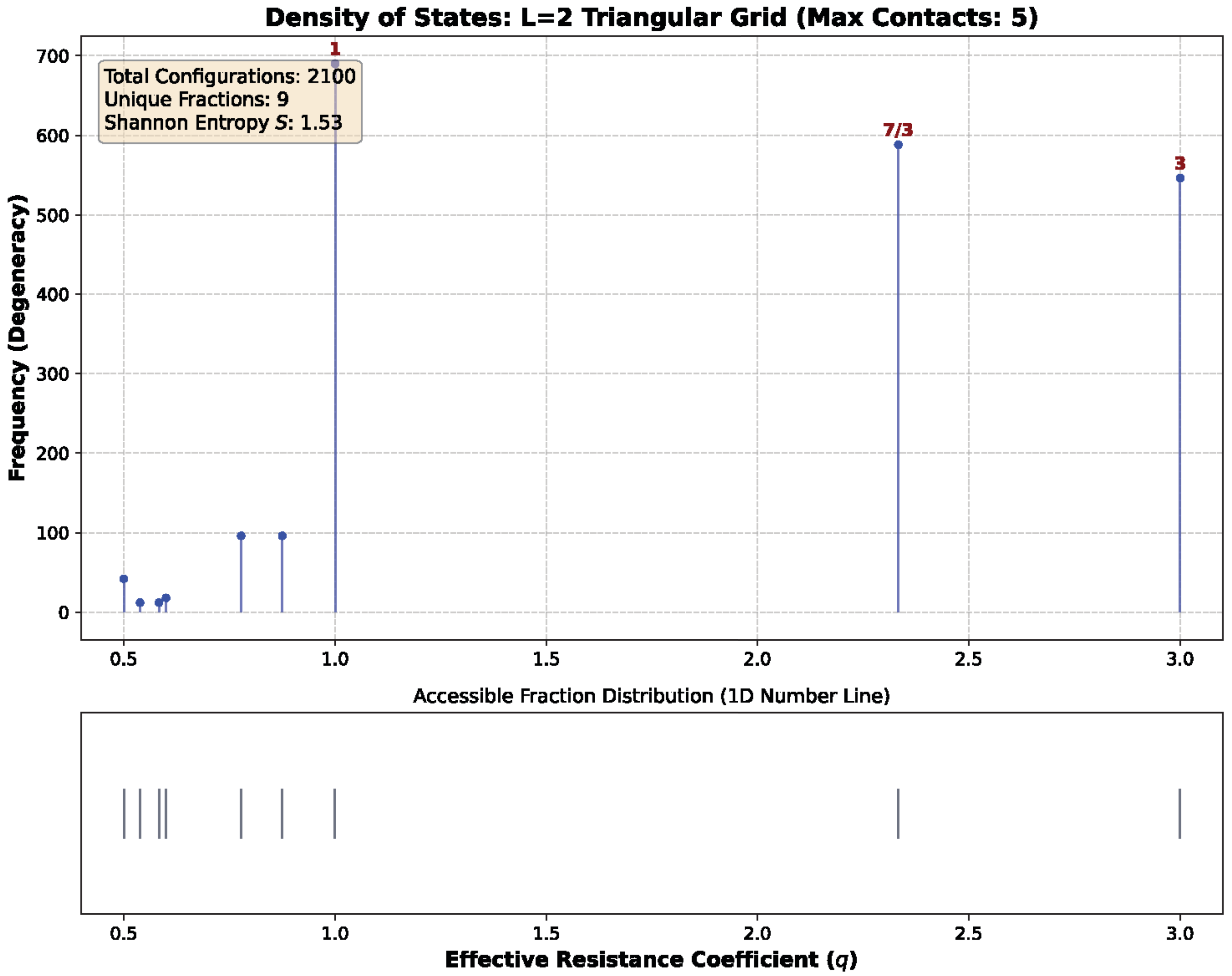
Density of States: L=2 Triangular Grid (Max Contacts: 5)
Total Configurations: 2100
Unique Fractions: 9
Shannon Entropy S: 1.53
1
7/3
3
Frequency (Degeneracy)
0
100
200
300
400
500
600
700
0.5
1.0
1.5
2.0
2.5
3.0
Accessible Fraction Distribution (1D Number Line)
Effective Resistance Coefficient (q)

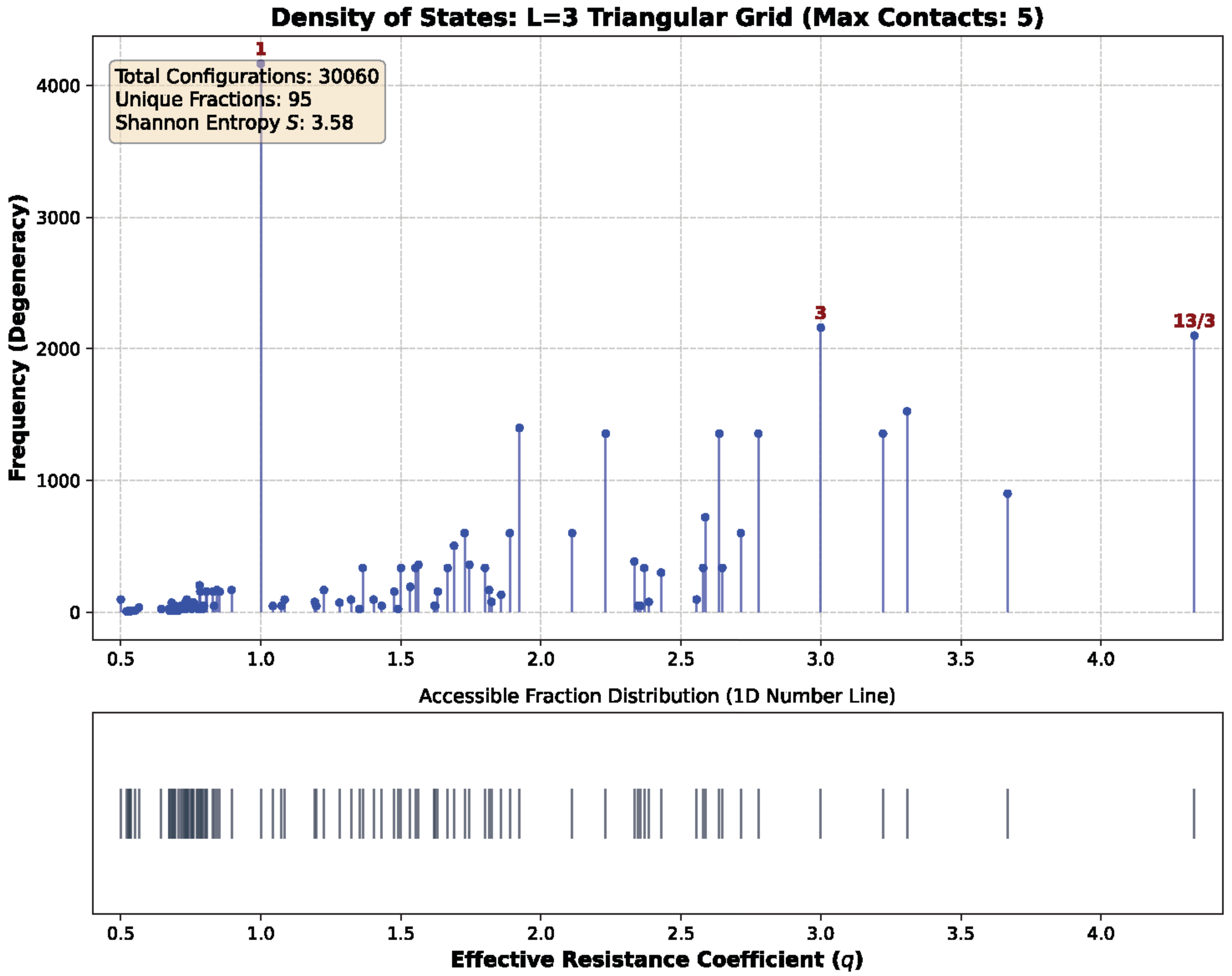
Density of States: L=3 Triangular Grid (Max Contacts: 5)
Total Configurations: 30060
Unique Fractions: 95
Shannon Entropy S: 3.58
1
3
13/3
Frequency (Degeneracy)
0
1000
2000
3000
4000
0.5
1.0
1.5
2.0
2.5
3.0
3.5
4.0
Accessible Fraction Distribution (1D Number Line)
Effective Resistance Coefficient (q)

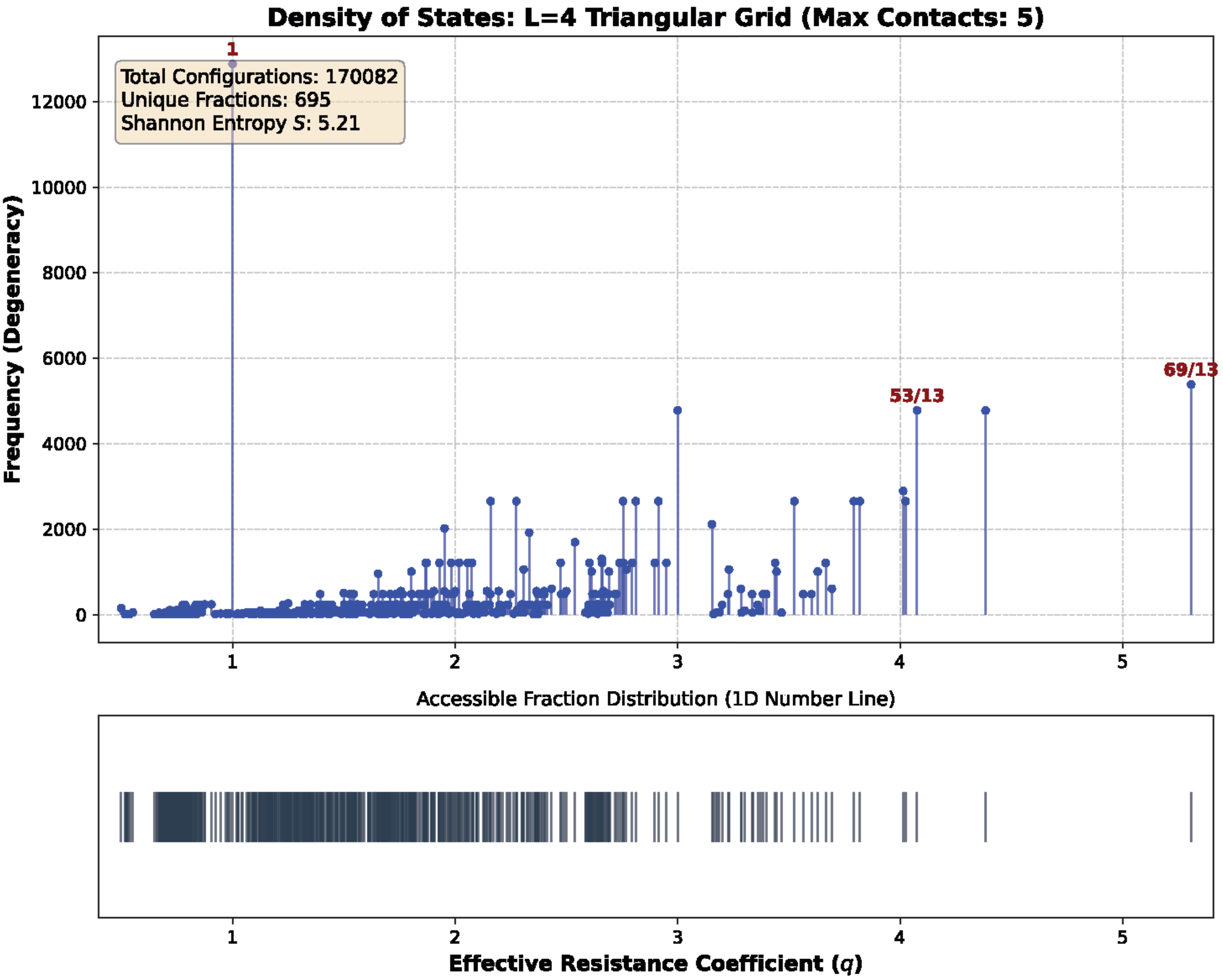

Density of States: L=4 Triangular Grid (Max Contacts: 5)
Total Configurations: 170082
Unique Fractions: 695
Shannon Entropy S: 5.21
1
53/13
69/13
Frequency (Degeneracy)
0
2000
4000
6000
8000
10000
12000
1
2
3
4
5
Accessible Fraction Distribution (1D Number Line)
Effective Resistance Coefficient (q)

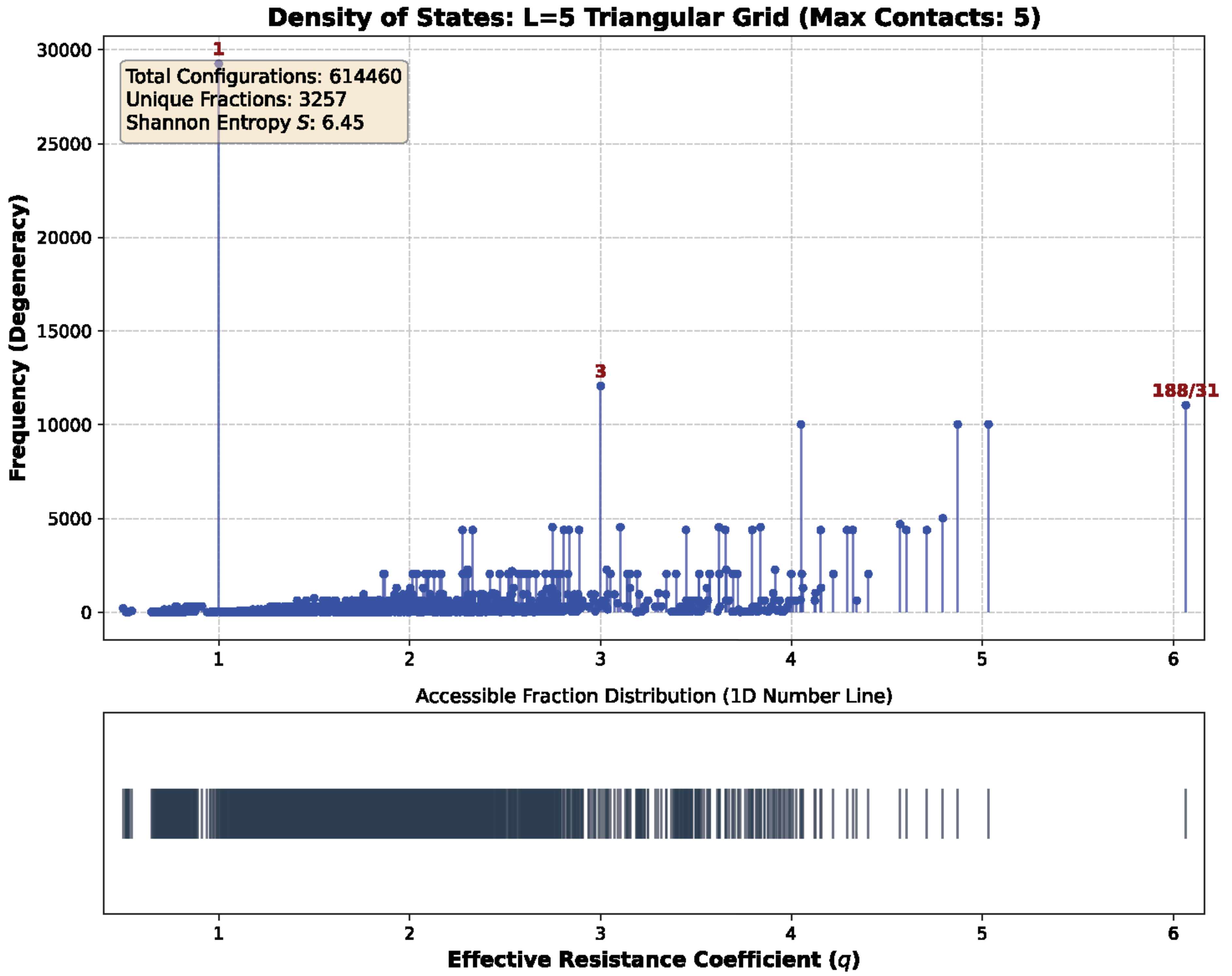
Density of States: L=5 Triangular Grid (Max Contacts: 5)
Total Configurations: 614460
Unique Fractions: 3257
Shannon Entropy S: 6.45
1
3
188/31
Frequency (Degeneracy)
0
5000
10000
15000
20000
25000
30000
1
2
3
4
5
6
Accessible Fraction Distribution (1D Number Line)
Effective Resistance Coefficient (q)

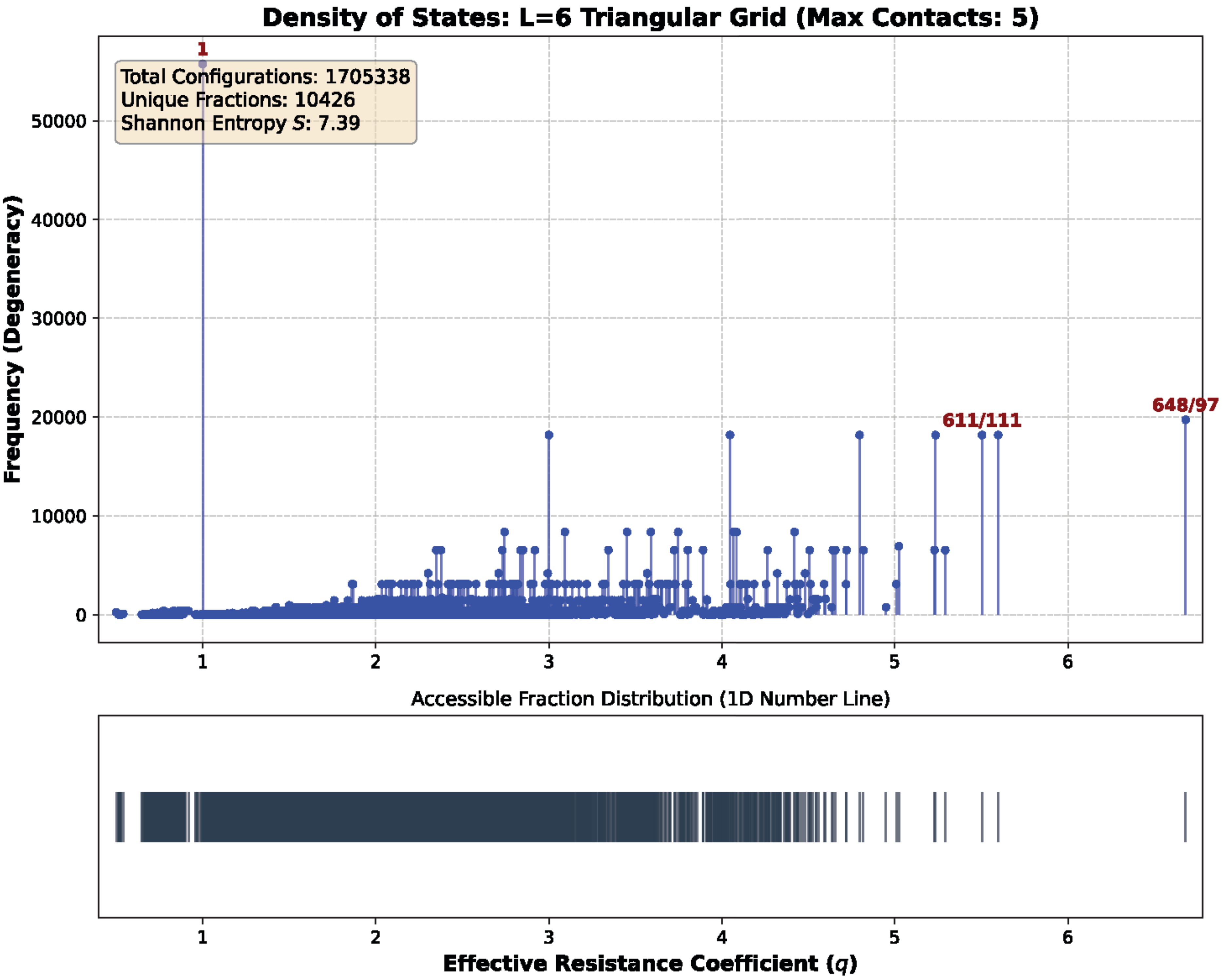

Density of States: L=6 Triangular Grid (Max Contacts: 5)
Total Configurations: 1705338
Unique Fractions: 10426
Shannon Entropy S: 7.39
1
611/111
648/97
Frequency (Degeneracy)
0
10000
20000
30000
40000
50000
1
2
3
4
5
6
Accessible Fraction Distribution (1D Number Line)
Effective Resistance Coefficient (q)

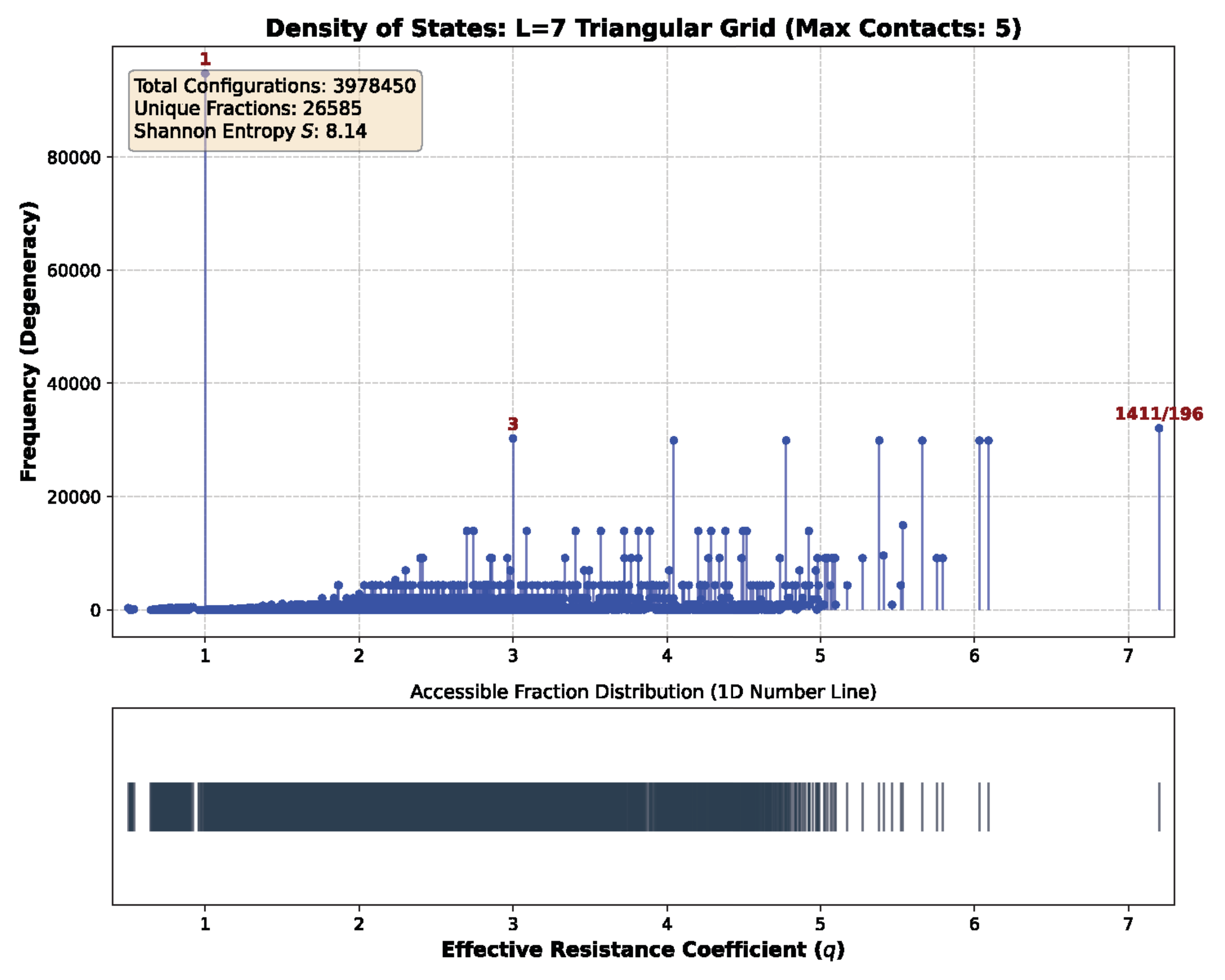

Density of States: L=7 Triangular Grid (Max Contacts: 5)
Total Configurations: 3978450
Unique Fractions: 26585
Shannon Entropy S: 8.14
1
3
1411/196
Frequency (Degeneracy)
0
20000
40000
60000
80000
1
2
3
4
5
6
7
Accessible Fraction Distribution (1D Number Line)
Effective Resistance Coefficient (q)